\documentclass[10pt,a4paper,twoside]{book}
\usepackage{amsmath}
\usepackage{amssymb}
\usepackage{graphicx}
\usepackage{float}
\usepackage{pictex}
\usepackage[dvipsnames,usenames]{color}
\usepackage{fancyhdr}
\usepackage{cite}
\usepackage{longtable}
\usepackage{epsfig}
\usepackage{psfrag}
\usepackage{bm}
\usepackage{color}

\usepackage{dcolumn}

\definecolor{Conf}{named}{Black}
\definecolor{GBE}{named}{Black}
\definecolor{OGE}{named}{Black}
\definecolor{Text-normal}{named}{Black}
\definecolor{Bonn}{named}{Black}

\begin{document}

\begin{titlepage}
\centering
\begin{large} Elmar P. Biernat  \end{large}                      \\[4cm]
 \begin{LARGE}\textbf{Electromagnetic Properties of}\end{LARGE}\\[4pt]
\begin{LARGE}\textbf{Few-Body Systems Within a}                                  \end{LARGE}
\\[4pt]
\begin{LARGE}\textbf{Point-Form Approach}                                                                                                        \end{LARGE}
\\[8cm]
{\Large Dissertation}\\[5pt]
zur Erlangung des Doktorgrades der Naturwissenschaften\\
an der
Karl-Franzens-Universit\"at Graz\\[1cm]
Eingereicht am Institut f\"ur Physik\\[1cm]
Betreuer:\\
Ao.~Univ.-Prof.~Mag.~Dr.~Wolfgang Schweiger\\
[1cm]
Graz, 2011
\thispagestyle{empty}
\end{titlepage}
\newpage\thispagestyle{empty}
\begin{titlepage}\thispagestyle{empty}
 \begin{large}

\textbf{Pr\"{a}ambel} \end{large} \vspace*{1cm}

Ich, Elmar P. Biernat, best\"{a}tige, dass es sich bei der hier vorgelegten
Dissertation um eine Originalarbeit handelt, die von mir selbst\"{a}ndig
angefertigt und abgefasst wurde. 
\thispagestyle{empty}\end{titlepage}
\newpage\thispagestyle{empty}
\begin{titlepage}
\thispagestyle{empty}
\centering \vspace*{8cm}
{\begin{small} {\textit{Para Nika}}\end{small} }\thispagestyle{empty}
\end{titlepage}\thispagestyle{empty}
\tableofcontents
\chapter{\label{chap:introduction} Introduction}
\section{Motivation}
With the discovery of the non-point-like nature of hadrons and nuclei from electron scattering experiments~\cite{Hofstadter:1957wk,Hofstadter:1963} the investigation of the structure of hadronic systems has become a vital field of research in modern nuclear and particle physics. According to quantum chromodynamics (QCD), which is accepted as the fundamental theory of the strong interaction, hadrons are bound states formed by strongly interacting quarks. The details of this mechanism are, however, not completely understood due to calculational difficulties that arise with a strongly-interacting quantum field theory in the low energy regime (see for example Ref.~\cite{Aitchison:2003}). The nuclear force acting between nucleons is regarded as a residual force of the underlying strong interaction and is responsible for the binding of nucleons to form nuclei. 

Ab initio calculations of hadron masses and the structure of hadrons require a non-perturbative solution of the QCD bound-state problem. This can be done by lattice as well as continuum methods, but only through enormous computational efforts. Therefore many effective models have been developed with the purpose providing a phenomenological description of hadronic properties. It is clear that such models, describing phenomena at length scales of nuclei (and smaller) and at energies and momenta used in electron-nuclei scattering experiments, should incorporate both, quantum theory and special relativity. A particular model that does not exhibit the difficulties of a strongly interacting quantum field theory at low energies and at the same time unifies quantum theory and special relativity is \textit{relativistic quantum mechanics}. 
The foundation of such a model has been laid by Wigner~\cite{Wigner:1939cj} and Bargmann~\cite{Bargmann:1954gh}. Contemporaneously, with his work about the \textit{forms of relativistic dynamics}~\cite{Dirac:1949cp}, Dirac found 3 preferred ways of how to include interactions into a relativistic theory. He called them \textit{instant}, \textit{front} and \textit{point} \textit{form} according to the hypersurfaces in Minkowski space that are left invariant under the action of maximal sets of Poincar\'{e} transformations which are not affected by interactions (these are called \textit{kinematic}). 
The point form, in particular, is characterized by the kinematic nature of Lorentz boosts and rotations. 

The point form of dynamics has already been considered in the context of quantum field theory (see, e.g., Refs.~\cite{Fubini:1972mf,Gromes:1974yu,Biernat:2007sz,Klink:2008qt}). 
How to put the idea of a point-form quantum field theory into practical use is currently under investigation using the analytically solvable Schwinger model~\cite{PhysRev.82.914,PhysRev.91.713}. The long-term goal is then to end up with a point-form formulation of QCD (for a recent review, cf. Ref.~\cite{Biernat:2010tp}).
    
The point form can also be applied to quantum systems with a finite number of degrees of freedom which generates a \textit{point-form relativistic quantum mechanics} (PFRQM). A particularly simple Poincar\'{e} invariant prescription of how to set up an interacting quantum mechanical model is the \textit{Bakamjian-Thomas construction}~\cite{Bakamjian:1953kh}. 
What one loses, however, in a Bakamjian-Thomas framework (for more than 2 particles) is cluster separability~\cite{Sokolov:1977ym,Coester:1982vt}~(for a detailed discussion of this problem see Ref.~\cite{Keister:1991sb}). 
 For the three-particle case a solution to the cluster problem (formulated in terms of S-operators) has been given by Coester~\cite{Coester:1965zz}. A general solution for an arbitrary number of particles has been proposed by Sokolov by introducing unitary operators, so-called \textit{packing operators}, that restore cluster separability~\cite{Sokolov:1977ym}. This solution is, however, rather formal and cumbersome for practical purposes. 

PFRQM has already been quite successfully applied to various problems of relativistic few-body physics, such as the calculation of masses~\cite{Glozman:1997ag,Glantschnig:2004mu,Melde:2008yr}, strong decays of hadrons~\cite{Melde:2005hy,Melde:2006yw,Sengl:2007yq}, strong form factors~\cite{Melde:2008dg}, axial charges~\cite{Choi:2010ty,Choi:2009pw} and for a classification of baryons~\cite{Melde:2008yr}. 

In the present work we concentrate on the study of electromagnetic properties of few-body systems. The phenomenological quantities associated with the electromagnetically probed structure of a hadronic bound system are the \textit{electromagnetic form factors}. They encode, in a relativistic invariant manner, all effects coming from the underlying QCD that influence the electromagnetic vertex of the bound state. 

Finding a microscopic description of the form factors amounts to finding an expression for the electromagnetic bound-state current in terms of the electromagnetic constituent currents and bound-state wave function. Due to the requirement of covariance the current cannot be the simple sum of constituent currents. This was first realized by Siegert~\cite{Siegert:1937yt}.
Further constraints on the bound-state current come from the requirements of current conservation and cluster separability%
~\cite{Lev:1994au}.
%

There have been various attempts within relativistic few-body physics to calculate form factors. For instance, the form factors of the simplest hadronic systems like the $\pi$ meson, the $\rho$ meson and the deuteron have been investigated extensively in different approaches, most of them using light-front dynamics; see, for example, Refs.~\cite{Chung:1988my,Brodsky:1992px,Cardarelli:1995dc,Carbonell:1999pt,Allen:2000ge,Bakker:2002mt,Gilman:2001yh,Jaus:2002sv,Choi:2004ww,deMelo:2005cy,Brodsky:2007hb,Carbonell:2008tz}.
Also PFRQM has already been used to analyze the electromagnetic structure of such simple hadronic two-body systems; see, for example, Refs.~\cite{Klink:1998qf,Allen:1998hb,Allen:2000ge,Wagenbrunn:2000es,Boffi:2001zb,Melde:2006jn,Melde:2007zz}. 
In these approaches the point-form spectator model is used to define an electromagnetic current operator which already satisfies all the necessary requirements in order to be compatible with a particular interaction model~\cite{Lev:1994au}.
%

In the present work however we follow a different strategy. Rather than making an ansatz for the most general current on which the necessary constraints are imposed, we derive a microscopic bound-state current that exhibits the required properties. Our approach is based on a relativistic multi-channel framework proposed by Klink~\cite{Klink:2000pp}, 
using field theoretical vertex interactions that are implemented in the Bakamjian-Thomas construction. 
This multi-channel formalism 
has already been applied successfully to different relativistic few-body problems such asthe calculation of vector-meson masses and partial decay widths within the chiral constituent model~\cite{KrassniggDiss:2001,Krassnigg:2003gh,Kleinhappel:2010,Kleinhappel:2010hh}, the study of the hydrogen atom and positronium~\cite{Lechner:2003} and for the derivation of a mass operator for pion-nucleon and nucleon-nucleon scattering using chiral
perturbation theory~\cite{Girlanda:2007ds}. This formalism, adapted to electron-bound-state scattering, makes it possible to calculate electromagnetic bound-state currents.
It was first proposed in Ref.~\cite{Fuchsberger:2007} for the case of a pseudoscalar quark-antiquark bound state in which the quarks were treated, for simplicity, as spinless particles. Despite this simplification, the form factor results were already quite promising. In particular, they showed perfect numerical agreement with the results of the standard front-form approach by~\cite{Chung:1988mu}.
The deeper reason for this equivalence was, however, not yet evident. The present work, where the constituents' spins are fully taken into account, attempts to give an explanation to this question. Moreover, by investigating spin-0 and spin-1 form factors and addressing the cluster problem we will find quite remarkable similarities with the covariant light-front approach of Refs.~\cite{Karmanov:1994ck,Carbonell:1998rj}. The present work also contains part of a benchmark calculation of the deuteron form factors, whose aim is to define more accurately the meaning of \lq relativistic effects' in different approaches of few-body physics. As a common starting point, we propose a simple model for the nucleon-nucleon interaction in the spirit of the Walecka model~\cite{ref.01}.

The same coupled-channel formalism has also been applied very recently to calculate electromagnetic and weak form factors of heavy-light systems~\cite{Rocha:2010wm,Rocha:2010uu}. Therein, a simple analytical expression for the Isgur-Wise function~\cite{Isgur:1989vq,Isgur:1989ed} has been found in the heavy-quark limit.

In this thesis we restrict ourselves to instantaneous interactions between the constituents that form the bound state. The formalism offers, however, the possibility to include additional channels with dynamical particle-exchange interactions between the constituents. Such an extension is under investigation and will not be part of this thesis.

To sum up, the great advantage of our point-form approach is manifest Lorentz covariance which will be used to derive currents and form factors for spin-0 and spin-1 bound states of two spin-1/2 particles. Due to the Bakamjian-Thomas framework we expect, however, difficulties associated with cluster separability. How such difficulties manifest themselves and how they can be circumvented is the second important issue of this thesis.  
\section{Synopsis}
Chap.~\ref{chap:2} is an introduction to relativistic quantum mechanics. It provides the theoretical background on which the present work is based. We point out the relevance of the universal covering group of the Poincar\'{e} group for a relativistic quantum theory and derive the Casimir operators of the Poincar\'{e} group.  With the one-particle irreducible representations we construct multi-particle representations, one of them being the velocity-state representation which is most convenient for our purposes.

Chap.~\ref{chap:3} is devoted to interactions in a Poincar\'{e} invariant quantum theory. This brings up the notion of Dirac's forms of relativistic dynamics, where we focus in particular on the point form. A Poincar\'{e} invariant prescription for how to introduce interactions into a free theory is the Bakamjian-Thomas construction, which is then used to formulate the bound-state problem for a mass operator. In this context we address the non-trivial problem of cluster separability. The definition of appropriate electromagnetic vertex interactions in terms of field-theoretical interaction densities concludes this chapter. 

Subsequently, in Chap.~\ref{chap:4}, we set up a relativistic coupled-channel problem which describes center-of-mass electron-bound-state scattering in the one-photon-exchange approximation. From the associated optical potential we extract the electromagnetic bound-state current in terms of constituent currents and bound-state wave functions. Properties of the current such as hermiticity, continuity and covariance are discussed for the particular cases of spin-0 and spin-1 bound states. 

The subject of Chap.~\ref{chap:5} is the extraction of the electromagnetic form factors from the expression for the bound-state current, which requires a careful analysis of its Lorentz structure. In this context it will be necessary to come back to the problem of cluster separability. By comparing our expressions for the current and the form factors with those of light-front dynamics notable similarities are found. 

In Chap.~\ref{chap:6} we introduce specific models for bound states. For mesons, like the $\pi$ and $\rho$ meson, we model confinement between the quarks by a simple harmonic-oscillator potential. For the deuteron we propose a meson-exchange nucleon-nucleon interaction inspired by the Walecka model.

The numerical results for our form factors of the pion, the $\rho$ meson and the deuteron are then presented and discussed in Chap.~\ref{chap:7}.

Chap.~\ref{chap:8} finally summarizes the thesis and gives a short outlook to further investigations.

The appendices contain important definitions and conventions to be used throughout the work together with basic relations, longer derivations and involved calculations. App.~\ref{app:A} gives the Minkowski space notations and a matrix representation of the Poincar\'{e} group in Minkowski space. App.~\ref{app:B} gives the representations to be used for the Dirac spinors and the polarization vectors together with their most basic properties and relations. App.~\ref{app:C} contains some useful relations associated with the SL$(2,\mathbb C)$ and the Clebsch-Gordan coefficients. In App.~\ref{app:D} we give the rather lengthy calculations and derivations of the currents obtained in Chaps.~\ref{chap:3} and~\ref{chap:4}. In App.~\ref{app:1phoptpotmatrixelemnts} we detail the derivation of the microscopic bound state current from the optical potential. App.~\ref{app:F} contains a detailed analysis of the extraction of the form factors described in Chap.~\ref{chap:5}. Finally, App.~\ref{App.A} gives the derivation of the meson-exchange potentials in the static approximation used in Chap.~\ref{chap:6}.

\chapter{Relativistic Quantum Mechanics}\label{chap:2}
In this chapter we present and briefly discuss the fundamental principles and ideas on which the present work is built. It provides the basis for the following chapters. We will proceed roughly along the lines of Refs.~\cite{Keister:1991sb,Polyzou:JLab,KrassniggDiss:2001,Sengldiss:2006,Thaller:1992,Scheck:2001}.
\section{Introduction}\label{sec.IntroWignertheorem}
The basis of our investigations is \textit{relativistic quantum mechanics}. This means that we are dealing with quantum mechanical models for a \textit{finite} number of particles that are invariant under the transformations of the \textit{Poincar\'e group}, the symmetry group of special relativity. 

Any quantum mechanical model is formulated on a Hilbert space $\mathcal H$, i.e. a complete complex linear vector space with elements called \textit{state vectors} in the quantum mechanical context. A state vector $\vert\varPsi\rangle$ describes a particular state of a physical system. Experimentally measurable quantities, i.e. the observables of a physical system, correspond to expectation values and transition probabilities between such states. These quantities are determined by Hilbert-space scalar products. Hilbert-space scalar products are invariant under unitary transformations of the state vectors, which implies that probabilities are also unitarily invariant. Therefore, in order to ensure that probabilities are preserved under symmetry transformations of the system under consideration, one requires the action of a symmetry transformation on a state vector $\vert\varPsi\rangle$ to be represented by a unitary operator. 

According to the principles of special relativity the laws of physics are equivalent in all inertial coordinate systems. In other words, equivalent measurements done in different inertial coordinate systems lead to identical results. Inertial systems are connected by Poincar\'e transformations which form a group, the Poincar\'e group. In a theory where both relativity and quantum mechanics are incorporated, the principles of special relativity should be formulated in terms of quantum probabilities in such a way that quantum observables are invariant under the change of inertial reference systems. This is guaranteed if the effect of a Poincar\'{e} transformation on a state vector $\vert\varPsi\rangle$ is represented by a unitary operator. In this way the only measurable quantities, namely probabilities, cannot be used to distinguish between different inertial coordinate systems~\cite{Keister:1991sb}. This integration of special relativity into quantum theory was first realized by Wigner~\cite{Wigner:1939cj}. A refined version of this theorem by Bargmann~\cite{Bargmann:1954gh} reads: 
\begin{quote}
 \textit{A quantum mechanical model formulated on a Hilbert space preserves probabilities in all inertial coordinate systems if and only if the correspondence between states
in different inertial coordinate systems can be realized by a single-valued unitary representation of the covering group of the Poincar\'e group.}\footnote{This theorem is often referred to as \textit{the Wigner theorem}.}    
\end{quote}                                                
The purpose for requiring the covering group of the Poincar\'e group instead of the Poincar\'e group itself is  to avoid ambiguities that appear when rotating half-integer spin particles through an angle of $2\pi$~\cite{Keister:1991sb}. By taking the covering group of the Poincar\'e group these unwanted ambiguities are removed (for a thorough discussion of this problem we refer, e.g., to Ref.~\cite{Thaller:1992}).
\section{The Poincar\'e Group}
\label{sec:tPG}
In this section we summarize the most important features of the symmetry group of special relativity, the Poincar\'e group. The theory of special relativity postulates the existence of inertial coordinate systems. These inertial systems are related by transformations that
 preserve the proper time $\tau_{12}$ between two events in Minkowski space. If these two events are described by the four-vectors $x_1^\mu$ and $x_2^\mu$, $\tau_{12}$ is defined by
\begin{equation}
 \tau_{12}^2:=(x_1-x_2)^\mu \mathrm g_{\mu\nu}(x_1-x_2)^\nu\,.
\label{eq:propertime}
\end{equation}
For the Minkowski-space conventions used throughout this work we refer to App.~\ref{app:Minkowskispace}.
The transformations that leave Eq.~(\ref{eq:propertime}) invariant form a group, the, so-called, \textit{inhomogeneous Lorentz group }or \textit{Poincar\'e group}. This group consists of the continuous one-parameter transformations of time translations, space translations, rotations and Lorentz boosts and the discrete transformations of time reversal, space inversion and four-dimensional reflections. The Poincar\'e group is the semi-direct product of the group $\mathbb{R}^4$ of space-time translations with the (homogeneous) Lorentz-group O$(1,3)$ with elements $\varLambda$.\footnote{O$(1,3)$ is the linear group of Minkowski space $\mathbb R^{1+3}$ that leaves the scalar product, Eq.~(\ref{eq:propertime}), invariant.}
Consequently, a general Poincar\'e-group element is the pair $(\varLambda,a)$ labeled by a Lorentz transformation $\varLambda$ and a space-time translation $a$. The most general point transformation associated with $(\varLambda,a)$ that satisfies Eq.~(\ref{eq:propertime}) is given by \begin{equation}\label{eq:PtrafoMinkowski}
 x^\mu\stackrel{\left(\varLambda,a\right) }{\longrightarrow}x'^\mu=\varLambda^\mu_{\,\,\nu}x^\nu+a^\mu\,.
\end{equation}
This transformation defines a representation of the Poincar\'e group in Minkowski space where $a^\mu$ is a constant four-vector and $\varLambda^\mu_{\,\,\nu}$ is a constant $(4\times4)$-matrix.
In the following we restrict ourselves to the continuous 
transformations which themselves form a subgroup of the Poincar\'e group, the so-called \textit{proper, orthochronous} Poincar\'e group.
It is the component of the Poincar\'e group containing the identity and it is characterized by the conditions 
$\det\vert\varLambda\vert=1$ and $\varLambda^0_{\,\,0}\geq 1$, referring to \textit{proper} and \textit{orthochronous}, respectively.\footnote{We will use the terms \textit{Poincar\'e group}
 and \textit{Poincar\'e transformation} to refer only to the proper, orthochronous Poincar\'e group. Its homogeneous part is called the \textit{proper, orthochronous} or \textit{restricted} Lorentz group SO$^+(1,3)$.} 

A rotation can be parametrized by a rotation axis and an angle of rotation described by the direction and the norm of the three-vector $\boldsymbol{\theta}$, respectively. In the $(4\times 4)$-matrix representation $\varLambda^\mu_{\,\,\nu}$ of the Lorentz group in Minkowski space a rotation is given by
\begin{equation}\label{eq:rotMinkwoskispace}
 R\left(\boldsymbol{\theta}\right):=\left(%
\begin{array}{cc} 1&0\\
 0&\boldsymbol R\left(\boldsymbol{\theta}\right)
\end{array}
\right)%
\end{equation}
with the usual SO$\left(3\right)$ rotation matrix $\boldsymbol R\left(\boldsymbol{\theta}\right)$.\footnote{SO$\left(3\right)$ is the group of real $(3\times 3)$-matrices $\boldsymbol R$ with $\boldsymbol R^\mathrm T=\boldsymbol R^{-1}$ and $\det \boldsymbol R=1$.}  

A canonical (rotationless) boost can be parametrized by a velocity $\boldsymbol v$ and is given in the above $(4\times 4)$-matrix representation in Minkowski space by
\begin{eqnarray}\label{eq:B_c}
 B_\mathrm c\left(\boldsymbol v\right)
:=
\left(%
\begin{array}{cc}
  v^0&\boldsymbol{v}^\mathrm T\\
  \boldsymbol{v}&\boldsymbol{1}+\frac{v^0-1}{\boldsymbol{v}^2}\boldsymbol
{v}\, \boldsymbol{v}^\mathrm T\\
\end{array}
\right)%
\end{eqnarray}
with $v^0=\sqrt{1+\boldsymbol{v}^2}$. The velocity $\boldsymbol v$ can alternatively be expressed in terms of the rapidity $\boldsymbol{\rho}$ defined by
$
 \boldsymbol \rho:=\frac{\boldsymbol v}{\vert\boldsymbol v\vert} \sinh^{-1}\vert \boldsymbol v\vert
$.\footnote{Note that $\boldsymbol v:=\frac{\mathrm d\boldsymbol x }{\mathrm d\tau}=v^0\boldsymbol v_{\mathrm {ph}}$ with the \textit{physical velocity} $\boldsymbol v_{\mathrm {ph}}=\frac{\mathrm d\boldsymbol x }{\mathrm d t}$ (see, e.g., Ref.~\cite{Scheck:1999fg}). }

At this point we introduce another type of boost associated with the helicity. 
It consists of a canonical boost in the three-direction to a desired magnitude $\vert \boldsymbol v\vert $, followed by 
a rotation into the desired direction $\hat{\boldsymbol v}$:\footnote{Here the \lq hat' denotes a unit vector into the direction of the corresponding vector.}
\begin{eqnarray}\label{eq:helicityboost}
 B_\mathrm h\left(\boldsymbol v\right):= R\left(\boldsymbol{\theta}\right)B_\mathrm c\left(\vert\boldsymbol v\vert\hat{\boldsymbol x}_3\right)\quad\text{with}\quad
\boldsymbol{\theta}=\frac{\hat{\boldsymbol x}_3\times \boldsymbol v}{\vert\hat{\boldsymbol x}_3\times \boldsymbol v\vert}\cos^{-1}(\hat{\boldsymbol x}_3\cdot\hat{\boldsymbol v})\,.
\end{eqnarray}
By expressing the matrices $R$ and $B_\mathrm c$ in exponential form we can write down the most general proper Lorentz transformation, which can be decomposed into a rotation and a canonical boost according to the, so-called, \textit{decomposition theorem} (a proof of this theorem can be found, e.g., in Ref.~\cite{Scheck:1999fg}):
\begin{eqnarray}\label{eq:LreprMinkowski}
 \varLambda\left(\boldsymbol{\theta},\boldsymbol \rho\right)=\exp\left(-\mathrm i\,\boldsymbol{\theta}\cdot \boldsymbol J\right) \exp\left(-\mathrm i\,\boldsymbol \rho\cdot \boldsymbol K\right)
\end{eqnarray}
with the \textit{generators} $\boldsymbol J$ and $\boldsymbol K$ for infinitesimal Lorentz transformations defined by
\begin{eqnarray}
J^j:=\left.\mathrm i\,\frac{\partial \varLambda\left(\boldsymbol{\theta},\boldsymbol \rho\right)}{\partial \theta^j}\right\vert_{\boldsymbol\theta=\boldsymbol\rho=0 }\quad\text{and}\quad
K^j:=\left.\mathrm i\,\frac{\partial \varLambda\left(\boldsymbol{\theta},\boldsymbol \rho\right)}{\partial \rho^j}\right\vert_{\boldsymbol\theta=\boldsymbol\rho=0 }\,,\quad j=1,2,3\,.
\end{eqnarray}
These six infinitesimal Lorentz generators can be combined to six independent components of an antisymmetric tensor $M^{\mu\nu}$ of rank 2 which transforms covariantly under Lorentz transformations. 
This tensor is called the \textit{angular momentum tensor} and its non-vanishing components are defined by $M^{ij}:=\epsilon^{ijk}J^k$ and $M^{0j}:=K^j$. 

The inhomogeneous part of the Poincar\'e group, consisting of space-time translations, can be represented in (extended) Minkowski space by $(5\times5)$-matrices.
The infinitesimal generators for space-time translations are then given by
\begin{eqnarray}
P^\mu:=\left.\mathrm i\,\mathrm g^{\mu\nu}\frac{\partial \,T\left(a\right)}{\partial a^\nu}\right\vert_{a^\nu=0 }
\end{eqnarray}
with a space-time translation represented in (extended) Minkowski space by
\begin{eqnarray}
T(a):=
\left(%
\begin{array}{cc}
 \mathrm g^\mu_{\nu}&a^\mu\\
  0^{\mathrm T\nu}&1\\
\end{array}
\right)\,.
\end{eqnarray}
A detailed derivation can be found in App.~\ref{app:MinkowskispaceMatrixrepPmu} or, e.g., in Ref.~\cite{Scheck:2001}.

The 10 infinitesimal generators of the Poincar\'e group in the $(5\times5)$-matrix representation of Minkowski space satisfy a set of commutation relations, the, so-called.\textit{ Poincar\'e algebra}.\footnote{The Poincar\'e algebra is the Lie algebra of the corresponding 
Lie group, in our case the Poincar\'e group.}
Its manifest covariant form reads
\begin{eqnarray}\label{eq:la1}
\left[ P^{\mu},P^{\nu}\right]& =&0\,,\\
\left[M^{\mu\nu}, P^{\rho}\right]&=&\mathrm i\left(\mathrm g^{\nu\rho}P^{\mu}-\mathrm g^{\mu\rho}P^{\nu}\label{eq:la2}
\right)\,,\\
\left[M^{\mu\nu}, M^{\lambda\sigma}\right]&=&
-\mathrm i\left(\mathrm g^{\mu\lambda}M^{\nu\sigma}-\mathrm g^{\nu\lambda}M^{\mu\sigma}+\mathrm g^{\nu\sigma}M^{\mu\lambda}-\mathrm g^{\mu\sigma}M^{\nu\lambda}\label{eq:la3}
\right)\,.
\end{eqnarray}
Eq.~(\ref{eq:la1}) are the commutation relations for the Abelian group of space-time translations, Eq.~(\ref{eq:la2}) ensures that $P^\mu$ transforms 
covariantly under Lorentz transformations and Eq.~(\ref{eq:la3}) is the Lie algebra of the Lorentz group.

\section{The Covering Group of the Poincar\'e Group}
\subsection{Two-Spinor Representations}
Remembering the Wigner theorem, the aim is to find a single-valued unitary representation of the covering group of the Poincar\'e group which acts on states of the Hilbert space. The covering group of Poincar\'e group is the inhomogeneous SL$\left(2,\mathbb C\right)$, also sometimes denoted as ISL$\left(2,\mathbb C\right)$, with the SL$\left(2,\mathbb C\right)$ being the covering group of the Lorentz group $\mathrm{SO}^+(1,3)$. It is the group of ordered pairs of complex $(2\times2)$-matrices
$\left(\underline{\varLambda},\underline{a}\right)$ with $\det \underline{\varLambda}=1$ and $\underline{a}=\underline{a}^\dagger$.\footnote{The \lq underline' of a symbol denotes an element of the SL$\left(2,\mathbb C\right)$.}
The relation between the ISL$\left(2,\mathbb C\right)$ and the Poincar\'e group can be seen as follows. To this end we note that any four-vector $x^\mu$ can be represented by a $2\times2$ Hermitean matrix defined by~\cite{Keister:1991sb}
\begin{eqnarray}\label{eq:xSL2C}
\underline{x}:=x^\mu\sigma_\mu\quad \Leftrightarrow\quad x^\mu=\frac12\mathrm{tr}\left(\underline x\, \sigma_\mu\right)
\end{eqnarray}
with $\sigma_\mu:=(\sigma^0,\boldsymbol \sigma)$. Here $\sigma^0\equiv  1_{2\times2}$ is the $(2\times2)$-identity matrix and 
$\boldsymbol \sigma\equiv\{\sigma_i\}$ are the usual Pauli matrices defined in App.~\ref{app:standardrepresentation}.
Note that here the usual Pauli matrices are defined via lower (covariant) indices. The relation to the proper time between two events, cf. Eq.~(\ref{eq:propertime}), is then given by
\begin{eqnarray}
\det \vert\underline {x}_1- \underline {x}_2\vert=\tau^2_{12}\,.
\end{eqnarray}
The most general transformations that preserve this determinant (i.e. the proper time), the hermiticity of $\underline x$, the handedness of space 
and the direction of time read
\begin{eqnarray}
\underline x\stackrel{\left(\underline{\varLambda},\underline{a}\right)}{\longrightarrow}\underline x'=\underline{\varLambda}\,\underline x\,\underline\varLambda^\dag+\underline a\,.
\end{eqnarray}
Then the connection to the Poincar\'e group elements $\left(\varLambda,a\right)$ follows from Eq.~(\ref{eq:xSL2C}):
\begin{eqnarray}\label{eq:lambdaSL2C}
\varLambda\left(\underline\varLambda\right)^\mu_{\,\,\nu}=\frac12 \mathrm{tr}(\sigma_\mu\, \underline \varLambda\, \sigma_\nu\,\underline \varLambda^\dag )\quad\text{and}\quad a^\mu=\frac12\mathrm{tr}\left(\underline a\, \sigma_\mu\right)\,
.\end{eqnarray} 
At this point we notice that each of the pairs $\left(\underline{\varLambda},\underline{a}\right)$ and $\left(-\underline{\varLambda},\underline{a}\right)$ 
correspond to the \textit{same} Poincar\'e transformation $\left(\varLambda\left(\underline\varLambda\right),a\right)=\left(\varLambda\left(-\underline\varLambda\right),a\right)$. 
This means that there is a 2-to-1 correspondence between the ISL$\left(2,\mathbb C\right)$ and the
Poincar\'e group. In other words, the group ISL$\left(2,\mathbb C\right)$ covers the Poincar\'e group twice. By identifying the elements $\left(\underline{\varLambda},\underline{a}\right)$ and $\left(-\underline{\varLambda},\underline{a}\right)$ in ISL$\left(2,\mathbb C\right)$ 
one obtains a group that is isomorphic to the Poincar\'e group~\cite{Polyzou:JLab}. 
The most general $\underline\varLambda$ with $\det\underline\varLambda=1$ can be written by the decomposition theorem as~\cite{Keister:1991sb,Scheck:2001}
\begin{eqnarray}\label{eq:LambdaSL2C}
\underline\varLambda\left(\boldsymbol\theta,\boldsymbol\rho\right)=\exp\left(\frac12\boldsymbol \rho \cdot\boldsymbol \sigma\right)\exp\left(-\frac{\mathrm i}{2}\,\boldsymbol \theta\cdot\boldsymbol \sigma\right)\,.
\end{eqnarray}
 In this, so-called, \textit{spinor representation} of the SL$\left(2,\mathbb C\right)$ rotations are given by the unitary $(2\times2)$-matrices 
\begin{eqnarray}
 \underline{R}\left(\boldsymbol\theta\right)=\sigma^0\cos\frac{\theta}{2}-\mathrm i\,\boldsymbol \sigma\cdot \hat{\boldsymbol{\theta}}\,\sin \frac{\theta}{2}\,,
\end{eqnarray}
whereas canonical boosts are given by the Hermitean $(2\times2)$-matrices
\begin{eqnarray}\label{eq:boostSL2C}
 \underline{B}_{\mathrm c}\left(\boldsymbol\rho\right)=\sigma^0\cosh\frac{\rho}{2}+\boldsymbol \sigma\cdot \hat{\boldsymbol{\rho}}\,\sinh
 \frac{\rho}{2}\,.
\end{eqnarray}
In App.~\ref{app:boostSL2C} we have collected some useful relations for these SL$\left(2,\mathbb C\right)$ elements, which we will need for later purposes.
 According to the \textit{polar decomposition theorem} any element of SL$\left(2,\mathbb C\right)$ can be written as a product of a unitary and a 
Hermitean matrix~\cite{Simon:1972,Keister:1991sb}. This corresponds to the decomposition theorem for Lorentz transformations.
For the description of massive spin-$\frac12$ particles there exist 2 different spinor representations of the SL$\left(2,\mathbb C\right)$:
the representation $\underline\varLambda$ with generators \begin{eqnarray}
 \underline{\boldsymbol{J}}=\frac{\boldsymbol\sigma}{2} \quad\text{and}\quad \underline{\boldsymbol{K}}=\mathrm i\,\frac{\boldsymbol\sigma}{2} 
\end{eqnarray}
 which we have already introduced in Eq.~(\ref{eq:LambdaSL2C}) and another, inequivalent representation defined by
$\underline \varLambda ':=(\underline\varLambda^\dag)^{-1}$ with generators  
\begin{eqnarray}
\underline{\boldsymbol{J}}'=\frac{\boldsymbol\sigma}{2}\quad\text{and}\quad  \underline{\boldsymbol{K}}'=-\mathrm i\,\frac{\boldsymbol\sigma}{2}\,.
\end{eqnarray}
These two spinor representations $\underline \varLambda$ and $\underline \varLambda '$ are called the \textit{right-} and the \textit{left-handed representation}, respectively. They are not 
equivalent but connected by space inversion~\cite{Scheck:2001}. 
\subsection{Four-Spinor Representation}
The fact 
that there is no element of SL$\left(2,\mathbb C\right)$ corresponding to space inversion requires to double the dimension of the representation space in order to obtain a matrix representation of the Poincar\'e group. A direct sum (up to a similarity transformation) of the right- and left-handed representation gives a linear matrix representation on $\mathbb C^4$. Consequently, combining the matrices $\underline\varLambda$ and $(\underline\varLambda^\dag)^{-1}$ into a $(4\times4)$-matrix gives~\cite{Thaller:1992}
\begin{eqnarray}\label{eq:Slambda}
 S\left(\underline{\varLambda}\right)
=
\left(%
\begin{array}{cc}\underline{\varLambda}&0
  \\
 0&(\underline\varLambda^\dag)^{-1}\\
\end{array}
\right)\,.%
\end{eqnarray}
In this $(4\times4)$-matrix representation of the Lorentz group
a general Lorentz transformation is given by~\cite{Gross}
\begin{eqnarray}&&\label{eq:smatrixrepr}
S\left( \underline \varLambda\left(\boldsymbol\theta,\boldsymbol\rho \right) \right)=\mathrm{exp}\left[\frac12\left(-\mathrm i\,\gamma^5\boldsymbol\theta+\boldsymbol\rho\right)\cdot\boldsymbol{\alpha}\right]\,.
\end{eqnarray}
Then a rotation and a canonical boost read   
\begin{eqnarray}&&
S\left( \underline R\left(\boldsymbol\theta \right) \right)= 1_4\cos\frac{\theta}{2} -\mathrm i\,\gamma^5\boldsymbol \alpha\cdot\hat{\boldsymbol{\theta}} \sin\frac{\theta}{2}
\end{eqnarray}
and
\begin{eqnarray}&&\label{eq:boost4dimsl2c}
S\left(\underline B_{\mathrm c}\left(\boldsymbol\rho \right) \right)= 1_4\cosh \frac{\rho}{2} +\boldsymbol \alpha\cdot\hat{\boldsymbol{\rho}}  \sinh\frac{\rho}{2}\,, 
\end{eqnarray}
respectively, with
\begin{eqnarray}\label{eq:alpha}
 1_4:=  1_{4\times4}\quad\text{and}\quad\boldsymbol\alpha
:=
\left(%
\begin{array}{cc}\boldsymbol\sigma&0
  \\
0& \boldsymbol\sigma\\
\end{array}
\right)\,.%
\end{eqnarray}
The representation of $\boldsymbol\alpha$ given by Eq.~(\ref{eq:alpha}) was already determined by the choice of the representation of
$S\left(\underline{\varLambda}\right)$ given by Eq.~(\ref{eq:Slambda}). A similarity transformation transforms the representation of Eq.~(\ref{eq:alpha}) into 
another representation, the, so-called, \textit{standard representation} defined in App.~\ref{app:standardrepresentation}. We will use the standard representation throughout the present work. 
\subsection{Unitary Representation}
According to the Wigner theorem we have to find a single-valued unitary representation of the ISL$\left(2,\mathbb C\right)$ acting on $\mathcal H$, which will be denoted by
$\hat U$.\footnote{Here the \lq hat' on the top of a symbol denotes an \textit{operator} acting on $\mathcal H$ and \textit{not} a unit vector, but despite of this ambiguity in the notation the distinction between unit vectors and Hilbert-space operators should be clear from the context.} $\hat U$ is a mapping from the ISL$\left(2,\mathbb C\right)$ to the space of linear operators on the quantum mechanical Hilbert space $\mathcal H$ and can be written in an exponential form as~\cite{Keister:1991sb}  \begin{eqnarray}
   \hat U\left(\underline\varLambda\left(\boldsymbol{\theta},\boldsymbol{\rho}\right),\underline a\right)=
\exp(-\mathrm i\, \hat P\cdot  a)\,\exp(-\mathrm i \,\hat {\boldsymbol{J}}\cdot\boldsymbol \theta -\mathrm i \,\hat {\boldsymbol{K}}\cdot\boldsymbol \rho )\,.
  \end{eqnarray}
The infinitesimal generators of this unitary representation of the ISL$\left(2,\mathbb C\right)$ are self-adjoint operators given by
\begin{eqnarray}\label{eq:infinitgener}
&&\hat J^j=\left.\mathrm i\,\frac{\partial \hat U\left(\underline R\left(\boldsymbol \theta\right)\right)}{\partial \theta^j}\right\vert_{\boldsymbol\theta=0 }\,,\quad \hat K^j=\left.\mathrm i\,\frac{\partial  \hat U\left(\underline B_{\mathrm c}\left(\boldsymbol \rho\right)\right)}{\partial \rho^j}\right\vert_{\boldsymbol\rho=0 }\,,\\ &&\text{and}\quad
\hat P^\mu=\left.\mathrm i\,\mathrm g^{\mu\nu}\frac{\partial \hat U\left(\underline a\right)}{\partial a^\nu}\right\vert_{a^\nu=0 }\,.\label{eq:infinitgener2}
\end{eqnarray}
The generators  $\hat {\boldsymbol J}$, $\hat {\boldsymbol P}$ and $\hat { P}^0$ have the familiar interpretations as physical quantities of the system under consideration: the generator for rotations $\hat {\boldsymbol J}$ is the total angular momentum operator, the generator for space 
translations $\hat {\boldsymbol P}$ is the total linear momentum operator, and the generator for time translations 
$\hat { P}^0$ represents the total energy of the system (note that the boost generator $\hat {\boldsymbol K}$ has no physical interpretation as observable)~\cite{Keister:1991sb}.

Using the group representation property it can be shown that these operators satisfy the Poincar\'e algebra, Eqs.~(\ref{eq:la1})-(\ref{eq:la3}). 
Furthermore, they have the following 
transformation properties under Poincar\'e transformations~\cite{Polyzou:JLab}:
\begin{eqnarray}
\label{eq:transfproppmu}
 \hat U^\dag\left(\underline \varLambda,\underline a\right)\hat P^\mu \hat U\left(\underline \varLambda,\underline a\right)&=&\varLambda^\mu_{\,\,\nu}\hat P^\nu\,,\\
  \hat U^\dag\left(\underline \varLambda,\underline a\right)\hat M^{\mu\nu} \hat U\left(\underline \varLambda,\underline a\right)&=&\varLambda^\mu_{\,\,\sigma}\varLambda^\nu_{\,\,\tau}
\left(\hat M^{\sigma\tau}-a^\sigma\hat P^\tau + a^\tau\hat P^\sigma \right)\,.
\end{eqnarray}

\section[Casimir Operators and Newton-Wigner Position Operator]{Casimir Operators and\\
 Newton-Wigner Position Operator}

This section is devoted to the Hilbert-space representation of the Casimir operators of the Poincar\'e group and to the Newton-Wigner position operator. The Casimir operators (operator invariants) of the Poincar\'e group are defined as the only independent self-adjoint operators, being polynomials in the Poincar\'e generators, that commute with every Poincar\'e generator. The states of an irreducible representation of the Poincar\'e group must be eigenstates of the Casimir operators. Therefore, their eigenvalues can be used to classify different types of irreducible representations~\cite{Keister:1991sb,Scadron}. In particular, we will see that these quantum numbers are the mass and the spin of a system associated with the eigenvalues of a mass operator and a spin operator, respectively. 
For later purposes we will also define a position operator, i.e. the canonically conjugate operator to the three-momentum operator.
\subsection{The Mass Operator}
The mass operator squared $\hat M^2$ is defined as the square of the four-momentum operator $\hat P^\mu$:
\begin{eqnarray}\label{eq:massoperator2}
 \hat M^2:=\hat P^\mu\hat P_{\mu}\,.
\end{eqnarray}                                                                                        
It is self-adjoint, commutes with every generator and is therefore a Casimir operator of the Poincar\'e group. The, so-called, \textit{spectral condition} requires $\hat M^2$ to have only non-negative eigenvalues~\cite{Keister:1991sb}. Then the square root of this operator can be defined as the operator
\begin{eqnarray}\label{eq:massoperator}
 \hat M:=\sqrt{\hat P^\mu\hat P_{\mu}}\,.
\end{eqnarray}
The eigenvalue of $\hat M$ will be called the \textit{physical mass} of the system.
 \subsection{The Spin Operator }
In addition to the mass operator we can construct a second Casimir operator from the generators of the Poincar\'e group. To this aim we introduce the, so-called, \textit{Pauli-Lubanski operator} $\hat W^\mu$ by
\begin{eqnarray}\label{eq:PL}
 \hat W^\mu:=-\frac12 \epsilon^{\mu\nu\sigma\tau}\hat P_\nu \hat M_{\sigma\tau}\,,
\end{eqnarray}
which is again a self-adjoint operator. $\hat W^\mu$ satisfies the following commutation
relations~\cite{Keister:1991sb}:
\begin{eqnarray}\label{eq:PLfourv1}
\left[ \hat W^{\mu},\hat P^{\nu}\right]& =&0\,,\\ 
\left[\hat M^{\mu\nu}, \hat W^{\rho}\right]&=&\mathrm i\left(\mathrm g^{\nu\rho}\hat W^{\mu}-\mathrm g^{\mu\rho}\hat W^{\nu}
\right)\label{eq:PLfourv2}\,,\\
\label{eq:PLfour3}
\left[ \hat W^{\mu},\hat W^{\nu}\right]& =&\mathrm i\, \epsilon^{\mu\nu\sigma\tau}\hat W_{\sigma}\hat P_{\tau}\,.
\end{eqnarray}
Eq.~(\ref{eq:PLfourv2}) ensures that $\hat W^\mu$ transforms like a four-vector under Lorentz transformations, cf. Eq.~(\ref{eq:transfproppmu}).
Therefore, its square is a Lorentz invariant and given by
\begin{eqnarray}\label{eq:W^2}
\hat W_\mu \hat W^\mu=-\hat M^2\hat{\boldsymbol{J}}_\mathrm g^2\,, 
\end{eqnarray}
where $\hat {\boldsymbol{J}}_\mathrm g$ is the total \textit{intrinsic spin operator}.
$\hat{\boldsymbol{J}}_\mathrm g^2$ is independent of $\hat M^2$, commutes with all the Poincar\'e generators and is therefore (beside $\hat M^2$) the second Casimir operator of the Poincar\'e group.

From the definition of $\hat W^\mu$, cf. Eq.~(\ref{eq:PL}), we find the Lorentz invariant relation
\begin{eqnarray}\label{eq:PW0}
\hat P_\mu\, \hat W^{\mu}=0\,. 
\end{eqnarray}
 This equation implies that $\hat W^{\mu}$ must be a space-like vector since $\hat P^\mu$ is time-like by the spectral condition. Any space-like vector can be Lorentz transformed into a reference frame where its time component vanishes. Eq.~(\ref{eq:PW0}) implies that this transformation is a \textit{general Lorentz boost} into the rest frame of the system. This boost is denoted by $B^{-1}_\mathrm g(\hat {\boldsymbol{V}})$ where $\hat V^\mu:=\hat P^\mu /\hat M$ is the four-velocity operator. $B^{-1}_\mathrm g(\hat {\boldsymbol{V}})$ together with Eq.~(\ref{eq:W^2}) can be used to define a spin operator $\hat{\boldsymbol{J}}_\mathrm g$ by\footnote{The spin operator $\hat{\boldsymbol{J}}_\mathrm g$ can equivalently be defined via the angular momentum tensor operator $\hat M^{\mu\nu}$ by the relation $
\hat J_\mathrm g^i=\frac12\epsilon^{ikl}B^{-1}_\mathrm g(\hat{\boldsymbol{V}})^k_{\,\,\mu}B^{-1}_\mathrm g(\hat{\boldsymbol{V}})^l_{\,\,\nu} \hat M^{\mu\nu}$~\cite{Keister:1991sb}.}
\begin{eqnarray}\label{eq:spin}
(0,\hat{\boldsymbol J}_\mathrm g)^\mu=\frac{1}{\hat M}B^{-1}_\mathrm g(\hat{\boldsymbol{V}})^\mu_{\,\,\nu}\hat W^{\nu}\,. 
\end{eqnarray}
It should be mentioned that this expression does not transform like a four-vector under Lorentz transformations. It rather transforms with a, so-called, \textit{Wigner rotation} $R_\mathrm{W_{\!g}}$:
\begin{eqnarray} \label{eq:jtransform}
 \hat U^\dag(\underline \varLambda,\underline a) \,(0,\hat{\boldsymbol J}_\mathrm g)^\mu\,\hat U(\underline \varLambda,\underline a)=R_\mathrm{W_{\!g}}(\hat V,\hat \varLambda)^\mu_{\,\,\nu} (0,\hat{\boldsymbol J}_\mathrm g)^\nu 
\end{eqnarray}
with 
\begin{eqnarray}\label{eq.Wignerot}
 R_\mathrm{W_{\!g}}(V,\varLambda):=B^{-1}_\mathrm g(\boldsymbol \varLambda V)\,\varLambda\, B_\mathrm g (\boldsymbol V)\,.
\end{eqnarray}
The proof of Eq.~(\ref{eq:jtransform}) can be found, e.g., in Ref.~\cite{Polyzou:JLab}.

The definition of $\hat{\boldsymbol{J}}_\mathrm g$ in Eq.~(\ref{eq:spin}) is still not unique, since there exist in principle infinite many different boosts (combinations of rotationless boosts and rotations) that transform a system with a 
given momentum $\hat P^\mu$ into its rest frame with momentum $(\hat M,\boldsymbol 0)^\mu$. Note that Lorentz invariance of Eq.~(\ref{eq:W^2}) implies that the square of the spin operator is independent of the particular choice of the boost 
$B^{-1}_\mathrm g(\hat {\boldsymbol{V}})$ and therefore unique.
In case $B^{-1}_\mathrm g(\hat{\boldsymbol{V}})$ is a canonical (rotationless) boost $B_{\mathrm c}^{-1}(\hat {\boldsymbol{V}})$ (cf. Eq.~(\ref{eq:B_c})), then Eq.~(\ref{eq:spin}) defines the \textit{canonical spin operator} $\hat{\boldsymbol{J}}_{\mathrm c}$:  
 \begin{eqnarray}\label{eq:canspin}
\left(0,\hat{\boldsymbol J}_{\mathrm c}\right)^\mu:=\frac{1}{\hat M}B_{\mathrm c}^{-1}(\hat{\boldsymbol{V}})^\mu_{\,\,\nu}\hat W^{\nu}\,. 
\end{eqnarray}
Only this canonical spin operator $\hat{\boldsymbol J}_{\mathrm c}$ transforms like a three-vector under rotations, since Wigner rotations for canonical boosts have the special property that
\begin{eqnarray}\label{eq:propWc}
 R_\mathrm{W_{\!c}}\left(V,R\right)=R
\end{eqnarray}
where $R$ is a rotation.\footnote{Note that Eq.~(\ref{eq:propWc}) implies the relation
$\label{eq:wignerrotatboost}
 R\, B_{\mathrm c}(\boldsymbol V)\, R^{-1} =B_{\mathrm c}(\boldsymbol R V)\,.$ } Therefore, the canonical Wigner rotation of a rotation is the rotation itself.
The proof of Eq.~(\ref{eq:propWc}) can be found, e.g., in Refs.~\cite{Klink:1998zz,Polyzou:JLab}. 
If a helicity boost is used in Eq.~(\ref{eq:spin}) to define the spin we speak of a \textit{helicity spin operator} $\hat{\boldsymbol J}_\mathrm h$, because its third component is the familiar expression for the helicity:
\begin{eqnarray}\label{eq:helicity}
 \hat J_\mathrm h^3=\frac{\hat {\boldsymbol P}\cdot \hat{\boldsymbol J}}{\vert\hat {\boldsymbol P}\vert}\,.
\end{eqnarray}
The special property of the helicity is that it has a well-defined value in the limit of a vanishing mass.

Using Eq.~(\ref{eq:PLfour3}) together with the spectral condition on $\hat M^2$ we obtain the usual SU$(2)$ Lie algebra for the spin operators:\footnote{SU(2) is the group of $(2\times 2)$-matrices $\underline R$ with $\underline R^\dag=\underline R^{-1}$ and $\det\vert\underline R\vert=1$. It is the universal covering group of the rotation group SO(3).}
 \begin{eqnarray}\label{eq:SU(2)}
\left[\hat J_\mathrm g^i,\hat J_\mathrm g^k\right]=\mathrm i\epsilon^{ikl}\hat J_\mathrm g^l.
\end{eqnarray}
These relations imply that the spin eigenvalue $J$ can only be an integer or a half-integer number. In the case of the helicity spin the SU(2) commutation relations are satisfied by the components $\hat J^3_{\mathrm h}$ and $\hat {\boldsymbol J}_{\mathrm h}\times \boldsymbol x_3/ \vert\boldsymbol x
_3\vert$.
\subsection{Newton-Wigner Position Operator}\label{sec:NWposition operator}
For the case of canonical spin one finds by direct computation that $\hat {\boldsymbol J}_\mathrm c$ is related to the Lorentz generators by
\begin{eqnarray}\label{eq:Joperatorexplicit}
\hat {\boldsymbol J}=\hat{\boldsymbol {X}}_\mathrm c\times \hat {\boldsymbol P}+\hat {\boldsymbol J}_\mathrm c
\end{eqnarray}
and
\begin{eqnarray}\label{eq:boostgeneratorsexplicit}
\hat {\boldsymbol K}=-\frac12\{\hat P^0,\hat{\boldsymbol {X}}_\mathrm c \}-
\frac{\hat {\boldsymbol P}\times\hat{\boldsymbol {J}}_\mathrm c }{\hat P^0+\hat M}\,.
\end{eqnarray}
Here the Hermitean operator $\hat{\boldsymbol {X}}_\mathrm c$ is the canonically conjugate operator to $\hat{\boldsymbol P}$ and given by
\begin{eqnarray}
\hat{\boldsymbol {X}}_\mathrm c =-\frac12\{\frac{1}{\hat P^0},\hat {\boldsymbol K}\}-
\frac{\hat{\boldsymbol P}\times(\hat P^0\hat {\boldsymbol J}-\hat{\boldsymbol P}\times \hat {\boldsymbol K} )}{\hat P^0\hat M(\hat P^0+\hat M)}\,.
\end{eqnarray}
It was introduced by Newton and Wigner in Ref.~\cite{Newton:1949cq} and is called the \textit{Newton-Wigner position operator}. The role of this operator will become evident later when dealing with the construction of dynamical representations of the Poincar\'e group.

It should be mentioned that the above relation between the canonical spin operator $\hat {\boldsymbol J}_\mathrm c$ and the total angular momentum $\hat {\boldsymbol J}$ is equivalent to the corresponding relation in non-relativistic quantum mechanics. This is special to canonical spins and allows one to couple spin and orbital angular momentum to the total angular momentum using the standard addition rules of combining angular momenta in non-relativistic quantum mechanics~\cite{Keister:1991sb}.
\section{One-Particle Irreducible Representations}
 The aim of this section is to end up with a Poincar\'e-invariant quantum mechanical description of a single particle with mass $m$ and spin $j$. This amounts to finding an irreducible representation of the Poincar\'e group on a one-particle Hilbert space. The eigenvalues $m^2$ and $j(j+1)$ of the Casimirs $\hat m^2$ and $\hat{\boldsymbol j}^2$, respectively, together with the sign of the eigenvalue $p^0$ of the Hamiltonian $\hat p^0$ can be used to classify irreducible representations~\cite{Polyzou:JLab}. In the present work we will only consider the massive spin-$j$ case with $m^2>0$ and $p^0>0$ and the massless spin-1 case with $m^2=0$ and $p^0>0$.  
\subsection{Massive Representations}
We start with massive particles.
A suitable basis for a one-particle Hilbert-space representation is given by the set of simultaneous eigenstates $\vert m,j,\boldsymbol p,\sigma\rangle$ of the commuting Hermitean one-body operators 
  \begin{eqnarray}
\hat m^2,\quad \hat {\boldsymbol j}^2,\quad \hat{\boldsymbol p}\quad\text{and}\quad \hat j_\mathrm c^3\,.
\end{eqnarray}
For the states $\vert m,j,\boldsymbol p,\sigma\rangle$ short-hand denoted as $\vert\boldsymbol p,\sigma\rangle$ we use a covariant normalization given by
\begin{eqnarray}\label{eq:norm1massivepart}
\langle \boldsymbol  p',\sigma' \vert \boldsymbol  p,\sigma\rangle=(2\pi)^3\delta_{\sigma\sigma'}2p^0\delta^3\left(\boldsymbol p-\boldsymbol p'\right)\,.
\end{eqnarray}
The corresponding completeness relation reads
\begin{eqnarray}
\hat{1}=\int_{\mathbb R^3}\frac{\mathrm d^3 p}{(2\pi)^32p^0}\sum_{\sigma=-j}^j\vert \boldsymbol  p,\sigma\rangle\langle \boldsymbol  p,\sigma\vert\,.
\end{eqnarray}
The action of $\hat U\left(\underline \varLambda,\underline a\right)$ on $\vert \boldsymbol p,\sigma\rangle $ is given by
\begin{eqnarray}\label{eq:1ptrafoprop}
\hat U\left(\underline \varLambda,\underline a\right)\vert \boldsymbol  p,\sigma\rangle &=&
\exp\left(-\mathrm i \varLambda p\cdot a\right)\sum_{\sigma'=-j}^j
D_{\sigma'\sigma}^j\left[\underline R_\mathrm {W_{\!c}}\left(v,\varLambda\right)\right]\vert \boldsymbol \varLambda p,\sigma'\rangle\,,
\end{eqnarray}
with $v=p/m$. 
For a detailed derivation of Eq.~(\ref{eq:1ptrafoprop}) see, e.g., Ref.~\cite{Keister:1991sb}.
Here
\begin{eqnarray}\label{eq:wignerDfuntion}
 D_{\sigma'\sigma}^j(\underline R):=\langle j,\sigma'\vert \hat U\left(\underline R,0\right)\vert j,\sigma\rangle
\end{eqnarray}
are the familiar \textit{Wigner $D$-functions} being a ($2j+1$) dimensional unitary irreducible representation of the rotation group $\mathrm {SU(2)}$ in the basis $\vert j,\sigma\rangle\equiv\vert\sigma\rangle$ (see, e.g., Ref.~\cite{Varshalovich}).
Some useful relations of these Wigner $D$-functions which will be repeatedly used throughout this work are collected in App.~(\ref{app:WignerDf}).
\subsection{Massless Representations for Spin-1 Particles}
In this section we turn to the case of massless particles with $m^2=0$. For a detailed discussion on the covariant treatment of massless spin-1 particles, such as the photon $\gamma$, within the so-called \textit{Gupta-Bleuler formalism} see, e.g. Refs.~\cite{Bleuler:1950cy,Gupta:1949rh,Schweber:1961,Klink:2000pq}. 
A suitable basis for massless spin-1 states is given by the state vectors $\vert j, \boldsymbol p,\lambda\rangle\equiv\vert \boldsymbol p,\lambda\rangle$ with $\lambda$ denoting the 4 polarization degrees-of-freedom of a massless vector field (2 physical and 2 unphysical ones).
These states are simultaneous eigenstates of the operators
 \begin{eqnarray}
\hat {\boldsymbol j}^2,\quad \hat{\boldsymbol p}\quad \text{and}\quad\hat j_{\mathrm h}^3\,.
\end{eqnarray}
These massless states are normalized covariantly according to
\begin{eqnarray}
\langle \boldsymbol p',\lambda' \vert \boldsymbol p,\lambda\rangle=(2\pi)^3(-\mathrm g_{\lambda\lambda'})2\vert\boldsymbol p\vert\delta^3\left(\boldsymbol p-\boldsymbol p'\right)\,.
\end{eqnarray}
The corresponding completeness relation reads
\begin{eqnarray}
\hat{1}=\int_{\mathbb R^3}\frac{\mathrm d^3 p}{(2\pi)^32\vert\boldsymbol p\vert}\sum_{\lambda=0}^3(-\mathrm g_{\lambda\lambda})\vert \boldsymbol p,\lambda\rangle\langle \boldsymbol p,\lambda\vert\,.
\end{eqnarray}
The transformation properties of $\vert \boldsymbol p,\lambda\rangle$ under Poincar\'e transformations are given by
 \begin{eqnarray}\label{eq:trafpropphotonstates}
  \hat U(\underline{\varLambda},\underline a)\vert \boldsymbol p,\lambda\rangle=\exp\left(-\mathrm i \varLambda p\cdot a\right)\sum_{\lambda'=0}^3
R_\mathrm{W_{\!h}}\left(p,\varLambda\right)_{\lambda'\lambda}\vert \boldsymbol \varLambda p,\lambda'\rangle
 \end{eqnarray}
where
\begin{eqnarray}
R_\mathrm{W_{\!h}}\left(p,\varLambda\right):=B^{-1}_\mathrm h(\boldsymbol \varLambda p)\,\varLambda\, B_\mathrm h(\boldsymbol p)
\end{eqnarray}
is the massless analogue of a Wigner rotation.
\section{Multi-Particle Representations}
\label{sec:multiparticlerepre}
After having found the one-particle irreducible representations of the Poincar\'e group in the previous section we consider now systems with more than 1 particle. We will restrict our considerations to a finite number of degrees of freedom. Our aim is to end up with a multi-particle representation of the Poincar\'e group which is constructed from the one-particle irreducible representations. 
In particular, we will use two different kinds of multi-particle bases, the tensor-product-state and the velocity-state basis. 
\subsection{Tensor-Product States}
The first step towards a multi-particle representation of the Poincar\'e group is to consider a $n$-particle Hilbert space which is the $n$-fold tensor product of single-particle Hilbert spaces of the previous section. A free $n$-particle momentum state basis of this $n$-particle Hilbert space is then constructed from tensor products of $n$ single-particle basis states as follows:
  \begin{eqnarray}\label{eq:npartstate}
   \vert \boldsymbol p_1,\sigma_1; \boldsymbol p_2,\sigma_2;\ldots; \boldsymbol p_n,\sigma_n\rangle\equiv \vert \{\boldsymbol p_i,\sigma_i\}\rangle:=
\vert \boldsymbol p_1,\sigma_1\rangle\otimes\vert \boldsymbol p_2,\sigma_2\rangle\otimes\ldots\otimes\vert \boldsymbol p_n,\sigma_n\rangle
\,.\nonumber\\
  \end{eqnarray}
Here $p_i^\mu$ and $\sigma_i$ are the physical four-momentum and canonical spin projection of the $i^{\text{th}}$ particle, respectively. 
 We assume that all single particles are on their respective mass shells, i.e.
\begin{eqnarray}
p_i^0=\sqrt{m_i^2+\boldsymbol p_i^2}\,,\quad\forall \,i=1,\ldots,n\,.
  \end{eqnarray}
The kinematic variables
\begin{eqnarray}
P_{n}^\mu=\sum_{i=1}^n p_i^\mu \quad\text{and}\quad M_{n}=\sqrt{P_{n}^2}
  \end{eqnarray}
are the free total four-momentum and the free invariant mass of the $n$-particle system, respectively. 
A unitary representation of the Poincar\'e group acting on the tensor-product states~(\ref{eq:npartstate}) is given by a tensor product of $n$ irreducible single-particle representations:
\begin{eqnarray}
\label{eq:unitPtrafo}
\hat{U}_{12\cdots n}\left[\underline \varLambda, \underline a\right]:=\hat{U}_1\left[\underline \varLambda, \underline a\right]\otimes \hat{U}_2\left[\underline \varLambda, \underline a\right]\otimes\ldots \otimes
 \hat{U}_n\left[\underline \varLambda, \underline a\right]\,.
  \end{eqnarray}
Consequently, the Poincar\'e generators of the $n$-particle system are sums of the $n$ one-particle generators:
\begin{eqnarray}
\hat P_{n}^\mu&:=&\hat p_1^\mu\otimes \hat{ 1}_2 \otimes\cdots \otimes  \hat{ 1}_n+\hat{ 1}_1\otimes \hat p_2^\mu \otimes\cdots \otimes \hat {1}_n+\ldots+
\hat {  1}_1\otimes \hat { 1}_2 \otimes\cdots \otimes \hat p_n^\mu\,;\nonumber\\
\hat M_{n}^{\mu\nu}&:=&\hat m_1^{\mu\nu}\otimes \hat { 1}_2 \otimes\cdots \otimes \hat {  1}_n+\hat {  1}_1\otimes \hat m_2^{\mu\nu} \otimes\cdots \otimes \hat {  1}_n+\ldots\nonumber\\&&+
\hat {  1}_1\otimes \hat {  1}_2 \otimes\cdots \otimes \hat m_n^{\mu\nu}\,.\label{eq:multipartG}
  \end{eqnarray}
The transformation properties of the multi-particle tensor-product states under Poincar\'e transformations follow from the corresponding transformation properties of the single-particle states, cf.  Eqs.~(\ref{eq:1ptrafoprop}) and~(\ref{eq:trafpropphotonstates}):
\begin{eqnarray}
\label{eq:trafopropertiestensorprod}
\hat{U}_{12\cdots n}\left[\underline \varLambda,\underline a\right]\vert\{ \boldsymbol p_i,\sigma_i\}\rangle=
\mathrm{e}^{-\mathrm i\varLambda P_n\cdot a}
\sum_{\{\sigma_i\}}  \vert\{\boldsymbol \varLambda p_i,\sigma_i'\}\rangle
\prod_{i=1}^n D^{j_i}_{\sigma_i'\sigma_i}\left[\underline R_\mathrm{W_{\!c}}\left(v_i,\varLambda\right)\right]\,
  \end{eqnarray} 
where $v_i:=p_i/m_i$ is the four-velocity of the $i^{\text{th}}$ particle.
  \subsection{Velocity States}\label{sec:velocitystates}
In this section we will introduce another, alternative basis for multi-particle states which is associated with the overall and internal motion of the multi-particle system. To be more precise, we will define a multi-particle state which is specified by the overall free four-velocity of the $n$-particle system
\begin{eqnarray}
\label{eq:overallvelocity}
V^\mu:=\frac{P_{n}^\mu}{M_{n}}
\end{eqnarray}
and the individual center-of-mass momenta 
\begin{eqnarray}\label{eq:commom}
 k_i:=B_\mathrm c^{-1}\left(\boldsymbol V\right)p_i\,,\quad i=1,\ldots,n\,.
  \end{eqnarray}
 Note that the $n$ center-of-mass momenta $\boldsymbol k_1,\ldots,\boldsymbol k_n$ are not independent from each other but constrained by the relation 
\begin{eqnarray}\label{eq:comconstraint}
 \sum_{i=1}^n\boldsymbol k_i=0\,.
\end{eqnarray}
Thus, the center-of-mass momentum of the $j^{\text{th}}$ particle $\boldsymbol k_j$ can be expressed in terms of the remaining $n-1$ center-of-mass momenta.

A multi-particle state characterized by $V$ and by the $n-1$ $\{ \boldsymbol k_i\}$ will be called a \textit{velocity state}. It was introduced by Klink in Ref.~\cite{Klink:1998zz}. Such a velocity state is defined by applying a canonical boost $\hat{U}_{12\cdots n}\left[\underline B_\mathrm c\left(\boldsymbol V\right)\right]$ to the multi-particle rest state: 
\begin{eqnarray}\label{eq:npartvelocitystate}\lefteqn{\vert  V;\boldsymbol k_1,\mu_1;\boldsymbol k_2,\mu_2;\ldots;\boldsymbol k_n,\mu_n\rangle \equiv\vert  V;\{\boldsymbol k_i,\mu_i\}\rangle}\nonumber\\&:=& 
\hat{U}_{12\cdots n}\left[\underline B_\mathrm c\left(\boldsymbol V\right)\right]\vert \{\boldsymbol k_i,\mu_i\}\rangle=  
\sum_{\{\sigma_i\}} \vert\{ \boldsymbol p_i,\sigma_i\}\rangle
\prod_{i=1}^n D^{j_i}_{\sigma_i\mu_i}\left[\underline R_\mathrm{W_{\!c}}\left(w_i,B_\mathrm c(\boldsymbol V)\right)\right]\nonumber\\
  \end{eqnarray}
with the internal four-velocities defined by $w_i:=k_i/m_i$. $\mu_i$ denotes the canonical spin projection of the $i^{\text{th}}$ particle associated with the multi-particle state in the rest frame.

The action of the four momentum operator on this velocity state can be easily evaluated using the transformation properties of $\hat P^\mu$, Eq.~(\ref{eq:transfproppmu}):
\begin{eqnarray}\label{eq:pmuveloictystates}&&
\hat P^\mu_{n} \vert  V;\{\boldsymbol k_i,\mu_i\}\rangle=
M_n V^\mu \vert  V;\{\boldsymbol k_i,\mu_i\}\rangle\,,
  \end{eqnarray}                                                                                                                                            
where $M_n=\sum_{i=1}^n k_i^0$. Therefore, a velocity state $\vert  V;\{\boldsymbol k_i,\mu_i\}\rangle$ is an eigenstate of $\hat P^\mu_{n}$ with eigenvalue $M_n V^\mu$. This suggests writing $\hat P^\mu_{n}$ as
the product 
\begin{eqnarray}
\hat P^\mu_{n} =\hat M_n \hat V^\mu
  \end{eqnarray}
where
\begin{eqnarray}
\hat M_n=\sqrt{\hat P_n^\mu\hat P_{n\mu}}\quad\text{and}\quad \hat V^\mu:=\frac{\hat P^\mu_{n}}{\hat M_n}
  \end{eqnarray}
 are the free invariant mass operator and 
the free four-velocity operator, respectively. $\hat V^\mu$ transforms like a four-vector operator under Lorentz transformations and satisfies the relation $\hat V^\mu\hat V_\mu=\hat{ 1}$. A velocity state
$\vert  V;\{\boldsymbol k_i,\mu_i\}\rangle$
is a simultaneous eigenstate of $\hat M_n$ and $\hat V^\mu$ with eigenvalues $M_n=\sum_{i=1}^n k_i^0$ and $V^\mu=(\sqrt{1+\boldsymbol V^2 },\boldsymbol V)$, respectively.  $\vert  V;\{\boldsymbol k_i,\mu_i\}\rangle$ is also an eigenstate of the  center-of-mass momentum operators $\hat k_i^\mu$. These operators do not transform covariantly under Lorentz transformations; rather they transform under a Wigner rotation in the same way as the spin operator $(0,\hat{\boldsymbol J}_\mathrm g)^\mu$, cf. Eq.~(\ref{eq:jtransform}).

The transformation properties of the velocity states under Lorentz transformations are given by
\begin{eqnarray}\label{eq:velocitystatetransfprop}
 \lefteqn{\hat U_{12\cdots n}\left(\underline \varLambda\right)\vert  V;\{\boldsymbol k_i,\mu_i\}\rangle}\nonumber\\&&=
\sum_{\{\mu_i'\}}\vert \varLambda  V; \{\boldsymbol R_\mathrm{W_{\!c}}(V,\varLambda) k_i,\mu_i'\}\rangle\prod_{i=1}^n D^{j_i}_{\mu_i'\mu_i}
\left[\underline R_\mathrm{W_{\!c}}\left(V,\varLambda\right)\right]\,.
\end{eqnarray}
For each \textit{massless} spin-1 particle occurring in the velocity state the corresponding Wigner $D$-function has to be replaced by the $(4\times 4)$-dimensional Minkowski-space representation of the Lorentz group (cf. Eq.~(\ref{eq:rotMinkwoskispace})):
\begin{eqnarray}
 D^{j_i}_{\mu_i'\mu_i}
\left[\underline R_\mathrm{W_{\!c}}\left(V,\varLambda\right)\right]\rightarrow R_\mathrm{W_{\!h}}\left[k_i,R_\mathrm{W_{\!c}}(V, \Lambda)\right]_{\mu_i'\mu_i}\,.
\end{eqnarray}
A comparison of the transformation properties of the tensor-product states, Eq.~(\ref{eq:trafopropertiestensorprod}), with those of the velocity states, Eq.~(\ref{eq:velocitystatetransfprop}), leads to the following notable observation: 
for tensor-product states each Wigner rotation in the $D$-functions depends on a \textit{different} single-particle four-velocity $v_i$, whereas for velocity states all momenta and canonical spins rotate with the \textit{same} Wigner rotation as a consequence of Eq.~(\ref{eq:propWc}). This property of the velocity states makes it possible to couple canonical spins and orbital angular momenta according to the standard addition rules using SU(2) Clebsch-Gordan coefficients~\cite{Klink:1998zz}.

The transformation properties of a velocity state under a space-time translation $\hat{U}_{12\cdots n}\left(\underline a\right)$ are equivalent with those for a tensor-product state if one identifies $P^\mu_n$ with $M_n V^\mu$ in the exponent of Eq.~(\ref{eq:trafopropertiestensorprod}):
\begin{eqnarray}&&\label{eq:velocitystatetransfpropsttransl}
\hat{U}_{12\cdots n}\left(\underline a\right)\vert  V;\{\boldsymbol k_i,\mu_i\}\rangle=
\vert  V;\{\boldsymbol k_i,\mu_i\}\rangle \mathrm e^{-\mathrm i M_n V\cdot a}\,.
  \end{eqnarray}
Orthogonality and completeness relations for the velocity states can be derived from the ones for the usual tensor-product states. To this we have to transform the corresponding integral measures for a system of $n$ particles according
\begin{eqnarray}
 \int\prod_{i=1}^{n}\frac{\mathrm d^3 p_i}{2p_i^0}= \int \frac{\mathrm d^3 V}{V^0} \prod_{i=1}^{n-1}\frac{\mathrm d^3 k_i}{2k_i^0}\frac{(\sum_{j=1}^n k_j^0)^3}{2k_n^0}\,,
\end{eqnarray}
where
$k_i^0:= \sqrt{m_i^2+\boldsymbol{k}_i^2}$ for $\forall\,i=1,\ldots, n-1$ and $k_n^0:= \sqrt{m_n^2+(\sum_{j=1}^{n-1}\boldsymbol{k}_j)^2}\,.$ Here the $n^{\text{th}}$ momentum has been chosen to be redundant without loss of generality. Then the completeness relation for the velocity states follows from the one for the tensor-product states:
\begin{eqnarray}\label{eq:vcompl}
\hat{1}_{\{n\}}&=&\sum_{\{ \mu_i\}}
\int
\frac{\mathrm d^3 V}{(2\pi)^3 V_0} \left(
\prod_{i=1}^{n-1}\frac{\mathrm d^3k_i}{(2 \pi)^3 2 k^0_i}
\right)\frac{\left(\sum_{i=1}^n k^0_i\right)^3}{2
k^0_n}\nonumber\\ & & \times \vert V; \{\boldsymbol k_i,\mu_i\} \rangle \langle  V;\{\boldsymbol k_i,\mu_i\}\vert\, .\nonumber\\
\end{eqnarray}
The corresponding orthogonality relation reads:
\begin{eqnarray}\label{eq:vnorm}
\lefteqn{\langle V^\prime; \{\boldsymbol k_i',\mu_i'\}
\vert \, V;\{\boldsymbol k_i,\mu_i\}\rangle  \nonumber}\\ & & = V_0 \,
\delta^3(\boldsymbol{V}^\prime-\boldsymbol{V})\, \frac{(2 \pi)^3 2
k^0_n}{\left( \sum_{i=1}^n k_i^0\right)^3} \left(
\prod_{i=1}^{n-1} (2\pi)^3 2 k^0_i
\delta^3(\boldsymbol{k}_i^\prime-\boldsymbol{k}_i)\right) \left( \prod_{i=1}^{n}
\delta_{\mu_i^\prime \mu_i}\right)\, .\nonumber\\
\end{eqnarray}
A derivation of these relations can be found in Ref.~\cite{KrassniggDiss:2001}.
The completeness and orthogonality relations, Eqs.~(\ref{eq:vcompl}) and~(\ref{eq:vnorm}), have to be modified for each occurring photon according to the replacements $\sum_{\mu_\gamma}
\rightarrow \sum_{\mu_\gamma} (-\mathrm g^{\mu_\gamma \mu_\gamma})$ and $\delta_{\mu_\gamma \mu_\gamma^\prime} \rightarrow
(-\mathrm g^{\mu_\gamma \mu_\gamma^\prime})$.
\subsection{Clebsch-Gordan Coefficients}\label{par:Clebsch-Gordan Coefficients}
In the present work we restrict our considerations to systems consisting of $n=3$ and $n=4$ particles among which two particles (1 and 2) are coupled together forming a composite subsystem (12). The free two-body subsystem (12) is the subject of this section in which we proceed along the lines of Refs.~\cite{Polyzou:JLab,Keister:1991sb}. Since we will work in the velocity-state representation for 3 and 4 particles we adopt the notation~(\ref{eq:npartvelocitystate}) for the momenta and spins of the particles 1 and 2. It should be noted that $\boldsymbol k_1$ and $\boldsymbol k_2$ can take any value if they denote the momenta of particles 1 and 2 in the center of mass of the 3- or 4-particle system.

The aim of this section is to construct appropriate Clebsch-Gordan coefficients of the Poincar\'e group which define a two-particle basis on which $\hat U_{12}\left(\underline\varLambda,\underline a\right)$ acts irreducibly. This is equivalent to expressing tensor products of 
single-particle eigenstates in terms of linear combinations of two-particle eigenstates of the total four momentum such that this eigenstate at rest transforms with a $(2j+1)$-dimensional irreducible representation of the SU(2) with respect to rotations~\cite{Polyzou:JLab}:
\begin{eqnarray}\label{eq:resteigenstatetrafoprop}
 &&\hat U_{12}\left(\underline R\right)\vert(m_{12},j);\boldsymbol 0,\mu_j\rangle=\sum_{\mu_j'}\vert(m_{12},j);\boldsymbol 0,\mu_j'\rangle D^j_{\mu_j'\mu_j}\left(\underline R\right)\,.
\end{eqnarray}
Here $m_{12}$ is the free invariant mass of the (12)-subsystem given by \begin{eqnarray}
 m_{12}=\sqrt{k_{12}^\mu k_{12\mu}} \quad\text{with}\quad k_{12}^\mu:=k_1^\mu+k_2^\mu\,.
\end{eqnarray}
With the free velocity of the (12)-subsystem
\begin{eqnarray}
 w_{12}:=\frac{k_{12}}{m_{12}}
\end{eqnarray}
we can define the relative (internal) momentum $\tilde{\boldsymbol{k}}$ of the the (12)-subsystem:\footnote{Quantities with a \lq tilde' always refer to the center-of-momentum frame of the (12)-subsystem.}
\begin{eqnarray}\label{eq:ktilde1}
 \tilde k_{i}:=B_\mathrm c^{-1}(\boldsymbol{w}_{12})k_i \quad \text{with}\quad i=1,\,2\quad \Rightarrow\quad \tilde{\boldsymbol{k}}_1=-\tilde{\boldsymbol{k}}_2=:\tilde{\boldsymbol{k}}\,.
\end{eqnarray} 
In this representation the non-interacting mass operator $\hat m_{12}$  has the form
  \begin{equation}\label{eq:massoperfree12}
\hat m_{12}=\sqrt{m_1^2+\hat{\tilde{\boldsymbol k}}^2}+\sqrt{m_2^2+\hat{\tilde{\boldsymbol k}}^2}\,.
\end{equation}
The velocities of the individual particles in the center-of-momentum frame of the (12)-subsystem then read
\begin{eqnarray}
 \tilde w_{i}:=\frac{\tilde k_i}{m_i} \quad\text{with}\quad i=1,2\,.
\end{eqnarray}
The two-particle state associated with the system at rest is obtained by applying a canonical boost $\hat{U}_{12}^{-1}[B_\mathrm c(\boldsymbol w_{12})]$ on the state $\vert\boldsymbol  k_1,\mu_1; \boldsymbol k_2,\mu_2\rangle$ and multiplying the Wigner-D functions to the other side with the help of Eq.~(\ref{eq:D-functdelta}):
\begin{eqnarray}
\label{eq:coupl2states}
 \lefteqn{\vert\tilde{\boldsymbol{k}},\tilde \mu_{1}; -\tilde{\boldsymbol{k}},\tilde \mu_{2}\rangle}\nonumber\\&=& \sum_{\mu_1\mu_2} \hat{U}_{12}^{-1}[B_\mathrm c( \boldsymbol w_{12})]\vert\boldsymbol   k_1,\mu_1; \boldsymbol  k_2,\mu_2\rangle\nonumber\\&&\times
 D^{j_1}_{\mu_1\tilde \mu_1}\left[\underline R_\mathrm{W_{\!c}}\left(\tilde w_1, B_{\mathrm c}(\boldsymbol w_{12})\right)\right]
D^{j_2}_{\mu_2\tilde \mu_2}\left[\underline R_\mathrm{W_{\!c}}\left(\tilde w_2,B_{\mathrm c}(\boldsymbol w_{12})\right)\right]\,.
\end{eqnarray}
This rest eigenstate of the four-momentum transforms under a rotation in the same way as rest eigenstates do in non-relativistic quantum mechanics:
 \begin{eqnarray}
\label{eq:transproprotreststate}
 \hat{U}_{12}(\underline R)\vert\tilde{\boldsymbol{k}},\tilde \mu_{1}; -\tilde{\boldsymbol{k}},\tilde \mu_{2}\rangle=\sum_{\tilde \mu_1'\tilde \mu_2'}
\vert\boldsymbol R  \tilde k,\tilde \mu_1'; -\boldsymbol R \tilde k,\tilde \mu_2'\rangle
 D^{j_1}_{\tilde \mu_1'\tilde \mu_1}\left[\underline R\right]
D^{j_2}_{\tilde\mu_2'\tilde \mu_2}\left[\underline R\right]\,.
\end{eqnarray}
In the derivation of Eq.~(\ref{eq:transproprotreststate}) we have exploited the special properties of canonical Wigner rotations, cf. Eq.~(\ref{eq:propWc}). Eq.~(\ref{eq:transproprotreststate}) reveals
the benefit of using canonical boosts that permits to combine spins and angular momenta in the usual non-relativistic manner. We can couple the spins $j_1$ and $j_2$ of the particles 1 and 2 to the total spin $s$ and the total spin and orbital angular momentum $l$ to the total angular momentum $j$ by using standard SU(2) Clebsch-Gordan coefficients 
$C^{s\mu_s}_{j_1\tilde \mu_1j_2\tilde \mu_2}:=\langle j_1,\tilde \mu_1;j_2,\tilde \mu_2\vert s,\mu_s\rangle$ and  $C^{j\mu_j}_{s\mu_sl\mu_l}:=\langle l,\mu_l; s,\mu_s\vert j,\mu_j\rangle$. Keeping this in mind we introduce
\begin{eqnarray}\label{eq:reststatecoupled}
 \vert\boldsymbol 0;\tilde{k};(l,s),j,\mu_j\rangle:= \sum_{\tilde \mu_1\tilde \mu_2\mu_l\mu_s}
\int_{\Omega} \mathrm d\Omega(\hat{\tilde {\boldsymbol{ k}}})Y_{l\mu_l}(\hat{\tilde{\boldsymbol{ k}}})\vert\tilde{\boldsymbol{k}},\tilde \mu_{1}; -\tilde{\boldsymbol{k}},\tilde \mu_{2}\rangle 
C^{s\mu_s}_{j_1\tilde \mu_1j_2\tilde \mu_2}
C^{j\mu_j}_{s\mu_sl\mu_l}\,\nonumber\\
\end{eqnarray}
with $\tilde{k}:=\vert\tilde{\boldsymbol {k}}\vert$. $\hat{\tilde {\boldsymbol k}}:=\tilde{\boldsymbol{ k}}/\tilde{k}$ and $\mathrm d\Omega (\hat{\tilde {\boldsymbol{ k}}}) $ being the two-dimensional surface element for the angular integration.\footnote{In this case the \lq hat' denotes a \textit{unit vector} and \textit{not} an operator.} Here we have eliminated the angles in $\hat{\tilde {\boldsymbol k}}$ in favor of discrete quantum numbers using the usual spherical harmonics $Y_{l\mu_l}(\hat{\boldsymbol{ k}}):=\langle\hat{\boldsymbol{ k}}\vert l,\mu_l\rangle$. The quantum numbers
$s$ and $l$ corresponding to total spin and orbital angular momentum label degeneracies. 
The spherical harmonics $Y_{l\mu_l}(\hat{\boldsymbol{ k}})$ and the Clebsch-Gordan coefficients $C_{l\mu_ls\mu_s}^{j\mu_j}$ satisfy the orthogonality relations
\begin{eqnarray}\label{eq:sphharmotho}
\int_\Omega \mathrm d\Omega(\hat{\tilde {\boldsymbol{ k}}}')Y^{\ast}_{l'\mu_l'}(\hat{\tilde{\boldsymbol{ k}}}')Y_{l\mu_l}(\hat{\tilde{\boldsymbol{ k}}}')=\delta_{ll'}\delta_{\mu_l\mu_l'}
\end{eqnarray}
and~\cite{Varshalovich}
\begin{eqnarray}
\sum_{\mu_l\mu_s}C_{l\mu_ls\mu_s}^{j'\mu_j'}C_{l\mu_ls\mu_s}^{j\mu_j}=\delta_{jj'}\delta_{\mu_j\mu_j'}\,,\label{eq:cgortho}
\end{eqnarray}
respectively.
The rest eigenstate of Eq.~(\ref{eq:reststatecoupled}) has the desired transformation properties under rotations as given in Eq.~(\ref{eq:resteigenstatetrafoprop}).
Applying a canonical boost $\hat U_{12}[\underline B_\mathrm c(\boldsymbol w_{12})]$ on this state gives an irreducible representation:
\begin{eqnarray}\label{eq:irredrepr}\lefteqn{
\vert \boldsymbol k_{12};\tilde{k};(l,s),j,\mu_j\rangle}\nonumber\\&=&\hat U_{12}[\underline B_\mathrm c(\boldsymbol w_{12})]
\vert\boldsymbol 0;\tilde{k};(l,s),j,\mu_j\rangle\nonumber\\&=&\sum_{\tilde \mu_1\tilde \mu_2\mu_l\mu_s} \int_\Omega \mathrm d\Omega(\hat{\tilde {\boldsymbol{ k}}})Y_{l\mu_l}(\hat{\tilde{\boldsymbol{ k}}})
 \vert \boldsymbol k_1,\mu_1;  \boldsymbol k_2,\mu_2\rangle
C^{s\mu_s}_{j_1\tilde \mu_1j_2\tilde \mu_2}C^{j\mu_j}_{l\mu_ls\mu_s} 
\nonumber\\&&\times
 D^{j_1}_{\mu_1\tilde \mu_1}\left[\underline R_\mathrm{W_{\!c}}\left(\tilde w_1, B_{\mathrm c}(\boldsymbol w_{12})\right)\right]
D^{j_2}_{\mu_2\tilde \mu_2}\left[\underline R_\mathrm{W_{\!c}}\left(\tilde w_2,B_{\mathrm c}(\boldsymbol w_{12})\right)\right].\nonumber\\
\end{eqnarray}
 This state transforms irreducibly under the action of $\hat U_{12}\left(\underline\varLambda,\underline a\right)$, i.e. it has the transformation properties of a single-particle state, cf. Eq.~(\ref{eq:1ptrafoprop}).
The states $\vert \boldsymbol k_{12};\tilde{k};(l,s),j,\mu_j\rangle$ defined by Eq.~(\ref{eq:irredrepr}) are simultaneous eigenstates of the operators $\hat m_{12}$, $\hat{\boldsymbol j}_{\mathrm c}^2$, $\hat j_\mathrm c^3$ and $\hat {\boldsymbol k}_{12}$.
These operators together with the canonically conjugate operators of $\hat {\boldsymbol k}_{12}$ and  $\hat j_\mathrm c^3$ commute with $\hat m_{12}$~\cite{Polyzou:JLab}.

Finally we write down the Clebsch-Gordan coefficients which are determined from Eq.~(\ref{eq:irredrepr})~\cite{Keister:1991sb,Polyzou:JLab}:
\begin{eqnarray}\label{eq:CGCoeffPoin}
\lefteqn{\langle \boldsymbol k_{12};\tilde{k};(l,s),j,\mu_j\vert\boldsymbol k_1,\mu_1; \boldsymbol k_2,\mu_2\rangle}\nonumber\\&=& (2\pi)^6\int_\Omega \mathrm d\Omega(\hat{\tilde{\boldsymbol{  k}}})
2 k_{1}^0\delta^3(\boldsymbol k_{1}-\boldsymbol B_\mathrm c(\boldsymbol w_{12}) \tilde k_1) 2 k_{2}^0\delta^3(\boldsymbol k_{2}-\boldsymbol B_\mathrm c(\boldsymbol w_{12})\tilde k_2) Y^{\ast }_{l\mu_l}(\hat{\tilde{\boldsymbol k}})
\nonumber\\&&\times 
C^{s\mu_s}_{j_1\tilde \mu_1j_2\tilde \mu_2}C^{j\mu_j}_{l\mu_ls\mu_s} D^{j_1}_{\tilde \mu_1\mu_1}[\underline R_\mathrm{W_{\!c}}^{-1}(\tilde w_1, B_{\mathrm c}(\boldsymbol w_{12}))]
D^{j_2}_{\tilde \mu_2\mu_2}[\underline R_\mathrm{W_{\!c}}^{-1}(\tilde w_2, B_{\mathrm c}(\boldsymbol w_{12}))]
\nonumber\\&=& 
(2\pi)^6 \int_\Omega \mathrm d\Omega(\hat{\tilde{\boldsymbol{  k}}})
2k_{12}^0 \delta^3(\boldsymbol k_{12}-\boldsymbol k_1-\boldsymbol k_2)  \frac{2\tilde k_1^0 2\tilde k_1^0}{2m_{12}}
\delta^3(\tilde{\boldsymbol k}-\boldsymbol B_\mathrm c^{-1}(\boldsymbol w_{12}) k_1)
Y^{\ast }_{l\mu_l}(\hat{\tilde{\boldsymbol k}})
\nonumber\\&&\times C^{s\mu_s}_{j_1\tilde \mu_1j_2\tilde \mu_2}C^{j\mu_j}_{l\mu_ls\mu_s}D^{j_1}_{\tilde \mu_1\mu_1}[\underline R_\mathrm{W_{\!c}}^{-1}(\tilde w_1, B_{\mathrm c}(\boldsymbol w_{12}))]
D^{j_2}_{\tilde \mu_2\mu_2}[\underline R_\mathrm{W_{\!c}}^{-1}(\tilde w_2, B_{\mathrm c}(\boldsymbol w_{12}))]\,.\nonumber\\
\end{eqnarray} 
Here we have used the Jacobian determinant $\left\vert\frac{\partial(\boldsymbol{k}_{12}, \tilde{\boldsymbol{k}})}{\partial(\boldsymbol k_1,\boldsymbol k_2)}\right\vert=\frac{k_{12}^0\tilde{k}_1^0\tilde{k}_2^0}{m_{12}k_1^0 k_2^0}$ 
in order to account for the change of variables in the delta functions.
This Clebsch-Gordan coefficient in Eq.~(\ref{eq:CGCoeffPoin}) corresponds to the coupling of 2 particles with canonical spin to a superposition of irreducible representations with canonical spin. 

The state vectors in Eq.~(\ref{eq:irredrepr}) are eigenstates of commuting self-adjoint operators. Their eigenvalues are the quantum numbers labeling the states in Eq.~(\ref{eq:irredrepr}).
The Clebsch-Gordan coefficient $C^{j\mu_j}_{l\mu_ls\mu_s}$ implies that the canonical spin operator $\hat {\boldsymbol j_c}$ of Eq.~(\ref{eq:canspin}) is the sum of 2 other operators $\hat{\boldsymbol l}$ and $\hat{\boldsymbol s}$, i.e. 
   \begin{equation}\label{eq:j_cls}
\hat {\boldsymbol j_c}=\hat{\boldsymbol l}+\hat{\boldsymbol s}\,, 
\end{equation}
whose eigenvalues are $l$ and $s$, respectively.
The other Clebsch-Gordan coefficient $C^{s\mu_s}_{j_1\tilde \mu_1j_2\tilde \mu_2}$ implies that $\hat{\boldsymbol s}$ is a sum of 2 one-body spin operators, where each is rotated by a different Wigner rotation:
  \begin{equation}\label{eq:sj1j2}
\hat {\boldsymbol s}=\boldsymbol R_\mathrm{W_{\!c}}\left[w_1, B^{-1}_{\mathrm c}\left(\boldsymbol w_{12}\right)\right]\hat{ j}_1+\boldsymbol R_\mathrm{W_{\!c}}\left[w_2, B^{-1}_{\mathrm c}\left(\boldsymbol w_{12}\right)\right]\hat{ j}_2\,.
\end{equation}
Eq.~(\ref{eq:j_cls}) can be considered as defining equation for the operator $\hat{\boldsymbol l}$ with $\hat {\boldsymbol j_c}$ and $\hat{\boldsymbol s}$ defined by Eqs.~(\ref{eq:canspin}) and (\ref{eq:sj1j2}), respectively~\cite{Keister:1991sb}.

\chapter{Dynamical Representations}\label{chap:3}
In the previous section we were dealing with a free many-body theory. The aim of this chapter, in which we proceed partly along the lines of Refs.~\cite{Keister:1991sb,Polyzou:JLab,KrassniggDiss:2001,Krassnigg:2003gh,Biernat:2010tp}, is to construct a relativistically invariant quantum mechanical model with interactions.
\section{Introduction}
 The usual procedure of incorporating interactions into a free many-body theory is to add an interaction term to the free Hamiltonian $\hat P_n^0$. Simultaneously, the Poincar\'e commutation relations have to be retained in order to guarantee relativistic invariance. As we will see, the only consistent way of adding interactions to $\hat P_n^0$ and retaining the  Poincar\'e algebra at the same time is to include interaction terms also in, at least, some of the 9 remaining Poincar\'e generators. Encountering some minimal sets of these interaction dependent generators will bring up the notion of Dirac's \textit{relativistic forms of dynamics}. 
A Poincar\'e invariant prescription of building up an interacting quantum mechanical model from a free theory is the, so-called, \textit{Bakamjian-Thomas construction}. In particular, this construction will be applied to instantaneous two-body interactions and to vertex interactions.
The instantaneous interactions will be used to formulate a bound-state problem, whose solution provides a two-body wave function.
The vertex interactions, which are motivated by a local quantum field theory, will allow us to set up a relativistic coupled-channel problem in the following chapter. 
\section{The Forms of Relativistic Dynamics}
\label{sec:formsofrelativisticdyn}
Wigner's fundamental theorem of Sec.~\ref{sec.IntroWignertheorem} phrases the formal problem of constructing a relativistically invariant quantum theory. This amounts to finding a representation of the Poincar\'e generators in terms of self-adjoint operators that act
on the multi-particle Hilbert space of Sec.~\ref{sec:multiparticlerepre} and satisfy the Poincar\'e
algebra, Eqs.~(\ref{eq:la1})-(\ref{eq:la3}). For a many-body theory without interactions this is easily achieved. The multi-particle generators of Eq.~(\ref{eq:multipartG}), being tensor products of the one-body generators, trivially satisfy the Poincar\'e algebra. To set up an \textit{interacting} quantum mechanical model is, however, a non-trivial task. The complication can be seen most easily by taking a closer look at a particular commutation relation, namely Eq.~(\ref{eq:la2}),
\begin{equation}
[\hat{P}^j, \hat{K}^k]=\mathrm i\, \delta_{\!jk}\, \hat{P}^0\,,
\end{equation}
between the generators for space translations $\hat{P}^j$ and the
boost generators $\hat{K}^k$. This commutation relation already indicates that interaction terms must appear in more than just one of the Poincar\'e generators: if interaction terms are included in the Hamiltonian $\hat{P}^0$, which is the natural way of doing, i.e. $\hat{P}^0=
\hat{P}_n^0+\hat{P}_\mathrm{int}^0$, then $\hat{K}^k$ or $\hat{P}^j$ (or both) must be interaction dependent too. A particular \textit{form} of relativistic dynamics is then characterized by the minimal set of Poincar\'e generators that have to contain interactions. In Dirac's seminal paper~\cite{Dirac:1949cp} on the forms of classical relativistic
dynamics the interaction dependent (\textit{dynamical}) Poincar\'e
generators are called \textit{Hamiltonians}. The remaining interaction free generators are called \textit{kinematic} and they generate the, so-called, \textit{kinematic group}. The kinematic group of a form leaves a particular space-like (or light-like) hypersurface of Minkowski space invariant. In this way there is a hypersurface associated with each form of relativistic dynamics. 

\begin{figure}[t!]
\begin{center} 
\includegraphics[clip=7cm,width=10cm]{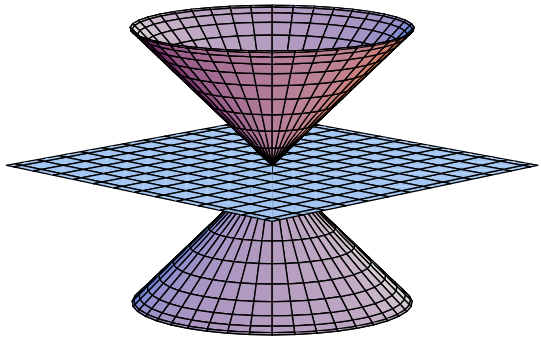}
\caption{\label{fig:if3d} The space-like hypersurface of Minkowski-space $x^0=0$, which is invariant under the instant-form kinematic group of spatial translations and rotations. Indicated is also the light cone.
}
\end{center} 
\end{figure}

\begin{figure}[t!]
\begin{center} 
\includegraphics[clip=7cm,width=10cm]{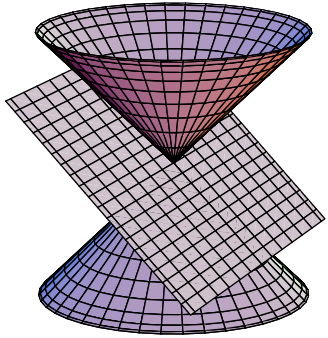}
\caption{\label{fig:ff3d} The light-like hypersurface $x^0+x^3=0$ associated with the front form (together with the light cone).
}
\end{center} 
\end{figure}

\begin{figure}[t!]
\begin{center} 
\includegraphics[clip=7cm,width=10cm]{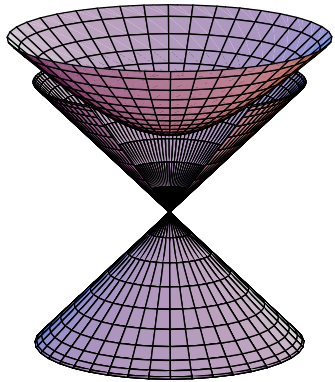}
\caption{\label{fig:pf3d} The forward hyperboloid $x^\mu x_\mu=\tau^2$, which is invariant under the Lorentz group, the kinematic group of the point form (together with the light cone).
}
\end{center} 
\end{figure}

The most prominent forms are the \textit{instant form}, the \textit{front form} and the \textit{point form} (for a thorough discussion on the relativistic forms of dynamics we refer to Refs.~\cite{Dirac:1949cp,Keister:1991sb,KrassniggDiss:2001}).
The instant form is the standard form characterized by the set of dynamical generators given by $\{\hat P^0,\hat K^1,\hat K^2,\hat K^3\}$. Its kinematic group consists of spatial translations and rotations leaving the  hypersurface of the \textit{instant} $x^0=0$ invariant. This hypersurface is depicted in Fig.~\ref{fig:if3d}. 
In the case of the front form the hypersurface left invariant under the action of the associated kinematic group is the light \textit{front} $x^0+x^3=0$, which is illustrated in Fig.~\ref{fig:ff3d}.
 A special feature about the front form is the fact that it has the largest kinematic group containing 7 generators. The remaining 3 dynamical generators are given by 
$\{\hat{P}^0-\hat{P^3},\hat{K}^1 - \hat{J}^2,\hat{K}^2 + \hat{J}^1\}$.
Among the three forms the
point form is the least known and, although it has definite
virtues, the least utilized. 
The kinematic group of the point form is the Lorentz group, which leaves the hyperboloid $x^\mu x_\mu=\tau^2$ (plotted in Fig.~\ref{fig:pf3d}) and in particular the \textit{point} $x^\mu=0$ invariant.
The point-form set of Hamiltonians is given by $\{\hat P^0,\hat P^1,\hat P^2,\hat P^3\}$, which provides a clean separation of Poincar\'e generators that are interaction dependent from those that are interaction free. The former are components of the
four-vector $\hat{P}^\mu$ whereas the latter can be combined into the
antisymmetric Lorentz tensor $\hat{M}^{\mu\nu}$, which has been already introduced in Sec.~\ref{sec:tPG}. This permits to express equations in point form in a manifestly Lorentz covariant form and makes for simple behavior under Lorentz transformations. The conditions for Poincar\'e invariance
can, e.g., be phrased in terms of the \textit{point-form equations}
\begin{eqnarray}
[\hat{P}^\mu,\hat{P}^\nu]=0\quad\text{and}\quad
\hat{U}(\underline \varLambda)\, \hat{P}^\mu \,\hat{U}^\dag(\underline \varLambda) = (
\varLambda^{-1} )^\mu_{\,\,\nu} \hat{P}^\nu \, .
\label{eq:pfeq}\end{eqnarray}
Here the Lorentz part of the Poincar\'e
commutation relations has been integrated out, which is possible
due to the kinematic nature of the rotation and boost generators.

\section{The Bakamjian-Thomas Construction}
\label{sec:BTconst}
By looking at the commutation relations involving the dynamical Poincar\'e generators it is evident that the interaction terms are, in general, subject to non-linear constraints. It has been already realized by Dirac in Ref.~\cite{Dirac:1949cp} that such conditions imposed by the Poincar\'e algebra are in general not easily fulfilled. This makes the problem of constructing a Poincar\'e-invariant few-body quantum theory quite intricate. A particular solution to the problem of finding a consistent set of the 10 Poincar\'e generators for an interacting quantum mechanical system with finitely many degrees of freedom is the, so-called, \textit{Bakamjian-Thomas construction} proposed by Bakamjian and Thomas in Ref.~\cite{Bakamjian:1953kh}. This construction is summarized as follows (for a detailed discussion we refer to Refs.~\cite{Keister:1991sb,KrassniggDiss:2001}): One starts with the non-interacting Poincar\'e generators, Eq.~(\ref{eq:multipartG}), which satisfy the Poincar\'e algebra trivially. From these generators one constructs a particular set of auxiliary operators which depends on the form of dynamics. The Poincar\'e algebra requires certain commutation relations between these auxiliary operators. One of these operators is always the free mass operator $\hat M_{n}$. In the next step an interaction term $\hat M_{\mathrm{int}}$ is added only to this mass operator to obtain an interacting mass operator
\begin{equation}\label{eq:totalinteractingmassoper}
\hat M=\hat M_{n}+\hat M_{\mathrm{int}}\,.
\end{equation}
 This interaction has to commute with all auxiliary operators such that the commutation relations between 
the auxiliary operators remain satisfied. This imposes linear constraints on the interaction terms which are, in general, easier to satisfy than the non-linear constraints imposed by the Poincar\'e algebra. In the final step the interacting Poincar\'e generators are (re-)constructed by simply replacing the free mass operator with the interacting mass operator in the expressions that define the generators in terms of the auxiliary operators. The commutation relations between the auxiliary operators imply that the Poincar\'e algebra is satisfied. This ensures relativistic invariance of the interacting quantum mechanical model resulting from this construction.
\subsubsection{Instant Form}
 It is clear that the set of auxiliary operators depends on the set of dynamical generators and is therefore different for each form of relativistic dynamics. In the instant form the auxiliary operators are given by $\hat M_{n},\hat {\boldsymbol P}_{n},\hat {\boldsymbol X}_{\mathrm c}, \hat {\boldsymbol J}_{\mathrm c}^2,\hat {J}_{\mathrm c}^3$ and the operator canonically conjugate to $\hat {J}_{\mathrm c}^3$. Equivalently, one can take the set $\{\hat M_{n},\hat {\boldsymbol P}_{n},\hat {\boldsymbol J}_{\mathrm c},\hat {\boldsymbol X}_{\mathrm c}\}$. In this case the only generators that are mass dependent and therefore become interaction dependent are the Hamiltonian $\hat {P}^0$  and the boost generators $\hat {\boldsymbol K}$, cf. Eqs.~(\ref{eq:massoperator}) and~(\ref{eq:boostgeneratorsexplicit}). The instant-form conditions on the interaction term ensuring Poincar\'e invariance read
\begin{equation}\label{eq:IFconstraints}
[\hat M_{\mathrm{int}},\hat {\boldsymbol P}_{n}]=
[\hat M_{\mathrm{int}},\hat {\boldsymbol J}_{\mathrm c}]=[\hat M_{\mathrm{int}},\hat {\boldsymbol X}_{\mathrm c}]=0\,.
\end{equation}
\subsubsection{Point Form}
If the Bakamjian-Thomas construction is carried out in the point form the set of auxiliary operators is given by 
$\{\hat M_n,\hat {\boldsymbol V},\hat{M}_n\hat {\boldsymbol X}_\mathrm c ,\hat {\boldsymbol J}_{\mathrm c}\}$ or  equivalently (cf. Eqs.~(\ref{eq:Joperatorexplicit}) and~(\ref{eq:boostgeneratorsexplicit})) by $\{\hat M_{n},\hat {\boldsymbol V},\hat {\boldsymbol K}_n,\hat {\boldsymbol J}_n\}$. Then the resulting interacting four-momentum
operator $\hat{P}^\mu$ is seen to separate into an interacting
mass operator $\hat{M}$ and the free four-velocity operator
$\hat{V}^\mu$,
\begin{equation}
\hat{P}^\mu = \hat{P}^\mu_{n} +
\hat{P}^\mu_{\mathrm{int}} =
\left(\hat{M}_{n}+\hat{M}_{\mathrm{int}} \right)
\hat{V}^\mu\, . \label{eq:PFBT}
\end{equation}
The point-form equations~(\ref{eq:pfeq}) imply that the interacting part of the mass operator $\hat{M}_\mathrm{int}$ must be a Lorentz scalar
that has to commute with the free velocity operator, i.e.
\begin{equation}\label{eq:commMV}
 [\hat{M}_{\mathrm{int}},\hat{V}^\mu]=[\hat{M}_{\mathrm{int}},\hat M_n^{\mu\nu}]=0\,.
\end{equation} 
Equivalently, these point-form conditions on the Bakamjian-Thomas type mass operator $\hat{M}_\mathrm{int}$ can be rewritten in terms of the operators $\hat{\boldsymbol X}_\mathrm c$ and $\hat {\boldsymbol J}_{\mathrm c}$ as
\begin{equation}\label{eq:commMValternativ}
 [\hat{M}_{\mathrm{int}},\hat{V}^\mu]=[\hat{M}_{\mathrm{int}},\hat{\boldsymbol X}_\mathrm c ]=[\hat{M}_{\mathrm{int}},\hat{\boldsymbol J}_\mathrm c ]=0\,.
\end{equation} 
 \subsubsection{Remarks}
One benefit of the Bakamjian-Thomas approach is that the operator for the total spin $\hat{\boldsymbol J}_{\mathrm g}$ of the interacting
system is not affected by interactions. 

Furthermore, the Bakamjian-Thomas framework allows one to construct Poincar\'e invariant models with instantaneous interactions, i.e. action at a distance. What one loses with instantaneous interactions as compared to a local quantum field theory is, however, \textit{microscopic locality}. Microscopic locality, which is sometimes also called \textit{Einstein causality} or \textit{microcausality}, roughly means that 2 observables commute if they are associated with 2 different regions of space-time that are separated by an arbitrarily small space-like distance~\cite{Streater:2000ff,Keister:1991sb}. It is possible to find infinitely many such independent observables for every finite space-time volume which implies that microscopic locality requires an infinite number of degrees of freedom. Therefore, in a relativistic quantum theory with a \textit{finite} number of degrees of freedom this property of microcausality is replaced by the weaker condition of \textit{macroscopic} locality, which is sometimes called \textit{cluster separability} or \textit{macrocausality}~\cite{Foldy:1960nb}. For a definition of macrocausality one simply replaces 'arbitrarily small' by 'sufficiently large' in the definition of microcausality. In other words, macrocausality means that disjunct subsystems of a quantum mechanical system should behave independently from each other, if they are separated by sufficiently large space-like distances. It turns out that it is the condition of macrocausality (and not microcausality) that can be tested by experiments~\cite{Keister:1991sb}.

As we will see, however, already for systems consisting of more than 2 particles, the Bakamjian-Thomas framework will violate not only microcausality, but also macrocausality. 

\section{Two-Body Interactions}
\label{sec:2binter}
Our aim is to find a two-body interaction defined on a three-particle Hilbert space within a point-form Bakamjian-Thomas framework. This means that we have conservation of the overall four-velocity of the three-particle system, i.e. the incoming and outgoing four velocities are equal. At the same time, the incoming and outgoing three-momenta of the non-interacting particles satisfy spectator conditions. These 2 requirements imply that the four-velocity of the interacting two-particle subsystem cannot be conserved. Therefore, we cannot directly use the point-form Bakamjian-Thomas construction on the two-particle Hilbert space to define the two-body interaction on the three-body Hilbert space.
One way to satisfy the above requirements is to define the two-body interaction on the two-particle Hilbert space in a way that resembles the instant-form Bakamjian-Thomas framework  and then imbed this two-body interaction consistent with the point-form Bakamjian-Thomas framework into the three-body Hilbert space.

\subsection{Two-Body Mass Operator}
It has been mentioned already that instantaneous interactions can be used within the Bakamjian-Thomas framework without losing Poincar\'e invariance. 
In this section we concentrate on the incorporation of instantaneous two-body interactions into a quantum mechanical model.

To this end we use an instant-form Bakamjian-Thomas construction formulated on a two-particle Hilbert space. In our rather brief discussion we follow closely the more detailed disquisitions of Refs.~\cite{Polyzou:JLab,Keister:1991sb}.
In Sec.~\ref{par:Clebsch-Gordan Coefficients} we have derived a basis for free two-particle states given by $\vert \boldsymbol k_{12};\tilde{k};(l,s),j,\mu_j\rangle$ which transforms irreducibly under 
the non-interacting unitary representation $\hat U_{12}\left(\underline\varLambda, \underline{a}\right)$. The free invariant mass operator $\hat m_{12}$ of the two-body system in this representation was given by Eq.~(\ref{eq:massoperfree12}). An interaction between particles 1 and 2 is introduced by adding an interaction operator $\hat m_{\mathrm{int}}$ to the free mass operator according to Eq.~(\ref{eq:totalinteractingmassoper}) to obtain a total mass operator for the two-particle cluster~C:
\begin{equation}\label{eq:mC}
\hat m_\mathrm{C}:=\hat m_{12}+\hat m_{\mathrm{int}}\,.
\end{equation}
Poincar\'e invariance requires that the instant-form set of the auxiliary operators $\{\hat m_\mathrm{C},\hat {\boldsymbol k}_{12},\hat{\boldsymbol j}_{\mathrm c},\hat{\boldsymbol x}_{\mathrm c} \}$ 
 satisfies the same commutation relations as $\{\hat m_{12},\hat {\boldsymbol k}_{12},\hat{\boldsymbol j}_{\mathrm c},\hat{\boldsymbol x}_{\mathrm c} \}$ which implies the instant-form constraints Eq.~(\ref{eq:IFconstraints}) on $\hat m_{\mathrm{int}}$:
\begin{eqnarray}
\label{eq:condmC}
\left[\hat {\boldsymbol k}_{12},\hat m_{\mathrm{int}}\right]=\left[\hat{\boldsymbol j}_{\mathrm c},\hat m_{\mathrm{int}}\right]=\left[\hat{\boldsymbol x}_{\mathrm c} ,\hat m_{\mathrm{int}}\right]=0\,.
\end{eqnarray}
From these conditions on $\hat m_{\mathrm {int}}$ we infer that matrix elements of $\hat m_{\mathrm{int}}$ in the non-interacting representation $\vert \boldsymbol k_{12};\tilde{k};l,s,j,\mu_j\rangle$ can be written in the form
\begin{eqnarray}\label{eq:2bodyinter}
\lefteqn{\langle \boldsymbol k_{12}';\tilde{k}';(l',s'),j',\mu_j'\vert \hat m_{\mathrm {int}}\vert\boldsymbol k_{12};\tilde{k};(l,s),j,\mu_j\rangle}\nonumber\\&=\delta_{jj'}\delta_{\mu_j\mu_j'}(2\pi)^3 2k_{12}^0\delta^3\left(\boldsymbol k_{12}-\boldsymbol k_{12}'
\right)\langle\tilde k';l',s'\vert\vert\hat m_{\mathrm {int}}^j\vert\vert \tilde k;l,s\rangle\,.
\end{eqnarray}
Here the reduced kernel $\langle\tilde k';l',s'\vert\vert\hat m_{\mathrm {int}}^j\vert\vert \tilde k;l,s\rangle$ is independent of the total two-particle momentum $\boldsymbol k_{12}$ and the total spin projection $\mu_j$.
 The eigenvalue problem for the mass operator $\hat m_{\mathrm C}$ is given by
\begin{eqnarray}\label{eq:eigenvalueproblem}
\left(\sqrt{m_1^2+\hat{\tilde{\boldsymbol k}}^2}+\sqrt{m_2^2+\hat{\tilde{\boldsymbol k}}^2}+\hat m_{\mathrm {int}}\right)\vert \varPsi_{\mathrm C}\rangle=m_\mathrm C\vert \varPsi_{\mathrm C}\rangle\,.
\end{eqnarray}
In the present work we concentrate on the study of bound states.\footnote{From now on we will also refer to the particles 1 and 2 as the constituents $\mathrm c_1$ and $\mathrm c_2$ of the bound state C.} Therefore, the eigenvalue equation~(\ref{eq:eigenvalueproblem}) is considered to be a bound-state equation leading to a discrete spectrum of $\hat m_{\mathrm C}$. The eigenvalues of $\hat m_{\mathrm C}$ will be denoted by $m_{n}$ where $n=0,1,2,\dots$ is the quantum number for radial excitations.  

We have constructed $\hat m_\mathrm{C}$ in such a way that it satisfies the same commutation relations as $\hat m_{12}$, i.e. it commutes with the remaining (free) auxiliary operators $\hat{\boldsymbol j}_{\mathrm c}^2$, $\hat j_\mathrm c^3$, $\hat {\boldsymbol k}_{12}$, $\hat{\boldsymbol x}_{\mathrm c}$ and the operator canonically conjugate to $\hat j_\mathrm c^3$~\cite{Polyzou:JLab}. Thus, there exist simultaneous eigenstates of $\hat m_\mathrm{C}$, $\hat{\boldsymbol j}_{\mathrm c}^2$, $\hat j_\mathrm c^3$ and $\hat {\boldsymbol k}_{12}$ denoted by $\vert \boldsymbol k_{12};m_{n},j,\mu_j\rangle\equiv\vert \boldsymbol k_{12};n,j,\mu_j\rangle$. These states transform irreducibly under the action of $\hat U_\mathrm C\left(\underline\varLambda,\underline a\right)$ and thus define a dynamical representation of the Poincar\'e group. It is simply obtained by replacing all eigenvalues $m_{12}$ by $m_{n}$ in the non-interacting representation defined by the action of $\hat U_{12}\left(\underline\varLambda,\underline a\right)$ on the non-interacting states $\vert \boldsymbol k_{12};\tilde k,(l,s),j,\mu_j\rangle$~\cite{Keister:1991sb}. Then, the four-momentum of the interacting two-particle cluster is given by
\begin{eqnarray}
 k_{\mathrm C}:= \left(
\begin{array}{c}
k_{\mathrm C}^0\\
\boldsymbol k_{12}
               \end{array}
               \right)\quad \text{with} \quad k_{\mathrm C}^0:=\sqrt{\boldsymbol k_{12}^2+m_{\mathrm C}^2}\,.
\end{eqnarray}
\subsubsection{Bound-State Wave Function}
Solving the bound-state problem Eq.~(\ref{eq:eigenvalueproblem}) leads to mass eigenfunctions of the form
 \begin{eqnarray}\label{eq:eigenfunctions}
\lefteqn{\langle \boldsymbol k_{12}';\tilde{k}';(l',s'),j',\mu_j'\vert \boldsymbol k_{12};n,j,\mu_j\rangle}\nonumber\\&=\bar N_2\delta_{jj'}\delta_{\mu_j\mu'_j} \delta^3\left(\boldsymbol k_{12}-\boldsymbol k'_{12}\right)u_{nl's'}^j(\tilde{k}')
\end{eqnarray}
where $\bar N_2$ is a normalization constant.
From Eq.~(\ref{eq:eigenfunctions}), using Eqs.~(\ref{eq:irredrepr}) and~(\ref{eq:CGCoeffPoin}) together with the relations Eqs.~(\ref{eq:RWprop1}),~(\ref{eq:sphharmotho}) and~(\ref{eq:cgortho}),
we obtain a representation of the eigenfunctions in terms of constituents' degrees of freedom
\begin{eqnarray}\label{eq:wavefunctions}
\langle \boldsymbol k_1',\mu_1';\boldsymbol k_2',\mu_2'\vert \boldsymbol k_{12};n,j,\mu_j\rangle=
\tilde N_2\delta^3\left(\boldsymbol k_{12}-\boldsymbol k_{1}'-\boldsymbol k_{2}'\right)\varPsi _{nj\mu_j\mu_1'\mu_2'}(\tilde{\boldsymbol k}')\,,\nonumber\\
\end{eqnarray}
where $\tilde N_2$ is a normalization constant and with the wave function defined by
\begin{eqnarray}\label{eq:2bodywf}
 \lefteqn{\varPsi _{nj\mu_j\mu_1'\mu_2'}(\tilde{\boldsymbol k}')}\nonumber\\&:=&
\sum_{ls\mu_l\mu_s\tilde \mu_1\tilde \mu_2}
Y_{l\mu_l}(\hat{\tilde{\boldsymbol{ k}}}')
C^{s\mu_s}_{j_1\tilde \mu_1j_2\tilde \mu_2}C^{j\mu_j}_{l\mu_ls\mu_s} u_{nls}^j(\tilde{k}')
\nonumber\\&&\times
 D^{j_1}_{\mu_1'\tilde \mu_1}\left[\underline R_\mathrm{W_{\!c}}\left(\tilde w_1',B_{\mathrm c}(\boldsymbol w_{12}')\right)\right]
D^{j_2}_{\mu_2'\tilde \mu_2}\left[\underline R_\mathrm{W_{\!c}}\left(\tilde w_2', B_{\mathrm c}(\boldsymbol w_{12}')\right)\right]\,.
\end{eqnarray}
We normalize this wave function to unity:
\begin{eqnarray}\label{eq:normeigenfucntions1}
\int \mathrm d^3 \tilde k'\sum_{\mu_1'\mu_2'} \varPsi^\ast _{nj\mu_j\mu_1'\mu_2'}(\tilde{\boldsymbol k}')\varPsi _{n'j'\mu_j'\mu_1'\mu_2'}(\tilde{\boldsymbol k}')
=\delta_{nn'}\delta_{jj'}\delta_{\mu_j\mu_j'}\,.
\end{eqnarray}


 \subsection{Embedding into a Three-Particle Hilbert Space}\label{sec:imbedding3Hilberts}
In the next step we embed the two-body interaction defined by Eq.~(\ref{eq:2bodyinter}) into a three-body Hilbert space of a system consisting of the two interacting particles 1 and 2 plus an additional spectator particle denoted by e.\footnote{\lq e' will stand later for \lq electron', not to be confused with the elementary charge. The distinction between them should, however, be clear from the context.} To this end we pursue a similar strategy as presented in Ref.~\cite{Keister:1991sb}.

In the velocity-state basis of Sec.~\ref{sec:velocitystates} the free three-particle system is represented by the state vector $\vert V;\boldsymbol k_1,\mu_1;\boldsymbol k_2,\mu_2;(\boldsymbol k_\mathrm e),\mu_\mathrm e\rangle$. The momentum in brackets is redundant since the momenta occurring here are subject to center-of-mass constraints, cf. Eq.~(\ref{eq:comconstraint}).  With the help of the Clebsch-Gordan coefficients for the Poincar\'e group, Eq.~(\ref{eq:CGCoeffPoin}), we can perform a basis change to $\vert V;\boldsymbol k_{12},\tilde k,(l,s),j,\mu_j;(\boldsymbol k_\mathrm e),\mu_\mathrm e\rangle$. In this basis we define the interaction on the three-body Hilbert space by simply including another Dirac delta function and 2 other Kronecker deltas in the definition of the two-body interaction in Eq.~(\ref{eq:2bodyinter}).
The additional delta function corresponds to overall-velocity conservation and is given by $V^0\delta^3(\boldsymbol V-\boldsymbol V')$. The 2 additional Kronecker deltas $\delta_{JJ'}$ and $\delta_{M_JM_J'}$ correspond to the total angular momentum of the three-particle system $|\boldsymbol J_\mathrm c^{(\prime)}|=|\boldsymbol j_\mathrm c^{(\prime)}+\boldsymbol j^{(\prime)}_\mathrm {c,e}|$ and its projection into the 3-direction $M_J^{(\prime)}=\mu_j^{(\prime)}+\mu_\mathrm e^{(\prime)}$, respectively. Therefore, due to the Kronecker delta $\delta_{\mu_j\mu_j'}$  occurring in Eq.~(\ref{eq:2bodyinter}) (implying $\mu_j=\mu_j'$), we can replace $\delta_{M_JM_J'}$ by $\delta_{\mu_\mathrm e\mu_\mathrm e'}$. Furthermore, due to the center-of-mass kinematics for velocity states, we can use $\delta^3(\boldsymbol k_{12}-\boldsymbol k_{12}')=\delta^3(\boldsymbol k_\mathrm e-\boldsymbol k_\mathrm e')$.
With these manipulations the definition of the two-body interaction on the three-body Hilbert space reads
 \begin{eqnarray}\label{eq:interactionBT3body}
\lefteqn{
\langle V';\boldsymbol k_{12}',\tilde k',(l',s'),j',\mu_j';(\boldsymbol k_\mathrm e'),\mu_\mathrm e'\vert \hat{ m}_{\mathrm {int }}
\vert  V;\boldsymbol k_{12},\tilde k,(l,s),j,\mu_j;(\boldsymbol k_\mathrm e),\mu_\mathrm e\rangle}\nonumber\\&&:=N 
V^0\delta^3(\boldsymbol V-\boldsymbol V')\delta_{jj'}\delta_{\mu_j\mu_j'} \delta_{\mu_\mathrm e\mu_\mathrm e'} \delta^3(\boldsymbol k_\mathrm e-\boldsymbol k_\mathrm e')
\langle\tilde k';l',s'\vert\vert \hat m_\mathrm{int} ^j\vert \vert\tilde  k;l,s\rangle\,,\nonumber\\
\end{eqnarray}
where $N$ is a normalization constant which is not fixed for the moment. The interaction $\hat { m}_{\mathrm {int }}$ satisfies the following commutation relations with the operators of the non-interacting system:
\begin{eqnarray}\label{eq:BT3bodycommrel}
[\hat { m}_{\mathrm {int }},\hat {\boldsymbol {V}}]=[\hat{ m}_{\mathrm {int }},\hat {\boldsymbol J}_{\mathrm c}]=
[\hat{ m}_{\mathrm {int }},\hat {\boldsymbol X}_{\mathrm c}]=
0\,,\\
\quad [\hat{ m}_{\mathrm {int }},\hat {\boldsymbol k}_{\mathrm e}]=[\hat{ m}_{\mathrm {int }},\hat {\boldsymbol j}_{\mathrm {c,e}}]=0\,.\label{eq:spectatorconditoins}
\end{eqnarray}
The first commutation relations Eq.~(\ref{eq:BT3bodycommrel}) are identical with the conditions of a point-form Bakamjian-Thomas construction (cf. Eq.~(\ref{eq:commMValternativ})) which means that Eq.~(\ref{eq:interactionBT3body}) defines a point-form Bakamjian-Thomas type mass operator. The commutation relations~(\ref{eq:spectatorconditoins}) follow from Eq.~(\ref{eq:condmC}) and can be viewed as \textit{spectator conditions} for the additional particle which has been assumed to be a non-interacting spectator. Note that Eqs.~(\ref{eq:BT3bodycommrel}) and (\ref{eq:spectatorconditoins}) imply that $[\hat{ m}_{\mathrm {int }},\hat {\boldsymbol j}_{\mathrm c}]=0$, which corresponds to the Kronecker deltas $\delta_{\mu_j\mu_j'}$ and $\delta_{jj'}$ in Eq.~(\ref{eq:interactionBT3body}). Furthermore, $[\hat{ m}_{\mathrm {int }},\hat {\boldsymbol k}_{\mathrm e}]=0$ is equivalent to $ [\hat{ m}_{\mathrm {int }},\hat {\boldsymbol k}_{12}]=0$ since $\hat {\boldsymbol k}_{12}=-\hat {\boldsymbol k}_{\mathrm e}$. Hence, the commutation relations on the two-body Hilbert-space, Eq.~(\ref{eq:condmC}), remain valid.
Eq.~(\ref{eq:interactionBT3body}) reveals that in a velocity-state representation of a Bakamjian-Thomas type model the overall velocity of the system can always be factored out as a velocity-conserving delta function, leaving the part for the pure internal motion.
\subsubsection{Four-Momentum Operator}
The next objective is to find a representation of the four-momentum operator corresponding to the interaction of Eq.~(\ref{eq:interactionBT3body}) on the three-body Hilbert space. To this end we define the interacting mass operator associated with the interacting pair and the spectator particle by
\begin{eqnarray}\label{eq:totalmassop}
\hat M_{\mathrm C\mathrm e}:=\sqrt{\hat {\boldsymbol k}_{\mathrm e}^2+\hat m_\mathrm C^2}+\sqrt{\hat {\boldsymbol k}_{\mathrm e}^2+m_\mathrm e^2}\,.
\end{eqnarray}
Then we can determine the two-body interaction on the three-body Hilbert space by subtracting the free mass operator $\hat M_{12\mathrm e}$ from Eq.~(\ref{eq:totalmassop}):
 \begin{eqnarray}\label{eq:MintBT}
 \hat M_\mathrm {int}:=\hat M_{\mathrm C\mathrm e}-\hat M_{12\mathrm e}\,.
\end{eqnarray}
This operator satisfies the same commutations relations as $\hat m_\mathrm {int}$, cf. Eqs.~(\ref{eq:BT3bodycommrel}) and~(\ref{eq:spectatorconditoins}). Finally, the interacting four-momentum operator associated with $\hat M_{e\mathrm C}$ of Eq.~(\ref{eq:totalmassop}) is easily obtained from Eq.~(\ref{eq:PFBT}):
\begin{eqnarray}\label{eq:BTfourmom}
\hat P^\mu = \hat M_{e\mathrm C}\hat V^\mu\,. 
\end{eqnarray}
\subsubsection{Bound-State Wave Function}
\label{sec:Two-BodyWaveFunction}
Next we look at the matrix elements of free and \textit{clustered}\footnote{With \textit{clustered velocity states} we mean velocity states which describe a $n$-particle system where particles 1 and 2 form a bound system.} velocity states $\langle V';\boldsymbol k_1',\mu_1';\boldsymbol k_2',\mu_2';(\boldsymbol k_\mathrm e'),\mu_\mathrm e'\vert V;\boldsymbol k_{12},n,j,\mu_j;(\boldsymbol k_\mathrm e),\mu_\mathrm e\rangle$ which define the two-body wave function on the three-body Hilbert space. As a consequence of Eq.~(\ref{eq:interactionBT3body}) they have the general form~\cite{Krassnigg:2003gh}
\begin{eqnarray}\label{eq:norm3partcl}
\lefteqn{\langle V';\boldsymbol k_1',\mu_1';\boldsymbol k_2',\mu_2';(\boldsymbol k_\mathrm e'),\mu_\mathrm e'\vert V;\boldsymbol k_{12},n,j,\mu_j;(\boldsymbol k_\mathrm e),\mu_\mathrm e \rangle}\nonumber\\&=
%
N_3V^0\delta ^3(\boldsymbol V-\boldsymbol V') \delta_{\mu_\mathrm e\mu_\mathrm e'} \delta^3\left(\boldsymbol k_{\mathrm e}-\boldsymbol k_{\mathrm e}'\right) \varPsi _{nj\mu_j\mu_1'\mu_2'}(\tilde{\boldsymbol k}')\,,
\end{eqnarray}
where we have used  Eq.~(\ref{eq:wavefunctions}) together with the relation $\delta^3\left(\boldsymbol k_{12}-\boldsymbol k_{1}'-\boldsymbol k_{2}'\right)=\delta^3\left(\boldsymbol k_{\mathrm e}-\boldsymbol k_{\mathrm e}'\right)$. $\tilde{\boldsymbol k}'$ is the momentum of particle 1 in the center of momentum of the free (12)-subsystem (cf. Eq.~(\ref{eq:ktilde1})). 
The normalization factor $N_3$ occurring in Eq.~(\ref{eq:norm3partcl}) is fixed by the normalization of the states $\vert V;\boldsymbol k_{12},n,j,\mu_j;(\boldsymbol k_\mathrm e),\mu_\mathrm e\rangle$, cf. Eq.~(\ref{eq:vnorm}), which reads
\begin{eqnarray}\label{eq:normeigenstates}
\lefteqn{\langle V'';\boldsymbol k_{12}'',n'',j'',\mu_j'';(\boldsymbol k_\mathrm e''),\mu_\mathrm e'' \vert V;\boldsymbol k_{12},n,j,\mu_j;(\boldsymbol k_\mathrm e),\mu_\mathrm e\rangle}\nonumber\\&=
(2\pi)^6V^0\delta^3(\boldsymbol V-\boldsymbol V'') \frac{2k_{\mathrm C}^02k_{\mathrm e}^0}{(k_{\mathrm C}^0+k_{\mathrm e}^0)^3}\delta^3(\boldsymbol k_{12}-\boldsymbol k_{12}'')\delta_{nn''}
\delta_{jj''}\delta_{\mu_j\mu_j''}\delta_{\mu_{\mathrm e}\mu_{\mathrm e}''}\,.\nonumber\\
\end{eqnarray}
$N_3$ is computed as follows: first we insert the three-particle velocity state unit $\hat 1'_{12\mathrm e}$ of Eq.~(\ref{eq:vcompl}) into the bracket of the left-hand side of Eq.~(\ref{eq:normeigenstates}) and use the Jacobian
\begin{eqnarray}
\frac{\mathrm d^3 k_1}{2k_1^0}\frac{\mathrm d^3 k_\mathrm e}{2k_\mathrm e^0}\frac{(k_1^0+k_2^0+k_\mathrm e^0)^3}{2k_2^0}=\frac{\mathrm d^3 k_{\mathrm e}}{2k_{\mathrm e}^0}\frac{\mathrm d^3 \tilde k}{2\tilde k_{1}^0}\frac{2(\tilde k_{1}^0+\tilde k_{2}^0)}{\tilde k_{2}^0}
\frac{(k_{12}^0+k_\mathrm e^0)^3}{2 k_{12}^0}\,.
\end{eqnarray}
 After performing the $\boldsymbol V'$- and $\boldsymbol k'_\mathrm e$-integrations together with the sum over $\mu_\mathrm e'$ the left-hand side of Eq.~(\ref{eq:normeigenstates}) becomes
\begin{eqnarray}
\lefteqn{\langle V'';\boldsymbol k_{12}'';n''j'',\mu_j'';(\boldsymbol k_\mathrm e''),\mu_\mathrm e'' \vert\hat 1'_{12\mathrm e}\vert V;\boldsymbol k_{12},n,j,\mu_j;(\boldsymbol k_\mathrm e),\mu_\mathrm e\rangle}
\nonumber\\&
=&
\frac {1}{(2\pi)^9} V^0\delta^3(\boldsymbol V-\boldsymbol V'') \delta^3(\boldsymbol k_\mathrm e -\boldsymbol k_\mathrm e'')\int \frac{\tilde k'^2\mathrm d \tilde k' }{2\tilde k_{1}'^0}\frac{2(\tilde k_{1}'^0+\tilde k_{2}'^0)}{2\tilde k_{2}'^0}\frac{(k_{12}'^0+k_\mathrm e'^0)^3}{2 k_{12}'^02k_{\mathrm e}'^0}
N_3^2
\nonumber\\&&\times\int \mathrm d \Omega(\hat{\tilde {\boldsymbol k}}')\sum_{\mu_1'\mu_2'}
\varPsi _{n''j''\mu_j''\mu_1'\mu_2'}^{\ast}(\tilde{\boldsymbol k}')\varPsi _{nj\mu_j\mu_1'\mu_2'}(\tilde{\boldsymbol k'})\,.
\end{eqnarray}
With the normalization of the wave function, Eq.~(\ref{eq:normeigenfucntions1}), and from comparison of the result with the right-hand side of Eq.~(\ref{eq:normeigenstates}), we finally infer that the normalization factor has the form
\begin{eqnarray}\label{eq:normalizationN_3}
N_3=(2\pi)^{15/2} \sqrt{\frac{2\tilde k'^0_12\tilde k'^0_2}{2(\tilde k'^0_1+\tilde k'^0_2)}}\sqrt{\frac{2 k^0_{\mathrm C}2k^0_{\mathrm e}}{(k^0_{\mathrm C}+k^0_{\mathrm e})^3}}
\sqrt{\frac{2 k'^0_{12}2k'^0_{\mathrm e}}{(k'^0_{12}+k'^0_{\mathrm e})^3}}\,.
\end{eqnarray}
\subsubsection{Transformation Properties}
In this subsection we investigate the transformation properties of the wave function $\varPsi _{nj\mu_j\mu_1'\mu_2'}(\tilde{\boldsymbol k})$ under Lorentz transformations. For simplicity, we restrict ourselves to a wave function with $l=0$, which is often referred to as \textit{s-wave}. In this case the Clebsch-Gordan coefficients and the spherical harmonics are simple expressions given by $C^{j\mu_j}_{00s\mu_s}=\delta_{sj}\delta_{\mu_s\mu_j}$ and $Y_{00} (\hat{\tilde{\boldsymbol{ k}}}')=1/\sqrt{4\pi}$. Then, from the normalization condition of the wave function, Eq.~(\ref{eq:normeigenfucntions1}), using the property of the Wigner $D$-functions, Eq.~(\ref{eq:RWprop1}), together with the orthogonality relations for the Clebsch-Gordan coefficients, Eq.~(\ref{eq:cgortho}), we find the following orthogonality relation for the radial wave function $u_{n0}(\tilde{k})\equiv u_{n0j}^j(\tilde{k})$:
\begin{eqnarray}\label{eq:normeigenfucntions}
\int_ {0}^\infty \mathrm d \tilde k\, \tilde k^2\, u^{\ast}_{n'0}(\tilde{k})u_{n0}(\tilde{k})=\delta_{nn'}\,.
\end{eqnarray}
 
We start our analysis by Lorentz transforming the matrix element of Eq.~(\ref{eq:norm3partcl}):
 \begin{eqnarray}
\lefteqn{\langle V';\boldsymbol k_1',\mu_1';\boldsymbol k_2',\mu_2';\mu_\mathrm e'\vert V;\boldsymbol k_{12},n,j,\mu_j;\mu_\mathrm e \rangle}\nonumber\\&
\stackrel{\varLambda}{\longrightarrow} &\langle \bar V';\bar{\boldsymbol k}_1',\bar \mu_1';\bar{\boldsymbol k}_2',\bar\mu_2';\bar\mu_\mathrm e'\vert \bar V;\bar{\boldsymbol k}_{12},n,j,\bar \mu_j;\bar\mu_\mathrm e \rangle\,.
\end{eqnarray}
Here we have dropped the dependences on the (redundant) momenta $\boldsymbol k_{\mathrm e}$ and $\boldsymbol k_{\mathrm e}'$. 
From the transformation properties of velocity states under Lorentz transformations, Eq.~(\ref{eq:velocitystatetransfprop}), we can rewrite the transformed matrix element in terms of the original one (using the short-hand notation $R_\mathrm{W_{\!c}}^{(\prime)}\equiv R_\mathrm{W_{\!c}}(V^{(\prime)},\varLambda)$): 
\begin{eqnarray}\label{eq:3partwftransf}
\lefteqn{\langle \bar V';\bar{\boldsymbol k}_1',\bar \mu_1';\bar{\boldsymbol k}_2',\bar\mu_2';\bar\mu_\mathrm e'\vert \bar V;\bar{\boldsymbol k}_{12},n,j,\bar \mu_j;\bar\mu_\mathrm e \rangle}
\nonumber\\&=&
\langle \varLambda V';\boldsymbol R_\mathrm{W_{\!c}}' k_1', \mu_1';\boldsymbol R_\mathrm{W_{\!c}}'  k_2', \mu_2';\mu_\mathrm e'\vert \varLambda V;\boldsymbol R_\mathrm{W_{\!c}}  k_{\mathrm C};n,j, \mu_j;\mu_\mathrm e \rangle
\nonumber\\&&\times D_{\mu_1'\bar\mu_1'}^{j_1\ast}\left[\underline R_\mathrm{W_{\!c}}^{-1}(V',\varLambda)\right] D_{\mu_2'\bar\mu_2'}^{j_2\ast}\left[\underline R_\mathrm{W_{\!c}}^{-1}(V',\varLambda)\right]
D_{\mu_\mathrm e'\bar\mu_\mathrm e'}^{j_\mathrm e\ast}\left[\underline R_\mathrm{W_{\!c}}^{-1}(V',\varLambda)\right] \nonumber\\&&\times D_{\mu_\mathrm e\bar\mu_\mathrm e}^{j_\mathrm e}\left[\underline R_\mathrm{W_{\!c}}^{-1}(V,\varLambda)\right]  
 D_{\mu_j\bar\mu_j}^{j}\left[\underline R_\mathrm{W_{\!c}}^{-1}(V,\varLambda)\right]
\nonumber\\&=&
(2\pi)^{15/2} \sqrt{\frac{2\tilde k'^0_12\tilde k'^0_2}{2(\tilde k'^0_1+\tilde k'^0_2)}}\sqrt{\frac{2 k^0_{\mathrm C}2k^0_{\mathrm e}}{(k^0_{\mathrm C}+k^0_{\mathrm e})^3}}
\sqrt{\frac{2 k'^0_{12}2k'^0_{\mathrm e}}{(k'^0_{12}+k'^0_{\mathrm e})^3}} \nonumber\\&&\times V^0\delta ^3(\boldsymbol V-\boldsymbol V') \delta_{ \mu_\mathrm e \mu_\mathrm e'} \delta^3\left(\boldsymbol k_{\mathrm e}-\boldsymbol k_{\mathrm e}'\right) 
\varPsi _{nj\mu_j\mu_1'\mu_2'}(\bar{\tilde{\boldsymbol k}}')
\nonumber\\&&\times 
D_{\mu_1'\bar\mu_1'}^{j_1\ast}\left[\underline R_\mathrm{W_{\!c}}^{-1}(V,\varLambda)\right] D_{\mu_2'\bar\mu_2'}^{j_2\ast}\left[\underline R_\mathrm{W_{\!c}}^{-1}(V,\varLambda)\right]
 \nonumber\\&&\times  
 D_{\mu_j\bar\mu_j}^{j}\left[\underline R_\mathrm{W_{\!c}}^{-1}(V,\varLambda)\right]\,.
\end{eqnarray}
Here we have used Lorentz invariance of the delta functions and the fact that energies are not affected by (Wigner) rotations of the momenta.
Note that under a Lorentz transformation the cluster center-of-mass momenta undergo the same Wigner rotation as the center-of-mass momenta of the three-particle system:
\begin{eqnarray}\label{eq:ktildebar1}
\lefteqn{\bar{\tilde k}_1'}&=&B^{-1}(\boldsymbol R_\mathrm{W_{\!c}} (V,\varLambda)w'_{12})  R_{\mathrm  {Wc}}(V,\varLambda) k'_1\nonumber\\&=&
B^{-1}(\boldsymbol R_\mathrm{W_{\!c}} (V,\varLambda)w'_{12})  R_\mathrm{W_{\!c}}(V,\varLambda)  B_\mathrm c(\boldsymbol  w'_{12})\tilde k'_1\nonumber\\&=&
R_\mathrm{W_{\!c}}(w'_{12},R_\mathrm{W_{\!c}}(V,\varLambda))\tilde k_1\nonumber\\&\stackrel{\text{(\ref{eq:propWc})}}{=}&R_\mathrm{W_{\!c}}(V,\varLambda)\tilde k'_1\,.
\end{eqnarray} 
Since a pure s-wave depends only on the magnitude of the three-momentum, which is not affected by a rotation, we have
 $\varPsi _{nj\mu_j\mu_1'\mu_2'}(\bar{\tilde{\boldsymbol k}}')=\varPsi _{nj\mu_j\mu_1'\mu_2'}(\tilde{\boldsymbol k}')$.
Comparing Eq.~(\ref{eq:3partwftransf}) with Eq.~(\ref{eq:norm3partcl}) we can read off the transformation properties of the wave function under
Lorentz transformations:
\begin{eqnarray}\label{eq:transfwavefunction}
 \lefteqn{\varPsi _{nj\mu_j \mu_1'\mu_2'}(\tilde{\boldsymbol k}')\stackrel{\varLambda}{\longrightarrow} \varPsi _{nj\bar \mu_j\bar \mu_1'\bar \mu_2'}(\bar{\tilde{\boldsymbol k}}')}
\nonumber\\&&= \varPsi _{nj\mu_j \mu_1'\mu_2'}(\tilde{\boldsymbol k}')  D_{\bar\mu_1'\mu_1'}^{j_1}\left[\underline R_\mathrm{W_{\!c}}(V,\varLambda)\right] D_{\bar\mu_2'\mu_2'}^{j_2}\left[\underline R_\mathrm{W_{\!c}}(V,\varLambda)\right]
 D_{\bar\mu_j\mu_j}^{j\ast}\left[\underline R_\mathrm{W_{\!c}}(V,\varLambda)\right]\,.\nonumber\\
\end{eqnarray} 
In App.~\ref{app:TrafopropCG} we also derive the transformation properties of the Clebsch-Gordan coefficients under Lorentz transformations which read
\begin{eqnarray}\label{eq:TrafopropCG}
\lefteqn{C^{j\mu_j}_{j_1\tilde \mu_1j_2\tilde \mu_2}\stackrel{\varLambda}{\rightarrow}
 C^{j\bar \mu_j}_{j_1\bar {\tilde \mu}_1j_2\bar{\tilde \mu}_2}}\nonumber\\&&=C^{j\mu_j}_{j_1\tilde \mu_1j_2\tilde \mu_2}D_{\bar\mu_j\mu_j}^{j\ast}\left[\underline R_\mathrm{W_{\!c}}(V,\varLambda)\right]
D^{j_1}_{\bar{\tilde\mu}_1\tilde\mu_1}[\underline R_\mathrm{W_{\!c}}(V,\varLambda)]D^{j_2}_{\bar{\tilde\mu}_2\tilde\mu_2}[\underline R_\mathrm{W_{\!c}}(V,\varLambda)]\,.
\nonumber\\
\end{eqnarray}
\subsection{Extension to Four-Body Hilbert Space}
The procedure of embedding two-body interactions into a three-body Hilbert space can be easily generalized to a $n$-body Hilbert space. For our purposes we consider a four-body Hilbert space of a system consisting of the interacting pair 1 and 2, the spectator e and another massless spectator $\gamma$ (\lq$\gamma$' stands for a photon). Instead of going explicitly through an analogous procedure as above we simply list the most important results which will be needed in the following sections. The total mass operator for the four-particle system is given by
\begin{eqnarray}\label{eq:totalmassop4body}
\hat M_{\mathrm C\mathrm e\gamma}:=\sqrt{\hat {\boldsymbol k}_{\mathrm C}^2+\hat m_\mathrm C^2}+\sqrt{\hat {\boldsymbol k}_{\mathrm e}^2+m_\mathrm e^2}+\vert\hat {\boldsymbol k}_{\gamma}\vert\,.
\end{eqnarray}
The two-body wave function is defined via the velocity-states matrix elements by
\begin{eqnarray}\label{eq:norm4partcl}
\lefteqn{\langle V';\boldsymbol k_1',\mu_1';(\boldsymbol k_2'),\mu_2';\boldsymbol k_\mathrm e',\mu_\mathrm e';\boldsymbol k'_\gamma,\mu_\gamma'\vert V;(\boldsymbol k_{12});n,j,\mu_j;\boldsymbol k_\mathrm e,\mu_\mathrm e;\boldsymbol k_\gamma,\mu_\gamma\rangle}\nonumber\\&=&
 (2\pi)^{21/2} \sqrt{\frac{2\tilde k'^0_12\tilde k'^0_2}{2(\tilde k'^0_1+\tilde k'^0_2)}}\sqrt{\frac{2 k^0_{\mathrm C}2k^0_{\mathrm e}2k^0_{\gamma}}{(k^0_{\mathrm C}+k^0_{\mathrm e}+k^0_{\gamma})^3}}
\sqrt{\frac{2 k'^0_{12}2k'^0_{\mathrm e}2k'^0_{\gamma}}{(k'^0_{12}+k'^0_{\mathrm e}+k'^0_{\gamma})^3}}V^0\nonumber\\&&\times\delta ^3(\boldsymbol V-\boldsymbol V') \delta_{\mu_\mathrm e\mu_\mathrm e'} \delta^3\left(\boldsymbol k_{\mathrm e}-\boldsymbol k_{\mathrm e}'\right) 
(-\mathrm g_{\mu_\gamma\mu_\gamma'}) \delta^3\left(\boldsymbol k_{\gamma}-\boldsymbol k_{\gamma}'\right)\varPsi _{nj\mu_j\mu_1'\mu_2'}(\tilde{\boldsymbol k}')\,.\nonumber\\
\end{eqnarray}
The orthogonality and completeness relations of this basis vectors are given by
\begin{eqnarray}\label{eq:normeigenstates4part}
\lefteqn{\langle V';\boldsymbol k_{12}';n',j',\mu_j';(\boldsymbol k_\mathrm e'),\mu_\mathrm e' ;\boldsymbol k'_\gamma, \mu'_\gamma \vert V;\boldsymbol k_{12},n,j,\mu_j;(\boldsymbol k_\mathrm e),\mu_\mathrm e;\boldsymbol k_\gamma, \mu_\gamma\rangle}\nonumber\\&=&
(2\pi)^9V^0\delta^3(\boldsymbol V-\boldsymbol V') \frac{2k_{\mathrm C}^02k_{\mathrm e}^0}{(k_{\mathrm C}^0+k_{\mathrm e}^0+k^0_\gamma)^3}\delta^3(\boldsymbol k_{12}-\boldsymbol k_{12}')\delta_{nn'}
\delta_{jj'}\delta_{\mu_j\mu_j'}\nonumber\\&&\times\delta_{\mu_{\mathrm e}\mu_{\mathrm e}'}(-\mathrm g^{\mu_{\gamma}\mu_{\gamma}'})2 k^0_\gamma\delta^3(\boldsymbol k_{\gamma}-\boldsymbol k_{\gamma}') 
\end{eqnarray} and
\begin{eqnarray}\label{eq:completenesseigenstates4part}
\hat{1}_{\mathrm {Ce}\gamma}&=&\frac{1}{(2\pi)^9}
\int \frac{\mathrm d^3 V}{V^0}\frac{\mathrm d^3 k_{12}}{2 k_{\mathrm C}^0}\frac{\mathrm d^3 k_{\gamma}}{2 k_{\gamma}^0}\frac{(k_{\mathrm C}^0+k_{\mathrm e}^0+k^0_\gamma)^3}{2k_{\mathrm e}^0}\nonumber\\&&\times\sum_{nj\mu_j\mu_{\mathrm e}\mu_\gamma}(-\mathrm g^{\mu_\gamma\mu_\gamma})
\vert V;\boldsymbol k_{12},n,j,\mu_j;(\boldsymbol k_\mathrm e),\mu_\mathrm e;\boldsymbol k_\gamma, \mu_\gamma\rangle\nonumber\\&&\times\langle V;\boldsymbol k_{12},n,j,\mu_j;(\boldsymbol k_\mathrm e),\mu_\mathrm e;\boldsymbol k_\gamma, \mu_\gamma\vert\,,
\end{eqnarray}
respectively.
\subsection{Cluster Properties}\label{sec:clusterproperties}
We have already mentioned that the concept of macrocausality (or equivalently cluster separability) becomes questionable within a
Bakamjian-Thomas construction for relativistic quantum mechanics of more than 2 particles. In this subsection we follow closely Ref.~\cite{Keister:1991sb} and discuss the most simple non-trivial case on this issue, the (2+1)-problem of a pair C of interacting particles 1 and 2 and a non-interacting spectator e. What follows applies equally to 2 or more non-interacting spectators. Cluster separability requires that, if the spectator is separated by a sufficient large space-like distance from the cluster, then both, the cluster and the spectator, should behave as independent subsystems. In other words, all relevant physical properties of the (12e)-system should also hold for the (12)-subsystem. With relevant physical properties in the context of cluster separability we mean the relativistic transformation laws~\cite{Keister:1991sb}. 

In order to find a mathematical 
formulation of the condition of cluster separability we consider a Hilbert space of the (2+1)-system, which is the tensor product 
of the Hilbert spaces of the cluster and the spectator. A unitary representation of the Poincar\'e group on this Hilbert space is then given by Eq.~(\ref{eq:unitPtrafo}):
 \begin{eqnarray}
\label{eq:unitPtrafo12e}
\hat{U}_{\mathrm C\mathrm e}\left(\underline \varLambda, \underline a\right):=\hat{U}_{\mathrm C}\left(\underline \varLambda, \underline a\right)\otimes \hat{U}_\mathrm e\left(\underline \varLambda, \underline a\right)\,.
  \end{eqnarray}
Here $\hat{U}_{\mathrm C}\left(\underline \varLambda, \underline a\right)$ and $\hat{U}_\mathrm e\left(\underline \varLambda, \underline a\right)$ are unitary representations
associated with the interacting pair and the spectator, respectively. Both these representations are assumed to have the same relativistic transformation properties as the whole (Ce)-system. 
From Eq.~(\ref{eq:unitPtrafo12e}) we can define translation operators by
\begin{eqnarray}
\hat{T}_{\mathrm C}\left(\underline a\right):=\hat{U}_{\mathrm C}\left(1_2, \underline a\right)\otimes \hat{1}_\mathrm e\quad\text{and}\quad
\hat{T}_{\mathrm e}\left(\underline{a}\right):=\hat{1}_{\mathrm C}\otimes \hat{U}_{\mathrm e}\left(1_2, \underline a\right) \end{eqnarray}
that translate the interacting pair and the spectator by $a^\mu$, respectively. Additionally, we consider an interacting representation $\hat{U}\left(\underline \varLambda, \underline a\right)$ of the Poincar\'e group associated with the interacting (Ce)-system.
This representation satisfies cluster separability if it satisfies the, so-called, \textit{cluster condition}~\cite{Keister:1991sb}:   
     \begin{eqnarray}\label{eq:clustercondition1}
\lim_{(b-c)^2\rightarrow-\infty}\langle\varPsi\vert \hat{T}_{\mathrm C}^\dag\left(\underline{\varLambda b}\right) \hat{T}_{\mathrm e}^\dag\left(\underline{\varLambda c}\right)
\left\lbrace\hat{U}\left(\underline \varLambda, \underline a\right)-\hat{U}_{\mathrm C\mathrm e}\left(\underline \varLambda, \underline a\right)\right\rbrace
\hat{T}_{\mathrm C}\left( \underline{b}\right) \hat{T}_{\mathrm e}\left(\underline{c}\right)\vert\varPsi\rangle=0\,.\nonumber\\
\end{eqnarray}
This condition can be interpreted in the sense that for infinitely large space-like separations of the subsystems C and e, $\hat{U}\left(\underline \varLambda, \underline a\right)$ becomes identical
to the tensor-product representation $\hat{U}_{\mathrm C\mathrm e}\left(\underline \varLambda, \underline a\right)$.
For infinitesimal Poincar\'e transformations the cluster condition Eq.~(\ref{eq:clustercondition1}) can be written as a \textit{cluster condition on the generators}~\cite{Keister:1991sb}:
  \begin{eqnarray}\label{eq:clustercondition2}
\lim_{(b-c)^2\rightarrow-\infty}\langle\varPsi\vert \hat{T}_{\mathrm C}^\dag\left(\underline  b\right) \hat{T}_{\mathrm e}^\dag\left(\underline  c\right)
\left\lbrace\hat{G}-\hat{G}_{\mathrm C\mathrm e}\right\rbrace
\hat{T}_{\mathrm C}\left( \underline  b\right) \hat{T}_{\mathrm e}\left(\underline  c\right)\vert\varPsi\rangle=0\,.\end{eqnarray}
Here the infinitesimal generators $\hat{G}_{\mathrm C\mathrm e} \equiv\{\hat P^\mu_{\mathrm C\mathrm e},\hat M^{\nu\lambda}_{\mathrm C\mathrm e}\}$ of the tensor-product representation, Eq.~(\ref{eq:unitPtrafo12e}), are given by Eq.~(\ref{eq:multipartG}):
\begin{eqnarray}\label{eq:Generatortensorprod}
 \hat P^\mu_{\mathrm C\mathrm e}=\hat p^\mu_{\mathrm C}\otimes \hat{1}_\mathrm e+ \hat{1}_{\mathrm C}\otimes \hat p^\mu_\mathrm e \quad\text{and}\quad
\hat M^{\nu\lambda}_{\mathrm C\mathrm e}=\hat m^{\nu\lambda}_{\mathrm C}\otimes  \hat{1}_\mathrm e+ \hat{1}_{\mathrm C}\otimes \hat m^{\nu\lambda}_\mathrm e\,.
\end{eqnarray}
In our case of the point form the generators of the cluster are
\begin{eqnarray}
 \hat p^\mu_{\mathrm C}=\hat p^\mu_{12}+\hat p^\mu_{\mathrm {int}}\quad\text{and}\quad \hat m^{\nu\lambda}_{\mathrm C}=\hat m^{\nu\lambda}_{12}=\hat m^{\nu\lambda}_{1}\otimes \hat{1}_{2}+ \hat{1}_{1}\otimes \hat m^{\nu\lambda}_{2} \,.
\end{eqnarray}
$\hat G\equiv\{\hat P^\mu,\hat M^{\nu\lambda}\}$ are the generators formally obtained from the derivative of $\hat{U}\left(\underline \varLambda, \underline a\right)$
with respect to the parameters for vanishing values of the parameters, cf. Eqs.~(\ref{eq:infinitgener}) and~(\ref{eq:infinitgener2}).

The physical interpretation of the cluster condition, Eq.~(\ref{eq:clustercondition2}), is quite obvious~\cite{Keister:1991sb}: 
cluster separability is satisfied if the total energy, linear and angular momentum of the (Ce)-system in the cluster limit is equal to the sum of the individual energies, linear and angular momenta of the cluster and the spectator, respectively. 

With a condition for macrocausality at hand we are now able to investigate the cluster properties of the representation of the Poincar\'e group associated with the point-form Bakamjian-Thomas construction derived in the previous sections. The generator of interest in our point-form formulation is the interacting four-momentum operator given by Eq.~(\ref{eq:BTfourmom}). For simplicity, we can restrict our considerations (without loss of generality) to separation operators being pure spatial translations $\hat T_i(\underline{\boldsymbol a})=\mathrm e^{\mathrm i \,\hat{\boldsymbol p}_i\cdot \boldsymbol a}$. Then the left-hand side of the cluster condition, Eq.~(\ref{eq:clustercondition2}), for the point-form Bakamjian-Thomas generators $\hat P^\mu$ of Eq.~(\ref{eq:BTfourmom}) reads
 \begin{eqnarray}\label{eq:clustercondition3}\lefteqn{
\lim_{\vert\boldsymbol b-\boldsymbol c\vert\rightarrow\infty}\langle\varPsi\vert \mathrm e^{-\mathrm i\,(\hat{\boldsymbol p}_{\mathrm C}\cdot\boldsymbol b+\hat{\boldsymbol p}_{\mathrm e}\cdot\boldsymbol c)} 
\left\lbrace\hat{P}^\mu-\hat{P}_{\mathrm C\mathrm e}^\mu\right\rbrace
 \mathrm e^{\mathrm i\,(\hat{\boldsymbol p}_{\mathrm C}\cdot\boldsymbol b+\hat{\boldsymbol p}_{\mathrm e}\cdot\boldsymbol c)}\vert\varPsi\rangle}\nonumber\\&&=
\lim_{\vert\boldsymbol b-\boldsymbol c\vert\rightarrow\infty}\langle\varPsi\vert \mathrm e^{-\mathrm i\,(\hat{\boldsymbol p}_{\mathrm C}\cdot\boldsymbol b+\hat{\boldsymbol p}_{\mathrm e}\cdot\boldsymbol c)} 
\left\lbrace\hat{P}_{\mathrm {int}}^\mu-\hat{P}_{12\mathrm {int},\mathrm e}^\mu\right\rbrace
 \mathrm e^{\mathrm i\,(\hat{\boldsymbol p}_{\mathrm C}\cdot\boldsymbol b+\hat{\boldsymbol p}_{\mathrm e}\cdot\boldsymbol c)}\vert\varPsi\rangle\,
\end{eqnarray}
where we have used that $\hat{P}^\mu=\hat{P}^\mu_{12\mathrm e}+\hat{P}_{\mathrm {int}}^\mu$ and $\hat{P}_{\mathrm C\mathrm e}^\mu=\hat{P}^\mu_{12\mathrm e}+\hat{P}_{12\mathrm {int},\mathrm e}^\mu$ from Eqs.~(\ref{eq:MintBT}) and~(\ref{eq:Generatortensorprod}), respectively. Since
$\hat{P}_{12\mathrm {int},\mathrm e}^\mu$ commutes with $\hat{\boldsymbol p}_{\mathrm C}$ and $\hat{\boldsymbol p}_{\mathrm e}$ by definition, the exponentials cancel in the second term of Eq.~(\ref{eq:clustercondition3}):
 \begin{eqnarray}\label{eq:clustercondition4}&&
\lim_{\vert\boldsymbol b-\boldsymbol c\vert\rightarrow\infty}\langle\varPsi\vert \mathrm e^{-\mathrm i\,(\hat{\boldsymbol p}_{\mathrm C}\cdot\boldsymbol b+\hat{\boldsymbol p}_{\mathrm e}\cdot\boldsymbol c)} 
\hat{P}_{12\mathrm {int},\mathrm e}^\mu
 \mathrm e^{\mathrm i\,(\hat{\boldsymbol p}_{\mathrm C}\cdot\boldsymbol b+\hat{\boldsymbol p}_{\mathrm e}\cdot\boldsymbol c)}\vert\varPsi\rangle=\langle\varPsi\vert 
\hat{P}_{12\mathrm {int},\mathrm e}^\mu
 \vert\varPsi\rangle\,.\nonumber\\
\end{eqnarray}
The first term of Eq.~(\ref{eq:clustercondition3}) can be rewritten in terms of the operators $\hat{\boldsymbol V}$ and $\hat{\boldsymbol k}_\mathrm e$ with $\hat k_{\mathrm C}=(\sqrt{\hat m_\mathrm C^2+\hat{\boldsymbol k}_\mathrm e ^2},-\hat{\boldsymbol k}_\mathrm e)$ as
\begin{eqnarray}\label{eq:clustercondition5}
\lim_{\vert\boldsymbol b-\boldsymbol c\vert\rightarrow\infty}\!\!\!\!\!\!\!\!\!\!\!\!&&\langle\varPsi\vert \mathrm e^{-\mathrm i\,(\boldsymbol B_\mathrm c(\hat{\boldsymbol V})\hat k_{\mathrm C}\cdot\boldsymbol b+\boldsymbol B_\mathrm c(\hat{\boldsymbol V})\hat k_{\mathrm e}\cdot\boldsymbol c)} 
\hat{M}_{\mathrm{int}}\hat{V}^\mu\,
 \mathrm e^{\mathrm i\,(\boldsymbol B_\mathrm c(\hat{\boldsymbol V})\hat k_{\mathrm C}\cdot\boldsymbol b+\boldsymbol B_\mathrm c(\hat{\boldsymbol V})\hat k_{\mathrm e}\cdot\boldsymbol c)}\vert\varPsi\rangle\nonumber\\&=&
\lim_{\vert\boldsymbol b\vert\rightarrow\infty}\langle\varPsi\vert\hat{V}^\mu \,\mathrm e^{-\mathrm i\,\boldsymbol B_\mathrm c(\hat{\boldsymbol V})\hat k_{\mathrm C}\cdot\boldsymbol b} 
\hat{M}_{\mathrm{int}}\,
 \mathrm e^{\mathrm i\,\boldsymbol B_\mathrm c(\hat{\boldsymbol V})\hat k_{\mathrm C}\cdot\boldsymbol b}\vert\varPsi\rangle\,.
\end{eqnarray}
Here we have used that $\hat V^\mu$ commutes with $\hat{\boldsymbol k}_\mathrm e$ and $\hat m_\mathrm C =\hat m_{12}+\hat m_{\mathrm {int}}$. We have further used that $\hat{M}_{\mathrm{int}}$ commutes with $\hat V^\mu$ and $\hat{\boldsymbol k}_\mathrm e$.
Note that from Eq.~(\ref{eq:B_c}) we have
\begin{eqnarray}
\boldsymbol B_\mathrm c(\hat{\boldsymbol V})\hat k_{\mathrm C}=\sqrt{\hat{\boldsymbol k}_\mathrm e^2+ \hat m_\mathrm C^2}\hat{\boldsymbol V}-\hat{\boldsymbol k}_\mathrm e-\frac{\hat{V}^0-1}{\hat{\boldsymbol{V}}^2}\hat{\boldsymbol
{V}} (\hat{\boldsymbol{V}}\cdot\hat{\boldsymbol k}_\mathrm e)\,.
\end{eqnarray}
We observe that only the first term of this expression does not commute with $\hat{M}_{\mathrm{int}}$ since it involves the operator $\hat{\tilde{\boldsymbol k}}$ which occurs in $\hat m_\mathrm C$, cf. Eq.~(\ref{eq:massoperfree12}). Using this, Eq.~(\ref{eq:clustercondition5}) can be further reduced to \begin{eqnarray}\label{eq:clustercondition5a}
      \lim_{\vert\boldsymbol b\vert\rightarrow\infty}\langle\varPsi\vert \hat{V}^\mu\,\mathrm e^{-\mathrm i \,\sqrt{\hat m_\mathrm C^2+ \hat{\boldsymbol k}_\mathrm e^2}\hat{\boldsymbol V}\cdot\boldsymbol b} 
\hat{M}_{\mathrm{int}}\,
 \mathrm e^{\mathrm i \,\sqrt{\hat m_\mathrm C^2+ \hat{\boldsymbol k}_\mathrm e^2}\hat{\boldsymbol V}\cdot\boldsymbol b}\vert\varPsi\rangle\,.                                                                                                                                                                                                                                                                                                                                                                                                                                                                                                                                                                     \end{eqnarray}
This expression has a similar structure as a corresponding expression in an instant-form analysis of Ref.~\cite{Keister:1991sb}. Therefore, we can use similar arguments as therein: under certain assumptions~(for details we refer to~\cite{Keister:1991sb}) about the regularity of the interaction and the wave function it can be shown that expression (\ref{eq:clustercondition5a}) vanishes, whereas in general Eq.~(\ref{eq:clustercondition4}) does not. This represents a violation of the cluster condition, Eq.~(\ref{eq:clustercondition2}), for the Bakamjian-Thomas generators $\hat P^\mu$ of Eq.~(\ref{eq:BTfourmom}).

The vanishing of (\ref{eq:clustercondition5a}) indicates the vanishing of the interaction between particles 1 and 2 in the limit where the non-interacting spectator is separated by an infinite large space-like distance. This means that the physical properties of the interacting pair are not independent of the non-interacting spectator. The problem is inherent in the definition of the cluster wave functions from velocity states for more than 2 particles, see, e.g., Eq.~(\ref{eq:norm3partcl}). In particular, the wave function depends on the energy of the spectator which is reflected in a $k_\mathrm e^0$-dependence of the normalization factors of Eq.~(\ref{eq:normalizationN_3}). Likewise the wave function changes in the presence of additional non-interacting
particles. This change is essentially proportional to the fraction
of the cluster binding energy over the invariant mass of the whole system. 
This indicates that, if the invariant mass of the whole system is made sufficiently large, the effects of wrong cluster separability properties should be minimized or in particular cases even be eliminated. 
\section[Quantum Field Theoretical Vertex Interactions]{Quantum Field Theoretical \\Vertex Interactions}
 Up to now we discussed only instantaneous interactions. In order to set up a relativistic multi-channel problem we need interactions which describe particle creation and annihilation. The definition of such quantum field theoretical vertex interactions within our point-form Bakamjian-Thomas framework is the subject of this section. It relies on the ideas of Refs.~\cite{Klink:2000pp,KrassniggDiss:2001}.
\subsection{Field Operators} 
The first step towards a field theoretical vertex interaction is the construction of field operators. In the present section we will consider only fermion-photon vertices with the fermion field being a (anti-)quark, nucleon or electron field.
\subsubsection{Dirac Fields}
The field operators for a charged massive spin-$1/2$ fermion field read
\begin{eqnarray}\hat \psi\left(x\right)=\frac{1}{(2\pi)^3}\int\frac{\mathrm d^3p}{2p^0}\sum_{\sigma=\pm\frac12}\left(\mathrm e^{\mathrm i p\cdot x}v_{\sigma}(\boldsymbol p)\hat d^\dag_{\sigma}(\boldsymbol p)+
\mathrm e^{-\mathrm i p\cdot x}u_{\sigma}(\boldsymbol p)\hat c_{\sigma}(\boldsymbol p)\right) \end{eqnarray}
and
\begin{eqnarray}
\hat {\bar{\psi}}\left(x\right)=\frac{1}{(2\pi)^3}\int\frac{\mathrm d^3p}{2p^0}\sum_{\sigma=\pm\frac12}\left(\mathrm e^{-\mathrm i p\cdot x}\bar v_{\sigma}(\boldsymbol p)\hat d_{\sigma}(\boldsymbol p)+
\mathrm e^{\mathrm i p\cdot x}\bar u_{\sigma}(\boldsymbol p)\hat c^\dag_{\sigma}(\boldsymbol p)\right)\,.
\end{eqnarray} 
$\hat \psi_{\alpha}\left(x\right)$ with $\alpha=1,\ldots,4$ transforms under Poincar\'e transformations with the four-dimensional matrix representation of the SL$(2,\mathbb C)$, which has been introduced in Eq.~(\ref{eq:smatrixrepr}):
\begin{eqnarray}
\label{eq:tranfpsi}&&\hat U(\underline \varLambda,\underline a)\,\hat \psi_{\alpha}\left(x\right)\,\hat U^\dag(\underline \varLambda,\underline a)=S_{\alpha\beta}(\underline \varLambda^{-1})\,\hat \psi_{\beta}\left(\varLambda x+a\right).
\end{eqnarray}
With the help of Eq.~(\ref{eq:Sgamm0}) the transformation properties of the adjoint field operators follow immediately:
\begin{eqnarray}\label{eq:tranfpsibar}
&&\hat U(\underline \varLambda,\underline a)\,\hat {\bar \psi}_{\alpha}\left(x\right)\,\hat U^\dag(\underline \varLambda,\underline a)=\hat {\bar \psi}_{\beta}\left(\varLambda x+a\right)S_{\beta\alpha}(\underline \varLambda)\,.
\end{eqnarray}
$u_{\sigma}(\boldsymbol p)_\alpha$ and $v_{\sigma}(\boldsymbol p)_\alpha$ are the  Dirac four-spinors defined by the action of a canonical boost on the spinors in the rest frame:
 \begin{eqnarray}
u_{\sigma}(\boldsymbol p)_{\alpha}&= &S_{\alpha\beta}[\underline B_{\mathrm c}(\boldsymbol v)] u_{\sigma}(\boldsymbol{0})_{\beta}\,,\\
v_{\sigma}(\boldsymbol p)_{\alpha}&= &S_{\alpha\beta}[\underline B_{\mathrm c}(\boldsymbol v)] v_{\sigma}(\boldsymbol{0})_{\beta}\,.
\end{eqnarray}
They correspond to fermion and anti-fermion solutions of the Dirac equation, respectively:
\begin{eqnarray}\label{eq:Diracequation}
\left(\gamma^\mu p_\mu-m\right) u_{\sigma}(\boldsymbol p)=0\quad\text{and}\quad\left(\gamma^\mu p_\mu+m\right) v_{\sigma}(\boldsymbol p)=0\,.
\end{eqnarray}
The explicit forms of the spinors are fixed by the representation used for the $\gamma$-matrices. In the standard representation, which will be used throughout this work, the spinors are given in App.~\ref{app:Diracrepresentation}. 
The adjoint Dirac spinors defined by $\bar u_{\sigma}(\boldsymbol p):=u_{\sigma}^\dag(\boldsymbol p)\gamma ^0$ and $\bar v_{\sigma}(\boldsymbol p):=v_{\sigma}^\dag(\boldsymbol p)\gamma ^0$ satisfy the adjoint Dirac equation:
\begin{eqnarray}
\label{eq:adjDiracequation}
\bar u_{\sigma}(\boldsymbol p)\left(\gamma^\mu p_\mu-m\right) =0\quad \text{and}\quad
\bar v_{\sigma}(\boldsymbol p)\left(\gamma^\mu p_\mu+m\right) =0\,.
\end{eqnarray}
The operators $\hat c^\dag_{\sigma}(\boldsymbol p)$ ($\hat c_{\sigma}(\boldsymbol p)$) and  $\hat d^\dag_{\sigma}(\boldsymbol p)$ ($\hat d_{\sigma}(\boldsymbol p)$) are creation (annihilation) 
operators for fermions and anti-fermions, respectively. They act on the, so-called, \textit{Fock space} which is the direct sum over all tensor products of single-particle Hilbert spaces.  The vacuum state $\vert0\rangle$ of this Fock space is defined via the action of the annihilation operators: \begin{eqnarray}\label{eq:vac}
\hat c_{\sigma}(\boldsymbol p)\vert0\rangle=\hat d_{\sigma}(\boldsymbol p)\vert0\rangle=0\,.
\end{eqnarray}
A $n$-particle tensor-product state containing (anti-)fermions is obtained by applying $n$ creation operators for (anti-)fermions to the vacuum state. Hence, for a $n$-fermion state, for example, we have
\begin{eqnarray}\label{eq:npartfockstate}
\hat c_{\sigma_1}^\dag (\boldsymbol p_1)\hat c_{\sigma_2}^\dag (\boldsymbol p_2)\cdots\hat c_{\sigma_n}^\dag (\boldsymbol p_n)\vert0\rangle=\vert \boldsymbol p_1,\sigma_1;\boldsymbol p_2,\sigma_2;\ldots;\boldsymbol p_n,\sigma_n\rangle\,
.\end{eqnarray}
For each anti-fermion occurring in Eq.~(\ref{eq:npartfockstate}) the corresponding fermion operator $\hat c_{\sigma_i}^\dag (\boldsymbol p_i)$ has to be replaced by $\hat d_{\sigma_i}^\dag (\boldsymbol p_i)$.
The creation and annihilation operators satisfy the following anti-commutation relations:
\begin{eqnarray}\label{eq:anticommrel}
\{\hat c_{\sigma} (\boldsymbol p),\hat c_{\sigma'}^\dag (\boldsymbol p')\}=
\{\hat d_{\sigma} (\boldsymbol p),\hat d_{\sigma'}^\dag (\boldsymbol p')\}=(2\pi)^32 p^0 \delta^3\left(\boldsymbol p-\boldsymbol p'\right)\delta_{\sigma\sigma'}\,.
\end{eqnarray} They are fixed by the normalization of the massive one-particle states Eq.~(\ref{eq:norm1massivepart}). All other anti-commutators vanish.
%
%
\subsubsection{Maxwell Field}
A field operator for a neutral massless spin-$1$ boson field, such as a photon, is given by 
\begin{eqnarray}\hat A^\mu(x)=\frac{1}{(2\pi)^3}\int\frac{\mathrm d^3p}{2\vert\boldsymbol p\vert}\sum_{\lambda=0}^3(-\mathrm g^{\lambda\lambda})\,\left(
\mathrm e^{\mathrm i p\cdot x}\epsilon_{\lambda}^\mu(\boldsymbol p)\hat a^\dag_{\lambda}(\boldsymbol p)+
\mathrm e^{-\mathrm i p\cdot x}\epsilon_{\lambda}^{\ast\mu}(\boldsymbol p)\hat a_{\lambda}(\boldsymbol p)\right)\,.\nonumber\\
\end{eqnarray}
The field operator $\hat A^\mu(x)$ transforms under a Poincar\'e transformation with the usual four-dimensional matrix representation of the Lorentz group Eq.~(\ref{eq:LreprMinkowski}):
\begin{eqnarray}
\label{eq:tranfAmu}
&&\hat U (\underline \varLambda,\underline a)\,\hat A^\mu (x)\,\hat U^\dag(\underline \varLambda,\underline a)=\left(\varLambda^{-1}\right)_{\,\,\nu}^\mu \hat A^{\nu}(\varLambda x+a).
\end{eqnarray}
The appropriately orthonormalized photon polarization vectors $\epsilon_{\lambda}^\mu(\boldsymbol p)$ are most conveniently expressed as the components of the helicity boost matrix 
$\epsilon_{\lambda}^\mu(\boldsymbol p)=B_\mathrm h(\boldsymbol p)^\mu_{\,\,\lambda}$~\cite{Klink:2000pq}, which
satisfy the completeness relation
\begin{equation}
 \label{eq:polcomp}
\sum_{\lambda=0}^3
\epsilon^{\mu}_{\lambda}(\boldsymbol {p})(-\mathrm g^{\lambda\lambda})\, \epsilon^{\ast \nu}_{\lambda}(\boldsymbol{p})
= - \mathrm g^{\mu \nu}\, .
\end{equation}
The operators $\hat a^\dag_{\lambda}(\boldsymbol p)$ and $\hat a_{\lambda}(\boldsymbol p)$ are photon creation and annihilation operators, respectively. They satisfy the commutation 
relations
\begin{eqnarray}\label{eq:commrel}
[\hat a_{\lambda} (\boldsymbol p),\hat a_{\lambda'}^\dag (\boldsymbol p')]=(2\pi)^32 p^0 \delta^3\left(\boldsymbol p-\boldsymbol p'\right)(-\mathrm g_{\lambda\lambda'})\,.
\end{eqnarray}
All other commutators vanish.
$\hat a_{\lambda}$ annihilates the vacuum state $\vert0\rangle$:
\begin{eqnarray}\label{eq:vac1}
\hat a_{\lambda}(\boldsymbol p)\vert0\rangle=0\,.
\end{eqnarray}

\subsection{Vertex Operators from Interaction Densities}
In a velocity-state basis the Bakamjian-Thomas type four-momentum
operator, Eq.~(\ref{eq:PFBT}), becomes diagonal in the total four-velocity $V^\mu$. This is a special feature of the Bakamjian-Thomas construction which, in general, does not hold for arbitrary interacting relativistic quantum theories. It is, in particular, not possible to factorize the four-momentum operator of an interacting point-form quantum field theory into a product of a four-velocity operator and an interacting mass operator, cf. Ref.~\cite{Biernat:2007sz}. Therefore, vertex operators responsible for photon emission and absorption cannot directly be taken from point-form quantum electrodynamics. One rather has to make the approximation that the total four-velocity of the system is conserved at the electromagnetic vertices to end up with a Bakamjian-Thomas type mass operator. In Ref.~\cite{Klink:2000pp} it has been demonstrated in some detail that this is a natural way to implement general field theoretical vertex interactions into a Bakamjian-Thomas type framework.
\subsubsection{Constituents}
 We start our discussion by looking at the interaction of a photon field with the constituents and the electron. To this end we consider a field theoretical interaction density $\hat{\mathcal L}_{\mathrm {int}}(x)$ for spinor quantum
electrodynamics~\cite{Bjorken:1964}. It is given by the normal ordered product of the field operators defined above:
\begin{eqnarray}\label{eq:lagrangeandens}
\hat{\mathcal L}_{\mathrm {int}}(x)&:=&-:\hat{J}_{1}^\mu(x) \hat A_{\mu}(x):-:\hat{J}_{2}^\mu(x) \hat A_{\mu}(x):\nonumber\\&&-\,|\,\mathrm e\,|\,
 Q_{\mathrm e}:\hat{\bar \psi}_{\mathrm e}(x)\gamma^\mu\hat \psi_{\mathrm e} (x) \hat A_{\mu}(x):\,.
\end{eqnarray}
$\hat{\mathcal L}_{\mathrm {int}}(x)$ describes the coupling of a photon field to the 2 spin-1/2 constituent fields and to the electron field. The $\hat{J}_{i}^\mu(x)$ are current operators of the constituents and are given for point-like constituents like, e.g. quarks, explicitly by $\hat{J}_{\mathrm q}^\mu(x)=\,|\,\mathrm e\,|\,Q_{\mathrm q}\hat{\bar \psi}_{\mathrm q}(x) \gamma^\mu\hat \psi_{\mathrm q}(x)$. 
The transformation properties of the fields, Eqs.~(\ref{eq:tranfpsi}) and~(\ref{eq:tranfpsibar}), together with Eq.~(\ref{eq:covgammamatr}) imply that the fermion current of the form $\hat{\bar \psi}(x)\gamma^\mu\hat \psi (x)$ transforms like a vector field (cf. Eq.~(\ref{eq:tranfAmu})). Furthermore, it satisfies current conservation:
\begin{eqnarray}\label{eq:continuity}
 \frac{\partial}{\partial x^\mu}\hat{J}_{\mathrm q}^\mu(x)=0\,.
\end{eqnarray}
We require a general current operator $\hat{J}_{i}^\mu(x)$ for non-point-like objects such as nucleons to have the same covariance and continuity properties as $\hat{J}_{\mathrm q}^\mu(x)$. Consequently, $\hat{\mathcal L}_{\mathrm {int}}(x)$ transforms like a Lorentz scalar:
\begin{eqnarray}\label{eq:tansfLagr}
 \hat U(\underline \varLambda,\underline a)\, \hat{\mathcal L}_{\mathrm {int}} (x)\,\hat U^\dag(\underline \varLambda,\underline a) = \hat{\mathcal L}_{\mathrm {int}} (\varLambda x+a)\,.
\end{eqnarray}
From the interaction Lagrangean density operator we can immediately construct the interaction part of the energy-momentum tensor operator, which reads $ \hat{\mathcal T}_{\mathrm{int}}^{\mu\nu}(x)=-\mathrm g^{\mu\nu}\hat{\mathcal L}_{\mathrm {int}} (x)$.
We have already seen in Sec.~\ref{sec:formsofrelativisticdyn} that the space-like hypersurface of Minkowski space that is left invariant under the point-form kinematic group is the forward hyperboloid $x^\mu x_\mu=\tau^2$. In a point-form quantum field theory obtained from canonical quantization this hyperboloid represents the field quantization surface on which the canonical (anti-)commutation relations on the field operators $\hat \psi\left(x\right)$, $\hat{ \bar\psi}\left(x\right)$ and $\hat A^\mu(x)$ are imposed~\cite{Biernat:2007sz}. Accordingly, the interaction four-momentum operator in a point-form quantum field theory is obtained by integrating $\hat{\mathcal T}_{\mathrm{int}}^{\mu\nu}(x)$ over the forward hyperboloid~\cite{Klink:2000pp,Biernat:2007sz}:
\begin{eqnarray}
\hat P_{\mathrm{int}}^\mu=-\int 2\mathrm d^4 x\,\delta(x^\mu x_\mu-\tau^2)\,\theta(x^0)\, x^\mu \hat{\mathcal L}_{\mathrm {int}} (x)\,.
\end{eqnarray}
The correct transformation properties of $\hat P_{\mathrm{int}}^\mu$ under Poincar\'e transformations, Eq.~(\ref{eq:transfproppmu}), follow immediately from Eq.~(\ref{eq:tansfLagr}) and the invariance of 
the hypersurface element $\mathrm d^4 x\,\delta(x^\mu x_\mu-\tau^2)\,\theta(x^0)$. Since $\hat{\mathcal L}_{\mathrm {int}} (x)$ is a local field operator that transforms like a Lorentz-scalar density it can be shown that the Lie algebra for translations, Eq.~(\ref{eq:la1}), is satisfied~\cite{Klink:2000pp}.

Now consider particular matrix elements of this interacting momentum operator between velocity states:
\begin{eqnarray}\label{eq:pmuintvelocitybasis}
\lefteqn{\langle V';\boldsymbol k_1',\mu_1'; \boldsymbol k_2',\mu_2';\boldsymbol k_\mathrm e',\mu_\mathrm e';\boldsymbol k_\mathrm \gamma',\mu_\gamma'\vert\hat P_{\mathrm{int}}^\mu\vert V;\boldsymbol k_1,\mu_1; \boldsymbol k_2,\mu_2;\boldsymbol k_\mathrm e,\mu_\mathrm e\rangle}\nonumber\\&
=&-\langle V';\boldsymbol k_1',\mu_1'; \boldsymbol k_2',\mu_2';\boldsymbol k_\mathrm e',\mu_\mathrm e';\boldsymbol k_\mathrm \gamma',\mu_\gamma'\vert\hat{\mathcal L}_{\mathrm {int}} (0)\vert V;\boldsymbol k_1,\mu_1; \boldsymbol k_2,\mu_2;\boldsymbol k_\mathrm e,\mu_\mathrm e\rangle\nonumber\\&&\times
\int 2\,\mathrm d^4 x\,\delta(x^\mu x_\mu-\tau^2)\,\theta(x^0) \,x^\mu\,\mathrm e^{-\mathrm i(M'_{12\mathrm e\gamma}V'-M_{12\mathrm e}V)\cdot  x}\,.
\end{eqnarray} 
Here we have used the transformation properties of the Lagrangean density and the velocity states under space-time translations, Eqs.~(\ref{eq:tansfLagr}) and~(\ref{eq:velocitystatetransfpropsttransl}), respectively. 

In order to find a Bakamjian-Thomas-type mass operator from $\hat P_{\mathrm{int}}^\mu$ that commutes with the velocity operator we have to assume the matrix elements in Eq.~(\ref{eq:pmuintvelocitybasis}) to be diagonal in $V$ and $V'$. For $\boldsymbol V=\boldsymbol V'$ the integral in Eq.~(\ref{eq:pmuintvelocitybasis}) can be evaluated and one finds
\begin{eqnarray}\label{eq:Lagvelocitybasis}
\lefteqn{
\langle V;\boldsymbol k_1',\mu_1'; \boldsymbol k_2',\mu_2';\boldsymbol k_\mathrm e',\mu_\mathrm e';\boldsymbol k_\mathrm \gamma',\mu_\gamma'\vert\hat{\mathcal L}_{\mathrm {int}} (0)\vert V;\boldsymbol k_1,\mu_1; \boldsymbol k_2,\mu_2;\boldsymbol k_\mathrm e,\mu_\mathrm e\rangle}\nonumber\\&&\times
\int 2\mathrm d^4 x\,\delta(x^\mu x_\mu-\tau^2)\,\theta(x^0)\, x^\mu\,\mathrm e^{-\mathrm i(M'_{12\mathrm e\gamma}-M_{12\mathrm e})V\cdot x}
\nonumber\\&
=&f_\mathrm v(\vert M'_{12\mathrm e\gamma}-M_{12\mathrm e}\vert)\,V^\mu
\nonumber\\&&\times\langle\boldsymbol k_1',\mu_1'; \boldsymbol k_2',\mu_2';\boldsymbol k_\mathrm e',\mu_\mathrm e';\boldsymbol k_\mathrm \gamma',\mu_\gamma'\vert\hat{\mathcal L}_{\mathrm {int}} (0) \vert \boldsymbol k_1,\mu_1; \boldsymbol k_2,\mu_2;\boldsymbol k_\mathrm e,\mu_\mathrm e\rangle\,.
\end{eqnarray}
The vertex factor $f_\mathrm v(\vert M'_{12\mathrm e\gamma}-M_{12\mathrm e}\vert)$ is a known function (for its explicit form we refer to Ref.~\cite{Klink:2000pp}), but it might as well be replaced by a phenomenological function to compensate for the neglect of the contributions of terms that are off-diagonal in the four-velocity, as well as regulate the integrals, if necessary~\cite{KrassniggDiss:2001}. These are, however, not our primary objectives here. Hence, we set $f_\mathrm v(\vert M'_{12\mathrm e\gamma}-M_{12\mathrm e}\vert)=1$ in the following. By separating the four-velocity $V^\mu$ in Eq.~(\ref{eq:Lagvelocitybasis}), the remaining expression can be used to define the matrix elements of the \textit{electromagnetic vertex operators} $K^\dag$ and $K$ between velocity states:
\begin{eqnarray}\label{eq:vertexoper1}
\lefteqn{
\langle V';\boldsymbol k_1',\mu_1'; \boldsymbol k_2',\mu_2';\boldsymbol k_\mathrm e',\mu_\mathrm e';\boldsymbol k_\mathrm \gamma',\mu_\gamma'\vert\hat K^\dag \vert V; \boldsymbol k_1,\mu_1; \boldsymbol k_2,\mu_2;\boldsymbol k_\mathrm e,\mu_\mathrm e\rangle}\nonumber\\&\equiv&
\langle V;\boldsymbol k_\mathrm e,\mu_\mathrm e;\boldsymbol k_1,\mu_1;\boldsymbol k_2,\mu_2 \vert\hat K \vert V';\boldsymbol k_\mathrm e',\mu_\mathrm e';\boldsymbol k_1',\mu_1';\boldsymbol k_2',\mu_2';\boldsymbol k'_\gamma,\mu'_\gamma\rangle^\ast\nonumber\\&:=&
\frac{(-1)(2\pi)^3}{\sqrt{M'^3_{12\mathrm e\gamma} M_{12\mathrm e}^3}}V^0\delta^3(\boldsymbol V-\boldsymbol V') 
\nonumber\\&&\times\langle\boldsymbol k_1',\mu_1'; \boldsymbol k_2',\mu_2';\boldsymbol k_\mathrm e',\mu_\mathrm e';\boldsymbol k_\mathrm \gamma',\mu_\gamma'\vert\hat{\mathcal L}_{\mathrm {int}} (0) \vert \boldsymbol k_1,\mu_1; \boldsymbol k_2,\mu_2;\boldsymbol k_\mathrm e,\mu_\mathrm e\rangle\,
\end{eqnarray}
with
\begin{eqnarray}
\label{eq:vertexinteractionlagrangedens}
\lefteqn{
\langle\boldsymbol k_1',\mu_1'; \boldsymbol k_2',\mu_2';\boldsymbol k_\mathrm e',\mu_\mathrm e';\boldsymbol k_\mathrm \gamma',\mu_\gamma'\vert\hat{\mathcal L}_{\mathrm {int}} (0) \vert \boldsymbol k_1,\mu_1; \boldsymbol k_2,\mu_2;\boldsymbol k_\mathrm e,\mu_\mathrm e\rangle}\nonumber\\&=&
(2\pi)^6  \,|\,\mathrm e\,|\,Q_{\mathrm e}\,
\bar{u}_{\mu_{\mathrm e}^\prime}(\boldsymbol{k}_{\mathrm e}^\prime)\gamma_\nu
u_{\mu_{\mathrm e}}(\boldsymbol{k}_\mathrm e)\,
\epsilon_{\mu_{\gamma}^\prime}^\nu(\boldsymbol{k}_{\gamma}^\prime)\,  \delta_{\mu_1\mu_1'}2 k_1^0 \delta^3(\boldsymbol{k}_1^\prime - \boldsymbol{k}_1) \nonumber\\&&\;\;\;\;\times  \delta_{\mu_2\mu_2'}2 k_2^0 \delta^3(\boldsymbol{k}_2^\prime - \boldsymbol{k}_2) \nonumber\\
& &  + (2\pi)^6
J_{1\nu}(\boldsymbol{k}_1^\prime,\mu_1^\prime;\boldsymbol{k}_1,\mu_1)
\epsilon_{\mu_{\gamma}^\prime}^\nu(\boldsymbol{k}_{\gamma}^\prime)\, \delta_{\mu_{\mathrm e}\mu_{\mathrm e}'} 2 k_{\mathrm e}^0
 \delta^3(\boldsymbol{k}_e^\prime - \boldsymbol{k}_e)   \nonumber\\&&\;\;\;\;\times
 \delta_{\mu_2\mu_2'} 2  k_2^0 \delta^3(\boldsymbol{k}_2^\prime - \boldsymbol{k}_2)  \nonumber \\
& &   + (2\pi)^6J_{2\nu}(\boldsymbol{k}_2^\prime,\mu_2^\prime;\boldsymbol{k}_2,\mu_2)\,
\epsilon_{\mu_{\gamma}^\prime}^\nu(\boldsymbol{k}_{\gamma}^\prime)\,  \delta_{\mu_{\mathrm e}\mu_{\mathrm e}'} 2 k_{\mathrm e}^0 
\delta^3(\boldsymbol{k}_e^\prime - \boldsymbol{k}_e)  \nonumber\\&&\;\;\;\;\times
 \delta_{\mu_1\mu_1'} 2  k_1^0 \delta^3(\boldsymbol{k}_1^\prime - \boldsymbol{k}_1)
\,. \nonumber\\
\end{eqnarray}
The factor $V^0/\sqrt{M'^3_{12\mathrm e\gamma} M_{12\mathrm e}^3}$ in Eq.~(\ref{eq:vertexoper1}) is derived from the requirement that an analogous procedure for the kinetic term in the Lagrangean provides the usual (kinetic) mass term~\cite{Lechner:2003}. The detailed derivation of Eq.~(\ref{eq:vertexinteractionlagrangedens}) is demonstrated in App.~\ref{app:vertexint}.
Here $J_{i}^\mu(\boldsymbol{k}_i^\prime,\mu_i^\prime;\boldsymbol{k}_i,\mu_i):=\langle\boldsymbol{k}_i^\prime,\mu_i^\prime\vert \hat J_{i}^\mu(0)\vert\boldsymbol{k}_i,\mu_i\rangle$ are the constituent currents which are given for the point-like (structureless) particle case of a quark and an antiquark by the expressions
\begin{eqnarray}
J_{\mathrm q}^{\mu}(\boldsymbol{k}_{\mathrm q}^\prime,\mu_{\mathrm q}^\prime;\boldsymbol{k}_{\mathrm q},\mu_{\mathrm q})=\,|\,\mathrm e\,|\, Q_{\mathrm q}
\bar{u}_{\mu_{\mathrm q}^\prime}(\boldsymbol{k}_{\mathrm q}^\prime)\gamma^\mu
u_{\mu_{\mathrm q}}(\boldsymbol{k}_{\mathrm q})
\end{eqnarray}
and
\begin{eqnarray}
J_{\bar {\mathrm q}}^{\mu}(\boldsymbol{k}_{\bar{\mathrm  q}}^\prime,\mu_{\bar {\mathrm q}}^\prime;\boldsymbol{k}_{\bar{\mathrm q}}, \mu_{\bar{\mathrm q}})=- \,|\,\mathrm e\,|\,Q_{\bar{\mathrm q}}
\bar{v}_{\mu_{\bar{\mathrm q}}}(\boldsymbol{k}_{\mathrm q})\gamma^\mu
v_{\mu_{\bar{\mathrm q}}^\prime}(\boldsymbol{k}_{\bar{\mathrm q}}^\prime)\,,
\end{eqnarray}
respectively.
For non-point-like constituents like nucleons the corresponding current matrix elements are given by
\begin{eqnarray}\label{eq:nuclcurrent}
 J_{\mathrm N}^{\mu}(\boldsymbol{k}_{\mathrm N}^\prime,\mu_{\mathrm N}^\prime;\boldsymbol{k}_{\mathrm N},\mu_{\mathrm N})=
\,|\,\mathrm e\,|\,\bar{u}_{\mu_{\mathrm N}^\prime}(\boldsymbol{k}_{\mathrm N}^\prime)\left[F^{\mathrm N}_1(Q_{\mathrm N}^2) \gamma^\mu+F^{\mathrm N}_2(Q_{\mathrm N}^2)\frac{\mathrm i\, q_{\mathrm N\nu} \sigma^{\mu\nu}}{2m_{\mathrm N}}\right]
u_{\mu_{\mathrm N}}(\boldsymbol{k}_{\mathrm N})\nonumber\\
\end{eqnarray}
with $\mathrm {N\equiv(p,\,n)}$ standing for \lq proton' and \lq neutron', respectively. The electromagnetic structure of the nucleons is described by Dirac and Pauli form factors $F^{\mathrm N}_1(Q_{\mathrm N}^2)$ and $F^{\mathrm N}_2(Q_{\mathrm N}^2)$, respectively, which are functions of the square of the four-momentum transfer on the constituents defined by\footnote{The four-momentum transfer squared $Q_i^2$ on the constituent should not be confused with the constituents charge (squared), but the distinction should be clear from the context.}
\begin{eqnarray}
 q_{\mathrm N}:=k_\mathrm N'-k_\mathrm N\,,\quad - q_{\mathrm N}^\mu q_{\mathrm N\mu}=:Q_{\mathrm N}^2\,.
\end{eqnarray}
Eq.~(\ref{eq:nuclcurrent}) is the most general expression for an electromagnetic current of an extended spin-$1/2$ object satisfying current conservation
\begin{eqnarray}
 q_{\mathrm N\mu}J_{\mathrm N}^{\mu}(\boldsymbol{k}_{\mathrm N}^\prime,\mu_{\mathrm N}^\prime;\boldsymbol{k}_{\mathrm N},\mu_{\mathrm N})=0\,
\end{eqnarray}
 and Lorentz covariance, cf. App.~\ref{app:covarianceproperties} (for a derivation of the expression~(\ref{eq:nuclcurrent}) we refer to, e.g., Ref.~\cite{Aitchison:2003}).
The operator $K$ ($K^\dag$) describes the annihilation (creation) of a photon by the constituents 1 and 2 and by the electron. In the following chapter we will see how a point-form Bakamjian-Thomas-type mass operator is constructed from these vertex operators.
The prescription, Eq.~(\ref{eq:vertexoper1}), is a very natural way to
introduce field theoretical vertex interactions into the
Bakamjian-Thomas framework. It preserves the
Lorentz structure of field theoretical vertex interactions. However, due to the {\em overall} velocity conserving delta function $\delta^3(\boldsymbol V-\boldsymbol V')$ and the kinematical factor $1/\sqrt{M_{\mathrm e 12}'^3 M_{\mathrm e 12\gamma}^3}$, the locality of the vertex is violated. These quantities do not
only depend on the kinematics at the vertex, but also on the
kinematics of the spectator particle(s).
\subsubsection{Bound State}
We can also define a Lagrangean density $\mathcal L_{\mathrm {int}}^{\mathrm C}(x)$ which describes the coupling of the photon field to the electron and to the bound state:
\begin{eqnarray}\label{eq:lagrangeandensbs}
\hat{\mathcal L}_{\mathrm {int}}^{\mathrm C}(x):=-\,|\,\mathrm e\,|\,
 Q_{\mathrm e}:\hat{\bar \psi}_{\mathrm e}(x)\gamma^\mu\hat \psi_{\mathrm e} (x) \hat A_{\mu}(x):-:\hat{J}_{\mathrm C}^\mu(x) \hat A_{\mu}(x):\,.
\end{eqnarray}
In a similar way as above we can define vertex operators that describe the emission and absorption of a photon by the bound state. To
distinguish them from the vertex operators for the constituents we will denote them by
$\hat K^{\mathrm C\dag}$ and $\hat K^{\mathrm C}$. They are defined by 
\begin{eqnarray}\label{eq:vertexoperKbs}
\lefteqn{\langle V';\boldsymbol k_\mathrm e',\mu_\mathrm e';\boldsymbol k_\mathrm C',\mu_j';\boldsymbol k'_\gamma,\mu'_\gamma\vert\hat K^{\mathrm C\dag} \vert V;\boldsymbol k_\mathrm e,\mu_\mathrm e;\boldsymbol k_\mathrm C,\mu_j\rangle}\nonumber\\&\equiv&
\langle V;\boldsymbol k_\mathrm e,\mu_\mathrm e;\boldsymbol k_\mathrm C,\mu_j\vert\hat K^\mathrm C \vert V';\boldsymbol k_\mathrm e',\mu_\mathrm e';\boldsymbol k_\mathrm C',\mu_j';\boldsymbol k'_\gamma,\mu'_\gamma\rangle^\ast\nonumber\\&:=&
\frac{(-1)(2\pi)^3}{\sqrt{M_{\mathrm {e C}}'^3 M_{\mathrm {e C}\gamma}^3}}V^0\delta^3(\boldsymbol V-\boldsymbol V') 
\nonumber\\&&\times\langle \boldsymbol k_\mathrm e',\mu_\mathrm e';\boldsymbol k_\mathrm C',\mu_j';\boldsymbol k'_\gamma,\mu'_\gamma\vert \hat{\mathcal L}^\mathrm C_{\mathrm {int}} (0)\vert\boldsymbol k_\mathrm e,\mu_\mathrm e;\boldsymbol k_\mathrm C,\mu_j\rangle
\nonumber\\&=&\frac{(-1)(2\pi)^6}{\sqrt{M_{\mathrm {e C}}'^3 M_{\mathrm {e C}\gamma}^3}}V^0\delta^3(\boldsymbol V-\boldsymbol V') 
\nonumber\\&&\times
 \left[ \,|\,\mathrm e\,|\,Q_{\mathrm e}\,
\bar{u}_{\mu_{\mathrm e}^\prime}(\boldsymbol{k}_{\mathrm e}^\prime)\gamma_\nu
u_{\mu_{\mathrm e}}(\boldsymbol{k}_\mathrm e)\,
\epsilon_{\mu_{\gamma}^\prime}^\nu(\boldsymbol{k}_{\gamma}^\prime)\,  \delta_{\mu_j\mu_j'}2 k_\mathrm C^0 \delta^3(\boldsymbol{k}_\mathrm C^\prime - \boldsymbol{k}_\mathrm C) \right. \nonumber\\
& & \left. \;\; + \,
J_{\mathrm C\nu}(\boldsymbol k_\mathrm C^\prime,\mu_j^\prime;\boldsymbol k_\mathrm C,\mu_j;K_{\mathrm e})\,
\epsilon_{\mu_{\gamma}^\prime}^\nu(\boldsymbol{k}_{\gamma}^\prime)\, \delta_{\mu_{\mathrm e}\mu_{\mathrm e}'} 2 k_{\mathrm e}^0
 \delta^3(\boldsymbol{k}_e^\prime - \boldsymbol{k}_e)  
 \right]\,. 
\end{eqnarray}
Here
$J_{\mathrm C}^\mu(\boldsymbol k_\mathrm C^\prime,\mu_j^\prime;\boldsymbol k_\mathrm C,\mu_j;K_{\mathrm e})$ is the bound-state current. It is the most general current that one can write down for a bound state with spin $j$ and may contain phenomenological form factors that account for the substructure of the bound state.
Beside the incoming and outgoing bound-state momenta and spins we have also written, as a further dependence, the sum of the incoming and outgoing electron momenta 
$K_{\mathrm e}^\mu:=k_{\mathrm e}^\mu+k_{\mathrm e}'^\mu$ into the argument of the current. The latter dependence can be understood as an effect of the violation of cluster separability in the Bakamjian-Thomas framework, which has been already discussed in Sec.~\ref{sec:clusterproperties}. As a consequence, the physical properties of the bound state are not independent of the electron. In particular, as we will see, the bound-state current contains, so-called, \textit{unphysical} contributions that exhibit a dependence on $K_{\mathrm e}$. 
Finding an explicit form of $J_{\mathrm C}^\mu(\boldsymbol k_\mathrm C^\prime,\mu_j^\prime;\boldsymbol k_\mathrm C,\mu_j;K_{\mathrm e})$ in terms of bound-state wave functions and constituent currents and investigating its Lorentz structure is one of the central objectives of this work.
\subsubsection{Covariance Properties}
At first glance it might seem that the currents for the electron $\bar{u}_{\mu_{\mathrm e}^\prime}(\boldsymbol{k}_{\mathrm e}^\prime)\gamma^\mu
u_{\mu_{\mathrm e}}(\boldsymbol{k}_\mathrm e)$, for the constituents $J_{i}^\mu(\boldsymbol{k}_i^\prime,\mu_i^\prime;\boldsymbol{k}_i,\mu_i)$ and for the bound-state
 $J_{\mathrm C}^\mu(\boldsymbol k_\mathrm C^\prime,\mu_j^\prime;\boldsymbol k_\mathrm C,\mu_j;K_{\mathrm e})$ defined in Eqs.~(\ref{eq:vertexinteractionlagrangedens})
and (\ref{eq:vertexoperKbs}) transform like four-vectors under Lorentz transformations. This is, however, not the case due to the center-of-mass momenta of the electron-bound-state(-photon) system that appear in the currents. The effect of a Lorentz transformation on such center-of-mass momenta is a Wigner rotation, which can be seen from the transformation properties of velocity states, Eq.~(\ref{eq:velocitystatetransfprop}). In order to investigate the transformation properties of the currents we look at velocity-state matrix elements of a Lorentz transformed vertex operator $\hat {\bar K}^\dag=\hat U(\underline\varLambda)\hat K^\dag\hat U^\dag(\underline\varLambda)$
 (using the short-hand notation $R^{(\prime)}_\mathrm{W_{\!c}}\equiv R^{(\prime)}_\mathrm{W_{\!c}}(V^{(\prime)},\varLambda)$):
\begin{eqnarray}\label{eq:vertexoperK1}
\lefteqn{
\langle V';\boldsymbol k_\mathrm e',\mu_\mathrm e';\boldsymbol k_1',\mu_1';\boldsymbol k_2',\mu_2';\boldsymbol k'_\gamma,\mu'_\gamma\vert\hat U(\underline\varLambda)\hat K\hat U^\dag(\underline\varLambda)\vert V;\boldsymbol k_\mathrm e,\mu_\mathrm e;\boldsymbol k_1,\mu_1;\boldsymbol k_2,\mu_2\rangle}\nonumber\\&=&
\langle \varLambda V';\boldsymbol R^{\prime}_\mathrm{W_{\!c}} k_\mathrm e', \bar \mu_\mathrm e';\boldsymbol R^{\prime}_\mathrm{W_{\!c}} k_1',\bar \mu_1';\boldsymbol R^{\prime}_\mathrm{W_{\!c}} k_2',\bar \mu_2';\boldsymbol R^{\prime}_\mathrm{W_{\!c}} k'_\gamma,\bar \mu'_\gamma\vert\nonumber\\&&\times\hat K\vert \varLambda V;\boldsymbol R_\mathrm{W_{\!c}} k_\mathrm e, \bar \mu_\mathrm e;\boldsymbol R_\mathrm{W_{\!c}} k_1,\bar \mu_1;\boldsymbol R_\mathrm{W_{\!c}}k_2, \bar \mu_2\rangle
D^{\frac12\ast}_{\bar \mu_\mathrm e'\mu_\mathrm e'}(\underline R_\mathrm{W_{\!c}}') 
 D^{\frac12\ast}_{\bar \mu_\mathrm 1'\mu_\mathrm 1'}(\underline R_\mathrm{W_{\!c}}')\nonumber\\&&\times
D^{\frac12\ast}_{\bar \mu_\mathrm 2'\mu_\mathrm 2'}(\underline R_\mathrm{W_{\!c}}')
  R_\mathrm{W_{\!h}}\left(k'_\gamma, R_\mathrm{W_{\!c}}\right)_{\bar \mu_\gamma'\mu_\gamma'}
D^{\frac12}_{\bar \mu_\mathrm e\mu_\mathrm e}(\underline R_\mathrm{W_{\!c}}) 
D^{\frac12}_{\bar \mu_\mathrm 1\mu_\mathrm 1}(\underline R_\mathrm{W_{\!c}})
D^{\frac12}_{\bar \mu_\mathrm 2\mu_\mathrm 2}(\underline R_\mathrm{W_{\!c}})
\nonumber\\&=&
\frac{(-1)(2\pi)^9}{\sqrt{M_{\mathrm e 12}'^3 M_{\mathrm e 12\gamma}^3}}V^0\delta^3(\boldsymbol V-\boldsymbol V') \epsilon_{ \bar \mu_{\gamma}^\prime}^\nu( \boldsymbol R_\mathrm{W_{\!c}}k_{\gamma}^\prime) 
R_\mathrm{W_{\!h}}\left(k'_\gamma, R_\mathrm{W_{\!c}}\right)_{\bar \mu_\gamma'\mu_\gamma'}  
\nonumber\\&&\times
 \left[ \,|\,\mathrm e\,|\,Q_{\mathrm e}\,
\bar{u}_{ \bar \mu_{\mathrm e}^\prime}(\boldsymbol R_\mathrm{W_{\!c}} k_{\mathrm e}^\prime)\gamma_\nu
u_{\bar \mu_{\mathrm e}}(\boldsymbol R_\mathrm{W_{\!c}} k_\mathrm e)
D^{\frac12\ast}_{\bar \mu_\mathrm e'\mu_\mathrm e'}(\underline R_\mathrm{W_{\!c}}) D^{\frac12}_{\bar \mu_\mathrm e\mu_\mathrm e}(\underline R_\mathrm{W_{\!c}}) 
\,
\right. \nonumber \\
& & \left.\;\;\;\times\delta_{ \mu_1 \mu_1'}2 k_1^0 \delta^3(\boldsymbol{k}_1^\prime - \boldsymbol{k}_1) \,  \delta_{ \mu_2 \mu_2'}2 k_2^0 \delta^3(\boldsymbol{k}_2^\prime - \boldsymbol{k}_2)\right. \nonumber\\
& & \left. \;\; + \,
J_{1\nu}(\boldsymbol R_\mathrm{W_{\!c}}  k_1^\prime, \bar \mu_1^\prime;\boldsymbol R_\mathrm{W_{\!c}}  k_1, \bar \mu_1)
D^{\frac12\ast}_{\bar \mu_\mathrm 1'\mu_\mathrm 1'}(\underline R_\mathrm{W_{\!c}})D^{\frac12}_{\bar \mu_\mathrm 1\mu_\mathrm 1}(\underline R_\mathrm{W_{\!c}})
 \right. \nonumber \\
& & \left.\;\;\;\times
 \delta_{\mu_{\mathrm e}\mu_{\mathrm e}'} 2 k_{\mathrm e}^0
 \delta^3(\boldsymbol{k}_e^\prime - \boldsymbol{k}_e)  
 \delta_{\mu_2\mu_2'} 2  k_2^0 \delta^3(\boldsymbol{k}_2^\prime - \boldsymbol{k}_2) \right. \nonumber \\
& & \left. \;\; + 
J_{2\nu}(\boldsymbol R_\mathrm{W_{\!c}}  k_2^\prime,\bar  \mu_2^\prime;\boldsymbol R_\mathrm{W_{\!c}}  k_2, \bar \mu_2)
D^{\frac12\ast}_{\bar \mu_\mathrm 2'\mu_\mathrm 2'}(\underline R_\mathrm{W_{\!c}})
D^{\frac12}_{\bar \mu_\mathrm 2\mu_\mathrm 2}(\underline R_\mathrm{W_{\!c}}) 
\right. \nonumber \\
& & \left.\;\;\;\times
 \delta_{\mu_{\mathrm e}\mu_{\mathrm e}'} 2 k_{\mathrm e}^0 
\delta^3(\boldsymbol{k}_e^\prime - \boldsymbol{k}_e) \,
 \delta_{\mu_1\mu_1'} 2  k_1^0 \delta^3(\boldsymbol{k}_1^\prime - \boldsymbol{k}_1)
\right]\,. 
\end{eqnarray}
From the covariant transformation property of the polarization vectors, Eq.~(\ref{eq:phpolarvectortransfprop}), we find 
\begin{eqnarray}
  \epsilon_{\mu_{\gamma}^\prime}^\mu(\boldsymbol k_{\gamma}^\prime) \stackrel{\varLambda}{\longrightarrow} \epsilon_{\bar \mu_{\gamma}^\prime}^\nu(\boldsymbol R_\mathrm{W_{\!c}}  k_{\gamma}^\prime)
= R_\mathrm{W_{\!c}}( V,\underline \varLambda)^\mu_{\,\,\nu} \epsilon_{\mu_{\gamma}^\prime}^\nu(\boldsymbol k_{\gamma}^\prime)
R_\mathrm{W_{\!h}}^{-1}\left(k'_\gamma, R_\mathrm{W_{\!c}}\right)_{\mu_\gamma'\bar \mu_\gamma'}\,.\nonumber\\
\end{eqnarray}
Demanding Lorentz invariance of the vertex by comparing Eq.~(\ref{eq:vertexoperK1}) with Eq.~(\ref{eq:vertexoper1}) we find the transformation properties of the currents under Lorentz transformations:
\begin{eqnarray}\label{eq:transformpropcomcurrents}
 \lefteqn{J_{i}^\mu( \boldsymbol k_i^\prime, \mu_i^\prime; \boldsymbol k_i,\mu_i) \stackrel{\varLambda}{\longrightarrow} J_{i}^\mu(\boldsymbol R_\mathrm{W_{\!c}}  k_i^\prime,\bar \mu_i^\prime;\boldsymbol R_\mathrm{W_{\!c}}  k_i,\bar \mu_i)}
\nonumber\\&=&R_\mathrm{W_{\!c}}(V,\varLambda)_{\,\,\nu}^\mu
J_{i}^\nu(\boldsymbol k_i^\prime,\mu_i';\boldsymbol  k_i,\mu_i)
D^{j_i\ast}_{\mu_i'\bar \mu_i '}[\underline R_\mathrm{W_{\!c}}^{-1}(V,\varLambda)]  
D^{j_i}_{\mu_i\bar \mu_i}[\underline R_\mathrm{W_{\!c}}^{-1}(V,\varLambda)]\nonumber\\
\end{eqnarray}
with $i=\mathrm {e,\,q,\,N}$.
For the bound-state current $J_{\mathrm C}^\mu(\boldsymbol k_\mathrm C^\prime,\mu_j^\prime;\boldsymbol k_\mathrm C,\mu_j;K_{\mathrm e})$ one finds the same transformation properties as in Eq.~(\ref{eq:transformpropcomcurrents}).
We see that our currents do not behave like four-vectors under Lorentz transformations. Instead, they transform by a Wigner rotation $R_\mathrm{W_{\!c}}(V,\varLambda)$. Currents with the correct covariance
properties, however, are obtained by going back to the physical
particle momenta $p_i^{(\prime)}$ and spin projections $\sigma_i^{(\prime)}$ of Eq.~(\ref{eq:npartstate}) by means of a canonical boost with velocity $\boldsymbol{V}$ (cf. Eq.~(\ref{eq:commom})) and a Wigner rotation $R_\mathrm{W_{\!c}}^{-1}(w_i^{(\prime)}, B_{\mathrm c}(\boldsymbol V))$, respectively.
Such \textit{covariant currents} are defined by~\cite{Biernat:2009my}
\begin{eqnarray}\label{eq:physicalphenocurrent}
 J_{i}^\mu( \boldsymbol p_i^\prime, \sigma_i^\prime; \boldsymbol p_i,\sigma_i)&:=&B_{\mathrm c}(\boldsymbol V)_{\,\,\nu}^\mu
J_{i}^\nu(\boldsymbol k_i^\prime,\mu_i';\boldsymbol  k_i,\mu_i)\nonumber\\&&\times 
D^{j_i\ast}_{\mu_i '\sigma_i' }[\underline R_\mathrm{W_{\!c}}^{-1}(w_i',B_{\mathrm c}(\boldsymbol V))]  
D^{j_i}_{\mu_i \sigma_i }[\underline R_\mathrm{W_{\!c}}^{-1}(w_i,B_{\mathrm c}(\boldsymbol V))]\nonumber\\
\end{eqnarray}
with $i=\mathrm {e,\,q,\,N}$. It is checked in App.~\ref{app:covphenomenocurrents} that these currents exhibit the correct covariance properties under Lorentz transformations.
Similarly, the bound-state current $J_{\mathrm C}^\mu(\boldsymbol p_\mathrm C^\prime,\sigma_j^\prime;\boldsymbol p_\mathrm C,\sigma_j;P_{\mathrm e})$ with $P_{\mathrm e}=p_{\mathrm e}+p_{\mathrm e}'$ is obtained according to the prescription of Eq.~(\ref{eq:physicalphenocurrent}).\footnote{The term \lq physical current' for $J_{\mathrm C}^\mu(\boldsymbol p_\mathrm C^\prime,\sigma_j^\prime;\boldsymbol p_\mathrm C,\sigma_j;P_{\mathrm e})$ is avoided due to the \textit{unphysical} contributions proportional to $P_{\mathrm e}$ which may occur in the expression for this current. Therefore, we rather speak in this context of \textit{covariant currents}. We use the same notation both for the covariant current $J_{\mathrm C}^\mu(\boldsymbol p_\mathrm C^\prime,\sigma_j^\prime;\boldsymbol p_\mathrm C,\sigma_j;P_{\mathrm e})$ and for the current $J_{\mathrm C}^\mu(\boldsymbol k_\mathrm C^\prime,\mu_j^\prime;\boldsymbol k_\mathrm C,\mu_j;K_{\mathrm e})$. The difference is indicated by the arguments.}

\chapter{Relativistic Coupled-Channel Problem}\label{chap:4}
\section{Introduction}
In this chapter we look at elastic electron-bound-state scattering in the center-of-momentum frame. This process is treated in the one-photon-approximation as a coupled-channel problem for a Bakamjian-Thomas-type mass operator. The corresponding system of coupled equations can be reduced to an eigenvalue problem for the elastic channel with a one-photon-exchange optical potential. \lq On-shell' matrix elements of this optical potential in the velocity state basis are sufficient for the identification and extraction of the electromagnetic bound-state current. Corresponding analytical expressions in terms of bound-state wave functions and constituent currents will be derived. The following formalism can also be found in Ref.~\cite{Biernat:2009my} for the particular case of the bound state being a pseudoscalar meson.

\section{Electron-Bound-State Scattering}
\subsection{Coupled-Channel Problem}
The kinematic group of the point form is the Lorentz group which makes a manifest Lorentz covariant formulation feasible. It is easy in the point form to obtain a relativistic generalization of the Schr\"odinger equation. This is achieved by simply replacing the non-relativistic Hamiltonian, i.e. the energy operator, by the 
four-momentum operator, whose components contain all interactions of the system. The resulting dynamical equations to be solved are given by the eigenvalue problem for the four-momentum operator~\cite{Klink:2000pp}:
\begin{eqnarray}\label{eq:dyneq}
\hat P^\mu\vert\varPsi\rangle=P^\mu\vert\varPsi\rangle\,.
\end{eqnarray}
We have already pointed out in Sec.~\ref{sec:BTconst} that a point-form Bakamjian-Thomas construction results in a total four-momentum operator that is the product of the free four-velocity operator times the total mass operator, cf. Eq.~(\ref{eq:PFBT}). The latter contains all information about the dynamics of the system. As a consequence it suffices to study the eigenvalue problem for the mass operator. Therefore Eq.~(\ref{eq:dyneq}) can be reduced to
\begin{eqnarray}\label{eq:masseq}
\hat M\vert\varPsi\rangle=M\vert\varPsi\rangle\,.
\end{eqnarray}
Our aim is to construct a Bakamjian-Thomas type mass operator $\hat M$ that describes photon exchange in such a way that retardation effects are fully taken into account. This can be achieved within a coupled-channel framework that allows for the creation and annihilation of a photon by means of the vertex operators defined in the previous chapter. To this end we treat the electromagnetic scattering of an electron by a bound two-body system as a two-channel problem for an appropriately defined Bakamjian-Thomas type mass operator. Such a mass operator acts on a Hilbert space that is the direct sum of the $12\mathrm e$ and $12\mathrm e\gamma$ Hilbert spaces. It can be written as a ($2\times2$)-matrix operator~\cite{Klink:2000pp}
\begin{equation}\label{eq:coupledchannelmassoperator}
 \hat M:=\hat M_{\mathrm Ce(\gamma)}+\hat M_{\mathrm {int}}=\left( \begin{array}{ll} \hat{M}_{\mathrm C\mathrm e } & 0 \\
0 & \hat{M}_{\mathrm C\mathrm e \gamma}
\end{array}\right) +\left( \begin{array}{ll} 0 & \hat{K} \\
\hat{K}^\dag & 0
\end{array}\right)\,.
\end{equation}
The diagonal elements of this mass operator, $\hat{M}_{ \mathrm C\mathrm e}$ and $\hat{M}_{ \mathrm C\mathrm e\gamma}$ of Eqs.~(\ref{eq:totalmassop}) and~(\ref{eq:totalmassop4body}), respectively,
are the invariant mass operators of the uncoupled $\mathrm C\mathrm e$- and
$\mathrm C\mathrm e \gamma$-systems. They may contain instantaneous two-body interactions between the constituents 1 and 2 (in the way introduced in Sec.~\ref{sec:2binter}). The off-diagonal terms $\hat{K}^\dag$ and $\hat{K}$ are the vertex operators defined by Eq.~(\ref{eq:vertexoper1}) and they account for the creation and annihilation of the photon at the electron and the constituents. In this way they couple the $12\mathrm e$ Hilbert space to the $12\mathrm e\gamma$ Hilbert space. The mass operator of Eq.~(\ref{eq:coupledchannelmassoperator}) is of point-form Bakamjian-Thomas type (cf. Eq.~(\ref{eq:commMV})) since it is defined to commute with $\hat V^\mu$ due to the delta functions in Eqs.~(\ref{eq:interactionBT3body}) and (\ref{eq:vertexoper1}).
Decomposing an arbitrary mass eigenstate of the system $\vert \varPsi \rangle$ into components belonging to the 2 channels $12\mathrm e$ and $12\mathrm e\gamma$,
\begin{equation} \vert \varPsi \rangle:=\left( \begin{array}{l} \vert \varPsi_{ \mathrm C\mathrm e}\rangle \\ \vert
\varPsi_{\mathrm C\mathrm e \gamma}\rangle
\end{array}\right),\label{eq:psi2channels}
\end{equation}
and inserting the definitions~(\ref{eq:coupledchannelmassoperator}) and~(\ref{eq:psi2channels}) into the eigenvalue equation~(\ref{eq:masseq}) we have
\begin{equation}
\left( \begin{array}{ll} \hat{M}_{\mathrm C\mathrm e } & \hat{K} \\
\hat{K}^\dag & \hat{M}_{ \mathrm C\mathrm e\gamma}
\end{array}\right) \left( \begin{array}{l} \vert \varPsi_{ \mathrm C\mathrm e}\rangle \\ \vert
\varPsi_{\mathrm C\mathrm e \gamma}\rangle
\end{array}\right)= M \left( \begin{array}{l} \vert \varPsi_{\mathrm C\mathrm e }\rangle \\ \vert
\varPsi_{\mathrm C\mathrm e \gamma}\rangle
\end{array}\right)\,.\label{eq:massev}
\end{equation} 
This is a linear system of 2 coupled equations for the
respective components:
\begin{eqnarray}\label{eq:coupchann1}
\hat{M}_{\mathrm C\mathrm e }\vert \varPsi_{ \mathrm C\mathrm e} \rangle + \hat{K} \vert \varPsi_{\mathrm C\mathrm e 
\gamma}
\rangle & = & M \vert \varPsi_{\mathrm C\mathrm e } \rangle\,,\\
\hat{K}^\dag \vert \varPsi_{\mathrm C\mathrm e } \rangle + \hat{M}_{\mathrm C\mathrm e \gamma} \vert
\varPsi_{\mathrm C\mathrm e \gamma} \rangle & = & M \vert \varPsi_{\mathrm C \mathrm e \gamma}\rangle\,\label{eq:coupchann2}
.
\end{eqnarray}
\subsection{One-Photon-Exchange Optical Potential}
\subsubsection{Constituents}
For our further purposes it is useful to apply a Feshbach
reduction to Eqs.~(\ref{eq:coupchann1}) and~(\ref{eq:coupchann2})  which means to solve the second equation for $\vert \varPsi_{\mathrm C\mathrm e  \gamma}\rangle$ and insert the resulting expression into the
first one to end up with an equation for $\vert \varPsi_{\mathrm C\mathrm e }\rangle$:
\begin{equation}
\label{eq:DynamicalEquationM} \left(\hat{M}_{\mathrm C \mathrm e }-M\right)\vert
\varPsi_{\mathrm C\mathrm e } \rangle = \underbrace{\hat{K}{ \left(\hat{M}_{\mathrm e \mathrm C\gamma}
-M\right) }^{-1} \hat{K}^\dag}_{=:\hat{
V}_\mathrm{opt}(M)}  \vert \varPsi_{\mathrm C\mathrm e } \rangle \, .
\end{equation}
The right-hand side of this equation describes the action of the
one-photon-exchange on the state $\vert \varPsi_{\mathrm C\mathrm e }\rangle$. This equation has the structure of an eigenvalue equation, but with the mass eigenvalue $M$
also appearing in the optical potential $\hat{V}_{\mathrm{opt}}(M)$. $\hat{V}_{\mathrm{opt}}(M)$ consists of
all possibilities to exchange the photon between the
electron and the constituents. It also includes loop contributions, i.e.
absorption by the emitting particle. $(M-\hat{M}_{ \mathrm C\mathrm e\gamma})^{-1}$ describes the propagation of the $\mathrm C\mathrm e \gamma$ intermediate
state and is thus responsible for retardation effects.\\
\subsubsection{Bound State}
On the bound-state level, where the structure of the bound state is not resolved in terms of constituents, the corresponding one-photon-exchange optical potential is simply obtained by replacing $\hat{K}$ with $\hat{K}^\mathrm C$ of Eq.~(\ref{eq:vertexoperKbs}) in $\hat{V}_\mathrm{opt}(M)$:
 \begin{equation}
\label{eq:DynamicalEquationM2}\hat{
V}_\mathrm{opt}^\mathrm C(M):= \hat{K}^{\mathrm C} { \left(\hat{M}_{\mathrm C\mathrm e \gamma}
-M\right) }^{-1} \hat{K}^ {\mathrm C\dag} \,,
\end{equation}
with $\hat{
V}_\mathrm{opt}^\mathrm C(M)$ acting now on a two-particle $\mathrm {Ce}$ Hilbert space.

\section{Calculation of the Bound-State Current}\label{sec:optpotcalc}
\subsection{Matrix Elements of the Optical Potential}
Our objective of this section is to derive an expression for the bound-state current $J_{\mathrm C}^\mu(\boldsymbol k_\mathrm C^\prime,\mu_j^\prime;\boldsymbol k_\mathrm C,\mu_j;K_{\mathrm e})$ in terms of bound-state wave functions and constituent currents. To this end we consider matrix elements of the optical potential $\hat{ V}_\mathrm{opt}(M)$ of Eq.~(\ref{eq:DynamicalEquationM}) between velocity states of the  electron-bound-state system. The electromagnetic current of a bound state is usually extracted from the elastic electron-bound-state scattering amplitude calculated in the one-photon-exchange approximation. This means that we can restrict our considerations to \lq on-shell' (os) matrix elements of $\hat{V}_\mathrm{opt}(M)$ for which the total invariant mass of the incoming and outgoing electron and bound state has to be the same, i.e.
\begin{eqnarray}\label{eq:onshell}
 M=k_{\mathrm e}^0+k_{\mathrm C}^0 =
k_{\mathrm e}'^0+k_{\mathrm C}'^0
\end{eqnarray}
since 
\begin{eqnarray} k_{\mathrm e}^0=k_{\mathrm e}'^0\,\quad\text{and}\quad
k_{\mathrm C}^0=k_{\mathrm C}'^0\,.\label{eq:elasticscattering}
\end{eqnarray}
Note that the velocity-state representation is associated with center-of-mass
kinematics, cf. Eq.~(\ref{eq:comconstraint}).

\begin{figure}
\begin{center}
\psfrag{e}{$\mathrm e$                }
\psfrag{C}{$\mathrm C$          }
\psfrag{c1}{$\mathrm c_1$         }
\psfrag{c2}{$\mathrm c_2$          }
\psfrag{gamma}{\;$\gamma$           }
\includegraphics[clip=7cm,width=50mm]{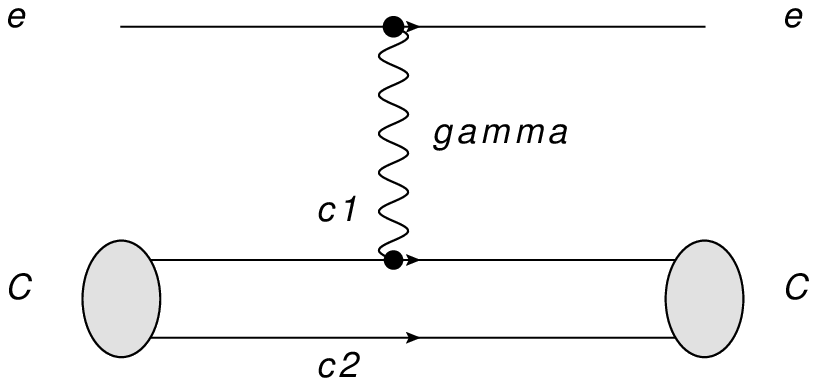}
\qquad\quad
\includegraphics[clip=7cm,width=50mm]{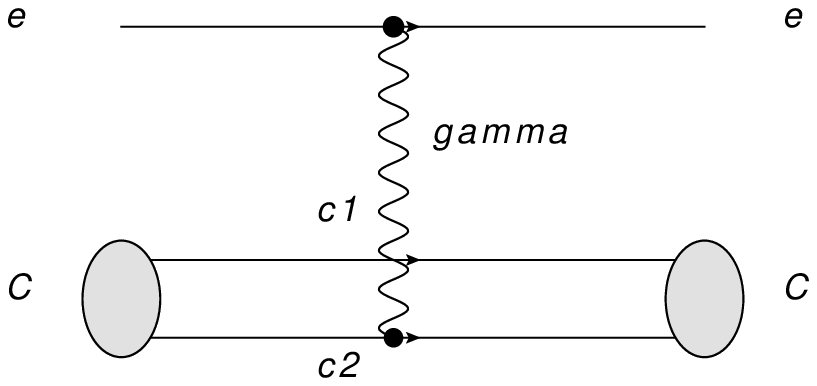}
\caption{\label{fig:graphcov}
The diagrammatic representation of the 2 covariant contributions to the one-photon-optical potential corresponding to the 2 terms in Eq.~(\ref{eq:1gammaamplitres}).  
}
\end{center} 
\end{figure}

We start our calculation by looking at on-shell matrix elements of the optical potential between clustered electron-bound-state velocity states of Sec.~\ref{sec:Two-BodyWaveFunction}:\footnote{Here we introduce the \lq underlining' of spins, momenta and velocities in order to have a further mean (beside multiple primes) for distinguishing quantum numbers in clustered from those in free velocity states. This underline should not be confused with the underline denoting elements of the SL$(2,\mathbb C)$, but the distinction should be clear from the context.}
\begin{eqnarray}\label{eq:1gammaamplit}
\lefteqn{\langle  \underline{V}^\prime;
\underline{\boldsymbol k}_{\mathrm e}^\prime, \underline{\mu}_{\mathrm e}^\prime;
\underline{\boldsymbol k}_{12}^\prime, n, j,\mu_j' \vert\,
\hat{V}_{\mathrm{opt}}(M) \vert\, \underline{V};
\underline{\boldsymbol k}_{\mathrm e}, \underline{\mu}_e; \underline{\boldsymbol k}_{12}, n,
j,\mu_j \rangle_{\mathrm{os}}} \nonumber \\
&=&  \langle  \underline{V}^\prime; \underline{\boldsymbol k}_{\mathrm e}^\prime,
\underline{\mu}_{\mathrm e}^\prime; \underline{\boldsymbol k}_{12}^\prime, n,j,\mu_j' \vert\, \hat{K} \left(\hat{M}_{\mathrm {e C} \gamma}
-M\right)^{-1} \hat{K}^\dag
\nonumber \\
&&\times \vert\, \underline{V};
\underline{\boldsymbol k}_{\mathrm e}, \underline{\mu}_{\mathrm e};\underline{\boldsymbol k}_{12}, n,
j,\mu_j \rangle_{\mathrm{os}} \,.
\end{eqnarray}
The first step towards an evaluation of Eq.~(\ref{eq:1gammaamplit}) is a multiple insertion of the completeness relations for free and clustered velocity states, cf. Eqs.~(\ref{eq:vcompl}) and~(\ref{eq:completenesseigenstates4part}), at the appropriate places on the right-hand side of Eq.~(\ref{eq:1gammaamplit}):
\begin{eqnarray}\label{eq:optpotonshell}
\lefteqn{
 \langle  \underline{V}^\prime; \underline{\boldsymbol k}_{\mathrm e}^\prime,
\underline{\mu}_{\mathrm e}^\prime; \underline{\boldsymbol k}_{12}^\prime, n,j,\mu_j' \vert\,\hat 1_{12\mathrm e}' \,\hat{K}\,\hat 1_{12\mathrm e\gamma}'''\, \left(\hat{M}_{\mathrm {e C} \gamma}
-M\right)^{-1} \hat 1_{\mathrm {Ce}\gamma}'' \nonumber }\\&&\times \hat 1_{12\mathrm e\gamma}''\,\hat{K}^\dag\,\hat 1_{12\mathrm e}\, \vert\, \underline{V};
\underline{\boldsymbol k}_{\mathrm e}, \underline{\mu}_{\mathrm e};\underline{\boldsymbol k}_{12}, n,
j,\mu_j \rangle_{\mathrm{os}} \,.
\end{eqnarray}
Next we use the orthogonality relations for the free and clustered velocity states, Eqs.~(\ref{eq:vnorm}) and~(\ref{eq:normeigenstates4part}), respectively, together with the expressions for the matrix elements of the vertex operators, Eq.~(\ref{eq:vertexoper1}), and the definitions of the wave functions, Eqs.~(\ref{eq:norm3partcl}) and~(\ref{eq:norm4partcl}). Then the necessary integrations and sums can be done by means of the appropriate Dirac and Kronecker deltas, respectively. 
In the rather lengthy calculation, which is presented in detail in App.~\ref{app:1phoptpotmatrixelemnts}, we neglect the 5 self-energy contributions in which the photon is emitted and absorbed by the same particle.\footnote{Photon exchange between particles 1 and 2 is also excluded as it is a self-energy contribution to the bound state.} The remaining 4 time-ordered contributions can be combined to 2 covariant contributions corresponding to the photon exchange between the electron and the two constituents. The result has the structure of a contraction of the electron current with the sum of 2 integrals (involving the wave functions and the constituent currents) multiplied with the covariant photon propagator:
\begin{eqnarray} \label{eq:1gammaamplitres} \lefteqn{\langle  \underline{V}^\prime;
\underline{\boldsymbol k}_{\mathrm e}^\prime, \underline{\mu}_{\mathrm e}^\prime;
\underline{\boldsymbol k}_{12}^\prime, n, j,\mu_j' \vert\,
\hat{V}_{\mathrm{opt}}(M) \vert\, \underline{V};
\underline{\boldsymbol k}_{\mathrm e}, \underline{\mu}_e; \underline{\boldsymbol k}_{12}, n,
j,\mu_j\rangle_{\mathrm{os}}} \nonumber \\
&=& 
(2\pi)^3\underline V^0\delta ^3(\underline {\boldsymbol V}-\underline {\boldsymbol V}')
\sqrt{\frac{2 k'^0_{\mathrm C}}{(k'^0_{\mathrm C}+\underline k'^0_{\mathrm e})^3}}\sqrt{\frac{2 k^0_{\mathrm C}}{(k^0_{\mathrm C}+\underline k^0_{\mathrm e})^3}}
 \nonumber\\&&\times
\sum_{\mu_1'\mu_2'}\left[\int \frac{\mathrm d^3 k_{1}'}{2 k_{1}'^0}\frac{1}{2k_{2}'^0}\frac{1}{2 k_{1}^0}\sqrt{\frac{2\tilde k'^0_12\tilde k'^0_2}{2(\tilde k'^0_1+\tilde k'^0_2)}}\sqrt{\frac{2\tilde k^0_12\tilde k^0_2}{2(\tilde k^0_1+\tilde k^0_2)}}
\sqrt{2 k'^0_{12}}\sqrt{2 k^0_{12}}\right.\nonumber\\&&\;\;\;\;\times\left.
\sum_{\mu_1}\varPsi^\ast _{nj\mu_j'\mu_1'\mu_2'}(\tilde{\boldsymbol k}')\varPsi _{nj\mu_j\mu_1\mu_2'}(\tilde{\boldsymbol k})
 J_{1\nu}(\boldsymbol{k}_1',\mu_1';\boldsymbol{k}_1,\mu_1)\right.\nonumber\\&&\;\;+\left.
\int \frac{\mathrm d^3 k_{2}'}{2 k_{2}'^0}\frac{1}{2k_{1}'^0}\frac{1}{2 k_{2}^0}\sqrt{\frac{2\tilde k'^0_12\tilde k'^0_2}{2(\tilde k'^0_1+\tilde k'^0_2)}}\sqrt{\frac{2\tilde k^0_12\tilde k^0_2}{2(\tilde k^0_1+\tilde k^0_2)}}
\sqrt{2 k'^0_{12}}\sqrt{2 k^0_{12}}\right.\nonumber\\&&\;\;\;\;\times\left.
\sum_{\mu_2}\varPsi^\ast _{nj\mu_j'\mu_1'\mu_2'}(\tilde{\boldsymbol k}')\varPsi _{nj\mu_j\mu_1'\mu_2}(\tilde{\boldsymbol k})
 J_{2\nu}(\boldsymbol{k}_2',\mu_2';\boldsymbol{k}_2,\mu_2) 
\right]\nonumber\\&&\times
\frac{(-\mathrm g^{\mu\nu})}{Q^2}\,|\,\mathrm e\,|\,
Q_{\mathrm e}\,
\bar{u}_{\underline{\mu}_{\mathrm e}'}(\boldsymbol{k}_{\mathrm e}^\prime)\gamma_\mu
u_{\underline{\mu}_{\mathrm e}}(\boldsymbol{k}_e)\,.
\end{eqnarray}
A diagrammatic representation of these contributions is depicted in Fig.~\ref{fig:graphcov}. Here the blobs that connect the $\mathrm c_1$ and $\mathrm c_2$ lines symbolize integrals over wave functions of the incoming and outgoing bound state.
The denominator of the covariant photon propagator, which
includes both time orderings, is given by $-q_\mu q^\mu=Q^2$ with $q^\mu = k_{\mathrm C}'^\mu
- k_{\mathrm C}^\mu$ being the four-momentum transfer between electron and bound state.\footnote{Note that $\boldsymbol{q}=\boldsymbol{k}_\gamma$, but $q^0
\neq k_\gamma^0 = \vert \boldsymbol{k}_\gamma \vert$.} 
Momenta with and without prime are related by $\boldsymbol{k}_i^{ \prime} =
\boldsymbol{k}_i + \boldsymbol{q}
= \boldsymbol{k}_i + \boldsymbol{k}_\gamma$ with $i=1,\,2$ and $\boldsymbol{k}_\mathrm e^{ \prime} =
\boldsymbol{k}_\mathrm e - \boldsymbol{k}_\gamma$. This means that we have three-momentum conservation
at the electromagnetic vertices, a property which one would not expect
in point-form quantum mechanics. Here it results from the
velocity-state representation, in particular the associated
center-of-mass kinematics and the spectator conditions for the inactive particles. The energy is, however, not conserved at the vertices. For the physical momenta, i.e. the
center-of-mass momenta boosted by $B_\mathrm c(\boldsymbol V)$, none of the four-momentum
components is, in
general (if $\boldsymbol V \neq 0$), conserved at the electromagnetic vertices.\footnote{Note that boosts mix spatial and temporal components.}\\
\subsection{Matrix Elements of the Optical Potential on the Bound-State Level}
In a similar way as in the previous section on the constituent level we can also calculate on-shell velocity-state matrix elements of the optical potential $\hat V_{\mathrm{opt}}^\mathrm C(M)$ given by Eq.~(\ref{eq:DynamicalEquationM2}). We work on the bound-state level which means that the structure of the bound state is not resolved in terms of constituents. The purpose of this calculation is the identification of the bound-state current from Eq.~(\ref{eq:1gammaamplitres}). In particular, it serves for the identification of the kinematical factors belonging to the current, which stand in front of the integral of Eq.~(\ref{eq:1gammaamplitres}).
The explicit calculation will not be demonstrated here, instead we refer to the thorough derivation in Ref.~\cite{Fuchsberger:2007}. It is briefly summarized as follows: one starts with the insertion of a velocity-state completeness relation at the appropriate place of matrix elements of $\hat V_{\mathrm{opt}}^\mathrm C(M)$. Then one uses the matrix elements of the vertex operators between velocity states, Eq.~(\ref{eq:vertexoperKbs}), in order to carry out the necessary integrals and sums
 by means of the corresponding Dirac and Kronecker deltas, respectively. If self-energy contributions
are again neglected the optical potential consists of 2 terms corresponding to the 2 possible time orderings. Restricting ourselves to on-shell matrix elements, cf. Eq.~(\ref{eq:onshell}), we can combine the 2 time orderings to the covariant photon propagator, cf. Eq.~(\ref{eq:covphotonprop}).
The final result is then given by~\cite{Fuchsberger:2007,Biernat:2009my}
\begin{eqnarray}\label{eq:1gammaamplitbslevel}
\lefteqn{\langle  V^\prime;
\boldsymbol k_{\mathrm e}^\prime,\mu_{\mathrm e}^\prime;
\boldsymbol k_{12}^\prime,\mu_j'\vert\,
\hat{V}_{\mathrm{opt}}^{\mathrm{C}}(M) \vert\,V;
\boldsymbol k_{\mathrm e}, \mu_{\mathrm e}; \boldsymbol k_{12},\mu_j \rangle_{\mathrm{os}}  }\nonumber \\
&=&(2\pi)^3V^0\delta^3(\boldsymbol V-\boldsymbol V')\frac{1}{\sqrt{(k_{\mathrm e}'^0+k_{\mathrm C}'^0)^3}}\frac{1}{\sqrt{(k_{\mathrm e}^0+k_{\mathrm C}^0)^3}}
\nonumber\\
&&\times \,|\,\mathrm e\,|\,Q_{\mathrm e}\bar{u}_{\mu_{\mathrm e}'}(\boldsymbol{k}_{\mathrm e}^\prime)\gamma_\mu
u_{\mu_{\mathrm e}}(\boldsymbol{k}_e)\frac{(-\mathrm g^{\mu\nu})}{Q^2}J_{\mathrm C\nu}(\boldsymbol k_{\mathrm C}',\mu_j';\boldsymbol k_{\mathrm C},\mu_j;K_\mathrm e)\,.
\end{eqnarray}
The expression on the right-hand side reveals that on-shell matrix elements of the optical
potential can be expressed as a contraction of the electromagnetic
electron current $|\,\mathrm e\,|\,Q_{\mathrm e}\bar{u}_{\mu_{\mathrm e}'}(\boldsymbol{k}_{\mathrm e}^\prime)\gamma_\mu u_{\mu_{\mathrm e}}$
with an electromagnetic bound-state current
$J_{\mathrm C\nu}(\boldsymbol k_{\mathrm C}',\mu_j';\boldsymbol k_{\mathrm C},\mu_j;K_\mathrm e)$
multiplied with the covariant photon propagator $-g^{\mu\nu}/Q^2$
(and a kinematical factor). Apart from the kinematical factor in front, the right-hand side of Eq.~(\ref{eq:1gammaamplitbslevel}) corresponds to the familiar one-photon exchange amplitude for elastic electron-bound-state scattering (calculated in the electron-bound-state center-of-mass system).
 
It should be mentioned that, in general, the four-momentum transfer between incoming and outgoing (active) constituent $q_1^\mu:=(k_1^\prime-k_1)^\mu$ deviates from the four-momentum transfer between incoming and outgoing cluster $q^\mu:=(k_\mathrm C^\prime-k_\mathrm C)^\mu$.
This can be seen as follows: whereas the three-momentum transfers are the same on the bound-state and
on the constituent level, i.e.
$\boldsymbol{q}=\boldsymbol{q}_1$, the
zero components differ, i.e. $q^0\neq q_1^0$. Due to the
center-of-mass kinematics associated with the velocity states
$k_\mathrm C^0=k_\mathrm C'^0$ and
hence $q^0=0$. On the other hand it can be shown that
\begin{eqnarray}
 (k_1'^0)^{2}=(k_1^0)^2-2 \boldsymbol{q}\cdot
\boldsymbol{k}_{2}^{\, \prime}\, \quad \text{so that} \quad q_1^0\neq 0\,.
\end{eqnarray}
 Accordingly, not all of the four-momentum transferred to the cluster is also transferred to the active constituent. \\
\subsection{Identifying the Bound-State Current}
By comparison of Eq.~(\ref{eq:1gammaamplitbslevel}) with Eq.~(\ref{eq:1gammaamplitres}) we can identify the bound-state current as
\begin{eqnarray} \label{eq:bscurrent3}
\lefteqn{J^\mu_{\mathrm C}(\boldsymbol k_{\mathrm C}',\mu_j';\boldsymbol k_{\mathrm C},\mu_j;K_\mathrm e)}\nonumber\\&=&
\sqrt{2 k^0_{\mathrm C}}\sqrt{2 k'^0_{\mathrm C}}
\sum_{\mu_1'\mu_2'}\left[\int \frac{\mathrm d^3 k_{1}'}{2 k_{1}'^0}\frac{1}{2k_{2}'^0}\frac{1}{2 k_{1}^0}\sqrt{\frac{2\tilde k'^0_12\tilde k'^0_2}{2(\tilde k'^0_1+\tilde k'^0_2)}}\sqrt{\frac{2\tilde k^0_12\tilde k^0_2}{2(\tilde k^0_1+\tilde k^0_2)}}
\right.\nonumber\\&&\;\;\;\;\times\left.
\sqrt{2 k'^0_{12}}\sqrt{2 k^0_{12}}\sum_{\mu_1}\varPsi^\ast _{nj\mu_j'\mu_1'\mu_2'}(\tilde{\boldsymbol k}')\varPsi _{nj\mu_j\mu_1\mu_2'}(\tilde{\boldsymbol k})
 J_{1}^\mu(\boldsymbol{k}_1'\mu_1';\boldsymbol{k}_1,\mu_1)\right.\nonumber\\&&\;\;+\left.
\int \frac{\mathrm d^3 k_{2}'}{2 k_{2}'^0}\frac{1}{2k_{1}'^0}\frac{1}{2 k_{2}^0}\sqrt{\frac{2\tilde k'^0_12\tilde k'^0_2}{2(\tilde k'^0_1+\tilde k'^0_2)}}\sqrt{\frac{2\tilde k^0_12\tilde k^0_2}{2(\tilde k^0_1+\tilde k^0_2)}}
\sqrt{2 k'^0_{12}}\sqrt{2 k^0_{12}}\right.\nonumber\\&&\;\;\;\;\times\left.
\sum_{\mu_2}\varPsi^\ast _{nj\mu_j'\mu_1'\mu_2'}(\tilde{\boldsymbol k}')\varPsi _{nj\mu_j\mu_1'\mu_2}(\tilde{\boldsymbol k})
 J_{2}^\mu(\boldsymbol{k}_2',\mu_2';\boldsymbol{k}_2,\mu_2) 
\right]\,.
\end{eqnarray}
\subsubsection{Simplifications for Equal Mass and Spin of the Constituents}
In the this work we restrict our considerations to bound systems consisting of 2 equal-mass equal-spin constituents. For the treatment of systems with unequal constituent masses within the present formalism, like heavy-light bound systems, we refer to Refs.~\cite{Rocha:2010uu,Rocha:2010wm}. According to our case, we assume the constituents to have equal masses, i.e. $m_1=m_2=:m$ and same spins, i.e. $j_1=j_2=\frac12$, which simplifies the expression for the current, Eq.~(\ref{eq:bscurrent3}), even further. First we manipulate the second term of Eq.~(\ref{eq:bscurrent3}) by simply renaming momenta and spins (which are integrated and summed over). We make the replacements 
 \begin{eqnarray}
  k_1'\leftrightarrow  k_2'\,,\quad k_1\leftrightarrow  k_2 \quad\text{and}\quad \mu_1'\leftrightarrow \mu_2'\,,\quad \mu_1\leftrightarrow \mu_2\,.
 \end{eqnarray}
Consequently, using the symmetry of the Clebsch-Gordan coefficients, i.e. $C^{s\mu_s}_{j_1\mu_1j_2\mu_2}=C^{s\mu_s}_{j_2\mu_2j_1\mu_1}$,
we can write Eq.~(\ref{eq:bscurrent3}) in a more compact form:
\begin{eqnarray}\label{eq:bscurrentequalmasses}
\lefteqn{J^\mu_{\mathrm C}(\boldsymbol k_{\mathrm C}',\mu_j';\boldsymbol k_{\mathrm C},\mu_j;K_\mathrm e)}\nonumber\\&=&
\sqrt{2 k^0_{\mathrm C}}\sqrt{2 k'^0_{\mathrm C}}
\sum_{\mu_1'\mu_2'\mu_1}\int \frac{\mathrm d^3 k_{1}'}{2 k_{1}'^0}\frac{1}{2k_{2}'^0}\frac{1}{2 k_{1}^0}\sqrt{\frac{2\tilde k'^0_12\tilde k'^0_2}{2(\tilde k'^0_1+\tilde k'^0_2)}}\sqrt{\frac{2\tilde k^0_12\tilde k^0_2}{2(\tilde k^0_1+\tilde k^0_2)}}
\nonumber\\&&\times\sqrt{2 k'^0_{12}}\sqrt{2 k^0_{12}}\varPsi^\ast _{nj\mu_j'\mu_1'\mu_2'}(\tilde{\boldsymbol k}')\varPsi _{nj\mu_j\mu_1\mu_2'}(\tilde{\boldsymbol k})\nonumber\\&&\times\left[
 J_{1}^\mu(\boldsymbol{k}_1',\mu_1';\boldsymbol{k}_1,\mu_1)+
 J_{2}^\mu(\boldsymbol{k}_1',\mu_1';\boldsymbol{k}_1,\mu_1) 
\right]
\,.
\end{eqnarray}
A further simplification of this expression is achieved by going from the $k_{1}'$-integration to a $\tilde k_{1}'$-integration by means of
the appropriate Jacobian, cf. Eq.~(\ref{eq:momchangekktilde}). Note that for equal masses we have $\tilde k^0_1=\tilde k^0_2=m_{12}/2$ and $\tilde k'^0_1=\tilde k'^0_2=m_{12}'/2$. Moreover, it can be shown that the spinor product for anti-particles, $\bar v_{\mu_1}(\boldsymbol k_1)\gamma^\mu v_{\mu_1'}(\boldsymbol k_1')$, can be replaced by the one for particles, $\bar u_{\mu_1'}(\boldsymbol k_1')\gamma^\mu u_{\mu_1}(\boldsymbol k_1)$, cf. Eq.~(\ref{eq:ugammauevgammav}). Writing out the wave functions defined in Eq.~(\ref{eq:wavefunctions}) we can combine the Wigner $D$-functions for the inactive (spectator) constituent 2 by means of Eq.~(\ref{eq:RWprop1}) due to the spectator condition $\boldsymbol k_2=\boldsymbol k_2'$:
\begin{eqnarray}
 \underline R_\mathrm{W_{\!c}}^{-1}\left(\tilde w_2', B_{\mathrm c}(\boldsymbol w_{12}')\right)\underline R_\mathrm{W_{\!c}}\left(\tilde w_2, B_{\mathrm c}(\boldsymbol w_{12})\right)=
\underline R_\mathrm{W_{\!c}}\left(\tilde w_2, B^{-1}_{\mathrm c}(\boldsymbol w_{12}') B_{\mathrm c}(\boldsymbol w_{12})\right)\,.\nonumber\\
\end{eqnarray}
With these manipulations the expression for the bound-state current finally reads
\begin{eqnarray}\label{eq:bscurrentequalmasses2} 
\lefteqn{J^\mu_{\mathrm C}(\boldsymbol k_{\mathrm C}',\mu_j';\boldsymbol k_{\mathrm C},\mu_j;K_\mathrm e)}\nonumber\\&=&
\sqrt{2 k^0_{\mathrm C}}\sqrt{2 k'^0_{\mathrm C}}
\sum\int \frac{\mathrm d^3 \tilde k_{1}'}{ k_{1}^0}
\sqrt{\frac{m_{12}}{m_{12}'}}
\sqrt{\frac{ k^0_{12}}{k'^0_{12}}}\nonumber\\&&\times
Y_{l'\mu_l'}^\ast(\hat{\tilde{\boldsymbol{ k}}}')
C^{s'\mu_s'\ast}_{\frac12\tilde \mu_1'\frac12\tilde \mu_2'}C^{j\mu_j'\ast}_{l'\mu_l's'\mu_s'} u^{j\ast}_{nl's'}(\tilde{k}')
Y_{l\mu_l}(\hat{\tilde{\boldsymbol{ k}}})
C^{s\mu_s}_{\frac12\tilde \mu_1\frac12\tilde \mu_2}C^{j\mu_j}_{l\mu_ls\mu_s} u_{nls}^j(\tilde{k})
\nonumber\\&&\times
D^{\frac12}_{\tilde \mu_1'\mu_1'}\left[\underline R_\mathrm{ Wc}^{-1}\left(\tilde w_1', B_{\mathrm c}(\boldsymbol w_{12}')\right)\right]
D^{\frac12}_{\mu_1\tilde \mu_1}\left[\underline R_\mathrm { Wc}\left(\tilde w_1, B_{\mathrm c}(\boldsymbol w_{12})\right)\right]
\nonumber\\&&\times D^{\frac12}_{\tilde \mu_2'\tilde \mu_2}\left[\underline R_\mathrm { Wc}\left(\tilde w_2, B^{-1}_{\mathrm c}(\boldsymbol w_{12}') B_{\mathrm c}(\boldsymbol w_{12})\right)\right] 
\nonumber\\&&\times\,|\,\mathrm e\,|\,\bar u_{\mu_1'}(\boldsymbol k_1')(\varGamma_1+\varGamma_2)^\mu u_{\mu_1}(\boldsymbol k_1)
 \,.
\end{eqnarray}
Here $\varGamma^\mu_i$ is the $\mathrm {c}_i\gamma\mathrm {c}_i$-vertex. It is given for point-like and non-point-like constituents like quarks and nucleons by the expressions
\begin{eqnarray}
 \varGamma^\mu_\mathrm q=Q_\mathrm q \gamma^\mu\quad\text{and} \quad
 \varGamma^\mu_\mathrm N=F^{\mathrm N}_1(Q_{\mathrm N}^2) \gamma^\mu+F^{\mathrm N}_2(Q_{\mathrm N}^2)\frac{\mathrm i\, q_{\mathrm N\nu} \sigma^{\mu\nu}}{2m_{\mathrm N}}\,,
\end{eqnarray}
respectively.
 \section{Current Properties} 
In this section we investigate hermiticity, covariance and continuity properties of our bound-state current given by Eq.~(\ref{eq:bscurrentequalmasses2}). The two former can be discussed without specifying the spin of the bound state. For the latter we consider the particular cases of a pseudoscalar and a vector bound state.
\subsection{Hermiticity}
\label{sec:hermiticity}
Hermiticity of the current means that
\begin{eqnarray}\label{eq:hermiticitybsc}
 J^\mu_{\mathrm C}(\boldsymbol k_{\mathrm C}',\mu_j';\boldsymbol k_{\mathrm C},\mu_j;K_\mathrm e)&=& \left[J^\mu_{\mathrm C}(\boldsymbol k_{\mathrm C}',\mu_j';\boldsymbol k_{\mathrm C},\mu_j;K_\mathrm e)\right]^\dag\nonumber\\&=&J^{\mu\ast}_{\mathrm C}(\boldsymbol k_{\mathrm C},\mu_j;\boldsymbol k_{\mathrm C}',\mu_j';K_\mathrm e)\,,
\end{eqnarray}
which amounts to showing invariance under interchange of all primed with unprimed momenta and spins and complex conjugation. To this we 
show hermiticity of the bound-state current Eq.~(\ref{eq:bscurrentequalmasses2}) as follows: we write down $J^{\mu\ast}_{\mathrm C}(\boldsymbol k_{\mathrm C},\mu_j;\boldsymbol k_{\mathrm C}',\mu_j';K_\mathrm e)$ from Eq.~(\ref{eq:bscurrentequalmasses2}) by interchanging all primed and unprimed momenta and spins and perform complex conjugation:
\begin{eqnarray} \label{eq:bscurrentherm}
\lefteqn{J^{\mu\ast}_{\mathrm C}(\boldsymbol k_{\mathrm C},\mu_j;\boldsymbol k_{\mathrm C}',\mu_j';K_\mathrm e)}\nonumber\\&=&
\sqrt{ k^0_{\mathrm C} k'^0_{\mathrm C}}\sum\int \frac{\mathrm d^3 \tilde k_{1}}{ k_{1}'^0}
\sqrt{\frac{m_{12'}}{m_{12}}}
\sqrt{\frac{ k'^0_{12}}{k^0_{12}}}\nonumber\\&&\times
Y_{l'\mu_l'}^\ast(\hat{\tilde{\boldsymbol{ k}}}')
C^{s'\mu_s'\ast}_{\frac12\tilde \mu_1'\frac12\tilde \mu_2'}C^{j\mu_j'\ast}_{l'\mu_l's'\mu_s'} u^{j\ast}_{nl's'}(\tilde{k}')
Y_{l\mu_l}(\hat{\tilde{\boldsymbol{ k}}})
C^{s\mu_s}_{\frac12\tilde \mu_1\frac12\tilde \mu_2}C^{j\mu_j}_{l\mu_ls\mu_s} u_{nls}^j(\tilde{k})
\nonumber\\&&\times
D^{\frac12\ast }_{\tilde \mu_1\mu_1}\left[\underline R_\mathrm{W_{\!c}}^{-1}\left(\tilde w_1, B_{\mathrm c}(\boldsymbol w_{12})\right)\right]
D^{\frac12\ast}_{\mu_1'\tilde \mu_1'}\left[\underline R_\mathrm{W_{\!c}}\left(\tilde w_1', B_{\mathrm c}(\boldsymbol w_{12}')\right)\right]
\nonumber\\&&\times D^{\frac12\ast}_{\tilde \mu_2\tilde \mu_2'}\left[\underline R_\mathrm{W_{\!c}}\left(\tilde w_2', B^{-1}_{\mathrm c}(\boldsymbol w_{12}) B_{\mathrm c}(\boldsymbol w_{12}')\right)\right] 
\nonumber\\&&\times\,|\,\mathrm e\,|\,\left[\bar u_{\mu_1}(\boldsymbol k_1)(\varGamma_1+\varGamma_2)^\mu u_{\mu_1'}(\boldsymbol k_1')\right]^\ast
 \,.
\end{eqnarray} 
Next we transform the integration measure in Eq.~(\ref{eq:bscurrentherm}) according to Eq.~(\ref{eq:momchangekktilde})
\begin{eqnarray}\label{eq:integrationmeausremanip}
\mathrm{d}^3 \tilde{k}_1=\mathrm{d}^3 \tilde{k}'_1
\frac{m_{12}}{m_{12}' }\frac{k_{12}^0}{k_{12}'^0}
\frac{k_1'^0}{k_1^0}
\end{eqnarray}
since $\mathrm{d}^3 k'_1=\mathrm{d}^3 k_1$.
Further the Wigner $D$-functions can be rewritten in their original form by using Eq.~(\ref{eq:RWprop1}). Finally we use the hermiticity of the constituent current, Eq.~(\ref{eq:nuclcurrenthermiticity}), to rewrite it in its original form. This shows that the right-hand sides of Eq.~(\ref{eq:bscurrentherm}) and Eq.~(\ref{eq:hermiticitybsc}) are equal.
\subsection{Covariance}   
In this section we discuss the covariance properties of the current given by Eq.~(\ref{eq:bscurrentequalmasses2}). For simplicity,
we restrict our investigation to the case of a pure s-wave with $l=l'=0$, which has already been discussed in Sec.~\ref{sec:imbedding3Hilberts}. In this case the spherical harmonics become a constant, $Y_{00}^\ast(\hat{\tilde{\boldsymbol{ k}}}^{\prime})=Y_{00}(\hat{\tilde{\boldsymbol{ k}}})=1/\sqrt{4\pi}$ and the Clebsch-Gordans become Kronecker deltas,
$C^{j\mu_j^{\prime}\ast}_{00s^{\prime}\mu_s^{\prime}}=\delta_{s^{\prime}j}\delta_{\mu_s^{\prime}\mu_j^{\prime}}$ and $C^{j\mu_j}_{00s\mu_s}=\delta_{sj}\delta_{\mu_s\mu_j}$.
Then the expression of the bound-state current, after summation and reordering, reads
\begin{eqnarray} \label{eq:bscurrent2}
\lefteqn{J^\mu_{\mathrm C}(\boldsymbol k_{\mathrm C}',\mu_j';\boldsymbol k_{\mathrm C},\mu_j;K_\mathrm e)}\nonumber\\&=&\frac{1}{4\pi}
\sqrt{ k^0_{\mathrm C} k'^0_{\mathrm C}}\sum\int \frac{\mathrm d^3 \tilde k_{1}'}{ k_{1}^0}
\sqrt{\frac{m_{12}}{m_{12}'}}
\sqrt{\frac{ k^0_{12}}{k'^0_{12}}}
u^\ast_{n0}(\tilde{k}')
 u_{n0}(\tilde{k})
\nonumber\\&&\times
D^{\frac12}_{\tilde \mu_1'\mu_1'}\left[\underline R_\mathrm{W_{\!c}}^{-1}\left(\tilde w_1', B_{\mathrm c}(\boldsymbol w_{12}')\right)\right]
C^{j'\mu_j'\ast}_{\frac12\tilde \mu_1'\frac12\tilde \mu_2'} \nonumber\\&&\times D^{\frac12}_{\tilde \mu_2'\tilde \mu_2}\left[\underline R_\mathrm{W_{\!c}}\left(\tilde w_2, B^{-1}_{\mathrm c}(\boldsymbol w_{12}') B_{\mathrm c}(\boldsymbol w_{12})\right)\right] 
\nonumber\\&&\times C^{j\mu_j}_{\frac12\tilde \mu_1\frac12\tilde \mu_2}D^{\frac12}_{\mu_1\tilde \mu_1}\left[\underline R_\mathrm{W_{\!c}}\left(\tilde w_1, B_{\mathrm c}(\boldsymbol w_{12})\right)\right]
\nonumber\\&&\times\,|\,\mathrm e\,|\,\bar u_{\mu_1'}(\boldsymbol k_1')(\varGamma_1+\varGamma_2)^\mu u_{\mu_1}(\boldsymbol k_1)
 \,.
\end{eqnarray} 
In App.~\ref{app:covmicrocurrents} we prove that this bound-state current associated with the center-of-mass system has the same transformation properties under Lorentz transformations as the center-of-mass currents of Eq.~(\ref{eq:transformpropcomcurrents}):
  \begin{eqnarray}\label{eq:trafopropcombscurrent}
    \lefteqn{J^\mu_{\mathrm C}(\boldsymbol k_{\mathrm C}',\mu_j';\boldsymbol k_{\mathrm C},\mu_j;K_\mathrm e)}\nonumber\\&\stackrel{\varLambda}{\longrightarrow}&J^\mu_{\mathrm C}(\boldsymbol R_\mathrm{W_{\!c}} (V,\varLambda)k_{\mathrm C}',\bar \mu_j';\boldsymbol R_\mathrm{W_{\!c}}(V,\varLambda)k_{\mathrm C},\bar \mu_j;R_\mathrm{W_{\!c}}(V,\varLambda)K_\mathrm e)\nonumber\\&=&
  R_\mathrm{W_{\!c}}(V,\varLambda)^\mu_{\,\,\nu}J^\nu_{\mathrm C}(\boldsymbol k_{\mathrm C}',\mu_j';\boldsymbol k_{\mathrm C},\mu_j;K_\mathrm e) D_{\mu_j\bar\mu_j}^{j}(\underline R_\mathrm{W_{\!c}}^{-1}(V,\varLambda)) D_{\mu_j'\bar\mu_j'}^{j\ast }(\underline R_\mathrm{W_{\!c}}^{-1}(V,\varLambda))\,.
 \nonumber\\ \end{eqnarray}
This serves as a reasonableness check of our microscopic bound-state current, cf. Eq.~(\ref{eq:bscurrent3}), expressed in terms of wave functions and constituent current. The bound-state current with the correct covariance properties is obtained in the same way as for the phenomenological currents, cf. Eq.~(\ref{eq:physicalphenocurrent}), namely by going to the physical particle momenta and corresponding spins by means of a canonical boost~\cite{Biernat:2009my}:
\begin{eqnarray}\label{eq:physicalmicrocurrent}
 \lefteqn{J_{\mathrm C}^\mu( \boldsymbol p_\mathrm C^\prime, \sigma_j^\prime; \boldsymbol p_\mathrm C,\sigma_j;P_\mathrm e)}\nonumber\\&:=&B_{\mathrm c}(\boldsymbol V)_{\,\,\nu}^\mu
J_{\mathrm C}^\nu(\boldsymbol k_\mathrm C^\prime,\mu_j';\boldsymbol  k_\mathrm C,\mu_j;K_\mathrm e)\nonumber\\&&\times 
D^{j\ast}_{\mu_j '\sigma_j' }(\underline R_\mathrm{W_{\!c}}^{-1}(w_{\mathrm C}',B_{\mathrm c}(\boldsymbol V)))  
D^{j}_{\mu_j \sigma_j }(\underline R_\mathrm{W_{\!c}}^{-1}(w_{\mathrm C},B_{\mathrm c}(\boldsymbol V)))
\nonumber\\&=&\frac{1}{4\pi}
\sqrt{ k^0_{\mathrm C} k'^0_{\mathrm C}}\sum\int \frac{\mathrm d^3 \tilde k_{1}'}{ k_{1}^0}
\sqrt{\frac{m_{12}}{m_{12}'}}
\sqrt{\frac{ k^0_{12}}{k'^0_{12}}}
u^\ast_{n0}(\tilde{k}')
 u_{n0}(\tilde{k})
\nonumber\\&&\times
D^{\frac12\ast}_{\sigma_1'\tilde \mu_1'}\left[\underline R_\mathrm{W_{\!c}}\left(\tilde w_1',B_{\mathrm c}(\boldsymbol V)B_{\mathrm c}(\boldsymbol w_{12}')\right)\right]
C^{j'\mu_j'\ast}_{\frac12\tilde \mu_1'\frac12\tilde \mu_2'} \nonumber\\&&\times D^{\frac12}_{\tilde \mu_2'\tilde \mu_2}\left[\underline R_\mathrm{W_{\!c}}\left(\tilde w_2, B^{-1}_{\mathrm c}(\boldsymbol w_{12}')B_{\mathrm c}(\boldsymbol w_{12})\right)\right] 
\nonumber\\&&\times C^{j\mu_j}_{\frac12\tilde \mu_1\frac12\tilde \mu_2}D^{\frac12}_{\sigma_1\tilde\mu_1}\left[\underline R_\mathrm{W_{\!c}}\left(\tilde w_1,B_{\mathrm c}(\boldsymbol V)B_{\mathrm c}(\boldsymbol w_{12})\right)\right]
\nonumber\\&&\times\,|\,\mathrm e\,|\,\bar 
u_{\sigma_1'}(\boldsymbol p_1')(\varGamma_1+\varGamma_2)^\mu u_{\sigma_1}(\boldsymbol p_1)
\nonumber\\&&\times  
D^{j\ast}_{\mu_j '\sigma_j' }(\underline R_\mathrm{W_{\!c}}^{-1}(w_{\mathrm C}',B_{\mathrm c}(\boldsymbol V)))  
D^{j}_{\mu_j \sigma_j }(\underline R_\mathrm{W_{\!c}}^{-1}(w_{\mathrm C},B_{\mathrm c}(\boldsymbol V)))
 \,,
\end{eqnarray}
with $w_\mathrm C^{(\prime)}:=k_\mathrm C^{(\prime)}/m_\mathrm C^{(\prime)}$. Here we have used the transformation properties of the constituent current, Eq.~(\ref{eq:transformpropcomcurrents}). The covariant current $J_{\mathrm C}^\mu( \boldsymbol p_\mathrm C^\prime, \sigma_j^\prime; \boldsymbol p_\mathrm C,\sigma_j;P_\mathrm e)$ of Eq.~(\ref{eq:physicalmicrocurrent}) transforms like a four-vector under Lorentz transformations. The corresponding proof is carried out in a similar manner as for the constituent currents in App.~\ref{app:covphenomenocurrents}. 
 
\subsection{Pseudoscalar Current}
In this section we consider our bound-state current of Eq.~(\ref{eq:bscurrentequalmasses}) for the case of the bound state being a charged, massive pseudoscalar particle (PS) with $j=0$ such as a $\pi^{\pm}$ meson (charged pion). In the constituent quark model a pseudoscalar meson is described by a confined quark-antiquark ($\mathrm q\bar {\mathrm q}$) pair. In particular, $\pi^{+}$ is a bound state of a $\mathrm u$-quark and a $\bar{\mathrm d}$-quark. 
\subsubsection{Simplifications}
In the case of $j=0$ and a pure s-wave the Clebsch-Gordan coefficients in the wave functions of Eq.~(\ref{eq:bscurrentequalmasses}) can be written as 
\begin{eqnarray}
 C^{00\ast}_{\frac12\tilde{\mu}_1' \frac12\tilde{\mu}_{2}'}=\frac{(-1)^{\frac12-\tilde{\mu}_1'}}{\sqrt{2}}\delta_{\tilde{\mu}_2'-\tilde{\mu}_{1}'}\,,\quad
C^{00}_{\frac12\tilde{\mu}_1 \frac12\tilde{\mu}_{2}}=\frac{(-1)^{\frac12-\tilde{\mu}_1}}{\sqrt{2}}\delta_{\tilde{\mu}_2-\tilde{\mu}_{1}}\,.
\end{eqnarray}
With these Kronecker deltas the sums over $\tilde{\mu}_2$ and $\tilde{\mu}_2'$ can be carried out. Then the product of the Clebsch-Gordan coefficients with the Wigner $D$-function for the inactive constituent 2 becomes
\begin{eqnarray}
\label{eq:CDC}
\lefteqn{
\frac12 (-1)^{1-\tilde{\mu}_1-\tilde{\mu}_1'}D^{\frac12}_{-\tilde \mu_1'-\tilde \mu_1}\left[\underline R_\mathrm{W_{\!c}}\left(\tilde w_2, B^{-1}_{\mathrm c}(\boldsymbol w_{12}') B_{\mathrm c}(\boldsymbol w_{12})\right)\right]}
\nonumber\\&
=&\frac12 D^{\frac12}_{\tilde \mu_1\tilde \mu_1'}\left[\underline R_\mathrm{W_{\!c}}^{-1}\left(\tilde w_2, B^{-1}_{\mathrm c}(\boldsymbol w_{12}') B_{\mathrm c}(\boldsymbol w_{12})\right)\right]\,.
\end{eqnarray}
Here we have made use of the properties of the Wigner $D$-functions, Eqs.~(\ref{eq:RWprop1}) and~(\ref{eq:RWprop4}). Using this expression allows us to combine all three $D$-functions to a single one. With these findings the current for a pseudoscalar particle reads
 \begin{eqnarray} \label{eq:pscurrent}
\lefteqn{J^\mu_{\mathrm {PS}}(\boldsymbol k_{\mathrm C}',\boldsymbol k_{\mathrm C},K_\mathrm e)}\nonumber\\&=&\frac{1}{4\pi}
\frac{\sqrt{ k^0_{\mathrm C} k'^0_{\mathrm C}}}{2}\int \frac{\mathrm d^3 \tilde k_{1}'}{ k_{1}^0}
\sqrt{\frac{m_{12}}{m_{12}'}}
\sqrt{\frac{ k^0_{12}}{k'^0_{12}}}
u^\ast_{n0}(\tilde{k}')
 u_{n0}(\tilde{k})
\nonumber\\&&\times
\sum_{\mu_1'\mu_1}D^{\frac12}_{\mu_1\mu_1'}\left[\underline R_\mathrm{W_{\!c}}\left(\tilde w_1, B_{\mathrm c}(\boldsymbol w_{12})\right)
\right.\nonumber\\&&\quad\times \left.\underline R_\mathrm{W_{\!c}}^{-1}\left(\tilde w_2, B^{-1}_{\mathrm c}(\boldsymbol w_{12}') B_{\mathrm c}(\boldsymbol w_{12})\right)
\underline R_\mathrm{W_{\!c}}^{-1}\left(\tilde w_1', B_{\mathrm c}(\boldsymbol w_{12}')\right)\right]
\nonumber\\&&\times\,|\,\mathrm e\,|\,\bar u_{\mu_1'}(\boldsymbol k_1')(\varGamma_1+\varGamma_2)^\mu u_{\mu_1}(\boldsymbol k_1)
 \,.
\end{eqnarray} 
The argument of the single Wigner $D$-function in this expression can be simplified by using some relations for canonical boosts in the SL(2,$\mathbb C$) representation, which are collected in App.~\ref{app:boostSL2C}:
\begin{eqnarray}\lefteqn{
 \underline R_\mathrm{W_{\!c}}\left(\tilde w_1, B_{\mathrm c}(\boldsymbol w_{12})\right)
\underline R_\mathrm{W_{\!c}}^{-1}\left(\tilde w_2, B^{-1}_{\mathrm c}(\boldsymbol w_{12}') B_{\mathrm c}(\boldsymbol w_{12})\right)
\underline R_\mathrm{W_{\!c}}^{-1}\left(\tilde w_1', B_{\mathrm c}(\boldsymbol w_{12}')\right)}\nonumber\\&&=
\underline B_{\mathrm c}\left(\boldsymbol w_1\right)
   \sigma_\tau 
\left(
\begin{array}{c}
w_{12}^{0}\\
-\boldsymbol w_{12}
\end{array}
\right)^\tau
\underline w_{2}\,
  \sigma_\nu
\left(
\begin{array}{c}
w_{12}'^{0}\\
-\boldsymbol w_{12}'
\end{array}
\right)^\nu\underline B_{\mathrm c}\left(\boldsymbol w'_1\right)\,.
\end{eqnarray}
With this simplification and going back to physical momenta the covariant current for a pseudoscalar bound state finally reads
\begin{eqnarray} \lefteqn{\label{eq:pscurrent2}
J^\mu_{\mathrm {PS}}(\boldsymbol p_{\mathrm C}',\boldsymbol p_{\mathrm C},P_\mathrm e)}\nonumber\\&=&\frac{1}{4\pi}
\frac{\sqrt{ k^0_{\mathrm C} k'^0_{\mathrm C}}}{2}\int \frac{\mathrm d^3 \tilde k_{1}'}{ k_{1}^0}
\sqrt{\frac{m_{12}}{m_{12}'}}
\sqrt{\frac{ k^0_{12}}{k'^0_{12}}}
u^\ast_{n0}(\tilde{k}')
 u_{n0}(\tilde{k})
\nonumber\\&&\;\;\times
\sum_{\sigma_1'\sigma_1}D^{\frac12}_{\sigma_1\sigma_1'}\left[
\underline B_\mathrm c(\boldsymbol v_1)
   \sigma_\tau 
B_\mathrm c(-\boldsymbol V)^\tau_{\,\,\rho} \left(
\begin{array}{c}
v_{12}^{0}\\
-\boldsymbol v_{12}
\end{array}
\right)^\rho
\underline v_{2}
\right.\nonumber\\&&\left.\;\;\;\;\times
  \sigma_\nu
B_\mathrm c(-\boldsymbol V)^\nu_{\,\,\lambda}\left(
\begin{array}{c}
v_{12}'^{0}\\
-\boldsymbol v_{12}'
\end{array}
\right)^\lambda
\underline B_\mathrm c(\boldsymbol v_1') \right]
\nonumber\\&&\;\;\times\,|\,\mathrm e\,|\,\bar u_{\sigma_1'}(\boldsymbol p_1')(\varGamma_1+\varGamma_2)^\mu u_{\sigma_1}(\boldsymbol p_1)
 \,.
\end{eqnarray}
\subsubsection{Continuity}
The microscopic current for a pseudoscalar bound system of Eq.~(\ref{eq:pscurrent2}) satisfies the condition of current conservation, cf. Eq.~(\ref{eq:continuity}), which reads 
\begin{eqnarray}\label{eq:currentconservps}
(p_\mathrm C'-p_\mathrm C)_{\mu}J^\mu_{\mathrm {PS}}(\boldsymbol p_{\mathrm C}',\boldsymbol p_{\mathrm C},P_\mathrm e)=(k_\mathrm C'-k_\mathrm C)_{\mu}J^\mu_{\mathrm {PS}}(\boldsymbol k_{\mathrm C}',\boldsymbol k_{\mathrm C},K_\mathrm e)
=0\,.
\end{eqnarray}
 The rather lengthy
analytical proof of Eq.~(\ref{eq:currentconservps}) is given in App.~\ref{app:currentconservation}. It amounts to showing that
the integral in Eq.~(\ref{eq:pscurrent2}) vanishes if the spinor
product $\bar{u}_{\mu_1'}(\boldsymbol k_1')\gamma_\nu u_{\mu_1}(\boldsymbol k_1)$ is replaced by $\bar{u}_{\mu_1'}(\boldsymbol k_1')\boldsymbol {\gamma}\cdot
(\boldsymbol k_1-\boldsymbol k_1')u_{\mu_1}(\boldsymbol k_1)$. By a change of the integration
variables $\mathrm d^3\tilde{k}_1^\prime\rightarrow \mathrm d^3\tilde{k}_1$ it can
be shown that the integral over $\bar{u}_{\mu_1'}(\boldsymbol k_1')\boldsymbol {\gamma}\cdot
\boldsymbol k_1' u_{\mu_1}(\boldsymbol k_1)$ goes over into the integral over $\bar{u}_{\mu_1'}(\boldsymbol k_1')\boldsymbol {\gamma}\cdot
\boldsymbol k_1 u_{\mu_1}(\boldsymbol k_1)$
(and vice versa) so that the difference of them vanishes. 

\subsection{Vector Current}
In this work we also consider charged, massive vector bound states with $j=1$. Examples for such systems are, e.g., the $\rho^{\pm}$ meson and the deuteron D. In the constituent quark model the $\rho^{+}$ meson is, like the $\pi^{+}$, considered as a confined pair of a $\mathrm u$-quark and a $\bar{\mathrm d}$-quark. The deuteron, the simplest of all nuclei, is composed of a proton p and a neutron n. In the following we do not further resolve the proton and the neutron in terms of quarks and gluons. Instead, the structure of the nucleons is taken into account in terms of phenomenological electromagnetic nucleon form factors, cf. Eq.~(\ref{eq:nuclcurrent}). 
\subsubsection{Simplifications}
For $j=1$ and the case of a pure s-wave the Clebsch-Gordan coefficients in the wave functions of Eq.~(\ref{eq:bscurrentequalmasses}), $C_{\frac12\tilde\mu_1\frac12\tilde\mu_2}^{1\mu_j}$ and $C_{\frac12\tilde\mu_1'\frac12\tilde\mu'_2}^{1\mu_j'\ast}$, can be expressed in $(2\times2)$-matrix form as~\cite{Buck:1979ff} 
  \begin{eqnarray}\label{eq:CG}
   C_{\frac12\tilde\mu_1\frac12\tilde\mu_2}^{1\mu_j}&=&\epsilon_{\mu_j}^\mu(\boldsymbol 0)\left(\sigma_\mu\frac{\varepsilon}{\sqrt 2}\right)_{\tilde\mu_1\tilde\mu_2}\,,\\
\label{eq:CGprimeddag}
  C_{\frac12\tilde\mu_1'\frac12\tilde\mu'_2}^{1\mu_j'\ast}&=&-
\epsilon^{\ast\mu}_{\mu_j'}(\boldsymbol 0)\left(\frac{\varepsilon}{\sqrt 2}\sigma_\mu\right)_{\tilde\mu_2'\tilde\mu_1'}\,.
  \end{eqnarray}
Here $\varepsilon=\mathrm i\, \sigma_2$ is the metric spinor and $\epsilon_{\mu_j^{(\prime)}}^\mu(\boldsymbol 0)$ are the polarization vectors for massive spin-1 particles in the rest frame given in App.~\ref{app:massivespin1polarizvecs}.
Using Lorentz invariance, cf. Eq.~(\ref{eq:edotsigma}) and inserting for the Clebsch-Gordan coefficients into Eq.~(\ref{eq:bscurrent2}) gives
\begin{eqnarray} \label{eq:bscurrentVector}
\lefteqn{J^\mu_{\mathrm V}(\boldsymbol k_{\mathrm C}',\mu_j';\boldsymbol k_{\mathrm C},\mu_j;K_\mathrm e)}\nonumber\\&=&\frac12
\epsilon^{\ast\sigma}_{\mu_j'}(\boldsymbol k_{\mathrm C}')\epsilon^{\tau}_{\mu_j}(\boldsymbol k_{\mathrm C})B_\mathrm c(-\boldsymbol w_{\mathrm C}')_{\,\,\sigma}^{\lambda}
B_\mathrm c(-\boldsymbol w_{\mathrm C})_{\,\,\tau}^{\nu} \nonumber\\&&\times
\frac{1}{4\pi}\sqrt{ k^0_{\mathrm C} k'^0_{\mathrm C}}\sum\int \frac{\mathrm d^3 \tilde k_{1}'}{ k_{1}^0}
\sqrt{\frac{m_{12}}{m_{12}'}}
\sqrt{\frac{ k^0_{12}}{k'^0_{12}}}
u^\ast_{n0}(\tilde{k}')
 u_{n0}(\tilde{k})
\nonumber\\&&\times
D^{\frac12}_{\mu_1\tilde \mu_1}\left[\underline R_\mathrm{W_{\!c}}\left(\tilde w_1, B_{\mathrm c}(\boldsymbol w_{12})\right)\right](\sigma_\nu)_{\tilde \mu_1\tilde \mu_2}
  \nonumber\\&&\times D^{\frac12}_{\tilde{\mu}_2\tilde{\mu}'_2}\left[\underline R_\mathrm{W_{\!c}}\left(\tilde{w}_{2}', B_{\mathrm c}^{-1}(\boldsymbol w_{12}) B_{\mathrm c} (\boldsymbol w'_{12})\right)\right]
\nonumber\\&&\times (\sigma_\lambda)_{\tilde \mu_2'\tilde \mu_1'} D^{\frac12}_{\tilde \mu_1'\mu_1'}\left[\underline R_\mathrm{W_{\!c}}^{-1}\left(\tilde w_1', B_{\mathrm c}(\boldsymbol w_{12}')\right)\right] 
\nonumber\\&&\times\,|\,\mathrm e\,|\,\bar u_{\mu_1'}(\boldsymbol k_1')(\varGamma_1+\varGamma_2)^\mu u_{\mu_1}(\boldsymbol k_1)
 \,,
\end{eqnarray} where we have used
Eq.~(\ref{eq:sigam2Dsigma2}).
The current with the correct covariance properties is then obtained from Eq.~(\ref{eq:bscurrentVector}) according to Eq.~(\ref{eq:physicalmicrocurrent}):
\begin{eqnarray} \label{eq:bscurrentVectorphys}
\lefteqn{J^\mu_{\mathrm V}(\boldsymbol p_{\mathrm C}',\sigma_j';\boldsymbol p_{\mathrm C},\sigma_j;P_\mathrm e)}\nonumber\\&=&\frac12
B_\mathrm c(-\boldsymbol V)^\sigma_{\,\,\kappa} \epsilon^{\ast\kappa}_{\sigma_j'}(\boldsymbol p_{\mathrm C}')B_\mathrm c(-\boldsymbol V)^\tau_{\,\,\omega}\epsilon^{\omega}_{\sigma_j}(\boldsymbol p_{\mathrm C})B_\mathrm c(-\boldsymbol w_{\mathrm C}')_{\,\,\sigma}^{\lambda}
B_\mathrm c(-\boldsymbol w_{\mathrm C})_{\,\,\tau}^{\nu} \nonumber\\&&\times
\frac{1}{4\pi}\sqrt{ k^0_{\mathrm C} k'^0_{\mathrm C}}\sum\int \frac{\mathrm d^3 \tilde k_{1}'}{ k_{1}^0}
\sqrt{\frac{m_{12}}{m_{12}'}}
\sqrt{\frac{ k^0_{12}}{k'^0_{12}}}
u^\ast_{n0}(\tilde{k}')
 u_{n0}(\tilde{k})
\nonumber\\&&\times D^{\frac12}_{\sigma _1\tilde \mu_1}\left[\underline R_\mathrm{W_{\!c}}\left(\tilde w_1,B_{\mathrm c}(\boldsymbol V)B_{\mathrm c}(\boldsymbol w_{12})\right)\right]
(\sigma_\nu)_{\tilde \mu_1\tilde \mu_2}
  \nonumber\\&&\times D^{\frac12}_{\tilde{\mu}_2\tilde{\mu}'_2}\left[\underline R_\mathrm{W_{\!c}}\left(\tilde{w}_{2}', B_{\mathrm c}^{-1}(\boldsymbol w_{12}) B_{\mathrm c} (\boldsymbol w'_{12})\right)\right]
\nonumber\\&&\times (\sigma_\lambda)_{\tilde \mu_2'\tilde \mu_1'} D^{\frac12}_{\tilde \mu_1'\sigma_1'}\left[\underline R_\mathrm{W_{\!c}}\left(v_1', B_{\mathrm c}^{-1}(\boldsymbol w_{12}')B^{-1}_{\mathrm c}(\boldsymbol V)\right)\right]
\nonumber\\&&\times\,|\,\mathrm e\,|\,\bar u_{\sigma_1'}(\boldsymbol p_1')(\varGamma_1+\varGamma_2)^\mu u_{\sigma_1}(\boldsymbol p_1)\,,
 \,
\end{eqnarray}
where we have used Eq.~(\ref{eq:polphysunphy}). From Eq.~(\ref{eq:bscurrentVector}) using 
\begin{eqnarray}\label{eq:currenttensorfromcurrent}
 J^\mu_{\mathrm V}(\boldsymbol k_{\mathrm C}',\mu_j';\boldsymbol k_{\mathrm C},\mu_j;K_\mathrm e)=J_\mathrm V (\boldsymbol k_{\mathrm C}';\boldsymbol k_{\mathrm C};K_\mathrm e)^\mu_{\sigma\tau}\epsilon^{\ast\sigma}_{\mu_j'}(\boldsymbol k_{\mathrm C}')\epsilon^{\tau}_{\mu_j}(\boldsymbol k_{\mathrm C})
\end{eqnarray}
 we can read off the current tensor as
\begin{eqnarray} \label{eq:bscurrenttensor}
\lefteqn{J_\mathrm V (\boldsymbol k_{\mathrm C}',\boldsymbol k_{\mathrm C},K_\mathrm e)^\mu_{\sigma\tau}}\nonumber\\&:=&\frac12
B_\mathrm c(-\boldsymbol w_{\mathrm C}')_{\,\,\sigma}^{\lambda}
B_\mathrm c(-\boldsymbol w_{\mathrm C})_{\,\,\tau}^{\nu} \nonumber\\&&\times\frac{1}{4\pi}\sqrt{ k^0_{\mathrm C} k'^0_{\mathrm C}}\sum\int \frac{\mathrm d^3 \tilde k_{1}'}{ k_{1}^0}
\sqrt{\frac{m_{12}}{m_{12}'}}
\sqrt{\frac{ k^0_{12}}{k'^0_{12}}}
u^\ast_{n0}(\tilde{k}')
 u_{n0}(\tilde{k})
\nonumber\\&&\times
D^{\frac12}_{\mu_1\tilde \mu_1}\left[\underline R_\mathrm{W_{\!c}}\left(\tilde w_1, B_{\mathrm c}(\boldsymbol w_{12})\right)\right](\sigma_\nu)_{\tilde \mu_1\tilde \mu_2}
  \nonumber\\&&\times D^{\frac12}_{\tilde{\mu}_2\tilde{\mu}'_2}\left[\underline R_\mathrm{W_{\!c}}\left(\tilde{w}_{2}', B_{\mathrm c}^{-1}(\boldsymbol w_{12}) B_{\mathrm c} (\boldsymbol w'_{12})\right)\right]
\nonumber\\&&\times (\sigma_\lambda)_{\tilde \mu_2'\tilde \mu_1'} D^{\frac12}_{\tilde \mu_1'\mu_1'}\left[\underline R_\mathrm{W_{\!c}}^{-1}\left(\tilde w_1', B_{\mathrm c}(\boldsymbol w_{12}')\right)\right] 
\nonumber\\&&\times\,|\,\mathrm e\,|\,\bar u_{\mu_1'}(\boldsymbol k_1')(\varGamma_1+\varGamma_2)^\mu u_{\mu_1}(\boldsymbol k_1)
 \,.
\end{eqnarray} 
This current tensor is independent of the incoming and outgoing spin projections $\mu_j$ and $\mu_j'$. The covariant current tensor is then obtained analogously from Eq.~(\ref{eq:bscurrentVectorphys}) as
\begin{eqnarray} \label{eq:bscurrenttensorphys}
\lefteqn{J_{\mathrm V}(\boldsymbol p_{\mathrm C}',\boldsymbol p_{\mathrm C},P_\mathrm e)^\mu_{\kappa\omega}}\nonumber\\&:=&\frac12
B_\mathrm c(-\boldsymbol V)^\sigma_{\,\,\kappa} B_\mathrm c(-\boldsymbol V)^\tau_{\,\,\omega}B_\mathrm c(-\boldsymbol w_{\mathrm C}')_{\,\,\sigma}^{\lambda}
B_\mathrm c(-\boldsymbol w_{\mathrm C})_{\,\,\tau}^{\nu} \nonumber\\&&\times
\frac{1}{4\pi}\sqrt{ k^0_{\mathrm C} k'^0_{\mathrm C}}\sum\int \frac{\mathrm d^3 \tilde k_{1}'}{ k_{1}^0}
\sqrt{\frac{m_{12}}{m_{12}'}}
\sqrt{\frac{ k^0_{12}}{k'^0_{12}}}
u^\ast_{n0}(\tilde{k}')
 u_{n0}(\tilde{k})
\nonumber\\&&\times D^{\frac12}_{\sigma _1\tilde \mu_1}\left[\underline R_\mathrm{W_{\!c}}\left(\tilde w_1,B_{\mathrm c}(\boldsymbol V)B_{\mathrm c}(\boldsymbol w_{12})\right)\right]
(\sigma_\nu)_{\tilde \mu_1\tilde \mu_2}
  \nonumber\\&&\times D^{\frac12}_{\tilde{\mu}_2\tilde{\mu}'_2}\left[\underline R_\mathrm{W_{\!c}}\left(\tilde{w}_{2}', B_{\mathrm c}^{-1}(\boldsymbol w_{12}) B_{\mathrm c} (\boldsymbol w'_{12})\right)\right]
\nonumber\\&&\times (\sigma_\lambda)_{\tilde \mu_2'\tilde \mu_1'} D^{\frac12}_{\tilde \mu_1'\sigma_1'}\left[\underline R_\mathrm{W_{\!c}}\left(v_1', B_{\mathrm c}^{-1}(\boldsymbol w_{12}')B^{-1}_{\mathrm c}(\boldsymbol V)\right)\right]
\nonumber\\&&\times\,|\,\mathrm e\,|\,\bar u_{\sigma_1'}(\boldsymbol p_1')(\varGamma_1+\varGamma_2)^\mu u_{\sigma_1}(\boldsymbol p_1)
 \,.
\end{eqnarray}
\subsubsection{Continuity}\label{sec:currentconservviolation}
Current conservation, Eq.~(\ref{eq:continuity}) does, in general, not hold for the electromagnetic current we obtain for a vector bound system as given by Eq.~(\ref{eq:bscurrentVectorphys}). This means 
\begin{eqnarray}\label{eq:currentnonconserv}
 \lefteqn{(p_\mathrm C'-p_\mathrm C)_{\mu}J^\mu_{\mathrm {V}}(\boldsymbol p_{\mathrm C}',\sigma_j';\boldsymbol p_{\mathrm C},\sigma_j;P_{\mathrm e})}\nonumber\\&=&-(\boldsymbol k_\mathrm C'-\boldsymbol k_\mathrm C)\cdot \boldsymbol J_{\mathrm {V}}(\boldsymbol k_{\mathrm C}',\mu_j';\boldsymbol k_{\mathrm C},\mu_j;K_\mathrm e)
\nonumber\\&&\times
 D^{j\ast}_{\mu_j '\sigma_j' }[\underline R_\mathrm{W_{\!c}}^{-1}(w_{\mathrm C}',B_{\mathrm c}(\boldsymbol V))] 
D^{j}_{\mu_j \sigma_j }[\underline R_\mathrm{W_{\!c}}^{-1}(w_{\mathrm C},B_{\mathrm c}(\boldsymbol V))]\nonumber\\&\neq&0\,.
\end{eqnarray}
This issue is discussed in more detail in App.~\ref{app:currentnonconservation}. There we explain that the reason for the failure of proving continuity is the fact that the product of the three Wigner $D$-functions together with the two Clebsch-Gordan coefficients in Eq.~(\ref{eq:physicalmicrocurrent}) \textit{cannot} be written as one single Wigner $D$-function like in the pseudoscalar case. Consequently, the properties of the Wigner $D$-functions necessary for showing current conservation cannot be used. An analysis of the covariant structure of the current~(\ref{eq:bscurrentVectorphys}) together with numerical results support the non-vanishing of the left-hand side of Eq.~(\ref{eq:currentnonconserv}).
\chapter{Electromagnetic Form Factors}\label{chap:5}
 \section{Introduction}
The invariant electron-bound-state scattering amplitude in the
one-photon-exchange approximation is usually expressed as the
contraction of the electron current with a bound-state
current times the covariant photon propagator.
The bound-state current is a sum of independent Lorentz covariants which are multiplied by Lorentz invariant functions, the so-called \textit{electromagnetic form factors} of the bound-state. They are real functions of the four-momentum-transfer squared ($q^\mu q_\mu = -Q^2$) and they represent the observables that describe the electromagnetic structure of the (extended) bound-state. The aim of this chapter is to extract such electromagnetic form factors from the bound-state currents derived in the previous chapter and finally express them in terms of the bound-state wave functions. 
\section{Covariant Structure}
In this section we investigate the covariant structure and the number of form factors needed to parametrize the bound-state currents derived previously. We start with the simple pseudoscalar case and then turn to the more complicated example of a spin-1 bound system.
\subsection{Pseudoscalar Current}\label{sec:psbscovstructure}
From covariance, continuity and macrocausality arguments it is clear that a correct physical current $I^\mu_{\mathrm{PS}}(\boldsymbol p_\mathrm C^\prime,\boldsymbol p_\mathrm C)$ of a pseudoscalar bound system should depend on only one covariant which is the sum of incoming and outgoing bound-state momenta $P_\mathrm C:=p_\mathrm C+p_\mathrm C'$ times an electromagnetic form factor $F$ of the bound system, with $F$ being a function of Mandelstam $\mathrm t=-Q^2$. It turns out, however, that this is not the
case for our electromagnetic current of a pseudoscalar bound system $J^\mu_{\mathrm {PS}}(\boldsymbol p_\mathrm C^\prime, \boldsymbol p_\mathrm C,P_\mathrm e)$, which we have derived from the
one-photon-exchange optical potential. It does not have all the
properties it should have. It satisfies current conservation and
transforms like a four-vector, but it cannot be written as a sum of
covariants times Lorentz invariants ($Q^2$ and $m_\mathrm C^2$) which are solely built from the incoming and outgoing bound-state four-momenta ($p_\mathrm C^\mu$ and $p_\mathrm C^{\mu\,\prime}$).
Our current exhibits an additional dependence on $P_\mathrm e$ which has already been indicated.
The general covariant structure of our current $J^\mu_{\mathrm {PS}}(\boldsymbol p_{\mathrm C}',\boldsymbol p_{\mathrm C},P_\mathrm e)$ can be derived from the following considerations:
without spin one has, in general, four independent current components corresponding to the Lorentz indices $\mu=0,\ldots, 3$. Hence, one needs four independent covariant structures multiplied by four form factors to parametrize the current. Due to the condition of current conservation, which is proved analytically in App.~\ref{app:currentconservation}, one component of the current can be expressed in terms of the other three, which eliminates one covariant structure. Another structure can be eliminated by the fact that the current $J^\mu_{\mathrm {PS}}(\boldsymbol p_{\mathrm C}',\boldsymbol p_{\mathrm C},P_\mathrm e)$ transforms
covariantly under the Lorentz group. This means that the current can always be transformed into a frame where one of its components vanishes. Finally, we are left with just 2 independent components of the current. Therefore, the current can be expanded in terms of 2 conserved covariants times 2 form factors $f$ and $\tilde b$. The expansion in terms of Lorentz covariants times form factors has to satisfy hermiticity. This has already been proved for the current in Sec.~\ref{sec:hermiticity}. Thus, the first of 2 covariants available is simply the sum of the incoming and outgoing bound-state momenta $P_\mathrm C$. For the second covariant we have to recall that our derivation of the current is based on the Bakamjian-Thomas construction which is known to provide wrong cluster properties. This issue has already been discussed in Sec.~\ref{sec:clusterproperties}. As a consequence, the physical properties of the bound state are not independent of an additional particle, in our case the electron. Therefore, we cannot be sure that the bound-state current we get does not also depend on the electron momenta. We actually find that our current cannot be fully expressed in terms of Hermitean bound-state covariants. The second current conserving Hermitean covariant available is the sum of the incoming and outgoing electron momenta $P_\mathrm e$. This is the reason why we have included $P_\mathrm e$ as a further dependence in the argument of the bound-state current, cf. Eq.~(\ref{eq:vertexoperKbs}).

The only independent Lorentz invariants that can be built from the incoming and outgoing bound-state and electron momenta are Mandelstam $\mathrm t$ and Mandelstam $\mathrm s$, i.e. the four-momentum-transfer squared and the square of the invariant mass of the electron-bound-state system, respectively. Therefore, wrong cluster properties may also influence the invariant vertex form factors $f$ and $\tilde b$ such that they do not only depend on $\mathrm t=-Q^2$ but also on
\begin{eqnarray}
  \mathrm s=(p_\mathrm C+p_\mathrm e)^2=\left(\sqrt{m_{\mathrm C}^2+\boldsymbol k_\mathrm C^2}+\sqrt{m_{\mathrm e}^2+\boldsymbol k_\mathrm C^2}\right)^2\,.
 \end{eqnarray}
The $\mathrm s$-dependence can equivalently be expressed as a dependence on the magnitude of the particle momenta
\begin{eqnarray}\label{eq:kmagnitude}
 k:=|\boldsymbol k_\mathrm C^{\prime}|=|\boldsymbol k_\mathrm e^{\prime}|=|\boldsymbol k_\mathrm C|=|\boldsymbol k_\mathrm e|\,,
\end{eqnarray} which turns out to be more convenient for our purposes. Here we have used Eqs.~(\ref{eq:elasticscattering}) and~(\ref{eq:comconstraint}). At this point it should be mentioned that Poincar\'e invariance of our Bakamjian-Thomas type approach would not be spoiled by vertex form factors that depend even on the whole set of independent Lorentz invariants involved in the process.
A reasonable microscopic model for electromagnetic form
factors should, of course, only depend on the momentum transfer squared $\mathrm t$ and not on $\mathrm s$. Nevertheless, due to the non-locality of the vertex defined in Eq.~(\ref{eq:vertexoper1}) both form factors $f$ and $\tilde b$ exhibit also a $\mathrm s$-dependence. As we will see in the following, this dependence can, however, be eliminated in a certain limit.

With these findings we can write down our pseudoscalar bound-state current as a sum of Lorentz covariants times form factors\cite{Biernat:2010py,Biernat:2010tp}:\footnote{The covariant structure of our current resembles the corresponding one obtained in a covariant light-front approach of Refs.~\cite{Karmanov:1994ck,Carbonell:1998rj}. The relation between them will be discussed later.}
 \begin{eqnarray}\label{eq:formffPS}
 J^\mu_{\mathrm {PS}}(\boldsymbol p_{\mathrm C}',\boldsymbol p_{\mathrm C},P_{\mathrm e})&=&
|\,\mathrm e\,|\,[f(Q^2,k)P_\mathrm C^\mu+\tilde b(Q^2,k)P_\mathrm e^\mu ]\nonumber\\&=&
|\,\mathrm e\,|\,\left[f(Q^2,k)P_\mathrm C^\mu+b(Q^2,k)P_\mathrm e^\mu \frac{P_\mathrm C^2}{P_\mathrm C\cdot P_\mathrm e}\right]\,.
\end{eqnarray}
Here the factor $P_\mathrm C^2/P_\mathrm C\cdot P_\mathrm e$ has been separated from $\tilde b$ for convenience. The decomposition~(\ref{eq:formffPS}) holds for arbitrary values of the bound-state momenta
$p_\mathrm C$ and $p_\mathrm C^\prime$ with one exception: in the so-called \textit{Breit frame}, i.e. backward scattering ($\boldsymbol k_\mathrm C=-\boldsymbol k_\mathrm C'$) in the electron-bound-state center-of-mass system. In this frame the two covariants $P_\mathrm C^\mu$ and $P_\mathrm e^\mu$ become proportional which precludes the separation of the two form factors.
 
It is clear that the two covariants $P_\mathrm C$ and $P_\mathrm e$ are not orthogonal to each other, i.e. $P_\mathrm C \cdot P_\mathrm e\neq0$. 
If we choose, e.g., the second covariant as
\begin{eqnarray}
 P^\mu_\perp:= P_\mathrm e^\mu \frac{P_\mathrm C^2}{P_\mathrm C\cdot P_\mathrm e}-P_\mathrm C^\mu\,,
\end{eqnarray}
which is orthogonal to $P_\mathrm C$,
 then we have, in general, different form factors $\bar f$ and $\bar b$:
\begin{eqnarray}
 J^\mu_{\mathrm {PS}}(\boldsymbol p_{\mathrm C}',\boldsymbol p_{\mathrm C},P_{\mathrm e})=|\,\mathrm e\,|\,[\bar f(Q^2,k)P_\mathrm C^\mu+\bar b(Q^2,k)P^\mu_\perp]\,.
\end{eqnarray}
Thus, there seems to be an ambiguity how to define the form factors by expanding the current in terms of covariants. 
It turns out, however, that only the form factor $f(Q^2,k)$ defined via the expansion~(\ref{eq:formffPS}) provides the correct charge of the bound state at $Q^2=0$, 
as it is required for the physical form factor. This justifies to call $f(Q^2,k)$ defined in Eq.~(\ref{eq:formffPS}) the \textit{physical form factor} of the pseudoscalar bound state. The remaining structure in Eq.~(\ref{eq:formffPS}) that is proportional to $P_\mathrm e$ will then be referred to as \textit{non-physical} (or \textit{spurious}) contribution with $b(Q^2,k)$ being the \textit{spurious form factor}. Hence, only the expansion~(\ref{eq:formffPS}) provides a sensible separation of the physical from the spurious contribution. The separation of Eq.~(\ref{eq:formffPS}) leads to the following definition: spurious contributions are defined as structures that depend on $P_\mathrm e$~\cite{Carbonell:1998rj}. 

By applying a canonical boost $B_\mathrm c(-\boldsymbol V)$ to $J^\mu_{\mathrm {PS}}(\boldsymbol p_{\mathrm C}',\boldsymbol p_{\mathrm C},P_\mathrm e)$ we find, according to Eq.~(\ref{eq:physicalmicrocurrent}) for the pseudoscalar case, the covariant structure of $J^\mu_{\mathrm {PS}}(\boldsymbol k_{\mathrm C}',\boldsymbol k_{\mathrm C},K_\mathrm e)$:
 \begin{eqnarray}\label{eq:formffPS2}
J^\mu_{\mathrm {PS}}(\boldsymbol k_{\mathrm C}',\boldsymbol k_{\mathrm C},K_\mathrm e)&=&B_\mathrm c(-\boldsymbol V)^\mu_{\,\,\nu} J^\nu_{\mathrm {PS}}(\boldsymbol p_{\mathrm C}',\boldsymbol p_{\mathrm C},P_\mathrm e)\nonumber\\&=&
|\,\mathrm e\,|\,\left[f(Q^2,k)K_\mathrm C^\mu+b(Q^2,k)K_\mathrm e^\mu \frac{K_\mathrm C^2}{K_\mathrm C\cdot K_\mathrm e}\right]\nonumber\\&=&
|\,\mathrm e\,|\,[f(Q^2,k)K_\mathrm C^\mu+\tilde b(Q^2,k)K_\mathrm e^\mu] 
\,.
\end{eqnarray} 
\subsection{Vector Current}
\label{sec:vectorboundstatesystem}
A correct physical current $I^\mu_{\mathrm{V}}(\boldsymbol p_\mathrm C^\prime,\sigma_j'; \boldsymbol p_\mathrm C,\sigma_j)$ of a vector bound system
should depend on 3 form factors, $F_1$, $F_2$ and $G_\mathrm M$ which are functions of Mandelstam $\mathrm t=-Q^2$. Its covariant structure is usually obtained by constructing from the tensor $\epsilon^{\mu\ast}_{\sigma_j'}(\boldsymbol p_\mathrm C')\epsilon^{\nu}_{\sigma_j}(\boldsymbol p_\mathrm C)$ all Hermitean, conserved four-vectors by appropriate multiplication and contraction with $\mathrm g^{\mu\nu}$, the sum $P_\mathrm C^\mu$ and/or the difference $d^\mu:=p_\mathrm C'^\mu-p_\mathrm C^\mu$ of the incoming and outgoing bound-state four-momenta.

However, as in the pseudoscalar case, the covariant structure of our bound-state current $J^\mu_{\mathrm {V}}(\boldsymbol p_{\mathrm C}',\sigma_j';\boldsymbol p_{\mathrm C},\sigma_j;P_{\mathrm e})$ of Eq.~(\ref{eq:bscurrentVectorphys}), cannot be solely built from the incoming and outgoing momenta and spins of the bound state. Due to the violation of cluster separability in the Bakamjian-Thomas framework, we expect that it exhibits an additional dependence on the sum of the electron momenta $P_\mathrm e$.
Furthermore, unlike in the pseudoscalar case due to Eq.~(\ref{eq:currentnonconserv}), we cannot demand current conservation and therefore we have to allow for non-conserved Lorentz structures proportional to $d^\mu$. 

The explicit construction of the covariant structure of $J^\mu_{\mathrm {V}}(\boldsymbol p_{\mathrm C}',\sigma_j';\boldsymbol p_{\mathrm C},\sigma_j;P_{\mathrm e})$ is demonstrated in App.~\ref{app:covstructure}. The analysis reveals that one can find 11 Hermitean covariants by contracting and/or multiplying the tensor $\epsilon^{\mu\ast}_{\sigma_j'}(\boldsymbol p_\mathrm C')\epsilon^{\nu}_{\sigma_j}(\boldsymbol p_\mathrm C)$ with $\mathrm g^{\mu\nu}$ and/or the available four-vectors $P_\mathrm C^\mu$, $d^\mu$ and/or $P_\mathrm e^\mu$. Consequently, we can parametrize the current
$J^\mu_{\mathrm {V}}(\boldsymbol p_{\mathrm C}',\sigma_j';\boldsymbol p_{\mathrm C},\sigma_j;P_\mathrm e)$ in terms of 11 form factors, the 3 physical form factors $f_1$, $f_2$ and $g_\mathrm M$ and 8 spurious form factors denoted $b_1,\ldots, b_8$. The form factors exhibit, due to the non-locality of the vertex in the Bakamjian-Thomas framework, an additional dependence on Mandelstam $\mathrm s$. The expansion of the current in terms of the Hermitean covariants that are collected in~(\ref{eq:cov1}), (\ref{eq:cov2}) and~(\ref{eq:cov3}) times the form factors reads\footnote{Like in the pseudoscalar case the covariant structure we obtain for our current resembles the corresponding ones of Refs.~\cite{Karmanov:1994ck,Carbonell:1998rj}. Thus, for later comparison, we have adopted the notation and normalizations of the spurious form factors of these works.}
\begin{eqnarray}\lefteqn{
 \frac{1}{|\,\mathrm e\,|}\,J^\mu_{\mathrm {V}}(\boldsymbol p_{\mathrm C}',\sigma_j';\boldsymbol p_{\mathrm C},\sigma_j;P_{\mathrm e})}\nonumber\\&=&
\left\lbrace f_1(Q^2,k)\epsilon^\ast_{\sigma_j'}(\boldsymbol p'_\mathrm C)\cdot\epsilon_{\sigma_j}(\boldsymbol p_\mathrm C)+f_2(Q^2,k)\frac{[\epsilon^\ast_{\sigma'_j}(\boldsymbol p'_\mathrm C)\cdot d]
[\epsilon_{\sigma_j}(\boldsymbol p_\mathrm C)\cdot d]}{2m_{\mathrm C}^2 }\right\rbrace P_\mathrm C^\mu\nonumber\\&&\;\;\;\;+
g_{\mathrm M}(Q^2,k)\left\lbrace\epsilon^{\mu\ast}_{\sigma'_j}(\boldsymbol p'_\mathrm C)[\epsilon_{\sigma_j}(\boldsymbol p_\mathrm C)\cdot d]-
\epsilon_{\sigma_j}^\mu (\boldsymbol p_\mathrm C)[\epsilon^\ast_{\sigma'_j}(\boldsymbol p'_\mathrm C)\cdot d]\right\rbrace\nonumber\\&&+
\frac{ m_{\mathrm C}^2}{P_\mathrm e\cdot P_\mathrm C}\left\lbrace
b_1(Q^2,k) \epsilon^\ast_{\sigma_j'}(\boldsymbol p'_\mathrm C)\cdot\epsilon_{\sigma_j}(\boldsymbol p_\mathrm C)+b_2(Q^2,k)\frac{[\epsilon^\ast_{\sigma'_j}(\boldsymbol p'_\mathrm C)\cdot d]
[\epsilon_{\sigma_j}(\boldsymbol p_\mathrm C)\cdot d]}{m_{\mathrm C}^2 }\right.\nonumber\\&&\left.\;\;\;\;+
b_3(Q^2,k)\,4 m_{\mathrm C}^2 \frac{[\epsilon^\ast_{\sigma'_j}(\boldsymbol p'_\mathrm C)\cdot P_\mathrm e][\epsilon_{\sigma_j}(\boldsymbol p_\mathrm C)\cdot P_\mathrm e]}{(P_\mathrm e\cdot P_\mathrm C)^2}
\right.\nonumber\\&&\left.\;\;\;\;+b_4(Q^2,k) \frac{ [\epsilon^\ast_{\sigma_j'}(\boldsymbol p'_\mathrm C)\cdot d][\epsilon_{\sigma_j}(\boldsymbol p_\mathrm C)\cdot P_\mathrm e]- 
[\epsilon^\ast_{\sigma_j'}(\boldsymbol p'_\mathrm C)\cdot P_\mathrm e][\epsilon_{\sigma_j}(\boldsymbol p_\mathrm C)\cdot d] }{P_\mathrm e\cdot P_\mathrm C}
\right\rbrace P_\mathrm e^\mu+\nonumber
\end{eqnarray}

\begin{eqnarray}
&&+
b_5(Q^2,k)\,4m_{\mathrm C}^2 \frac{[\epsilon^\ast_{\sigma'_j}(\boldsymbol p'_\mathrm C)\cdot P_\mathrm e][\epsilon_{\sigma_j}(\boldsymbol p_\mathrm C)\cdot P_\mathrm e]}{(P_\mathrm e\cdot P_\mathrm C)^2}P_\mathrm C^\mu
\nonumber\\&&+
b_6(Q^2,k) \frac{[\epsilon^\ast_{\sigma_j'}(\boldsymbol p'_\mathrm C)\cdot d][\epsilon_{\sigma_j}(\boldsymbol p_\mathrm C)\cdot P_\mathrm e]- 
[\epsilon^\ast_{\sigma_j'}(\boldsymbol p'_\mathrm C)\cdot P_\mathrm e][\epsilon_{\sigma_j}(\boldsymbol p_\mathrm C)\cdot d] }{P_\mathrm e\cdot P_\mathrm C} P_\mathrm C^\mu
\nonumber\\&&+
b_7(Q^2,k) \,2 m_{\mathrm C}^2\frac{\epsilon^{\mu\ast}_{\sigma'_j}(\boldsymbol p'_\mathrm C)[\epsilon_{\sigma_j}(\boldsymbol p_\mathrm C)\cdot P_\mathrm e]+
\epsilon_{\sigma_j}^\mu (\boldsymbol p_\mathrm C)[\epsilon^\ast_{\sigma'_j}(\boldsymbol p'_\mathrm C)\cdot P_\mathrm e]}{P_\mathrm e\cdot P_\mathrm C} 
\nonumber\\&&+b_8(Q^2,k)
\frac{[\epsilon^\ast_{\sigma_j'}(\boldsymbol p'_\mathrm C)\cdot d][\epsilon_{\sigma_j}(\boldsymbol p_\mathrm C)\cdot P_\mathrm e]+ 
[\epsilon^\ast_{\sigma_j'}(\boldsymbol p'_\mathrm C)\cdot P_\mathrm e][\epsilon_{\sigma_j}(\boldsymbol p_\mathrm C)\cdot d] }{P_\mathrm e\cdot P_\mathrm C}\,d^\mu\,.\nonumber\\
\label{eq:covstructurephysdeutcurr}
\end{eqnarray}
As in the pseudoscalar case there seems to be an ambiguity how to separate the physical from unphysical contributions. Again it turns out, however, that only the above decomposition with the associated definition of form factors gives the correct charge of the bound state at zero momentum transfer, as it is required for the physical charge form factor $G_\mathrm C$. Again this justifies to define unphysical (or spurious) contributions as structures proportional to first or higher degrees of $P_\mathrm e$, i.e. all structures multiplied by the spurious form factors $b_1,\,\ldots,\,b_8$ in Eq.~(\ref{eq:covstructurephysdeutcurr}). 

By separation of the polarization vectors we find the covariant structure of the current tensor $J_{\mathrm V}(\boldsymbol p_{\mathrm C}',\boldsymbol p_{\mathrm C}, P_\mathrm e)^\mu_{\sigma\tau}$:
 \begin{eqnarray}
\label{eq:currentphneospin1cov}
\lefteqn{
 \frac{1}{|\,\mathrm e\,|}\,J_{\mathrm V}(\boldsymbol p_{\mathrm C}',\boldsymbol p_{\mathrm C},P_{\mathrm e})^\mu_{\sigma\tau}}\nonumber\\&=&
\left\lbrace f_1(Q^2,k)\mathrm g_{\sigma\tau}+f_2(Q^2,k)\frac{d_\sigma d_\tau}{2m_{\mathrm C}^2 }\right\rbrace P_\mathrm C^\mu
+g_{\mathrm M}(Q^2,k)\left\lbrace\mathrm g^\mu_{\sigma}d_\tau-\mathrm g^\mu_{\tau}d_\sigma\right\rbrace\nonumber\\&&+
\left\lbrace
b_1(Q^2,k) \mathrm g_{\sigma\tau}+b_2(Q^2,k)\frac{d_\sigma d_\tau}{m_{\mathrm C}^2 }
+b_3(Q^2,k)4 m_{\mathrm C}^2 \frac{P_{\mathrm e \sigma}P_{\mathrm e \tau}}{(P_\mathrm e\cdot P_\mathrm C)^2}
\right.\nonumber\\&&\left.\;\;\;\;+b_4(Q^2,k) \frac{ P_{\mathrm e \tau}d_\sigma-  P_{\mathrm e \sigma}d_\tau}{P_\mathrm e\cdot P_\mathrm C}
\right\rbrace\frac{ m_{\mathrm C}^2}{P_\mathrm e\cdot P_\mathrm C} P_\mathrm e^\mu\nonumber\\&&+
b_5(Q^2,k) \frac{P_{\mathrm e \sigma}P_{\mathrm e \tau}}{(P_\mathrm e\cdot P_\mathrm C)^2}\,4 m_{\mathrm C}^2P_\mathrm C^\mu
+b_6(Q^2,k) \frac{ P_{\mathrm e \tau}d_\sigma-  P_{\mathrm e \sigma}d_\tau}{ P_\mathrm e\cdot P_\mathrm C} P_\mathrm C^\mu
\nonumber\\&&+
b_7(Q^2,k) \,2 m_{\mathrm C}^2\frac{\mathrm g^\mu_\sigma P_{\mathrm e\tau}+\mathrm g^\mu_\tau P_{\mathrm e\sigma}}{P_\mathrm e\cdot P_\mathrm C} 
+b_8(Q^2,k)
\frac{d_\sigma P_{\mathrm e\tau}+d_\tau P_{\mathrm e\sigma} }{P_\mathrm e\cdot P_\mathrm C}\,d^\mu\,.
\end{eqnarray}
The covariant structure of the bound-state current $J^\mu_{\mathrm {V}}(\boldsymbol k_{\mathrm C}',\mu_j';\boldsymbol k_{\mathrm C},\mu_j;K_{\mathrm e})$ is then obtained by inverting Eq.~(\ref{eq:physicalmicrocurrent}), which means canonically boosting
$J^\mu_{\mathrm {V}}(\boldsymbol p_{\mathrm C}',\sigma_j';\boldsymbol p_{\mathrm C},\sigma_j;P_{\mathrm e})$ with $B_\mathrm c(-\boldsymbol V)$ and multiplying with
$D^{1\ast}_{\sigma_j'\mu_j ' }[\underline R_\mathrm{W_{\!c}}(w_{\mathrm C}',B_{\mathrm c}(\boldsymbol V))]  
D^{1}_{ \sigma_j\mu_j }[\underline R_\mathrm{W_{\!c}}(w_{\mathrm C},B_{\mathrm c}(\boldsymbol V))]$:
\begin{eqnarray}\label{eq:currentphneospin1}
\lefteqn{\frac{1}{|\,\mathrm e\,|}\,J^\mu_{\mathrm {V}}(\boldsymbol k_{\mathrm C}',\mu_j';\boldsymbol k_{\mathrm C},\mu_j;K_{\mathrm e})}\nonumber\\&=&
\frac{1}{|\,\mathrm e\,|}\,
B_{\mathrm c}(-\boldsymbol V)^\mu_{\,\,\nu}J^\nu_{\mathrm {V}}(\boldsymbol p_{\mathrm C}',\sigma_j';\boldsymbol p_{\mathrm C},\sigma_j;P_{\mathrm e})
\nonumber\\&&\times D^{1\ast}_{\sigma_j'\mu_j ' }[\underline R_\mathrm{W_{\!c}}(w_{\mathrm C}',B_{\mathrm c}(\boldsymbol V))]  
D^{1}_{ \sigma_j\mu_j }[\underline R_\mathrm{W_{\!c}}(w_{\mathrm C},B_{\mathrm c}(\boldsymbol V))]\nonumber\\&=&
\left\lbrace f_1(Q^2,k)\epsilon^\ast_{\mu_j'}(\boldsymbol k'_\mathrm C)\cdot\epsilon_{\mu_j}(\boldsymbol k_\mathrm C)+f_2(Q^2,k)\frac{[\epsilon^\ast_{\mu'_j}(\boldsymbol k'_\mathrm C)\cdot q]
[\epsilon_{\mu_j}(\boldsymbol k_\mathrm C)\cdot q]}{2m_{\mathrm C}^2 }\right\rbrace K_\mathrm C^\mu\nonumber\\&&+
g_{\mathrm M}(Q^2,k)\left\lbrace\epsilon^{\mu\ast}_{\mu'_j}(\boldsymbol k'_\mathrm C)[\epsilon_{\mu_j}(\boldsymbol k_\mathrm C)\cdot q]-
\epsilon_{\mu_j}^\mu (\boldsymbol k_\mathrm C)[\epsilon^\ast_{\mu'_j}(\boldsymbol k'_\mathrm C)\cdot q]\right\rbrace\nonumber\\&&+
\frac{ m_{\mathrm C}^2}{K_\mathrm e\cdot K_\mathrm C}\left\lbrace
b_1(Q^2,k) \epsilon^\ast_{\mu_j'}(\boldsymbol k'_\mathrm C)\cdot\epsilon_{\mu_j}(\boldsymbol k_\mathrm C)+b_2(Q^2,k)\frac{[\epsilon^\ast_{\mu'_j}(\boldsymbol k'_\mathrm C)\cdot q]
[\epsilon_{\mu_j}(\boldsymbol k_\mathrm C)\cdot q]}{m_{\mathrm C}^2 }\right.\nonumber\\&&\left.\;\;\;\;+
b_3(Q^2,k)\,4 m_{\mathrm C}^2 \frac{[\epsilon^\ast_{\mu'_j}(\boldsymbol k'_\mathrm C)\cdot K_\mathrm e][\epsilon_{\mu_j}(\boldsymbol k_\mathrm C)\cdot K_\mathrm e]}{(K_\mathrm e\cdot K_\mathrm C)^2}
\right.\nonumber\\&&\left.\;\;\;\;+b_4(Q^2,k) \frac{ [\epsilon^\ast_{\mu_j'}(\boldsymbol k'_\mathrm C)\cdot q][\epsilon_{\mu_j}(\boldsymbol k_\mathrm C)\cdot K_\mathrm e]- 
[\epsilon^\ast_{\mu_j'}(\boldsymbol k'_\mathrm C)\cdot K_\mathrm e][\epsilon_{\mu_j}(\boldsymbol k_\mathrm C)\cdot q] }{K_\mathrm e\cdot K_\mathrm C}
\right\rbrace K_\mathrm e^\mu\nonumber\\&&+
b_5(Q^2,k) \frac{[\epsilon^\ast_{\mu'_j}(\boldsymbol k'_\mathrm C)\cdot K_\mathrm e][\epsilon_{\mu_j}(\boldsymbol k_\mathrm C)\cdot K_\mathrm e]}{(K_\mathrm e\cdot K_\mathrm C)^2}4m_{\mathrm C}^2K_\mathrm C^\mu
\nonumber\\&&+
b_6(Q^2,k) \frac{[\epsilon^\ast_{\mu_j'}(\boldsymbol k'_\mathrm C)\cdot q][\epsilon_{\mu_j}(\boldsymbol k_\mathrm C)\cdot K_\mathrm e]- 
[\epsilon^\ast_{\mu_j'}(\boldsymbol k'_\mathrm C)\cdot K_\mathrm e][\epsilon_{\mu_j}(\boldsymbol k_\mathrm C)\cdot q] }{K_\mathrm e\cdot K_\mathrm C} K_\mathrm C^\mu
\nonumber\\&&+
b_7(Q^2,k) \,2 m_{\mathrm C}^2\frac{\epsilon^{\mu\ast}_{\mu'_j}(\boldsymbol k'_\mathrm C)[\epsilon_{\mu_j}(\boldsymbol k_\mathrm C)\cdot K_\mathrm e]+
\epsilon_{\mu_j}^\mu (\boldsymbol k_\mathrm C)[\epsilon^\ast_{\mu'_j}(\boldsymbol k'_\mathrm C)\cdot K_\mathrm e]}{K_\mathrm e\cdot K_\mathrm C} 
\nonumber\\&&+b_8(Q^2,k)
\frac{[\epsilon^\ast_{\mu_j'}(\boldsymbol k'_\mathrm C)\cdot q][\epsilon_{\mu_j}(\boldsymbol k_\mathrm C)\cdot K_\mathrm e]+ 
[\epsilon^\ast_{\mu_j'}(\boldsymbol k'_\mathrm C)\cdot K_\mathrm e][\epsilon_{\mu_j}(\boldsymbol k_\mathrm C)\cdot q] }{K_\mathrm e\cdot K_\mathrm C}\,q^\mu\,,
\end{eqnarray}
where we have used Eq.~(\ref{eq:polphysunphy}).
The covariant structure of the current tensor $J_{\mathrm V}(\boldsymbol k_{\mathrm C}';\boldsymbol k_{\mathrm C};K_{\mathrm e})^\mu_{\sigma\tau}$ can be obtained from $J_{\mathrm V}(\boldsymbol p_{\mathrm C}';\boldsymbol p_{\mathrm C};P_{\mathrm e})^\mu_{\sigma\tau}$ with the help of the canonical boosts $B_\mathrm c (-\boldsymbol V)$:
\begin{eqnarray}
\lefteqn{
\frac{1}{|\,\mathrm e\,|}\,J_{\mathrm V}(\boldsymbol k_{\mathrm C}';\boldsymbol k_{\mathrm C};K_{\mathrm e})^\mu_{\sigma\tau}}\nonumber\\&=&\frac{1}{|\,\mathrm e\,|}\,
B_\mathrm c (-\boldsymbol V)^{\,\,\lambda}_\sigma B_\mathrm c(-\boldsymbol V)^{\,\,\rho}_\tau
B_\mathrm c(-\boldsymbol V)^\mu_{\,\,\nu} J_{\mathrm V}(\boldsymbol p_{\mathrm C}';\boldsymbol p_{\mathrm C};P_{\mathrm e})^\nu_{\lambda\rho}
\nonumber\\
&=&\left\lbrace f_1(Q^2,k)\mathrm g_{\sigma\tau}+f_2(Q^2,k)\frac{q_\sigma q_\tau}{2m_{\mathrm C}^2 }\right\rbrace K_\mathrm C^\mu
+g_{\mathrm M}(Q^2,k)\left\lbrace\mathrm g^\mu_{\sigma}q_\tau-\mathrm g^\mu_{\tau}q_\sigma\right\rbrace\nonumber\\&&+
\left\lbrace
b_1(Q^2,k) \mathrm g_{\sigma\tau}+b_2(Q^2,k)\frac{q_\sigma q_\tau}{m_{\mathrm C}^2 }
+b_3(Q^2,k)4 m_{\mathrm C}^2 \frac{K_{\mathrm e \sigma}K_{\mathrm e \tau}}{(K_\mathrm e\cdot K_\mathrm C)^2}
\right.\nonumber\\&&\left.\;\;\;\;+b_4(Q^2,k) \frac{ K_{\mathrm e \tau}q_\sigma-  K_{\mathrm e \sigma}q_\tau}{K_\mathrm e\cdot K_\mathrm C}
\right\rbrace\frac{ m_{\mathrm C}^2}{K_\mathrm e\cdot K_\mathrm C} K_\mathrm e^\mu\nonumber+
\end{eqnarray}
\begin{eqnarray}
\label{eq:currenttensordecomp}
&&+b_5(Q^2,k) \frac{K_{\mathrm e \sigma}K_{\mathrm e \tau}}{(K_\mathrm e\cdot K_\mathrm C)^2}\,4 m_{\mathrm C}^2K_\mathrm C^\mu
+b_6(Q^2,k) \frac{ K_{\mathrm e \tau}q_\sigma-  K_{\mathrm e \sigma}q_\tau}{ K_\mathrm e\cdot K_\mathrm C} K_\mathrm C^\mu
\nonumber\\&&+
b_7(Q^2,k)\, 2 m_{\mathrm C}^2\frac{\mathrm g^\mu_\sigma K_{\mathrm e\tau}+\mathrm g^\mu_\tau K_{\mathrm e\sigma}}{K_\mathrm e\cdot K_\mathrm C} 
+b_8(Q^2,k)
\frac{q_\sigma K_{\mathrm e\tau}+q_\tau K_{\mathrm e\sigma} }{K_\mathrm e\cdot K_\mathrm C}\,q^\mu\,,\nonumber\\
\end{eqnarray}
where we have used Eq.~(\ref{eq:lambdaortho}).

\section{Extracting the Form Factors}
The aim of this section is to extract the physical form factors defined by Eqs.~(\ref{eq:formffPS2}) and~(\ref{eq:currentphneospin1}) from the corresponding expressions for the bound-state currents, Eqs.~(\ref{eq:pscurrent}) and~(\ref{eq:bscurrentVector}), respectively. 
\subsection{Kinematics}\label{sec:kinematics}
 In order to find explicit expressions for the form factors as overlap integrals of the bound-state wave functions we have to specify our kinematics. For convenience (without loss of generality) we choose the (1,3)-plane as the scattering plane and the incoming bound-state three-momenta and the momentum transfer as~\cite{Fuchsberger:2007,Biernat:2009my}
\begin{equation}\label{eq:kM}
\boldsymbol{k}_\mathrm C=\left(%
\begin{array}{c} -\frac{Q}{2}\\0\\
\sqrt{k^2-\frac{Q^2}{4}}
\end{array}
\right)\, , \quad
\boldsymbol{q}=\left(%
\begin{array}{c} Q\\0\\
0
\end{array}
\right)\,, \qquad \boldsymbol{k}_\mathrm C^\prime = \boldsymbol{k}_\mathrm C + \boldsymbol{q}\,, 
\end{equation}
with $k$ given by Eq.~(\ref{eq:kmagnitude}).
It should be noted that $Q$ is restricted by $Q<2k$ in order to keep the bound-state momenta real. In the following we will consider two particular choices of frames, which are characterized by the value of $k$: the standard Breit frame with $k=Q/2$ and the \textit{infinite-momentum frame} with $k\rightarrow\infty$.
\subsubsection{Limit $k\rightarrow\infty$}
In the infinite-momentum frame, where $k\rightarrow\infty$, we have for the incoming bound-state and electron four-momenta
\begin{equation}\label{eq:kMinfinity}
k_\mathrm C^{\mu}\stackrel{k \rightarrow 
\infty}{\longrightarrow}\left(%
\begin{array}{c} k\\
-\frac{Q}{2}\\0\\
k
\end{array}
\right)^\mu \quad\text {and}\quad
k_\mathrm e^{\mu}\stackrel{k \rightarrow 
\infty}{\longrightarrow}\left(%
\begin{array}{c}k\\
\frac{Q}{2}\\0\\
-k
\end{array}
\right)^\mu\, ,
\end{equation} 
respectively. For the internal cluster center-of-mass momentum and the free invariant cluster mass we
find in this limit~\cite{Fuchsberger:2007,Biernat:2009my}
\begin{equation}\label{eq:kinf}
\tilde{\boldsymbol{k}}_1\stackrel{k \rightarrow
\infty}{\longrightarrow} \left(%
\begin{array}{c} \tilde{k}_1^{1\prime}+\left(
\frac{\tilde{k}_1^{3\prime}}{m_{12}^{\prime }}-\frac12 \right)Q \\
\tilde{k}_1^{2\prime}\\ \tilde{k}_1^{3\prime}\frac{m_{12}}{m_{12}^{\prime }}
\end{array}
\right)\, ,
 \end{equation}
and
\begin{equation}
m_{12}\stackrel{k \rightarrow
\infty}{\longrightarrow} \sqrt{m_{12}^{\prime
2}-\frac{4\tilde{k}_1^{1\prime}
m_{12}^{\prime}}{2\tilde{k}_1^{3\prime}+m_{12}^{\prime}}Q+\frac{\left(
m_{12}^{\prime}-2\tilde{k}_1^{3\prime}\right)
}{2\tilde{k}_1^{3\prime}+m_{12}^{\prime}}Q^2}\, ,
\end{equation}
respectively. Further, the kinematical factors of the microscopic bound-state current, Eq.~(\ref{eq:bscurrent2}), become
   \begin{eqnarray}\label{eq:kinfactorsinf}
   \sqrt{ k^0_{\mathrm C} k'^0_{\mathrm C}}\stackrel{k \rightarrow
\infty}{\longrightarrow} k\quad \text{and} \quad\sqrt{\frac{ k^0_{12}}{k'^0_{12}}}\stackrel{k \rightarrow
\infty}{\longrightarrow}1\,.
   \end{eqnarray}
Finally, we give the expression for the constituent current over $k_i^0$ expanded in a series around $k=\infty$ and keeping only terms up to $\mathcal O(1/k)$:
\begin{eqnarray}
\label{eq:constcurrentexpansion}
\lefteqn{\frac{J_i^{\mu}\left(\boldsymbol{k}_i',\mu_i';\boldsymbol{k}_i,\mu_i \right)}{k_i^0}}\nonumber\\&=&
\,|\,\mathrm e\,|\,F_1^i(Q_i^2)\left[\delta_{\mu_i\mu_i'}
\right.\nonumber\\&&\left.\;\;\times
\left(2,\frac{4 m_{12}' \tilde k_i'^1 - m_{12}' Q_i + 
 2  \tilde k_i'^3 Q_i}{k (m_{12}' + 2\tilde k_i'^3)}, -\frac{2 m_{12}'(-2 \tilde k_i'^2 -2\mu_i \, \mathrm i \,Q_i)}{k ( m_{12}' + 2  \tilde k_i'^3)},2\right)^\mu\right.\nonumber\\
&&\;\;+\left.\delta_{\mu_i'-\mu_i}\left(-\frac{2 m_{12}' Q_i}{k (m_{12}' +2 \tilde k_i'^3)}, 0, 0, -\frac{2 m_{12}' Q_i}{k (m_{12}' +2 \tilde k_i'^3)}\right)^\mu\right]
\nonumber\\&&+
(-1)^{\mu_i-\frac12}\sqrt{\tau} \,|\,\mathrm e\,|\,F_2^i(Q_i^2)\left[\frac{\delta_{\mu_i'\mu_i}}{
  k (m_{12}' + 2 \tilde k_i'^3)} 
\right.\nonumber\\&&\left.\;\;\times
\left(-2 \mathrm i \, m_{12}' (- 2 \tilde k_i'^2  -2\mu_i \, \mathrm i \, Q_i), 0, 4 \mathrm i \,m_i m_{12}' , -
  2 (- 2\mathrm i\, m_{12}' \tilde k_i'^2 + 2\mu_i m_{12}' Q_i)\right)^\mu\right.\nonumber\\&&\left.\;\;+\delta_{\mu_i'-\mu_i} \left(2, \frac{4 m_{12}' \tilde k_i'^1 - 
  m_{12}' Q_i + 2 \tilde k_i'^3 Q_i}{
 k ( m_{12}' + 2 \tilde k_i'^3 )}, \frac{4 m_{12}'\tilde k_i'^2}{
  k ( m_{12}' + 2 \tilde k_i'^3)}, 2\right)^\mu\right]\nonumber\\&&+\mathcal O(1/k^2)
\end{eqnarray}
with $i=\mathrm{\mathrm q,\mathrm N}$.
In the limit $k \rightarrow\infty$ the above expression becomes 
\begin{eqnarray}\label{eq:nuclcurrlim}
 \lefteqn{\frac{J_i^{\mu}\left(\boldsymbol{k}_i',\mu_i';\boldsymbol{k}_i,\mu_i \right)}{k_i^0}}\nonumber\\
&&\stackrel{k \rightarrow
\infty}{\longrightarrow}\,|\,\mathrm e\,|\,\left[F_1^i(Q^2)\delta_{\mu_i'\mu_i}+(-1)^{\mu_i-\frac12}\sqrt{\tau} F_2^i(Q^2)\delta_{\mu_i'-\mu_i} \right](2,0,0,2)^\mu\,,\nonumber\\
\end{eqnarray}
with $\tau=Q^2/(4m_i^2)$. 
In addition, we note that in the limit $k \rightarrow
\infty$ the four-momentum transfer to the (active) constituent, in particular its zero-component, goes over to the four-momentum transfer to the cluster:
\begin{eqnarray}\label{eq:qiequivq}
 q_i^\mu=k_i'^\mu-k_i^\mu\stackrel{k \rightarrow
\infty}{\longrightarrow}  q^\mu=k_\mathrm C'^\mu-k_\mathrm C^\mu\,.
\end{eqnarray}
\subsection{Pseudoscalar Bound States}
\subsubsection{Coupled Equations for the Form Factors}
The analysis in Sec.~\ref{sec:psbscovstructure} revealed that the current for pseudoscalar bound systems 
$J^\mu_{\mathrm {PS}}(\boldsymbol k_\mathrm C^\prime;\boldsymbol k_\mathrm C;K_\mathrm e)$ has 2 vanishing components due to current conservation and covariance. With the chosen kinematics given in Sec.~\ref{sec:kinematics} the two, in general non-vanishing components are $J^0_{\mathrm {PS}}(\boldsymbol k_\mathrm C^\prime; \boldsymbol k_\mathrm C;K_\mathrm e)$ and 
$J^3_{\mathrm {PS}}(\boldsymbol k_\mathrm C^\prime; \boldsymbol k_\mathrm C;K_\mathrm e)$. This can be easily seen from Eq.~(\ref{eq:formffPS2}) by inserting~(\ref{eq:kM}) for the momenta. Therefore, we have 
 2 coupled equations for the form factors:
\begin{eqnarray}
 2\sqrt{k^2+m_{\mathrm C}^2}f(Q^2,k)+2\sqrt{k^2+m_\mathrm e^2}\,\tilde b(Q^2,k)&=&\frac{1}{|\,\mathrm e\,|}J^0_{\mathrm {PS}}(\boldsymbol k_\mathrm C^\prime;\boldsymbol k_\mathrm C;K_\mathrm e)\,,\nonumber\\
2\sqrt{k^2-\frac{Q^2}{4}}f(Q^2,k)-2\sqrt{k^2-\frac{Q^2}{4}}\tilde b(Q^2,k)&=&\frac{1}{|\,\mathrm e\,|}J^3_{\mathrm {PS}}(\boldsymbol k_\mathrm C^\prime; \boldsymbol k_\mathrm C;K_\mathrm e)\,.\nonumber\\
\end{eqnarray}
Solving for the form factors we obtain the expressions
\begin{eqnarray}\label{eq:f1}
 f(Q^2,k)&=&\frac{1}{|\,\mathrm e\,|\,\left(1+\frac{\sqrt{k^2+m_{\mathrm C}^2}}{\sqrt{k^2+m_\mathrm e^2}}\right)}\left(\frac{J^0_{\mathrm {PS}}(\boldsymbol k_\mathrm C^\prime; \boldsymbol k_\mathrm C;K_\mathrm e)}{2\sqrt{k^2+m_\mathrm e^2}}+\frac{J^3_{\mathrm {PS}}(\boldsymbol k_\mathrm C^\prime; \boldsymbol k_\mathrm C;K_\mathrm e)}{2\sqrt{k^2-\frac{Q^2}{4}}}\right)\,,\nonumber\\
\\
\label{eq:f2}
\tilde b(Q^2,k)&=&\frac{1}{|\,\mathrm e\,|\,\left(1+\frac{\sqrt{k^2+m_\mathrm e^2}}{\sqrt{k^2+m_{\mathrm C}^2}}\right)}\left(\frac{J^0_{\mathrm {PS}}(\boldsymbol k_\mathrm C^\prime; \boldsymbol k_\mathrm C;K_\mathrm e)}{2\sqrt{k^2+m_{\mathrm C}^2}}-\frac{J^3_{\mathrm {PS}}(\boldsymbol k_\mathrm C^\prime; \boldsymbol k_\mathrm C;K_\mathrm e)}{2\sqrt{k^2-\frac{Q^2}{4}}}\right)\,.\nonumber\\
\end{eqnarray}
A numerical analysis of these expressions, which will be presented in Sec.~\ref{sec:empionff}, confirms that both, the physical form factor $f$ and the spurious form factor
$\tilde b$ depend indeed on $k$. However,
this $k$-dependence of $f$ vanishes rather quickly with increasing $k$ (or equivalently increasing invariant bound-state-electron
mass $M_{\mathrm {Ce}}^{(\prime)}$ or increasing Mandelstam $\mathrm s$). At the same time the spurious form factor $\tilde b$ is seen to vanish. It is thus suggestive to take the limit $k\rightarrow
\infty$ to get a sensible result for the physical form factor $f$ that only depends on $Q^2$. In addition, this limit removes all (unwanted) spurious contributions in the current of~(\ref{eq:formffPS}) due to the vanishing of $\tilde b$~\cite{Biernat:2010py,Biernat:2010tp}.   
\subsubsection{Limit $k\rightarrow\infty$}
\label{sec:imfPS}
Before we examine the analytic expressions for the form factors $f$ and $\tilde b$, Eqs.~(\ref{eq:f1}) and~(\ref{eq:f2}), in the limit $k \rightarrow
\infty$, we note that the point-like constituent current over $k_i^0$ becomes, for $k \rightarrow
\infty$, equal to $|\,\mathrm e\,|\,\delta_{{\mu_i'\mu_i}}(2,0,0,2)^\mu$, as can be seen from Eq.~(\ref{eq:nuclcurrlim}). Thus, for point-like constituents, the only expression occurring under the integral of the pseudoscalar bound-state current, Eq.~(\ref{eq:pscurrent}), that has not yet been investigated in this limit is the sum over the Wigner $D$-function $D^{\frac12}_{\mu_1\mu_1'}$ multiplied by the constituent current. Evidently, this product becomes the trace over the Wigner $D$-function and is given with our standard kinematics, Eq.~(\ref{eq:kM}), by\footnote{Due to the complexity of the left-hand side this equation has been computed using {\sc Mathematica}$^{\begin{scriptsize}\textcopyright                                                                                                                                                                                                                                                                                                                                              \end{scriptsize}                                                                                                                                                                                                                                                                                                                                                                                                                                                                                                                                                                                                                                                                                                                                                                                                                                         }$.}~\cite{Biernat:2009my}
\begin{eqnarray}&&
 \frac12\sum_{\mu_1} D^{\frac12}_{\mu_1\mu_1}\left[\underline B_{\mathrm c}\left(\boldsymbol w_1\right)
   \sigma_\tau 
\left(
\begin{array}{c}
w_{12}^{0}\\
-\boldsymbol w_{12}
\end{array}
\right)^\tau
\underline w_{2}\,
  \sigma_\nu
\left(
\begin{array}{c}
w_{12}'^{0}\\
-\boldsymbol w_{12}'
\end{array}
\right)^\nu\underline B_{\mathrm c}\left(\boldsymbol w'_1\right)\right]\nonumber\\&&\;\;\;\;\stackrel{k \rightarrow
\infty}{\longrightarrow} \frac{m_{12}^\prime}{m_{12}}-\frac{2
\tilde{k}_1^{\prime 1} \, Q}{m_{12} (m_{12}^\prime + 2
\tilde{k}_1^{\prime 3})}\nonumber\\&&\;\;\;\;=:\mathcal{S} \,.\label{eq:spinrot}
\end{eqnarray}
Note that the factor 1/2 in front comes from the two Clebsch-Gordan coefficients that have been included in $D^{\frac12}_{\mu_1\mu_1'}$, cf. Eq.~(\ref{eq:CDC}).
$\mathcal S$ of Eq.~(\ref{eq:spinrot}) is obviously a finite expression. Eq.~(\ref{eq:nuclcurrlim}) together with the expressions for the kinematical factors, Eqs.~(\ref{eq:kinfactorsinf}) and~(\ref{eq:spinrot}), give rise to the conclusion that
\begin{eqnarray}
 \lim_{k \rightarrow
\infty}J^\mu_{\mathrm {PS}}(\boldsymbol k_\mathrm C^\prime; \boldsymbol k_\mathrm C;K_\mathrm e)&=:& |\,\mathrm e\,|\,F(Q^2)\lim_{k \rightarrow
\infty} K_\mathrm C^\mu\,\label{eq:formfactorF}\\
\Rightarrow \quad \lim_{k \rightarrow
\infty}J^0_{\mathrm {PS}}(\boldsymbol k_\mathrm C^\prime;\boldsymbol  k_\mathrm C;K_\mathrm e)&=&\lim_{k \rightarrow
\infty}J^3_{\mathrm {PS}}(\boldsymbol k_\mathrm C^\prime; \boldsymbol k_\mathrm C;K_\mathrm e)\,.
\end{eqnarray}
We see that the microscopic current factorizes explicitly into the  covariant $K_\mathrm C^\mu$ times a $Q^2$-dependent integral which can be identified as the physical form factor of the pseudoscalar bound system. We finally end up with a quite simple analytic expression for $F$ defined by Eq.~(\ref{eq:f1}) for $k \rightarrow
\infty$ (or equivalently by Eq.~(\ref{eq:formfactorF}))~\cite{Biernat:2009my}:
\begin{eqnarray}
 F(Q^2)&:=&\lim_{k \rightarrow
\infty} f(Q^2,k)\nonumber\\&=&\frac{1}{|\,\mathrm e\,|}\lim_{k \rightarrow
\infty} \frac{1}{2k}J^0_{\mathrm {PS}}(\boldsymbol k_\mathrm C^\prime;\boldsymbol  k_\mathrm C;K_\mathrm e)\nonumber\\&=&
\frac{1}{4\pi}\int\mathrm{d}^3\tilde{k}^\prime_1
\sqrt{\frac{m_{12}}{m'_{12}}}\, \mathcal{S}\, u_{n
0}^\ast\,(\tilde{k}_1^\prime)\, u_{n 0}\,(\tilde{k}_1)\,. \label{eq:pionformfactor}
\end{eqnarray}
This expression is clearly independent of $k$. Further, we note that $F(Q^2)$ is independent of the reference frame since our Bakamjian-Thomas type approach ensures Poincar\'{e} invariance. The integrand
on the right-hand side of Eq.~(\ref{eq:pionformfactor}) depends only
on the momentum transfer $Q$ and the internal constituent momentum
$\tilde{\boldsymbol k}_1^\prime$, which is integrated over. Details of the dynamics enter solely via the form of the bound-state
wave function $ u_{n 0}\,(\tilde{k}_1)$ and not via the mass of the bound cluster. What we have achieved with
Eq.~(\ref{eq:pionformfactor}) is an impulse approximation to the
electromagnetic bound-state form factor. In the limit $k \rightarrow \infty$ the whole photon momentum is
transferred to one of the constituents due to Eq.~(\ref{eq:qiequivq}), whereas the other
one acts as a spectator. It should also be noted that the constituents' spins described by the spin-rotation factor $\mathcal{S}$ have a substantial effect on the
electromagnetic form factor over nearly the whole momentum-transfer
range and thus must not be neglected~\cite{Biernat:2009my}.

The spurious form factor vanishes in the limit $k \rightarrow \infty$~\cite{Biernat:2010py,Biernat:2010tp}:\begin{eqnarray}
 \lim_{k \rightarrow
\infty} \tilde b(Q^2,k)=0\,.
\end{eqnarray} Furthermore, the whole spurious part of the current is seen to vanish in the limit $k \rightarrow
\infty$:
\begin{eqnarray}
 \lim_{k \rightarrow
\infty} \tilde b(Q^2,k)K_\mathrm e^\mu=0\,.
\end{eqnarray} 
Thus the electromagnetic current for a pseudoscalar bound system $J^\mu_{\mathrm {PS}}(\boldsymbol p_{\mathrm C}';\boldsymbol p_{\mathrm C})$, Eq.~(\ref{eq:pscurrent2}), exhibits the correct cluster-separability properties in the limit $k\rightarrow \infty$. We can also turn this around
and say that we have found a reference frame for the $\gamma^\ast
\mathrm C \rightarrow \mathrm C$ subprocess in which a one-body constituent current already provides the correct cluster-separability properties for the electromagnetic
current of a pseudoscalar bound system~\cite{Biernat:2009my}.
\subsubsection{Projection Vector}
 Another prescription to extract the form factor is similar to the one proposed in Refs.~\cite{Karmanov:1994ck,Carbonell:1998rj}. Contracting the pseudoscalar bound-state current with the four-vector $K^\mu_\mathrm e/K_\mathrm C\cdot K_\mathrm e$ in the limit $k \rightarrow
\infty$ projects out the form factor:
 \begin{eqnarray}\label{eq:projoutF}
 F(Q^2)=\frac{1}{|\,\mathrm e\,|}\lim_{k \rightarrow
\infty}\frac{K_{\mathrm e\mu}}{K_\mathrm C\cdot K_\mathrm e}J^\mu_{\mathrm {PS}}(\boldsymbol k_\mathrm C^\prime;\boldsymbol  k_\mathrm C;K_\mathrm e)\,.
\end{eqnarray}
This prescription for the extraction of the form factors is equivalent to solving the system of equations for the form factors. Both results for $F(Q^2)$, Eqs.~(\ref{eq:projoutF}) and~(\ref{eq:pionformfactor}), coincide.
\subsubsection{Physical Current}
With a prescription for eliminating the unphysical contributions we are now able to define a physical covariant current of a pseudoscalar bound system that satisfies all required properties including macrocausality.
It is given by
\begin{eqnarray}
 I^\mu_{\mathrm {PS}}(\boldsymbol V):=B_\mathrm c(\boldsymbol V)^\mu_{\,\,\nu} \lim_{k\rightarrow\infty}J_{\mathrm {PS}}^\nu(\boldsymbol k_\mathrm C^\prime; \boldsymbol k_\mathrm C;K_\mathrm e)\,.
\end{eqnarray}

\subsubsection{Comparison with the Standard Front-Form Approach}\label{sec:comparestandardff}                                                                                                      
We extract the electromagnetic form factor of a pseudoscalar bound system in the limit $k\equiv \vert \boldsymbol{k}_\mathrm C\vert \rightarrow \infty$. It means that we consider the subprocess where the photon is absorbed or emitted by the the bound state in the
infinite-momentum frame of the bound state~\cite{Biernat:2009my}. This offers the
possibility of making a direct comparison with form factor analyses
done in the front form of relativistic dynamics. The $k \rightarrow \infty$ limit with our standard kinematics, as chosen
in Eq.~(\ref{eq:kM}), implies in particular that we work in a
reference frame in which the plus component of the four-momentum
transfer $q^+:=q^0+q^3$ vanishes. $q^{+}=0$ frames
are also popular in the form-factor studies in front
form~\cite{Chung:1988mu,Keister:1991sb,Simula:2002vm,Coester:2005cv}. One reason is that the impulse approximation can
be formulated consistently in any $q^+=0$ frame (for the plus
component of the current operator)~\cite{Keister:1991sb}. The
other benefit is that, so-called, \textit{Z-graphs} are
suppressed in such frames~\cite{Simula:2002vm}.

Our point-form calculation can be related to front form results by
an appropriate change of variables. To show this relation we define the
longitudinal momentum fractions $z$ as
\begin{equation}
z^{(\prime)}:=\frac{\tilde{k}_1^{3(\prime)}}{m_{12}^{(\prime)
}}+\frac{1}{2}
\end{equation}
and introduce the short-hand notation
\begin{equation}
\boldsymbol k_\perp^{(\prime)}:=\left(
\begin{array}{c}\tilde{k}_1^{1(\prime)}\\\tilde{k}_1^{2(\prime)}
\end{array}\right)
\end{equation}
for the intrinsic transverse momentum of the active incoming
(outgoing) constituent. From Eq.~(\ref{eq:kinf}) we infer that
\begin{equation}
z=z^\prime \quad \mathrm{and}\quad
\boldsymbol k_\perp^\prime=\boldsymbol k_\perp+(1-z)\, \boldsymbol q_\perp \quad
\mathrm{with}\quad \boldsymbol q_\perp=\left(
\begin{array}{c}Q\\0
\end{array}\right)\, .
\end{equation}
The free invariant mass of the incoming (outgoing) $12$-system
can then be written as
\begin{equation}\label{eq:invmlc}
m_{12}^{(\prime)}=\sqrt{\frac{m^2+\boldsymbol k_\perp^{(\prime)
2}}{z (1-z)}}\, .
\end{equation}
With these relations the Jacobian for the variable transformation
$\{\tilde{k}_1^{1\prime},\tilde{k}_1^{2\prime},\tilde{k}_1^{3\prime}\}
\rightarrow \{z,k_\perp^1,k_\perp^2\}$ becomes
\begin{equation}
\frac{\partial(\tilde{k}_1^{1\prime},\tilde{k}_1^{2\prime},\tilde{k}_1^{3\prime})}
{\partial(z,k_\perp^1,k_\perp^2)}=\frac{m_{12}^{\prime}}{4 z (1-z)}\,.
\end{equation}
Then the integral for the electromagnetic form factor takes on the
form~\cite{Biernat:2009my}
\begin{equation}\label{eq:ffff}
F(Q^2) =\frac{1}{4 \pi }\int_0^1 \mathrm{d}z \int_{\mathbb{R}^2}
\mathrm{d}^2\!k_\perp \frac{\sqrt{m_{12}\,
m'_{12}}}{4z(1-z)}\, \mathcal{M}\, u_{n 0}^\ast\,(\tilde{k}_1^\prime)\, u_{n 0}\,(\tilde{k}_1)
 \, .
\end{equation}
The argument of the wave functions is easily expressed in terms of
$z$ and $\boldsymbol k_\perp$ if one uses $\tilde{\boldsymbol k}_1^{(\prime)
2}=m^2+{m_{12}^{(\prime)\,2}}/{4}$ and
Eq.~(\ref{eq:invmlc}). By the change of variables the
spin-rotation factor $\mathcal{S}$, cf. Eq.~(\ref{eq:spinrot}),
goes over into the Melosh-rotation factor of Ref.~\cite{Chung:1988mu}
\begin{equation}\label{eq:melrot}
\mathcal{M}=\frac{m_{12}}{m_{12}^{\prime}}\left(
1+\frac{(1-z)\, (\boldsymbol q_\perp \cdot
\boldsymbol k_\perp)}{m^2+\boldsymbol k_\perp^2}\right)\, .
\end{equation}
$\mathcal{S}$ and $\mathcal{M}$ describe the effect of the constituents'
spin onto the electromagnetic form factor in point form and in
front form, respectively. Eqs.~(\ref{eq:ffff}) and
(\ref{eq:melrot}) are identical to the corresponding formulae in
Refs.~\cite{Chung:1988mu,Simula:2002vm}.  This is a remarkable
result. Starting from two different forms of relativistic dynamics
and applying completely different procedures to identify the
electromagnetic form factor of a pseudoscalar bound system the outcome is the same. It
means that relativity is treated in an equivalent way and the
physical ingredients are alike. Since the infinite-momentum frame
we use is just a particular $q^+=0$ frame, Z-graph contributions
to the electromagnetic form factor are also suppressed in
our point-form approach. This is a welcome feature, because
Z-graphs can play a significant role in $q^+\neq 0$
frames~\cite{Simula:2002vm} and one should have control on them when
form-factor predictions are compared with experiment.
\subsubsection{Comparison with Covariant Light-Front Dynamics}
The above standard front form of relativistic dynamics, which is characterized by the light-front plane $x^0+x^3=0$ (cf.~Sec.~\ref{sec:formsofrelativisticdyn}) is not manifestly (explicitly) covariant due to the dynamic nature of some Lorentz transformations that change the orientation of the light front. This problem can, however, be overcome by introducing an arbitrary light-like four-vector $\omega^\mu$ that defines the orientation of the light front which is then described by the equality $\omega\cdot x=0$. Such an explicit covariant light-front approach has been formulated in the works of Refs.~\cite{Karmanov:1994ck,Carbonell:1998rj}. The particular choice of $\omega=(1,0,0,-1)$ then corresponds to the standard (non-covariant) front-form approach.

The Lorentz structure of the current $J^\mu_{\mathrm {PS}}(\boldsymbol k_{\mathrm C}';\boldsymbol k_{\mathrm C};K_{\mathrm e})$ for a pseudoscalar bound state described by Eq.~(\ref{eq:formffPS}) resembles the corresponding current obtained within the covariant light-front dynamics of Refs.~\cite{Karmanov:1994ck,Carbonell:1998rj}. In these works the authors encounter a spurious (unphysical) contribution to the current which is associated with $\omega^\mu$. Their spurious contribution is comparable to our spurious contribution if $\omega^\mu$ is identified with $K_\mathrm e^\mu$. It has already been mentioned that spurious $K_\mathrm e$-dependent terms in the current of our point-form Bakamjian-Thomas approach can be traced back to the violation of
cluster separability. However, by taking the limit $k \rightarrow
\infty$ we can eliminate these cluster separability violating effects~\cite{Biernat:2010py}. In the covariant light-front formalism the spurious $\omega$-dependent contribution is rather the consequence of the most general ansatz for a current of a pseudoscalar bound-state that includes
the orientation of the light front.

In the standard light front approach, i.e. $\omega=(1,0,0,-1)$, the electromagnetic form factor of a pseudoscalar bound state is usually extracted from matrix elements of the plus component of the current operator. Due to $\omega^+=0$ in the standard case, the second (spurious) part of the current proportional to $\omega^\mu$ does not contribute~\cite{Karmanov:1994ck}. Therefore, it is not surprising that standard light-front and covariant light-front dynamics give the same results for the form factor of a pseudoscalar bound-state, which is some way a special case due to the simplicity of spin-0 systems. Consequently, taking the plus component of the current to extract the form factor in the standard light-front approach plays a similar role to taking the limit $k\rightarrow\infty$ in our approach in the sense that in both cases one gets rid of the spurious contribution in the current.
\subsubsection{Comparison with the Point-Form Spectator Model}\label{sec:PFSM}
Relativistic point-form quantum mechanics has also been applied in
Ref.~\cite{Wagenbrunn:2000es,Boffi:2001zb} to calculate
electroweak baryon form factors within a constituent quark model.
The strategy for the extraction of electromagnetic form factors, however,
differs from the one in the present work. We apply the Bakamjian-Thomas
type framework to the full electron-bound-state system
in order to derive the electromagnetic bound-state current. In their works the Bakamjian-Thomas framework for the bound-state system is only used to obtain the bound-state wave function. This wave function is then plugged into an ansatz for the electromagnetic
bound-state current. The ansatz is constrained by the requirements of continuity and covariance. It is shown that these constraints can be satisfied by a spectator current if not all of the photon momentum is transferred to the active constituent. The momentum transfer to the active constituent $q_1$ is uniquely determined by total
four-momentum conservation for the $\gamma^\ast \mathrm C\rightarrow \mathrm C $
subprocess and by the spectator conditions. An ambiguity in
defining such a spectator current, however, enters through a
normalization factor which has to be introduced in order to recover the bound-state charge from the electric form factor
in the limit $Q^2\rightarrow 0$~\cite{Melde:2004qu}. Since both
quantities, $q_1$ and the normalization factor depend effectively on
all quark momenta and not only on those of the active ones, the model
current constructed in this way cannot be considered as a pure
one-body current~\cite{Melde:2007zz}. It has therefore been termed
\textit{point-form spectator model} to distinguish it
from the usual impulse approximation.

A comparison of our result for the form factor with the one of the point-form spectator model reveals that the dynamics enter into the former solely via the bound-state wave function, whereas the latter also exhibits an explicit dependence on the bound-state mass $m_\mathrm C$. Within the point-form spectator model the eigenvalue spectrum of the mass operator is thus directly connected with the electromagnetic
structure of its eigenstates~\cite{Biernat:2009my}. We shall come back on this issue when doing a numerical comparison between both approaches.
\subsection{Vector Bound States}
\subsubsection{Current Matrix Elements}
By a numerical analysis, whose dynamical ingredients will be discussed in detail later, using the standard kinematics of Eq.~(\ref{eq:kM}) we observe that our vector bound-state current $J^\mu_{\mathrm {V}}(\boldsymbol k_{\mathrm C}',\mu_j';\boldsymbol k_{\mathrm C},\mu_j;K_{\mathrm e})$ of Eq.~(\ref{eq:bscurrentVector}) has, in general, 11 independent, non-vanishing matrix elements. They are given by (using the short-hand notation $J^\mu_{\mathrm {V}}(\boldsymbol k_{\mathrm C}',\mu_j';\boldsymbol k_{\mathrm C},\mu_j;K_{\mathrm e})\equiv J^\mu_{\mathrm {V}}(\mu_j',\mu_j)$): 
\begin{eqnarray}&&\label{eq:indepmatrixelements}
J^0_{\mathrm {V}}(1,-1)\,,\quad
J^3_{\mathrm {V}}(1,-1)\,,\quad
J^0_{\mathrm {V}}(0,0)\,,\quad
J^3_{\mathrm {V}}(0,0)\,,\quad
J^0_{\mathrm {V}}(1,0)\,,\quad
J^1_{\mathrm {V}}(1,0)\,,\nonumber\\&&
J^2_{\mathrm {V}}(1,0)\,,\quad
J^3_{\mathrm {V}}(1,0)\,,\quad
J^0_{\mathrm {V}}(1,1)\,,\quad
J^2_{\mathrm {V}}(1,1)\,,\quad
J^3_{\mathrm {V}}(1,1)\,.
\end{eqnarray}
This confirms numerically the outcome of the formal analysis in Sec.~\ref{sec:vectorboundstatesystem}. Note that the remaining non-vanishing matrix elements are related to the above ones by the following relations:
\begin{eqnarray}
 J^0_{\mathrm {V}}(1,-1)&=&J^0_{\mathrm {V}}(-1,1)\,,\\
J^3_{\mathrm {V}}(1,-1)&=&J^3_{\mathrm {V}}(-1,1)\,,\\
J^0_{\mathrm {V}}(1,1)&=&J^0_{\mathrm {V}}(-1,-1)\,,\\
J^3_{\mathrm {V}}(1,1)&=&J^3_{\mathrm {V}}(-1,-1)\,,\\
J^2_{\mathrm {V}}(1,1)&=&-J^2_{\mathrm {V}}(-1,-1)\,,\\
J^0_{\mathrm {V}}(1,0)&=&-J^0_{\mathrm {V}}(-1,0)=J^0_{\mathrm {V}}(0,-1)=-J^0_{\mathrm {V}}(0,1)\,,\\
J^3_{\mathrm {V}}(1,0)&=&-J^3_{\mathrm {V}}(-1,0)=J^3_{\mathrm {V}}(0,-1)=-J^3_{\mathrm {V}}(0,1)\,,\\
J^1_{\mathrm {V}}(1,0)&=&-J^1_{\mathrm {V}}(-1,0)=J^1_{\mathrm {V}}(0,1)=-J^1_{\mathrm {V}}(0,1)\,,\\
J^2_{\mathrm {V}}(1,0)&=&J^2_{\mathrm {V}}(-1,0)=-J^2_{\mathrm {V}}(0,1)=-J^2_{\mathrm {V}}(0,-1)\,.
\end{eqnarray}   
\subsubsection{Limit $k\rightarrow\infty$}
Of the 11 form factors only the 3 physical form factors $f_1$, $f_2$ and $g_\mathrm M$ are of interest. We extract them from Eq.~(\ref{eq:bscurrentVector}) by using the decomposition~(\ref{eq:currentphneospin1}). As in the pseudoscalar case, we take the limit $k \rightarrow\infty$ where the form factors become independent of $k$. 
 Furthermore, using our standard kinematics, we observe that the zeroth and third component of the current become identical in this limit, i.e.
\begin{eqnarray} 
 J^0_{\mu_j'\mu_j}:=\lim_{k \rightarrow
\infty}J^0_{\mathrm V}(\mu_j',\mu_j)=\lim_{k \rightarrow
\infty}J^3_{\mathrm V}(\mu_j',\mu_j)=J^3_{\mu_j'\mu_j}\,,
\end{eqnarray}
 which reduces the number of independent matrix elements from 11 to 7. This means, however, that 4 of the 8 spurious contributions cannot be eliminated by simply taking the limit $k \rightarrow
\infty$. This complication occurs due to the complexity of spin-1 bound systems as compared to the previous spin-0 case.
 Nonetheless, a careful analysis (cf. App.~\ref{app:ifinitemomeframe}) of the covariant structure of the current reveals that the 3 current matrix elements $J^0_{11}, J^0_{1-1}$ and $J^2_{11}$ do not contain leading order spurious contributions (for a similar analysis in the light-front formalism see Ref.~\cite{Melikhov:2001pm}). These \lq\lq good'' matrix elements are therefore appropriate for the extraction of the physical form factors, which are then given by the following rather lengthy expressions (for a derivation we refer to App.~\ref{app:ifinitemomeframe}): 
\begin{eqnarray}
 \label{eq:ff1me}
 F_1(Q^2)&:=&\lim_{k\rightarrow\infty}f_1(Q^2,k)\nonumber\\&=& -\frac{1}{|\,\mathrm e\,|}\lim_{k\rightarrow\infty} \frac{1}{2k}\left[J^0_{\mathrm {V}}(1,1)+J^0_{\mathrm {V}}(1,-1)\right]\nonumber\\&=&-\frac{1}{4\pi}\int\mathrm{d}^3\tilde{k}'_1\sqrt{\frac{m_{12}}{m'_{12}}}u_{n0}^\ast\left(\tilde{k}'_1 \right)u_{n0}\left(\tilde{k}_1 \right)
\nonumber\\&&\times\left\lbrace\left[F_1^1(Q^2)+F_1^2(Q^2)\right](\mathcal S^{11}_{1}+\mathcal S^{1-1}_{1})\right.\nonumber\\&&\;\;\;\;+
\left.\sqrt{\tau}\left[F_2^1(Q^2)+F_2^2(Q^2)\right](\mathcal S^{11}_{2}+\mathcal S^{1-1}_{2}) \right\rbrace\,,\\ \nonumber\\
\label{eq:ff2me}
F_2(Q^2)&:=&\lim_{k\rightarrow\infty}f_2(Q^2,k)\nonumber\\&=&-\frac{1}{|\,\mathrm e\,|\,\eta}\lim_{k\rightarrow\infty} \frac{1}{2k}J^0_{\mathrm {V}}(1,-1)\nonumber\\&=&-\frac{1}{\eta}\frac{1}{4\pi}\int\mathrm{d}^3\tilde{k}'_1\sqrt{\frac{m_{12}}{m'_{12}}}u_{n0}^\ast\left(\tilde{k}'_1 \right)u_{n0}\left(\tilde{k}_1 \right)
\nonumber\\&&\times\left\lbrace\left[F_1^1(Q^2)+F_1^2(Q^2)\right]\mathcal S^{1-1}_1+\sqrt{\tau}\left[F_2^1(Q^2)+F_2^2(Q^2)\right]\mathcal S^{1-1}_2 \right\rbrace\,,\nonumber\\\\ \nonumber\\
\label{eq:ffGMme}
 G_\mathrm M(Q^2)&:=&\lim_{k\rightarrow\infty}g_\mathrm M(Q^2,k)\nonumber\\&=&-\frac{\mathrm i}{|\,\mathrm e\,|\,Q} J^2_{11}\nonumber\\&=&
-\frac{\mathrm i}{Q}\frac{1}{4\pi}\int\mathrm{d}^3\tilde{k}'_1\sqrt{\frac{m_{12}}{ m'_{12}}}u_{n0}^\ast\left(\tilde{k}'_1 \right)u_{n0}\left(\tilde{k}_1 \right)
\frac{m_{12}'}{ ( m_{12}' + 2  \tilde k_1'^3)}
\nonumber\\&&\times\left\lbrace\left[F_1^1(Q^2) +F_1^2(Q^2)\right]\left(
\tilde k_1'^2 \mathcal S^{11}_{1}+  \frac{\mathrm i \,Q}{2} 
\mathcal S^{11}_{3}\right)\right.\nonumber\\&&\;\;\;\;+
\left.
\sqrt{\tau}\left[F_2^1(Q^2)
 +F_2^2(Q^2)\right]\left(\mathrm i\, m \,\mathcal S^{11}_{3}+\tilde k_1'^2\mathcal S^{11}_{2}\right)
\right\rbrace\,, 
\end{eqnarray}
with $\eta=Q^2/(4 m_{\mathrm C}^2)$. Here $\mathcal S^{\mu_j'\mu_j}_{1}$, $\mathcal S^{\mu_j'\mu_j}_{2}$ and $\mathcal S^{\mu_j'\mu_j}_{3}$ are the spin rotation factors given by
\begin{eqnarray}\label{eq:S1}
\mathcal S_1^{\mu_j'\mu_j}&:=&\lim_{k\rightarrow \infty}\sum D^{\frac12}_{\mu_1\tilde{\mu}_1}\left[\underline R_\mathrm{W_{\!c}}\left(\tilde{w}_1,B_\mathrm c\left(\boldsymbol w_{12} \right)\right)\right]C^{1\mu_j}_{\frac12\tilde{\mu}_1 \frac12\tilde{\mu}_{2}}
\nonumber\\&&\times
D^{\frac12}_{\tilde{\mu}_{2}'\tilde{\mu}_{2}}\left[\underline R_\mathrm{W_{\!c}}\left(\tilde{w}_{2},B_\mathrm c^{-1}\left(\boldsymbol w'_{12}\right)B_\mathrm c\left(\boldsymbol w_{12}\right)\right)\right]\nonumber\\&&\times C^{1\mu_j'}_{\frac12\tilde{\mu}_2' \frac12\tilde{\mu}_{1}'}D^{\frac12}_{\tilde{\mu}_1'\mu_1}\left[\underline R_\mathrm{W_{\!c}}\left(w'_{1},B_\mathrm c^{-1}\left(\boldsymbol w'_{12} \right)\right)\right]\,,
\end{eqnarray}
\begin{eqnarray}
\label{eq:S2}
\mathcal S^{\mu_j'\mu_j}_{2}&:=&\lim_{k\rightarrow \infty}\sum(-1)^{\mu_1-\frac12} D^{\frac12}_{\mu_1\tilde{\mu}_1}\left[\underline R_\mathrm{W_{\!c}}\left(\tilde{w}_1,B_\mathrm c\left(\boldsymbol w_{12} \right)\right)\right]C^{1\mu_j}_{\frac12\tilde{\mu}_1 \frac12\tilde{\mu}_{2}}
\nonumber\\&&\times
D^{\frac12}_{\tilde{\mu}_{2}'\tilde{\mu}_{2}}\left[\underline R_\mathrm{W_{\!c}}\left(\tilde{w}_{2},B_\mathrm c^{-1}\left(\boldsymbol w'_{12}\right)B_\mathrm c\left(\boldsymbol w_{12}\right)\right)\right]\nonumber\\&&\times
C^{1\mu_j'}_{\frac12\tilde{\mu}_2' \frac12\tilde{\mu}_{1}'}D^{\frac12}_{\tilde{\mu}_1'-\mu_1}\left[\underline R_\mathrm{W_{\!c}}\left(w'_{1},B_\mathrm c^{-1}\left(\boldsymbol w'_{12} \right)\right)\right]\,,
\\
\label{eq:S3}
\mathcal S^{\mu_j'\mu_j}_{3}&:=&\lim_{k\rightarrow \infty}\sum(-1)^{\mu_1-\frac12} D^{\frac12}_{\mu_1\tilde{\mu}_1}\left[\underline R_\mathrm{W_{\!c}}\left(\tilde{w}_1,B_\mathrm c\left(\boldsymbol w_{12} \right)\right)\right]C^{1\mu_j}_{\frac12\tilde{\mu}_1 \frac12\tilde{\mu}_{2}}
\nonumber\\&&\times D^{\frac12}_{\tilde{\mu}_{2}'\tilde{\mu}_{2}}\left[\underline R_\mathrm{W_{\!c}}\left(\tilde{w}_{2},B_\mathrm c^{-1}\left(\boldsymbol w'_{12}\right)B_\mathrm c\left(\boldsymbol w_{12}\right)\right)\right]\nonumber\\&&\times
C^{1\mu_j'}_{\frac12\tilde{\mu}_2' \frac12\tilde{\mu}_{1}'}D^{\frac12}_{\tilde{\mu}_1'\mu_1}\left[\underline R_\mathrm{W_{\!c}}\left(w'_{1},B_\mathrm c^{-1}\left(\boldsymbol w'_{12}\right)\right)\right]\,.
\end{eqnarray}
The expressions for the form factors Eqs. (\ref{eq:ff1me})-(\ref{eq:ffGMme}) cannot be simplified further like in the pseudoscalar case. However, for point-like constituents like quarks with $F_1^1(Q^2)+F_1^2(Q^2)=1$ and $F_2^1(Q^2)=F_2^2(Q^2)=0$, they become a little bit shorter.
\subsubsection{Projection Tensors}
Another, equivalent prescription to extract the physical form factors is proposed in Refs.~\cite{Karmanov:1994ck,Carbonell:1998rj}. 
To this end we define appropriate tensors $F_{1\mu}^{\sigma\tau}$, $F_{2\mu}^{\sigma\tau}$ and $G_{\mathrm M\mu}^{\sigma\tau}$ that project out the form factors $F_1$, $F_2$ and $G_\mathrm M$ from the current tensor $J_\mathrm V (\boldsymbol k_{\mathrm C}',\boldsymbol k_{\mathrm C},K_{\mathrm e})^\mu_{\sigma\tau}$. These projection tensors are fixed by Eq.~(\ref{eq:currenttensordecomp}) and given by~\cite{Karmanov:1994ck,Carbonell:1998rj} 
\begin{eqnarray}
F_{1\mu}^{\sigma\tau}&:=&
 \frac{K_{\mathrm e\mu}}{ K_\mathrm e\cdot K_\mathrm C}\left(\mathrm g^{\sigma\tau}-\frac{q^\sigma q^\tau}{q^2}-\frac{K_\mathrm C^\sigma K_\mathrm e^\tau+K_\mathrm C^\tau K_\mathrm e^\sigma}{K_\mathrm e\cdot K_\mathrm C}+K_\mathrm C^2
\frac{K_\mathrm e^\sigma K_\mathrm e^\tau}{(K_\mathrm e\cdot K_\mathrm C)^2}\right)\,,\nonumber\\\\
F_{2\mu}^{\sigma\tau}&:=&-\frac{K_{\mathrm e\mu}}{ (K_\mathrm e\cdot K_\mathrm C)q^2}\left(\mathrm g^{\sigma\tau}-\frac{q^\sigma q^\tau}{q^2}-
\frac{K_\mathrm C^\sigma K_\mathrm e^\tau+K_\mathrm C^\tau K_\mathrm e^\sigma}{K_\mathrm e\cdot K_\mathrm C}
\right.\nonumber\\&&\;\;\;\;+4 m_{\mathrm C}^2\frac{K_\mathrm e^\sigma K_\mathrm e^\tau}{ (K_\mathrm e\cdot K_\mathrm C)^2}-\left.\frac{q^\sigma K_\mathrm e^\tau-q^\tau K_\mathrm e^\sigma}{K_\mathrm e\cdot K_\mathrm C}\right)\,,\\
 G_{\mathrm M\mu}^{\sigma\tau}&:=&\frac12
\left[\frac{\mathrm g^\sigma_\mu q^\tau-\mathrm g^\tau_\mu q^\sigma}{q^2}+\frac{\mathrm g^\sigma_\mu K_\mathrm e^\tau+\mathrm g^\tau_\mu K_\mathrm e^\sigma}{K_\mathrm e\cdot K_\mathrm C}\right.\nonumber\\&&\left.
\;\;\;\;+\frac{K_{e\mu}}{K_e\cdot K_\mathrm C}\left(-K_\mathrm C^2\frac{q^\sigma K_\mathrm e^\tau-q^\tau K_\mathrm e^\sigma}{(K_\mathrm e\cdot K_\mathrm C)q^2}+
\frac{q^\sigma K_\mathrm C^\tau-q^\tau K_\mathrm C^\sigma}{q^2}\right.\right.\nonumber\\&&\left.\left.\;\;\;\;\;\;\;\;+
2K_\mathrm C^2\frac{K_\mathrm e^\sigma K_\mathrm e^\tau}{(K_\mathrm e\cdot K_\mathrm C)^2}-
\frac{K_\mathrm C^\sigma K_\mathrm e^\tau-K_\mathrm C^\tau K_\mathrm e^\sigma}{K_\mathrm e\cdot K_\mathrm C}\right)\right.\nonumber\\&&\left.\;\;\;\;+ 
K_{\mathrm C\mu}\left(\frac{q^\sigma K_\mathrm e^\tau-q^\tau K_e^\sigma}{q^2(K_\mathrm e\cdot K_\mathrm C)}-
2\frac{K_\mathrm e^\sigma K_\mathrm e^\tau}{(K_\mathrm e\cdot K_\mathrm C)^2}\right)
-q_{\mu} \frac{q^\sigma K_\mathrm e^\tau+q^\tau K_\mathrm e^\sigma}{(K_\mathrm e\cdot K_\mathrm C)q^2}
\right]\,.\nonumber\\
\end{eqnarray}
 Then the form factors are obtained by contraction of the projection tensors with the current tensor:
\begin{eqnarray}\label{eq:ff1projection}
 F_1(Q^2)&=&\frac{1}{|\,\mathrm e\,|}\lim_{k\rightarrow \infty}F_{1\mu}^{\sigma\tau} J_\mathrm V (\boldsymbol k_{\mathrm C}',\boldsymbol k_{\mathrm C},K_{\mathrm e})^\mu_{\sigma\tau}\,,\\
F_2(Q^2)&=&\frac{1}{|\,\mathrm e\,|}\lim_{k\rightarrow \infty}F_{2\mu}^{\sigma\tau} J_\mathrm V (\boldsymbol k_{\mathrm C}',\boldsymbol k_{\mathrm C},K_{\mathrm e})^\mu_{\sigma\tau}\,,\label{eq:ff2projection}\\
G_\mathrm M(Q^2)&=&\frac{1}{|\,\mathrm e\,|}\lim_{k\rightarrow \infty}G_{\mathrm M\mu}^{\sigma\tau} J_\mathrm V (\boldsymbol k_{\mathrm C}',\boldsymbol k_{\mathrm C},K_{\mathrm e})^\mu_{\sigma\tau}\,.\label{eq:ffGMprojection}
\end{eqnarray}
These are finite expressions and independent of $k$. It can be shown that these expressions are identical to the corresponding ones obtained from current matrix elements, Eqs.~(\ref{eq:ff1me}),~(\ref{eq:ff2me}) and (\ref{eq:ffGMme}) (for details, see App.~\ref{app:projecttensors}).
\subsubsection{Physical Current}\label{sec:spuriousconstrib}
We have already discussed in Sec.~\ref{sec:currentconservviolation} the failure of showing continuity for the bound-state current $J^\mu_{\mathrm {V}}(\boldsymbol p_{\mathrm C}',\sigma_j';\boldsymbol p_{\mathrm C},\sigma_j;P_{\mathrm e})$. Actually the current is not conserved due to the spurious contributions of $b_7$ and $b_8$ (cf. Eq.~(\ref{eq:currentphneospin1cov})). A numerical analysis using our standard kinematics shows that these contributions do not cancel such that there is a non-vanishing contribution to $J^1_{\mathrm {V}}(1,0)$. Even in the $k \rightarrow
\infty$ limit these continuity violating contributions survive. Nevertheless, with an unambiguous prescription at hand to separate the physical from the unphysical contributions we can define the physical part of the current by
\begin{eqnarray}\lefteqn{
  \frac{1}{|\,\mathrm e\,|}I_{\mathrm V}^\mu(\boldsymbol k_\mathrm C^\prime,\mu_j';\boldsymbol  k_\mathrm C,\mu_j)}\nonumber\\&:=&
\left[f_1(Q^2,k)\epsilon^\ast_{\mu_j'}(\boldsymbol k'_\mathrm C)\cdot\epsilon_{\mu_j}(\boldsymbol k_\mathrm C)+f_2(Q^2,k)\frac{[\epsilon^\ast_{\mu'_j}(\boldsymbol k'_\mathrm C)\cdot q]
[\epsilon_{\mu_j}(\boldsymbol k_\mathrm C)\cdot q]}{2m_{\mathrm C}^2 }\right] K_\mathrm C^\mu\nonumber\\&&+
g_{\mathrm M}(Q^2,k)\left[\epsilon^{\mu\ast}_{\mu'_j}(\boldsymbol k'_\mathrm C)[\epsilon_{\mu_j}(\boldsymbol k_\mathrm C)\cdot q]-
\epsilon_{\mu_j}^\mu (\boldsymbol k_\mathrm C)[\epsilon^\ast_{\mu'_j}(\boldsymbol k'_\mathrm C)\cdot q]\right]\,.
 \end{eqnarray}
The covariant physical current is then obtained from Eq.~(\ref{eq:physicalmicrocurrent}) by
\begin{eqnarray}
 I^\mu_{\mathrm V\sigma_j'\sigma_j}(\boldsymbol V)&:=&B_{\mathrm c}(\boldsymbol V)_{\,\,\nu}^\mu
\lim_{k\rightarrow \infty}
I_{\mathrm V}^\nu(\boldsymbol k_\mathrm C^\prime,\mu_j';\boldsymbol  k_\mathrm C,\mu_j)\nonumber\\&&\times 
D^{j\ast}_{\mu_j '\sigma_j' }[\underline R_\mathrm{W_{\!c}}^{-1}(w_{\mathrm C}',B_{\mathrm c}(\boldsymbol V))]  
D^{j}_{\mu_j \sigma_j }[\underline R_\mathrm{W_{\!c}}^{-1}(w_{\mathrm C},B_{\mathrm c}(\boldsymbol V))]\,.\nonumber\\
\end{eqnarray}
This current satisfies all required properties including current conservation and macrocausality.
\subsubsection{Angular Condition}
The four matrix elements $I^0_{11}, I^0_{1-1}, I^0_{10}$ and $I^0_{00}$  with $I^\mu_{\mathrm V\sigma_j'\sigma_j}(\boldsymbol 0)=:I^\mu_{\sigma_j'\sigma_j}$  depend on the 3 physical form factors and hence are not independent from each other.
They satisfy the, so-called, \textit{angular condition} (cf., e.g., Refs.~\cite{Grach:1983hd,Carbonell:1998rj,Bakker:2002aw})
\begin{eqnarray}\label{eq:angularcondition}
(1+2\eta)I^0_{11}+I^0_{1-1}-2\sqrt{2\eta}I^0_{10}-I^0_{00}=0\,,
\end{eqnarray}
where $\eta=\frac{Q^2}{4 m_{\mathrm C}^2}$.
For the matrix elements of the full current $J^0_{\mu_j'\mu_j}$ including spurious contributions the angular condition is, however, not satisfied due to the spurious contributions of $B_5$ and $B_7$ (cf. Eqs.~(\ref{eq:J000})-(\ref{eq:J0m11})):
  \begin{eqnarray}\label{eq:angularconditionviolation}&&
\lim_{k\rightarrow \infty}\frac{1}{2k}\left[(1+2\eta)J_\mathrm V^0(1,1)+J_\mathrm V^0(1,-1)-2\sqrt{2\eta}J^0_\mathrm V(1,0)-J^0_\mathrm V(0,0)\right]\nonumber\\&&\;\;\;\;=-\,|\,\mathrm e\,|\,\left[B_5(Q^2)+B_7(Q^2)\right]\,,
\end{eqnarray}
with $B_i(Q^2):=\lim_{k\rightarrow \infty}b_i(Q^2,k)$.

From Eq.~(\ref{eq:angularconditionviolation}) it becomes evident that
the 4 matrix elements $J^0_{11},J^0_{1-1},J^0_{10}$ and $J^0_{00}$ cannot be used to extract the 3 physical form factors in an unambiguous way. In the literature different triplets have been chosen to calculate the form factors~\cite{Grach:1983hd,Chung:1988my,Brodsky:1992px,Frankfurt:1993ut}. These different prescriptions lead to, in general, different results for the form factors due to non-physical contributions that enter the form factors (for a numerical and analytical comparison of different approaches, see Ref.~\cite{Cardarelli:1994yq} and Ref.~\cite{Karmanov:1996qc}, respectively). It is clear that if the angular condition were satisfied, different prescriptions would lead to the same form factor results~\cite{Keister:1993mg,Carbonell:1998rj}. However, the magnetic form factor obtained from the plus component current would still contain the spurious form factor $B_6$ (for the relations between the form factors obtained from different prescriptions cf. Ref.~\cite{Carbonell:1998rj}). Translating to our case this can be seen from Eqs.~(\ref{eq:J011}) and~(\ref{eq:J010}):
\begin{eqnarray}
 G_\mathrm M(Q^2)-B_6(Q^2)=\frac{1}{|\,\mathrm e\,|}\lim_{k\rightarrow\infty}\frac1k(J^0_\mathrm V(1,0)-\frac{m_\mathrm C\sqrt{2}}{Q}J^0_\mathrm V(1,1))\,.\label{eq:GMB6}
\end{eqnarray}
 Anyway, as we shall see later by a numerical analysis, the spurious contributions, that violate continuity and the angular condition together with $B_6$, are relatively small.
\subsubsection{$G_\mathrm C$, $G_\mathrm Q$ and Elastic Scattering Observables}
The charge and quadrupole form factors $G_\mathrm C$ and $G_\mathrm Q$, respectively, are expressed through $F_1$, $F_2$ and $G_\mathrm M$ by
 \begin{eqnarray}
 G_\mathrm C(Q^2)&=& -F_1(Q^2)-\frac{2\eta}{3}\left[F_1(Q^2)+G_\mathrm M(Q^2)-F_2(Q^2)(1+\eta)\right]\,,\nonumber\\\label{eq:rhoGC}
\\
G_\mathrm Q(Q^2)&=&-F_1(Q^2)-G_\mathrm M(Q^2)+F_2(Q^2)(1+\eta)\,.\label{eq:rhoGQ}
\end{eqnarray}
These form factors have the limits given by
\begin{eqnarray}
\lim_{Q^2\rightarrow0}G_\mathrm C(Q^2)&=&1\,,\\\label{eq:GCrho0}
\lim_{Q^2\rightarrow0}G_\mathrm M(Q^2)&=&\mu_\mathrm V\,,\label{eq:murho}\\
\lim_{Q^2\rightarrow0}G_\mathrm Q(Q^2)&=&Q_\mathrm V\,,\label{eq:qrho}
\end{eqnarray}
where $+1$ is the charge in units of the fundamental charge $|\,\mathrm e\,|$, $\mu_\mathrm V$ the magnetic dipole moment in units of $|\,\mathrm e\,|/2m_{\mathrm C}$ and $Q_\mathrm V$ the electric quadrupole moment in units of $|\,\mathrm e\,|/m_{\mathrm C}^2$.
For point-like spin-1 bound systems the magnetic dipole and the electric quadrupole moments are $\mu_\mathrm V=2$ and $Q_\mathrm V=-1$, respectively.

The usual observables of elastic scattering of an electron by a vector bound-state are the structure functions $A(Q^2)$, $B(Q^2)$ and the tensor polarization $T_{20}(Q^2)$. 
$A(Q^2)$, $B(Q^2)$ are
determined from the unpolarized laboratory frame differential cross section using the Rosenbluth formula.
They are given in terms of form factors by 
\begin{eqnarray}\label{eq:Aobservable}
A(Q^2)=G_\mathrm C^2(Q^2)+\frac{8}{9}\eta^2G_\mathrm Q^2(Q^2)+\frac23 \eta\, G_\mathrm M^2(Q^2)
\end{eqnarray}and
\begin{eqnarray}
B(Q^2)=\frac43 \eta (1+\eta)G_\mathrm M^2(Q^2)\,.\label{eq:Bobservable}
\end{eqnarray}
The observable $T_{20}(Q^2)$ for quadrupole polarization is extracted from the difference 
in the cross sections for target bound states having canonical spin polarizations $\mu_j=1$ and $\mu_j=0$.
In terms of form factors it reads \begin{eqnarray}\label{eq:T20}
\lefteqn{T_{20}(Q^2)}\nonumber\\&=&-\sqrt 2 \eta\,  \frac{\frac49 \eta \, G_\mathrm Q^2(Q^2)+\frac43 G_\mathrm Q(Q^2)G_\mathrm C(Q^2)+\frac13(\frac12+(1+\eta)\tan^2(\theta/2))G_\mathrm M^2(Q^2)}{A(Q^2)+B(Q^2)\tan^2(\theta/2)}\,.\nonumber\\
\end{eqnarray}
\subsubsection{Comparison with Covariant Light-Front Dynamics}
The covariant structure of the bound-state current $J^\mu_{\mathrm {V}}(\boldsymbol k_{\mathrm C}',\mu_j',\boldsymbol k_{\mathrm C},\mu_j,K_{\mathrm e})$ resembles the corresponding one obtained within the covariant
light-front approach of Refs.~\cite{Karmanov:1994ck,Carbonell:1998rj}. In these works the authors encounter 8 spurious contributions to the current that are associated with $\omega^\mu$. These $\omega$-dependent spurious contributions correspond to our $K_\mathrm e$-dependent spurious contributions. As in the pseudoscalar case, our spurious contributions to the spin-1 current can be traced back to the violation of
cluster separability (cf.~Sec~\ref{sec:clusterproperties}). The $\omega$-dependent contributions of
Ref.~\cite{Carbonell:1998rj} are rather the consequence of
the most general ansatz for a current of a vector bound state. 
The difference to the pseudoscalar case is, however, that we cannot completely get rid of all 8 spurious contributions by simply taking the $k\rightarrow \infty$ limit. This resembles the situation in standard light-front dynamics with $\omega=(1,0,0,-1)$, in which the extraction of the form factors is based on the plus component of the current operator. By restricting to the plus component in the standard approach the 4 spurious contributions of $B_1,\ldots,B_4$, proportional to $\omega^\mu$, are eliminated. However, the contributions of $B_5,\ldots,B_8$ survive which lead to the violation of the angular condition. In this sense the prescription of taking the limit $k\rightarrow \infty$ in our approach can be compared to taking the plus component of the current operator in standard light-front dynamics. 

The only consistent way to extract the physical form factors unambiguously in covariant light front dynamics is the method of projecting them out of the current tensor by means of appropriate tensors (cf. Refs.~\cite{Karmanov:1994ck,Carbonell:1998rj} and the above discussion). 
After elimination of the spurious contributions the spin matrix elements of the plus component of the remaining physical part satisfy the usual angular condition, Eq.~(\ref{eq:angularcondition}).
\subsubsection{Comparison with Point-Form Spectator Approximation}
The point-form of relativistic dynamics has also been used to calculate the elastic form factors of a vector system, cf. e.g. Ref.~\cite{Allen:2000ge}. Similar as in the point-form spectator model the Bakamjian-Thomas
type framework is only applied on the Hilbert space of the 2 constituents (and not on the electron-bound-state Hilbert space as in our case). This means that cluster separability is trivially satisfied. The point-form Bakamjian-Thomas construction is used to obtain the bound-state wave function and the bound-state mass which are then inserted into a general ansatz for a Lorentz covariant electromagnetic bound-state current for a spin-1 system. The independent matrix elements of this spectator current are defined in the Breit frame where the $J_\mathrm V^\mu(1,-1)$ contribution vanishes. The constraints from current conservation are also imposed in the Breit frame where they are most easily satisfied. Due to the kinematic nature of Lorentz transformations in the point form the current can be transformed into any arbitrary frame. This procedure ensures that the angular condition is satisfied. The 3 electromagnetic form factors are then extracted in the Breit frame from the 3 independent current matrix elements. As in the case of the spectator model, the momentum transfer on the active constituent is greater than the momentum transfer on the bound state~\cite{Allen:2000ge}.
\chapter{Models}\label{chap:6}
\section{Introduction}
In order to carry out a numerical analysis of the form factors derived in the previous chapter we have to specify the dynamics to obtain a bound-state wave function and a bound-state mass that can be inserted into the analytical expressions for the form factors. In this chapter we introduce the particular models we will be using in order to describe the interaction between the constituents. For simplicity we will restrict ourselves to instantaneous interactions which can be used within the Bakamjian-Thomas framework without losing Poincar\'{e} invariance.
According to Eq.~(\ref{eq:mC}) the term $\hat m_{\mathrm{int}}$ describing this instantaneous interaction is added to the free invariant two-particle mass operator $\hat m_{12}$. For hadrons, such as the $\pi$ meson (pion) or the $\rho$ meson, we need a confining quark-antiquark interaction which will be modeled in our case by a simple harmonic-oscillator potential~\cite{Keister:1991sb}. 
For nuclei, such as the deuteron, the interaction between the 2 nucleons is described by a simple Walecka-type two-meson-exchange model. In the static approximation each term of this potential becomes a usual (instantaneous) Yukawa interaction.  
\section{Harmonic-Oscillator Model for Mesons}\label{sec:homodelformesons}
\subsection{Wave Function}
The fact that particles with color, like quarks, are not observed as free particles is known as \textit{confinement}. The three-dimensional isotropic harmonic oscillator provides already a simple analytically solvable model for 2 quarks with a confining interaction. In the following we will restrict ourselves to the case of equal quark masses $m_\mathrm q$.
We start our rather brief discussion, which goes along the lines of Refs.~\cite{Keister:1991sb,KrassniggDiss:2001}, by squaring the mass operator in Eq.~(\ref{eq:mC}) to obtain \begin{eqnarray}\label{eq:mCsquared}
\hat m_{\mathrm C}^2=\hat m_{12}^2+\hat {\tilde m}_{\mathrm {int}}^2\,.
\end{eqnarray}
By setting 
\begin{eqnarray}
\hat {\tilde m}_{\mathrm {int}}^2=-4 a^2\boldsymbol \nabla_{\tilde {\boldsymbol k}}^2\,,
\end{eqnarray}
the square of the total mass operator $\hat m_{\mathrm C}^2$ in Eq.~(\ref{eq:mCsquared}) takes the form of a harmonic-oscillator Hamiltonian. Here the parameter $a$ is the oscillator length. The corresponding eigenvalue problem for $\hat m_{\mathrm C}^2$
 \begin{eqnarray}
 \hat m_{\mathrm C}^2\varPsi(\tilde{\boldsymbol k})=m_{\mathrm C}^2\varPsi(\tilde{\boldsymbol k})
 \end{eqnarray}
leads to the eigenfunctions $\varPsi(\tilde{\boldsymbol k})=\varPsi_{nl\mu_l}(\tilde{\boldsymbol k})=u_{nl}(\tilde{ k})Y_{l\mu_l}(\hat{\tilde{\boldsymbol k}})$.
Here the radial wave function is given by
\begin{eqnarray}\label{eq:unl}
 u_{nl}(\tilde{ k})=\frac{1}{\pi^{\frac14}a^{\frac32}}\sqrt{\frac{2^{n+l+2}n!}{(2n+2l+1)!!}}L_n^{l+\frac12}\left(\frac{\tilde k^2}{a^2}\right)\left(\frac{\tilde k}{a}\right)^l\mathrm e^{-\frac{\tilde k^2}{2a^2}}
 \end{eqnarray}
where $L_n^{l+\frac12}$ are the generalized Laguerre polynomials.
The eigenvalues of $\hat m_{\mathrm C}^2$ are given by
\begin{eqnarray}\label{eq:spectrumho}
m_{nl}^2=8a^2\left(2n+l+\frac32\right)+4m_\mathrm q.
 \end{eqnarray} 
\subsection{Parameters}
\label{sec:HOparameters}
 There are just 2 free parameters in this simple model, the
oscillator parameter $a$ and the quark mass $m_{\mathrm q}$.
For the simplest case of the ground state the wave function reads
\begin{equation}\label{eq:u00}
u_{00}(\tilde{ k}) = \frac{2}{\pi^{\frac14}a^{\frac32}}\mathrm e^{-\frac{\tilde k^2}{2a^2}}\,.
\end{equation}
The wave function is solely determined by the oscillator parameter
$a$ (cf. Eq.~(\ref{eq:unl})). We are going to use 2 sets of parameters which are fixed either by fitting the electromagnetic pion form factor or by fitting the mass spectra. For the latter we observe that a free constant $c_0$ can be
added to the confinement potential to shift the mass spectrum without changing the wave function. We have fixed the parameters $a$, $m_{\mathrm q}$ and $c_0$ for the harmonic-oscillator confinement potential through the
vector-meson spectrum as done in Refs.~\cite{KrassniggDiss:2001,Krassnigg:2003gh}. With the values for the
parameters $a=312~\mathrm{MeV}$, $c_0=-1.04~\mathrm{GeV}^2$ and
$m_{\mathrm q}=340~\mathrm{MeV}$, the masses $m_{00}$ and $m_{10}$ of the vector-meson ground states and
first excited states, respectively, are reasonably well reproduced~\cite{Krassnigg:2003gh}. Applying them to
the $\pi$ and $\rho$ meson and its excitations we observe that the first and second radial excitations are about 10\% too high as compared to
experiment and the $\pi$ and $\rho$ ground state have a mass of $770~\mathrm{MeV}$~\cite{KrassniggDiss:2001,Krassnigg:2003gh}.
These are quite reasonable values for a pure central confining potential
in view of the fact that spin-spin forces from an additional
hyperfine interaction can bring them close to the experimental
masses~\cite{Carlson:1983rw}.
\section{Walecka-Type Model for the Deuteron}
\label{sec:waleckatypemodel}
In this section we propose a simple nucleon-nucleon interaction potential that is inspired by the, so-called, \textit{Walecka model}~\cite{ref.01}. The derivation presented here can also be found in Ref.~\cite{Bakker:2010}. It is the first step of a benchmark calculation with the purpose to reach unanimity in different approaches of few-body physics regarding the definition of \lq relativistic effects'. The idea is to find a common ground on which the deuteron electromagnetic form factors are calculated in various approaches and then compared with each other. All numerical computations of this section have been carried out with {\sc Mathematica}$^{\begin{scriptsize}\textcopyright                                                                                                                                                                                                                                                                                                                                              \end{scriptsize}                                                                                                                                                                                                                                                                                                                                                                                                                                                                                                                                                                                                                                                                                                                                                                                                                                         }$.
\subsection{Introduction}
\label{Sect.I}
We define an interaction Lagrangean density
that is simple enough to derive a simple, effective nucleon-nucleon
interaction, but realistic enough in the sense that this interaction
supports an s-wave bound state that resembles the s-wave part of the
deuteron. The Lagrangean
chosen is inspired by the $\sigma-\omega$-model, also known as the Walecka
model \cite{ref.01}, which is also often referred to as \textit{quantum hadrodynamics}~\cite{Serot:1997xg}. The present model means a drastic simplification,
which, of course, does not meet the sophistication of realistic
nucleon-nucleon interactions found to be needed to achieve precise
agreement with the data. 
Below we apply this idea in the simplest possible context,
namely to write the interaction in the form of a static potential to
be used in the non-relativistic Schr\"{o}dinger equation. Our model is
compared throughout to the Malfliet-Tjon potential~\cite{ref.04}. 
The Walecka model requires a regularization procedure for which we take Pauli-Villars
regularization~\cite{ref.02}. This amounts to introducing for every
exchanged physical boson an unphysical one that regulates the boson
propagator in such a way that the interaction defined by boson exchange
becomes well-behaved in momentum space.
\subsection{Lagrangean Density of the Walecka Model}
\label{Sect.II}

We use the system of natural units $\mathrm c=\hbar=1$, which means that
masses and momenta are given in MeV and radii in 1/MeV. The translation to lengths in fm is given by $\hbar\, \mathrm c = 197.3~\mathrm{MeV\,fm}$.

The interaction Lagrangean densities in the
Walecka model  are given by~\cite{ref.01,ref.05,ref.06}
\begin{equation}
 \hat {\mathcal L}^{\text{N}\sigma}_{\mathrm {int}}(x) = g_{\sigma} \hat {\bar{\psi}}_\mathrm N(x)\hat \psi_\mathrm N(x)\hat \sigma(x)
\label{eq.01}
\end{equation}
and
\begin{equation}
 \hat {\mathcal L}^{\text{N}\omega}_{\mathrm {int}}(x) =\mathrm  i\, g_{\omega} \hat{\bar{\psi}}_\mathrm N(x)
 \gamma_{\mu}\hat \psi_\mathrm N(x)\hat \omega^{\mu}(x)+
 \frac{f_{\omega}}{4m_{\mathrm N}} \hat{\bar{\psi}}_\mathrm N(x)\sigma_{\mu\nu}\hat \psi_\mathrm N(x)
 (\partial^{\mu}\hat\omega^{\nu}(x)-\partial^{\nu}\hat \omega^{\mu}(x)).
\label{eq.02}
\end{equation}
Here $\psi_\mathrm N(x)$ denotes the nucleon field of mass $m_{\text N}$, $\sigma (x)$
denotes a neutral scalar meson field of mass $m_{\sigma}$ and $\omega^{\mu}(x)$
denotes a neutral vector meson field of mass $m_{\omega}$.  $\hat{\mathcal
L}^{\text{N}\sigma}_{\mathrm {int}}(x)$ describes the coupling of the scalar meson ($\sigma$) to the nucleons and
$\hat{\mathcal L}^{\text{N}\omega}_{\mathrm {int}}(x)$ describes the coupling of the vector meson ($\omega$)
to the nucleons. 
Using the Feynman rules the second-order amplitudes for scalar and vector
exchange become, respectively
\begin{eqnarray}
m^{\mathrm N\sigma}_{\mathrm {int}}(\tilde k'_1, \tilde k_1) &=&   g^2_{\sigma} \bar{\tilde u}_{\tilde \mu'_1}(\tilde {\boldsymbol k}_1') \tilde u_{\tilde \mu_1}(\tilde {\boldsymbol k}_1)
 \frac{1}{(\tilde k'_1 - \tilde k_1)^2 - m^2_\sigma}
 \bar{\tilde u}_{\tilde \mu_2'}(\tilde {\boldsymbol k}_2') \tilde u_{\tilde \mu_2}(\tilde {\boldsymbol k}_2)\, ,
\\
 m^{\mathrm N\omega}_{\mathrm {int}}(\tilde k'_1, \tilde k_1)  &=&  - g^2_{\omega} \bar{\tilde u}_{\tilde \mu'_1}(\tilde {\boldsymbol k}_1') \gamma_\mu \tilde u_{\tilde \mu_1}(\tilde {\boldsymbol k}_1)
 \frac{\mathrm g^{\mu\nu} + (\tilde k'^\mu_1 - \tilde k^\mu_1) (\tilde k'^\nu_2 - \tilde k^\nu_2)/m_{\omega}^2}
{(\tilde k'_1 - \tilde k_1)^2 - m^2_\omega} \nonumber\\&&\times
\bar{\tilde u}_{\tilde \mu_2'}(\tilde {\boldsymbol k}_2')\gamma_\nu \tilde u_{\tilde \mu_2}(\tilde {\boldsymbol k}_2) \,.
\label{eq.03}
\end{eqnarray}
Here we use Dirac spinors normalized to unity, i.e. $\tilde u_{\sigma}(\boldsymbol p):=\frac{1}{\sqrt{2m}} u_{\sigma}(\boldsymbol p)$.
Note that for simplicity we take only pure vector exchange for the
$\omega$ meson. This can be justified by looking at a comparison
of different models for the vector-meson exchange in Ref.~\cite{ref.03}. Realistic models favour no, or only very small tensor
coupling for the  $\omega$ meson.

For the Pauli-Villars particles, the same basic forms for the second-order
matrix elements are used with one difference: the amplitudes have opposite
sign. This sign difference is, of course, the reason why the Pauli-Villars
bosons provide regularization of the potentials.

\subsubsection{Static Potentials in Momentum Space}
\label{Sect.III}

In the static (non-relativistic) approximation the amplitudes turn out to be~(for details see App.~\ref{App.A})
\begin{equation}
 m^{\mathrm {N}\sigma}_{\mathrm {int}} (\tilde{\boldsymbol q}_1) = -
 \frac{g_{\sigma}^2}{\tilde{\boldsymbol q}_1^2+m_{\sigma}^2}
\label{eq.04}
\end{equation}
and 
\begin{equation}
 m^{\mathrm {N}\omega}_{\mathrm {int}} (\tilde{\boldsymbol q}_1)=\frac{g_{\omega}^2}{\tilde{\boldsymbol q}_1^2+
 m_{\omega}^2}\,,
\label{eq.05}
\end{equation}
where $\tilde{\boldsymbol q}_1:= \tilde {\boldsymbol{k}}_1'-\tilde {\boldsymbol{k}}_1$ denotes the exchanged three-momentum. The energies
occurring in the Dirac spinors have been approximated by the nucleon
masses and only the large components of the Dirac spinors are retained.

\subsubsection{Potentials in Coordinate Space}
\label{Sect.IV}

The potentials in configuration space are given by the Fourier transforms
of expressions~(\ref{eq.04}) and (\ref{eq.05}):
\begin{equation}
 V^{\mathrm {N}\sigma}(\boldsymbol{r}) =
 \int \frac{\mathrm d^3 \tilde q_1}{(2\pi)^3}\,m^{\mathrm {N}\sigma}_{\mathrm {int}} (\tilde{\boldsymbol q}_1)\,
 \mathrm e^{\mathrm i \tilde{\boldsymbol q}_1\cdot \boldsymbol r}\,.
\label{eq.06}
\end{equation}
This integral can be done by contour integration giving the usual Yukawa potential~\cite{PeskinSchroeder}
\begin{equation}
\label{eq.07}
 V^{\mathrm {N}\sigma}(r)=-\frac{g_{\sigma}^2}{4\pi} \frac{ \mathrm e^{-m_{\sigma}r }}{r}\,,
\end{equation}
where $r=\vert \boldsymbol r\vert$.  Similarly for the vector-meson-exchange potential:
\begin{equation}
  V^{\mathrm {N}\omega}(r)=\frac{g_{\omega}^2}{4\pi} \frac{ \mathrm e^{-m_{\omega}r }}{r}\,.
\label{eq.08}
\end{equation}

\subsubsection{Pauli-Villars Regularization}
\label{Sect.V}

In the Pauli-Villars regularization prescription one introduces
fictitious heavy particles in the Lagrangean density. Then the regularized
interaction Lagrangean density is given by\footnote{Note that we include only one Pauli-Villars field for either meson. This is sufficient for cutting off the kernels in momentum space if the effective interaction is obtained by taking matrix elements between on-mass-shell nucleons.
Otherwise, e.g. if one uses the Bethe-Salpeter equation, more Pauli-Villars fields are needed~\cite{Karmanov}.}
\begin{eqnarray}
 \hat{\mathcal L}_{\text{int}}(x)&=& g_{\sigma} \hat {\bar{\psi}}_\mathrm N(x)\hat \psi_\mathrm N(x)\hat \sigma(x)+
\mathrm i\, g_{\omega} \hat{\bar{\psi}}_\mathrm N(x)
 \gamma_{\mu}\hat \psi_\mathrm N(x)\hat \omega^{\mu}(x)\nonumber\\&&+
\mathrm i\,g_{\sigma} \hat {\bar{\psi}}_\mathrm N(x)\hat \psi_\mathrm N(x)\hat \sigma_{\mathrm{PV}}(x)+
 g_{\omega} \hat{\bar{\psi}}_\mathrm N(x)
 \gamma_{\mu}\hat \psi_\mathrm N(x)\hat \omega^{\mu}_{\mathrm{PV}}(x)\,
,
\label{eq.09}
\end{eqnarray} 
where $\sigma_{\mathrm{PV}}(x)$ and $\omega^{\mu}_{\mathrm{PV}}(x)$ are the Pauli-Villars fields with large
masses $\varLambda_\sigma$ and  $\varLambda_\omega$, respectively. Note that we
couple the Pauli-Villars bosons with the same strength as the physical
bosons but with an additional factor $\mathrm i$, which translates in second order
into a relative minus sign of the amplitudes.

This introduction of Pauli-Villars fields in the Lagrangean results in
making the following replacement for the $\sigma$-meson propagators in
Eq.~(\ref{eq.04}):
\begin{equation}
 \frac{1}{\tilde{\boldsymbol q}_1^2+m_{\sigma}^2}\longrightarrow
 \frac{1}{\tilde{\boldsymbol q}_1^2+m_{\sigma}^2}-
 \frac{1}{\tilde{\boldsymbol q}_1^2+\varLambda_\sigma^2}\,.
\label{eq.10}
\end{equation}
Similarly for the $\omega$-meson propagator in Eq.~(\ref{eq.05}):
\begin{equation}
 \frac{1}{\tilde{\boldsymbol q}_1^2+
 m_{\omega}^2}\longrightarrow 
 \frac{1}{\tilde{\boldsymbol q}_1^2+
 m_{\omega}^2}-\frac{1}{\tilde{\boldsymbol q}_1^2+\varLambda_{\omega}^2}\,.
\label{eq.11}
\end{equation}
Therefore, the Walecka potential in configuration space finally reads
\begin{equation}
  V(r) = -\frac{g_{\sigma}^2}{4\pi}
 \frac{ \mathrm e^{-m_{\sigma}r }}{r}+
 \frac{g_{\sigma}^2}{4\pi} \frac{ \mathrm e^{-\varLambda_{\sigma}r }}{r}+
 \frac{g_{\omega}^2}{4\pi}
 \frac{ \mathrm e^{-m_{\omega}r }}{r}-\frac{g_{\omega}^2}{4\pi}
 \frac{ \mathrm e^{-\varLambda_{\omega}r }}{r}\,.
\label{eq.12}
\end{equation}
\subsection{Coordinate Space}
\subsubsection{Solving the Schr\"odinger Equation and Fixing the Parameters}
\label{Sect.VI}

The two-body s-wave ($l=0$) Schr\"odinger equation for the reduced deuteron wave function 
$w_{0}(r) := r\, \tilde u_{0}(r)$ reads 
\begin{equation}
 \left[-\frac{\hbar^2}{2m_{\text{red}}}\frac{\mathrm {d}^2}{\mathrm dr^2} +
 V(r)\right] w_{0}(r)=E_\mathrm B\,  w_{0}(r)\,,
\label{eq.13}
\end{equation}
where
\begin{equation}
 2m_{\text{red}} = \frac{2m_\mathrm pm_\mathrm n}{m_\mathrm p+m_\mathrm n} = 938.92~\mathrm{MeV}/\mathrm c^2 \approx m_{\rm N}\,.
\label{eq.14}
\end{equation} %
The potential has the form, (re)expressing $r$ in fm and the $\lambda$'s
in $\mathrm {fm}^{-1}$
\begin{equation}
 V(r) = \hbar\, \mathrm c\left[-\frac{g_{\sigma}^2}{4\pi}
 \frac{ \mathrm e^{-r/\lambda_{\sigma} }}{r}+
 \frac{g_{\sigma}^2}{4\pi}
 \frac{ \mathrm e^{-r/\lambda_{\mathrm{PV}\sigma} }}{r}+
 \frac{g_{\omega}^2}{4\pi} \frac{ \mathrm e^{-r/\lambda_{\omega} }}{r}-
 \frac{g_{\omega}^2}{4\pi}
 \frac{ \mathrm e^{-r/\lambda_{\mathrm{PV}\omega} }}{r}\right]\,,
\label{eq.15}
\end{equation}
where the quantities $\lambda$ are the Compton wavelengths of the
corresponding meson masses, e.g. $\lambda_\sigma = \hbar/m_\sigma \mathrm c$.

We use the existing data on the recommended $\sigma$-meson mass, the
experimental $\omega$-meson mass, the deuteron binding energy, and the
triplet scattering length to fix the parameters. For the Pauli-Villars
particles we use the mass values (cut off values) $\varLambda_\sigma =
1000$~MeV and $\varLambda_\omega = 1500$~MeV, which are reasonable values
for these cut offs, although they are somewhat smaller than usual. We did
not vary them, because the model proposed here is simply not realistic
enough to warrant much effort in this direction. Yet, we wanted this
model to produce a reasonable s-wave deuteron wave function.  The values
of the parameters we found are given in Tab.~\ref{table1}.
\begin{table*} 

\begin{small}
\begin{tabular}{|l|r|r|r|}

\hline
 parameters (units)& present &present, no PV & MT~\cite{ref.04}\\
\hline
$E_\mathrm B\, (\mathrm {MeV})$ & -2.224575 & -3.36772 & -2.27203\\
\hline
 $a_\mathrm t\, (\mathrm {fm})$ & 5.4151 & 4.58658 & 5.4739\\
\hline
$g_{\sigma}^2/4\pi$& 6.31 & 6.31 &  3.22749\\
\hline
$g_{\omega}^2/4\pi$& 18.617 & 18.617 & 7.40758\\
\hline
 $\lambda_{\sigma}\, (\mathrm {fm})\quad (m_{\sigma} \,(\mathrm {MeV/c^2}) )$&0.493317 (400) &0.493317 (400) &0.643087 (306.8)\\
\hline
$\lambda_{\omega}\, (\mathrm {fm}) \quad (m_{\omega} \,(\mathrm {MeV/c^2}) ) $&0.25211 (782.7) &0.25211 (782.7) & 0.321543 (613.6)\\
\hline
 $\lambda_{\mathrm{PV}\sigma}\, (\mathrm {fm})\quad  (\varLambda_{\sigma}  \,(\mathrm {MeV/c^2}))$ & 0.19732 (1000) & $0$ ($\infty$)& $0$ ($\infty$)\\
\hline
$\lambda_{\mathrm{PV}\omega}\, (\mathrm {fm})\quad  (\varLambda_{\omega}  \,(\mathrm {MeV/c^2}))$ & 0.131551 (1500)  & $0$ ($\infty$)& $0$ ($\infty$)\\
\hline

\end{tabular}
\end{small}\caption{\label{table1} The values for the masses
and coupling constants are chosen such that the deuteron binding energy
$E_\mathrm B$ and the scattering length $a_t$ are well reproduced. These values
are also similar to those proposed in~\cite{ref.03}.}
\end{table*}

\begin{figure*}[h!]
\centering 
\psfrag{-}{\begin{small}\;-              \end{small}}
\psfrag{100}{\begin{small}\!\!
100                \end{small}}
\psfrag{50}{\begin{small}\!\!
50
           \end{small}}
\psfrag{1}{\begin{small}1              \end{small}}
\psfrag{2}{\begin{small}2            \end{small}}
\psfrag{3}{\begin{small}3           \end{small}}
\psfrag{4}{\begin{small}4      \end{small}}
\psfrag{r}{\begin{small}
$r$ (fm)        \end{small}}
\psfrag{V}{\!\!\!\!\!\!\!\!\!\!\!\!\!\begin{small}$V(r)$ (MeV)                                       \end{small}}
\psfrag{our}{\begin{small}
present model      \end{small}}
\psfrag{noPV}{\begin{small}
present model, no PV  \end{small}}
\psfrag{MT}{\begin{small}
Malfliet-Tjon III~\cite{ref.04}  \end{small}}
\vspace{8mm}
\includegraphics[width=100mm]{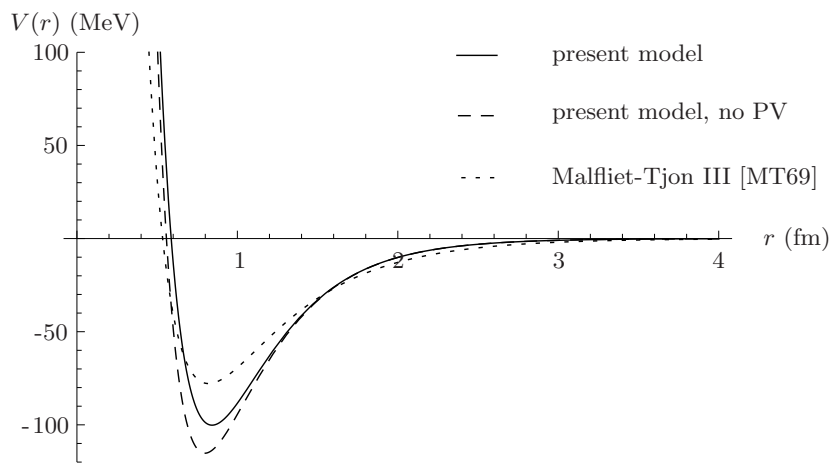}

\vspace{2mm}

\caption{\label{fig.01}The present potential with and without Pauli-Villars 
regularization compared with the Malfliet-Tjon 
potential~\cite{ref.04}. The parameters are given in Tab.~\ref{table1}.}
\end{figure*}

The potential is depicted in Fig.~\ref{fig.01} together with the simple
Malfliet-Tjon-III potential\cite{ref.04}.
 It is clear that the present
model has stronger attraction at intermediate range and stronger repulsion
at short range than the Malfliet-Tjon potential. This difference can
be expected to show up in the bound-state wave function at intermediate
range.

\begin{figure}[h!]
\centering 
\psfrag{0}{\begin{small}0
                \end{small}}
\psfrag{5}{\begin{small}5
           \end{small}}
\psfrag{10}{\begin{small}10              \end{small}}
\psfrag{20}{\begin{small}20          \end{small}}
\psfrag{15}{\begin{small}15          \end{small}}
\psfrag{0.1}{\begin{small}
\!\!\!0.1     \end{small}}
\psfrag{0.2}{\begin{small}
\!\!\!0.2     \end{small}}\psfrag{0.3}{\begin{small}
\!\!\!0.3    \end{small}}\psfrag{0.4}{\begin{small}
\!\!\!0.4     \end{small}}\psfrag{0.5}{\begin{small}
\!\!\!0.5     \end{small}}\psfrag{0.6}{\begin{small}
\!\!\!0.6    \end{small}}
\psfrag{r}{\begin{small}
$r$ (fm)        \end{small}}
\psfrag{u}{\!\!\!\!\!\!\begin{small}$w_{0}(r) (\mathrm{fm}^{-\frac12})$\end{small}}                                      
\psfrag{W}{\begin{small}
present model      \end{small}}
\psfrag{WnoPV}{\begin{small}
present model, no PV  \end{small}}
\psfrag{M}{\begin{small}
Malfliet-Tjon III~\cite{ref.04}  \end{small}}
\vspace{10mm}
\includegraphics[width=120mm]{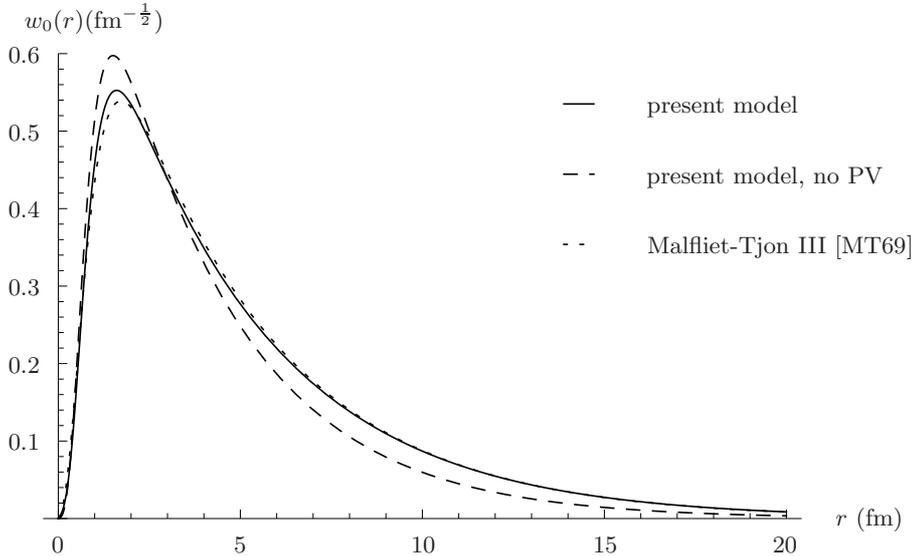}

\caption{\label{fig.02} Present model wave functions with and without 
Pauli-Villars regularization compared with the Malfliet-Tjon model wave 
function in units of $\mathrm{fm}^{-\frac{1}{2}}$. 
The Schr\"odinger equation for the Malfliet-Tjon potential is solved 
with the parameters given in Tab.~\ref{table1}.}
\end{figure}

The Schr\"odinger equation~(\ref{eq.13})
is solved numerically, with the solution $w_{0}(r)$ being subject to the boundary condition $w_{0}(0)=0$. 
The normalized bound state wave function $w_{0}(r)$ is depicted in 
Fig.~\ref{fig.02}.

\begin{figure}[h!]
\begin{center} 
\psfrag{-}{\begin{small}\;-              \end{small}}
\psfrag{2}{\begin{small}2
                \end{small}}
\psfrag{4}{\begin{small}4
           \end{small}}
\psfrag{6}{\begin{small}6             \end{small}}
\psfrag{8}{\begin{small}8          \end{small}}
\psfrag{10}{\begin{small}10          \end{small}}
\psfrag{r}{\begin{small}
$r$ (fm)        \end{small}}
\psfrag{u}{\!\!\!\!\!\!\begin{small}$w(r) (\mathrm{fm}^{-\frac12})$\end{small}}                                      
\psfrag{uw}{\begin{small}
$w_{0}(r)$    \end{small}}
\psfrag{uass}{\begin{small}
$w_{\mathrm{as}}(r)=5.421-r$ \end{small}}
\psfrag{unP}{\begin{small}
$w_{0}(r)$, no PV   \end{small}}
\psfrag{uassnP}{\begin{small}
$w_{\mathrm{as}}(r)=4.587-r$, no PV \end{small}}
\psfrag{M}{\begin{small}
Malfliet-Tjon III~\cite{ref.04}  \end{small}}
\vspace{8mm}

\includegraphics[width=110mm]{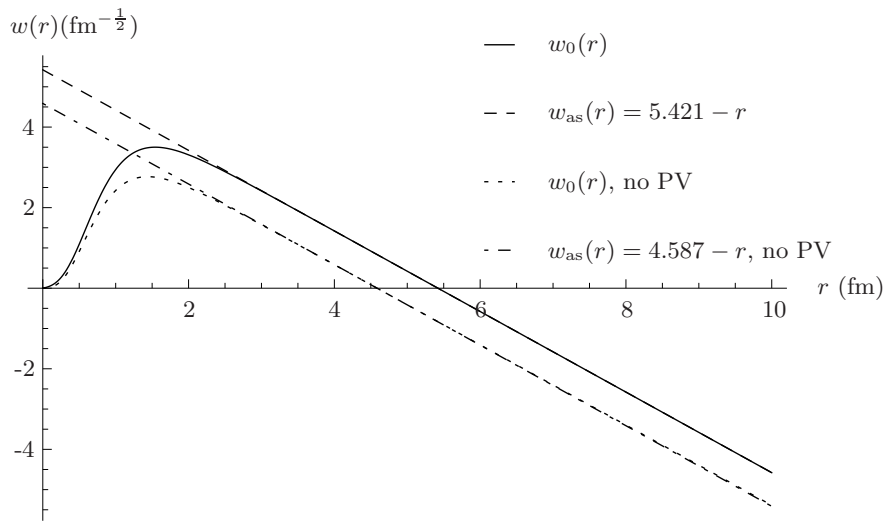}

\vspace{1ex}

\caption{\label{fig.03} Wave functions at energy $E_\mathrm B = 0$ for the present
model with and without Pauli-Villars regularization included in the
potential.  The scattering length is determined by a linear fit to the
wave function at large $r$.}

\vspace{1ex}

\end{center} 
\end{figure}

As a characteristic quantity that measures the quality of the wave function
non-locally, we  calculated the mean-square radius defined by~\cite{ref.08} 
\begin{equation}
 \langle r^2\rangle:=\frac{1}{4}\int_0^{\infty} {\mathrm dr}\, r^2 w_0^2(r)\,.
\label{eq.17}
\end{equation}
We find a value for the root-mean-squared radius $\sqrt{\langle
r^2\rangle}=1.95015$~fm close to Malfliet-Tjon's value $\sqrt{\langle
r^2\rangle}=1.97625$~fm. This is also in accordance with other results
for the deuteron root-mean-square radius~\cite{ref.08}.  The triplet
scattering length $a_\mathrm t$ is obtained from the solution $w_{0}(r)$ of the
Schr\"odinger equation Eq.~(\ref{eq.13}) corresponding to $E_\mathrm B=0$, see
Fig.~\ref{fig.03}. We find the value $a_\mathrm t = 5.421$~fm, compared to the
experimental value of $5.432$~fm \cite{ref.09}.
It is amusing to check the importance of the Pauli-Villars bosons in
the static approximation, where they are not needed for convergence. In
Table~\ref{table1} we give the values of the energy and the scattering
length in the situation that these bosons are omitted. Clearly, the values
of the  bound-state energy, $-3.368$~MeV, compared to $-2.225$~MeV,
and $4.587$~fm compared to $5.421$~fm, are quite reasonable, although
of course far outside the error bounds of the experimental data.

\subsubsection{Body Form Factor}
\label{Sect.VII}

\begin{figure}[h!]
\begin{center}
\psfrag{0.2}{\begin{small}
\!\!\!0.2
                \end{small}}
\psfrag{0.4}{\begin{small}
\!\!\!0.4
           \end{small}}
\psfrag{0.6}{\begin{small}
\!\!\!0.6             \end{small}}
\psfrag{0.8}{\begin{small}
\!\!\!0.8          \end{small}}
\psfrag{1.0}{\begin{small}
\!1         \end{small}}
\psfrag{200}{\begin{small}
\!\!\!200         \end{small}}
\psfrag{400}{\begin{small}
\!\!\!400         \end{small}}
\psfrag{600}{\begin{small}
\!\!\!600         \end{small}}
\psfrag{800}{\begin{small}
\!\!\!800         \end{small}}
\psfrag{1000}{\begin{small}
$10^3$         \end{small}}
\psfrag{q}{\begin{small}$\tilde q_1$ (MeV/c)        \end{small}}
\psfrag{F}{\!\!\!\begin{small}$F_\mathrm B(\tilde q_1)$\end{small}}                                      

\vspace{4ex}

\includegraphics[width=90mm]{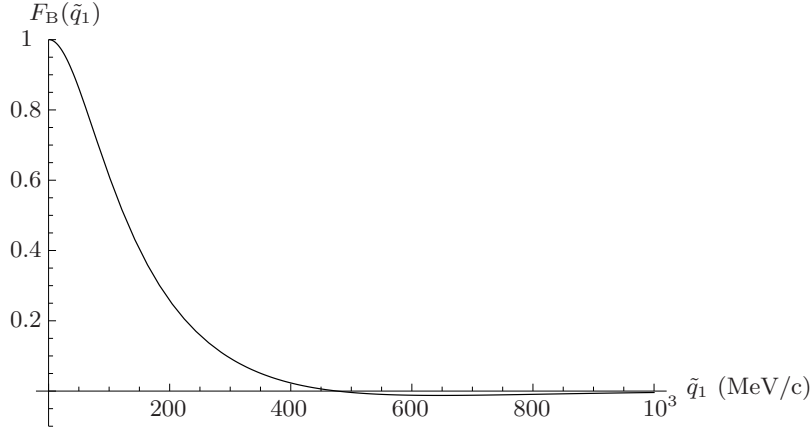}

\caption{\label{fig.04} The form factor in the present model including
Pauli-Villars regularization. The form factor for the Malfliet-Tjon
deuteron is indistinguishable on this scale.}

\end{center} 
\end{figure}

As we do not specify the charges of the fermions in the simple benchmark
model, we calculate the body form factor of the bound state only. For the
s-wave bound state found here, it is given by the well-known formula\footnote{$F_\mathrm B(\tilde q_1)$ corresponds to the electric charge form factor $G_\mathrm C(2 \tilde q_1)$ (cf. Eq.~\ref{eq:rhoGC}) for point-like nucleons and for the relativistic kinematical factors and spin-rotation factors set equal to one.}~\cite{BrownJackson}
\begin{equation}
 F_\mathrm B(\tilde q_1) = \int^\infty_0 {\rm d} r\, j_0 ( \tilde q_1 r)\, w_{0}^2 (r)\,,
\label{eq.18}
\end{equation}
where $j_0 ( \tilde q_1 r)=\sin  (\tilde q_1 r)/(\tilde q_1 r)$  is the first spherical Bessel function and
 $w_{0}(r)$ is normalized as
\begin{equation}
 \int^\infty_0 {\rm d} r\,w_{0}^2 (r) = 1\,.
\label{eq.19}
\end{equation}
This normalization guarantees $F_\mathrm B(0) = 1$. The resulting form factor for the
present model is depicted in Fig.~\ref{fig.04}.

\begin{figure}[h!]
\begin{center}

\psfrag{0.2}{\begin{small}
\!\!\!0.2
                \end{small}}
\psfrag{0.4}{\begin{small}
\!\!\!0.4
           \end{small}}
\psfrag{0.6}{\begin{small}
\!\!\!0.6             \end{small}}
\psfrag{0.8}{\begin{small}
\!\!\!0.8          \end{small}}
\psfrag{1.0}{\begin{small}
\!1         \end{small}}
\psfrag{200}{\begin{small}
\!\!\!200         \end{small}}
\psfrag{50}{\begin{small}
\!\!\!50         \end{small}}
\psfrag{100}{\begin{small}
\!\!\!100         \end{small}}
\psfrag{150}{\begin{small}
\!\!\!150         \end{small}}
\psfrag{1000}{\begin{small}
$10^3$         \end{small}}
\psfrag{q}{\begin{small}
$\tilde q_1$ (MeV/c)        \end{small}}
\psfrag{F}{\!\!\!\begin{small}$F(\tilde q_1)$\end{small}}                                      
\psfrag{ff}{\begin{small}$F_\mathrm B(\tilde q_1)$\end{small}}  
\psfrag{ffass}{\begin{small}$1-2\,\langle
r^2\rangle\,\tilde q_1^2/3$\end{small}}      
\includegraphics[width=90mm]{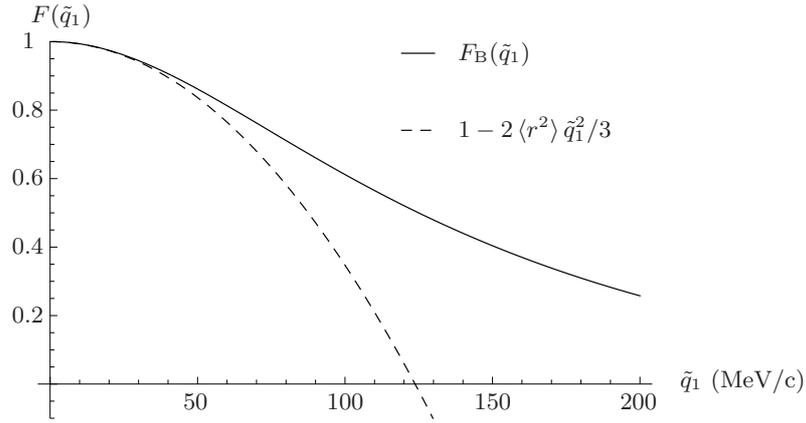}

\caption{\label{fig.05} The form factor together with the form for small $\tilde q_1$.}

\end{center} 
\end{figure}

The small-$\tilde q_1$ behaviour can be read off in the usual way from the small-$x$
behaviour of the spherical Bessel function, namely $j_0(x) \sim 1-x^2/6$,
which gives
\begin{equation}
  F_\mathrm B(\tilde q_1) \sim 1-  \frac{\tilde q_1^2}{6} \, \int^\infty_0 {\rm d} r\, r^2\,w_{0}^2 (r) \,.
\label{eq.20}
\end{equation}
In Fig.~\ref{fig.05} we show the form factor at small $\tilde q_1$ together
with the limiting form given in Eq.~(\ref{eq.20}). (Note the factor $4$
that comes from the factor $1/4$ in the definition of $\langle r^2 \rangle$,
Eq.~(\ref{eq.17}).)

\subsection{Momentum Space}
\label{Sect.VIII}
\subsubsection{Solving the Integral Equation}
\begin{figure}[h!]
\begin{center}

\psfrag{0.02}{\begin{small}
\!\!\!0.02
                \end{small}}
\psfrag{0.04}{\begin{small}
\!\!\!0.04
           \end{small}}
\psfrag{0.06}{\begin{small}
\!\!\!0.06             \end{small}}
\psfrag{0.08}{\begin{small}
\!\!\!0.08          \end{small}}
\psfrag{1.0}{\begin{small}
\!1         \end{small}}
\psfrag{200}{\begin{small}
\!\!\!200         \end{small}}
\psfrag{50}{\begin{small}
\!\!\!50         \end{small}}
\psfrag{100}{\begin{small}
\!\!\!100         \end{small}}
\psfrag{150}{\begin{small}
\!\!\!150         \end{small}}
\psfrag{250}{\begin{small}
\!\!\!250         \end{small}}
\psfrag{300}{\begin{small}
\!\!\!300         \end{small}}
\psfrag{0}{\begin{small}
\!\!\!0         \end{small}}
\psfrag{1000}{\begin{small}
$10^3$         \end{small}}
\psfrag{u}{\begin{small}\!\!\!\!\!\!\!\!\!$u_0(\tilde k) \,(\mathrm {MeV}^{-\frac32})$        \end{small}}
\psfrag{k}{\!\!\!\begin{small}$\tilde k$ (MeV/c)\end{small}}                                      
\psfrag{ff}{\begin{small}$F_\mathrm B(\tilde q_1)$\end{small}}  
\psfrag{ffass}{\begin{small}$1-2\,\langle
r^2\rangle\,\tilde q_1^2/3$\end{small}}      
\vspace{8ex}
\includegraphics[width=100mm]{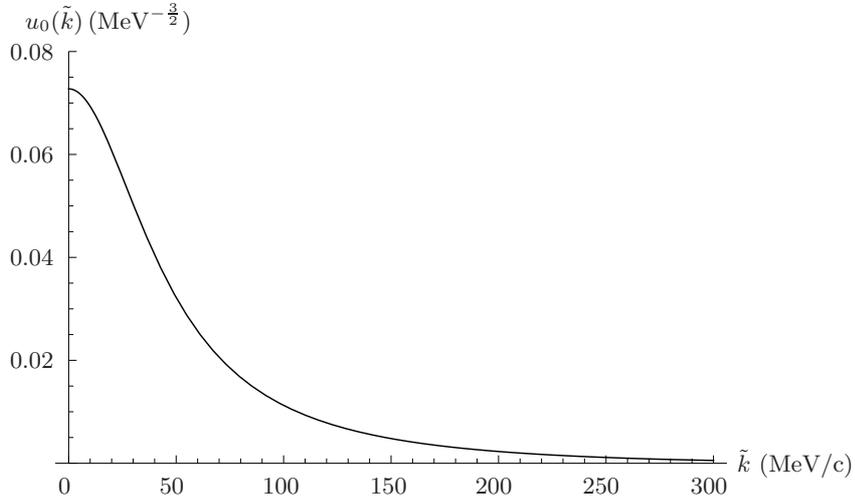}
 
\caption{\label{fig.06} The wave function $u_0(\tilde k) $ in momentum
space for the present model including Pauli-Villars regularization. The
wave function for the Malfliet-Tjon deuteron is indistinguishable on
this scale.}

\end{center} 
\end{figure}

\begin{figure}[h!]
\begin{center}

\psfrag{0.0005}{\begin{small}
\!\!\!\!\!\!\!\!0.0005
                \end{small}}
\psfrag{0.0010}{\begin{small}
\!\!\!\!\!0.001
           \end{small}}
\psfrag{0.0015}{\begin{small}
\!\!\!\!\!\!\!\!0.0015            \end{small}}
\psfrag{0.0020}{\begin{small}
         \end{small}}
\psfrag{1.0}{\begin{small}
\!1         \end{small}}
\psfrag{200}{\begin{small}
\!\!\!200         \end{small}}
\psfrag{50}{\begin{small}
\!\!\!50         \end{small}}
\psfrag{100}{\begin{small}
\!\!\!100         \end{small}}
\psfrag{150}{\begin{small}
\!\!\!150         \end{small}}
\psfrag{250}{\begin{small}
\!\!\!250         \end{small}}
\psfrag{300}{\begin{small}
\!\!\!300         \end{small}}
\psfrag{350}{\begin{small}
\!\!\!350         \end{small}}
\psfrag{400}{\begin{small}
\!\!\!400         \end{small}}
\psfrag{450}{\begin{small}
\!\!\!450         \end{small}}
\psfrag{u}{\begin{small}\!\!\!\!\!\!\!\!\!
\!\!\!\!\!$u_0(\tilde k) \,(\mathrm {MeV}^{-\frac32})$        \end{small}}
\psfrag{k}{\begin{small}$\tilde k$ (MeV/c)\end{small}}                                      
\psfrag{uW}{\begin{small}present model\end{small}}  
\psfrag{uMT}{\begin{small}Malfliet-Tjon III~\cite{ref.04}  \end{small}}      
\includegraphics[width=100mm]{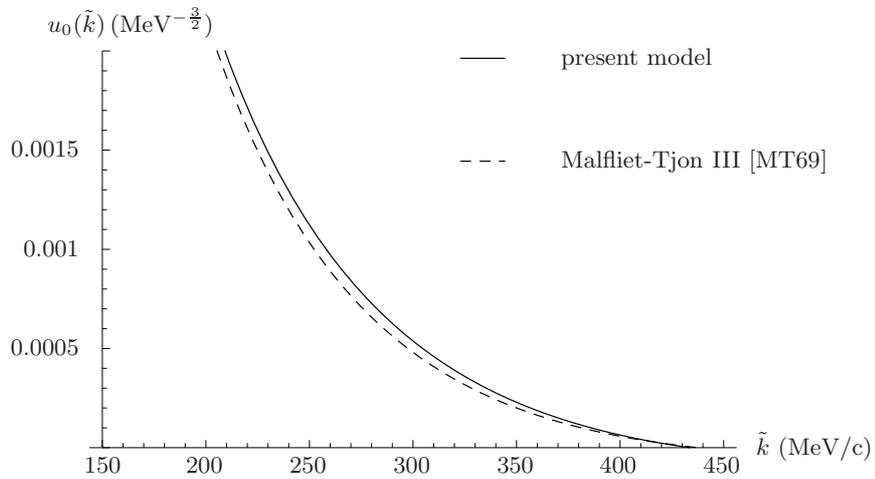}

\caption{\label{fig.07} The wave function $u_0(\tilde k) $ in momentum
space for the present model including Pauli-Villars regularization and
the wave function for the Malfliet-Tjon deuteron for large momenta.}

\end{center} 
\end{figure}

In order to facilitate the comparison of our model with calculations
restricted to momentum space, we transformed the Schr\"{o}dinger
equation to momentum space
which is the integral equation obtained by Fourier transforming Eq.~(\ref{eq.13}):
\begin{equation}
 \frac{\tilde k^2}{m_\mathrm N}\, u_{0}(\tilde k)+\int\frac{\mathrm d^3 \tilde k'}{(2 \pi)^3} \,m_{\mathrm {int}}(\vert\tilde{\boldsymbol k}_1-\tilde{\boldsymbol k}_1'\vert)\, u_{0}(\tilde k')=E_\mathrm B \,u_{0}(\tilde k)\,, 
\end{equation}
where 
\begin{eqnarray}
 m_{\mathrm {int}}(\vert\tilde{\boldsymbol k}-\tilde{\boldsymbol k}'\vert)&=&-\frac{g_\sigma^2}{(\tilde{\boldsymbol k}-\tilde{\boldsymbol k}')^2+m_{\sigma}^2}+
 \frac{g_\sigma^2}{(\tilde{\boldsymbol k}-\tilde{\boldsymbol k}')^2+\varLambda_{\sigma}^2}\nonumber\\&&+\frac{g_\omega^2}{(\tilde{\boldsymbol k}-\tilde{\boldsymbol k}')^2+m_{\omega}^2}-
 \frac{g_\omega^2}{(\tilde{\boldsymbol k}-\tilde{\boldsymbol k}')^2+\varLambda_{\omega}^2}\,.
\end{eqnarray}
Note that Eq.~(\ref{eq.19}) fixes the normalization of $u_0(\tilde k)$ as
 \begin{eqnarray}
 \int^\infty_0 {\rm d} \tilde k\, \tilde k^2\,u_0^2 (\tilde k) = (2\pi)^3\,.
\end{eqnarray} 
This defines the difference between $u_0 (\tilde k)$ and $u_{00} (\tilde k)$ from Eq.~(\ref{eq:normeigenfucntions}) by a factor of $(2\pi)^{\frac32}$:
\begin{eqnarray}
u_{00} (\tilde k)\equiv \frac{1}{(2\pi)^{\frac32}}\, u_0 (\tilde k)\,.
\end{eqnarray} 
Since there is only one bound-state for the deuteron, we have left away the index $n=0$ which also served as a distinction between $u_0 (\tilde k)$ and $u_{00} (\tilde k)$.\footnote{Note that we define $\tilde \varPsi_{00}(\boldsymbol{r}) =
Y_{00}(\hat{\boldsymbol r}) \tilde u_0(r)$ and its Fourier transform as  $\varPsi_{00}(\tilde{\boldsymbol k})
= Y_{00}(\hat{\tilde{\boldsymbol k}})u_0(\tilde k)$.}
The angular integrations can be done by expanding $m_{\mathrm {int}}(\vert\tilde{\boldsymbol k}_1-\tilde{\boldsymbol k}_1'\vert)$ in terms of partial waves (for a detailed discussion see, e.g. Ref.~\cite{vanIersel:2000}). 
The result is 
\begin{equation}\label{eq:momentumspacebseq}
 \frac{\tilde k^2}{m_\mathrm N}\,u_0(\tilde k)+\frac{1}{8 \pi^2}\int_{0}^{\infty}\mathrm d \tilde k'\,m_{\mathrm {int}}( \tilde k,\tilde k')\, u_0(\tilde k')=E_\mathrm B u_0(\tilde k)\,, 
\end{equation}
where
\begin{eqnarray}
m_{\mathrm {int}}( \tilde k,\tilde k')&=&\frac{\tilde k'}{\tilde k}\left\lbrace -g^2_{\sigma}\ln\left[ \frac{(\tilde k+\tilde k')^2+m_\sigma^2}{(\tilde k-\tilde k')^2+m_\sigma^2}\right]+
g^2_{\sigma}\ln\left[ \frac{(\tilde k+\tilde k')^2+\varLambda_\sigma^2}{(\tilde k-\tilde k')^2+\varLambda_\sigma^2}\right]\right.\nonumber\\&&\left.+g^2_{\omega}\ln\left[ \frac{(\tilde k+\tilde k')^2+m_\omega^2}{(\tilde k-\tilde k')^2+m_\omega^2}\right]-
g^2_{\omega}\ln\left[ \frac{(\tilde k+\tilde k')^2+\varLambda_\omega^2}{(\tilde k-\tilde k')^2+\varLambda_\omega^2}\right]\right\rbrace.\nonumber\\
\end{eqnarray}
The integral equation (\ref{eq:momentumspacebseq}) is solved for the wave function of the bound state with the numerical method of standard Gauss quadrature (for details see, e.g. Ref.~\cite{vanIersel:2000}).
The determination of the energy
served as a check of our numerical calculations. Indeed, we found the
same energy in momentum space and coordinate space. The momentum-space
wave function $u_0(\tilde k)$ is shown in Fig.~\ref{fig.06}. On this scale,
we cannot see the difference between the Malfliet-Tjon wave function and
the one obtained with the present model. Looking at Fig.~\ref{fig.02}
we see that in coordinate space the two models differ most below a few
fm, so we may expect the momentum-space wave functions to be different
at momenta around 200 MeV/c, which turns out to be true, as
Fig.~\ref{fig.07} shows.

\begin{figure}[h!]
\begin{center}

\psfrag{0.02}{\begin{small}
\!\!\!\!\!\!0.02
                \end{small}}
\psfrag{0.04}{\begin{small}
\!\!\!\!\!\!0.04
           \end{small}}
\psfrag{0.06}{\begin{small}
\!\!\!\!\!\!0.06             \end{small}}
\psfrag{0.08}{\begin{small}
\!\!\!\!\!\!0.08          \end{small}}
\psfrag{1.0}{\begin{small}
\!1         \end{small}}
\psfrag{20}{\begin{small}
\!\!\!20         \end{small}}
\psfrag{40}{\begin{small}
\!\!\!40         \end{small}}
\psfrag{60}{\begin{small}
\!\!\!60         \end{small}}
\psfrag{80}{\begin{small}
\!\!\!80         \end{small}}
\psfrag{100}{\begin{small}
\!\!\!100         \end{small}}
\psfrag{250}{\begin{small}
\!\!\!250         \end{small}}
\psfrag{300}{\begin{small}
\!\!\!300         \end{small}}
\psfrag{0}{\begin{small}
\!\!\!0         \end{small}}
\psfrag{1000}{\begin{small}
$10^3$         \end{small}}
\psfrag{u}{\begin{small}\!\!\!\!\!\!\!\!\!
\!\!\!$u(\tilde k) \,(\mathrm {MeV}^{-\frac32})$        \end{small}}
\psfrag{k}{\begin{small}$\tilde k$ (MeV/c)\end{small}}                                      
\psfrag{uW}{\begin{small}$u_0(\tilde k)$\end{small}}  
\psfrag{usma}{\begin{small}$\alpha^2 u_0(0)/(\tilde k^2+\alpha^2) $\end{small}}   
\includegraphics[width=100mm]{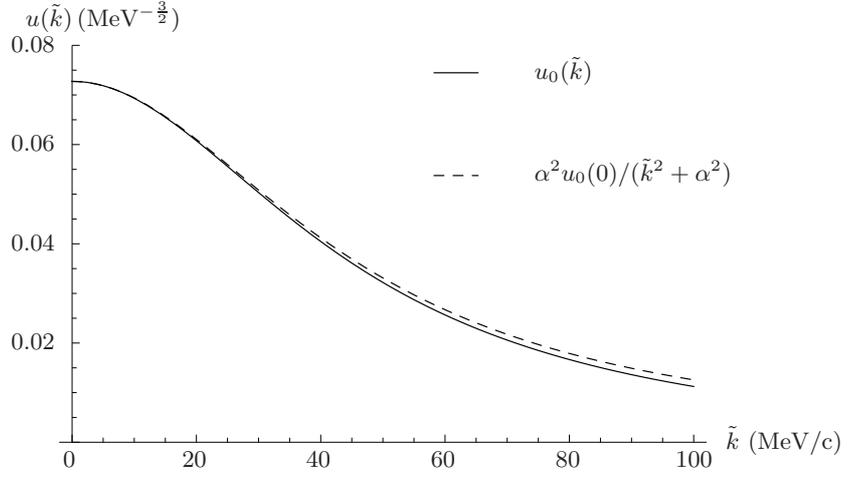}

\caption{\label{fig.08} The momentum-space wave function $u_0(\tilde k) $
for the present model including Pauli-Villars regularization and the
pole part of this wave function for small momenta.}

\end{center} 
\end{figure}

For small values of the momenta the momentum-space wave function is
dominated by the asymptotic behaviour of the coordinate-space wave
function, i.e. $\tilde u_0(r) \sim \exp (- \alpha r)$ with $\alpha^2=-E_\mathrm Bm_\mathrm N$. The corresponding
behaviour in momentum space is $u_0(\tilde k) \sim 1/(\alpha^2 + \tilde k^2)$. 
The latter is plotted in Fig.~\ref{fig.08}.

\subsubsection{Relativistic Kinetic Energy}
\label{Sect.IX}

The simplest modification of the present model replaces the
kinetic-energy operator $\tilde{\boldsymbol{k}}^2/m_{\mathrm N}$ by the operator
$2\sqrt{\tilde{\boldsymbol{k}}^2 + m^2_{\mathrm N}} - 2m_{\mathrm N}$. Then the bound-state equation Eq.~(\ref{eq:momentumspacebseq}) becomes a mass eigenvalue equation (cf. Eq.~(\ref{eq:eigenvalueproblem})) 
\begin{equation}\label{eq:schroedingermomspace}
 2\sqrt{\tilde{\boldsymbol{k}}^2 + m^2_{\mathrm N}} u_0(\tilde k)+\int\frac{\mathrm d^3 \tilde k'}{(2 \pi)^3} m_{\mathrm {int}}(\vert\tilde{\boldsymbol k}_1-\tilde{\boldsymbol k}_1'\vert) u_0(\tilde k')=(\underbrace{E_\mathrm B^\mathrm{rel}+2m_{\mathrm N}}_{=m_{\mathrm D}}) u_0(\tilde k)\,. 
\end{equation}
 It is well
known that this replacement effectively weakens the kinetic energy and
consequently lowers the energy of the bound state(s) if the potential
remains the same. This is indeed what we find. The eigenvalue of the
Hamiltonian drops to $E_\mathrm B^\mathrm{rel}=-2.73414~\mathrm {MeV}$ which gives a deuteron mass of $m_{00}=m_\mathrm D=1.87511~\mathrm {GeV}$.

\begin{figure}[t!]
\begin{center}

\psfrag{0.02}{\begin{small}
\!\!\!\!\!\!
                \end{small}}
\psfrag{0.04}{\begin{small}
\!\!\!\!\!\!
           \end{small}}
\psfrag{0.06}{\begin{small}
\!\!\!\!\!\!             \end{small}}
\psfrag{0.08}{\begin{small}
\!\!\!\!\!\!0.08          \end{small}}

\psfrag{0.01}{\begin{small}
\!\!\!\!\!\!0.01
                \end{small}}
\psfrag{0.03}{\begin{small}
\!\!\!\!\!\!0.03
           \end{small}}
\psfrag{0.05}{\begin{small}
\!\!\!\!\!\!0.05             \end{small}}
\psfrag{0.07}{\begin{small}
\!\!\!\!\!\!0.07          \end{small}}
\psfrag{1.0}{\begin{small}
\!1         \end{small}}
\psfrag{200}{\begin{small}
\!\!\!200         \end{small}}
\psfrag{400}{\begin{small}
\!\!\!400         \end{small}}
\psfrag{60}{\begin{small}
\!\!\!60         \end{small}}
\psfrag{80}{\begin{small}
\!\!\!80         \end{small}}
\psfrag{100}{\begin{small}
\!\!\!100         \end{small}}
\psfrag{250}{\begin{small}
\!\!\!250         \end{small}}
\psfrag{300}{\begin{small}
\!\!\!300         \end{small}}
\psfrag{0}{\begin{small}
\!\!\!0         \end{small}}
\psfrag{1000}{\begin{small}
$10^3$         \end{small}}
\psfrag{u}{\begin{small}\!\!\!\!\!\!
\!\!\!$u(\tilde k) \,(\mathrm {MeV}^{-\frac32})$        \end{small}}
\psfrag{k}{\begin{small}$\tilde k$ (MeV/c)\end{small}}                                      
\psfrag{uW}{\begin{small}$u_0(\tilde k)$, relativistic\end{small}}  
\psfrag{usma}{\begin{small}$\alpha_\mathrm{rel}^2 u_0(0)/(\tilde k^2+\alpha_\mathrm{rel}^2) $\end{small}}  
\includegraphics[width=100mm]{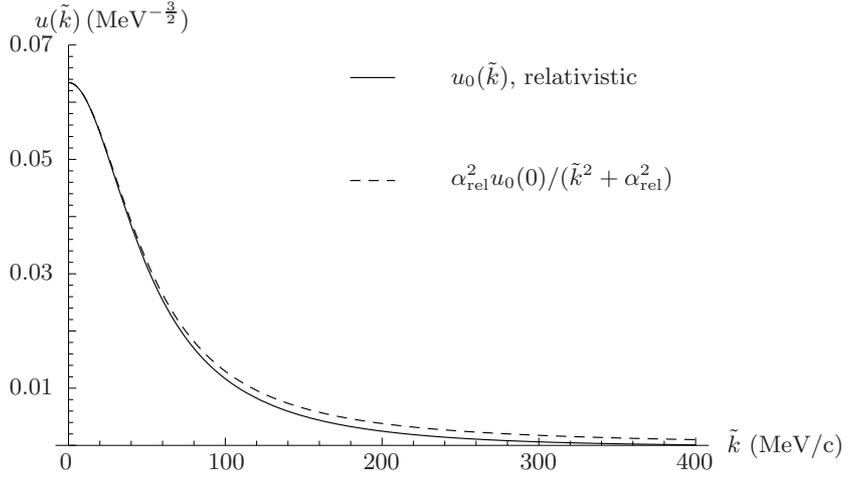}

\caption{\label{fig.09} The momentum-space wave function $u_0(\tilde k)$ for
the present model including Pauli-Villars regularization and relativistic
kinetic-energy operator. Also shown is the pole part.}

\end{center} 
\end{figure}
\begin{figure}[b!]
\begin{center}

\psfrag{0.02}{\begin{small}
\!\!\!\!\!\!0.02
                \end{small}}
\psfrag{0.04}{\begin{small}
\!\!\!\!\!\!0.04
           \end{small}}
\psfrag{0.06}{\begin{small}
\!\!\!\!\!\!0.06             \end{small}}
\psfrag{0.08}{\begin{small}
\!\!\!\!\!\!0.08          \end{small}}
\psfrag{200}{\begin{small}
\!\!\!200         \end{small}}
\psfrag{400}{\begin{small}
\!\!\!400         \end{small}}
\psfrag{60}{\begin{small}
\!\!\!60         \end{small}}
\psfrag{80}{\begin{small}
\!\!\!80         \end{small}}
\psfrag{100}{\begin{small}
\!\!\!100         \end{small}}
\psfrag{250}{\begin{small}
\!\!\!250         \end{small}}
\psfrag{300}{\begin{small}
\!\!\!300         \end{small}}
\psfrag{0}{\begin{small}
\!\!\!0         \end{small}}
\psfrag{1000}{\begin{small}
$10^3$         \end{small}}
\psfrag{u}{\begin{small}\!\!\!\!\!\!
\!\!\!$u(\tilde k) \,(\mathrm {MeV}^{-\frac32})$        \end{small}}
\psfrag{k}{\begin{small}$\tilde k$ (MeV/c)\end{small}}                                      
\psfrag{uWrel}{\begin{small}$u_0(\tilde k)$, relativistic\end{small}} 
\psfrag{uW}{\begin{small}$u_0(\tilde k)$, non-relativistic\end{small}} 
\psfrag{usma}{\begin{small}$\frac{-E_\mathrm {B}^\mathrm{rel} m_\mathrm N u_0(0)}{\tilde k^2-E_\mathrm B^\mathrm{rel}m_\mathrm N} $\end{small}}
\includegraphics[width=100mm]{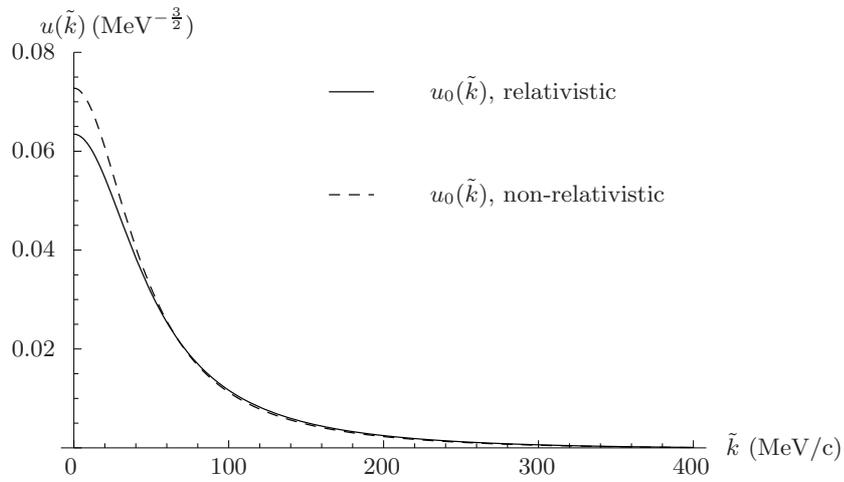}

\caption{\label{fig.10} The momentum-space wave functions $u_0(\tilde k)$ for
the present model including Pauli-Villars regularization and relativistic
 and non-relativistic kinetic-energy operator.}

\end{center} 
\end{figure}

In Fig.~\ref{fig.10} we show the momentum-space wave functions for the case
where the non-relativistic kinetic-energy operator is used and the one where
the relativistic operator is used.

\begin{figure}[b!]
\begin{center}

\psfrag{1.0}{\begin{small}
\!\!1
                \end{small}}
\psfrag{0.2}{\begin{small}
\!\!\!\!\!\!0.2
           \end{small}}
\psfrag{0.4}{\begin{small}
\!\!\!\!\!\!0.4
           \end{small}}
\psfrag{0.6}{\begin{small}
\!\!\!\!\!\!0.6             \end{small}}
\psfrag{0.8}{\begin{small}
\!\!\!\!\!\!0.8          \end{small}}
\psfrag{200}{\begin{small}
\!\!\!200         \end{small}}
\psfrag{400}{\begin{small}
\!\!\!400         \end{small}}
\psfrag{600}{\begin{small}
\!\!\!600         \end{small}}
\psfrag{800}{\begin{small}
\!\!\!800        \end{small}}
\psfrag{250}{\begin{small}
\!\!\!250         \end{small}}
\psfrag{300}{\begin{small}
\!\!\!300         \end{small}}
\psfrag{0}{\begin{small}
\!\!\!0         \end{small}}
\psfrag{1000}{\begin{small}
$10^3$         \end{small}}
\psfrag{k}{\begin{small}
$\tilde q_1$ (MeV/c)        \end{small}}
\psfrag{F}{\!\!\!\begin{small}$F_\mathrm B(\tilde q_1)$\end{small}}                                  
\psfrag{FWrel}{\begin{small}relativistic kinetic energy\end{small}} 
\psfrag{FW}{\begin{small}non-relativistic\end{small}} 
\psfrag{usma}{\begin{small}$\frac{-E_\mathrm {B}^\mathrm{rel} m_\mathrm N u_0(0)}{\tilde k^2-E_\mathrm B^\mathrm{rel}m_\mathrm N} $\end{small}}
\includegraphics[width=90mm]{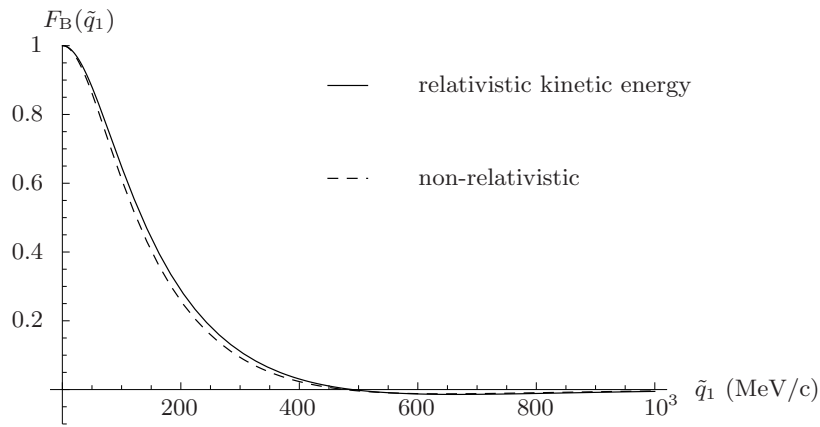}

\caption{\label{fig.11} A comparison of the form factors obtained for the
present model including Pauli-Villars regularization with the relativistic
kinetic-energy operator and the non-relativistic one.}

\end{center} 

\end{figure}

\begin{figure}[t!]
\begin{center}

\psfrag{1.0}{\begin{small}
\!\!1
                \end{small}}
\psfrag{0.1}{\begin{small}
\!\!\!\!0.1
           \end{small}}
\psfrag{0.01}{\begin{small}
\!\!\!\!0.01
           \end{small}}
\psfrag{0.001}{\begin{small}
\!\!\!$10^{-3}$          \end{small}}
\psfrag{0.0001}{\begin{small}
$10^{-4}$         \end{small}}
\psfrag{200}{\begin{small}
\!\!\!200         \end{small}}
\psfrag{400}{\begin{small}
\!\!\!400         \end{small}}
\psfrag{600}{\begin{small}
\!\!\!600         \end{small}}
\psfrag{800}{\begin{small}
\!\!\!800        \end{small}}
\psfrag{250}{\begin{small}
\!\!\!250         \end{small}}
\psfrag{300}{\begin{small}
\!\!\!300         \end{small}}
\psfrag{0}{\begin{small}
\!\!\!0         \end{small}}
\psfrag{1000}{\begin{small}
$10^3$         \end{small}}
\psfrag{k}{\begin{small}
$\tilde q_1$ (MeV/c)        \end{small}}
\psfrag{F}{\!\!\!\!\!\!\begin{small}$\vert F_\mathrm B(\tilde q_1)\vert$\end{small}}                                  
\psfrag{FWrel}{\begin{small}relativistic kinetic energy\end{small}} 
\psfrag{FW}{\begin{small}non-relativistic\end{small}}
 
\includegraphics[width=90mm]{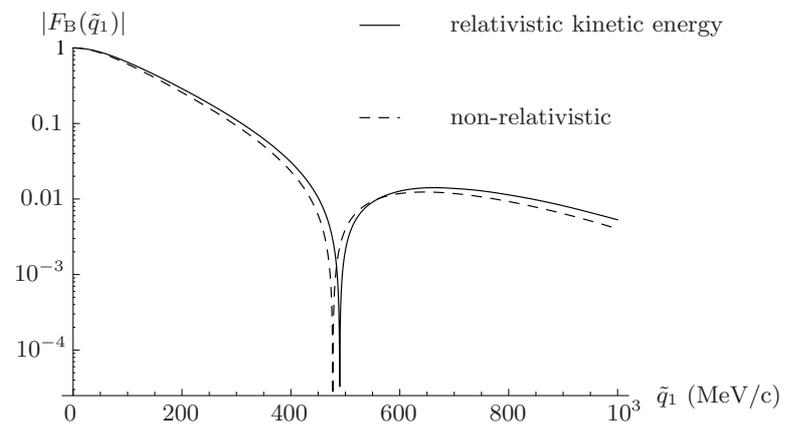}
\caption{\label{fig.12} The same in semi-log representation.}

\end{center} 

\end{figure}

The body form factor expression, Eq.~(\ref{eq.18}), is given in momentum space by
\begin{equation}
 F_\mathrm B(\tilde q_1) = \frac{1}{4\pi}\int{\rm d}^3 \tilde k\,u_0(\tilde k)u_0(|\tilde {\boldsymbol k}+\tilde {\boldsymbol q}_1|)\,.
\end{equation}
In Figs.~\ref{fig.11} and \ref{fig.12} we show again the form factor
for the Walecka-type model, but now together with the one where the
relativistic kinetic energy is used.  The differences between the two
are hardly noticeable.

\newpage
\chapter{Numerical Results}\label{chap:7}
Having solved the bound-state problem for the harmonic-oscillator potential and the Walecka-type model in the previous chapter we are now ready to make a numerical analysis of the form factor expressions for pseudoscalar and vector bound-states. All computations of this chapter have been done using {\sc Mathematica}$^{\begin{scriptsize}\textcopyright                                                                                                                                                                                                                                                                                                                                              \end{scriptsize}                                                                                                                                                                                                                                                                                                                                                                                                                                                                                                                                                                                                                                                                                                                                                                                                                                         }$.
\section{Electromagnetic $\pi$-Meson Form Factor}\label{sec:empionff}
In the present work confinement between the quark and the antiquark of a charged pion $\pi^{\pm}$ is modeled by the simple harmonic-oscillator potential introduced in Sec.~\ref{sec:homodelformesons}. There are just 2 free parameters in this simple model, the oscillator parameter $a$ and the constituent-quark mass $m_\mathrm q$. Using Eq.~(\ref{eq:pionformfactor}) for the form factor a reasonable fit of the experimental data~\cite{Huber:2008id,Bebek:1977pe,Bebek:1974ww,Bebek:1974iz,Brown:1973wr,Amendolia:1986wj} can be achieved by choosing the parameter values $a=350$~MeV$ $ and $m_\mathrm q=210$~MeV, which are taken from Ref.~\cite{Chung:1988mu}. At this point we shall emphasize that our goal is not to end up with an optimal fit of the pion form factor, but we rather want to exhibit the virtues of our point-form approach and compare it to other attempts to develop microscopic models for hadron form factors.
\subsection{Numerical Study}
We start with a numerical study of the influence of cluster-separability violating effects on the pion current and on the pion form factors. To this end we take the simple harmonic-oscillator wave function, Eq.~(\ref{eq:u00}), to calculate the pion current, Eq.~(\ref{eq:pscurrent}).\footnote{We have chosen for the 2 parameters, $m_\mathrm q=340$~MeV and
$a=312$~MeV. A justification for these values has already been given in Sec.~\ref{sec:HOparameters} or can also be found in Ref.~\cite{KrassniggDiss:2001}. Their precise value, however, is not important for this qualitative study.} Using our standard kinematics of Eq.~(\ref{eq:kM}) we find that the current $J_{\pi}^\mu(\boldsymbol k_\mathrm C',k_\mathrm C)$ has actually 2 vanishing components as expected from the theoretical analysis of Sec.~\ref{sec:psbscovstructure}: $J_{\pi}^1(\boldsymbol k_\mathrm C',\boldsymbol k_\mathrm C)=J_{\pi}^2(\boldsymbol k_\mathrm C',\boldsymbol k_\mathrm C)=0$.
There are, in general, 2 non-vanishing non-equal components, $J_{\pi}^0(\boldsymbol k_\mathrm C',\boldsymbol k_\mathrm C)$ and $J_{\pi}^3(\boldsymbol k_\mathrm C',\boldsymbol k_\mathrm C)$, except in the Breit frame ($k=Q/2$) and in the infinite-momentum frame ($k\rightarrow\infty$). 
In the Breit frame the third component is seen to vanish: $J_{\pi}^3(\boldsymbol k_\mathrm C',\boldsymbol k_\mathrm C)=0$.
This confirms exactly our findings of Sec.~\ref{sec:psbscovstructure}, namely that it is not possible to separate the 2 form factors $f$ and $\tilde b$ in the Breit frame, which essentially means that one is left with just one form factor (a linear combination of $f$ and $\tilde b$). 

\begin{figure}[t!]
\begin{center}

\psfrag{0}{\begin{small}0
                \end{small}}
\psfrag{0.0}{\begin{small}0
                \end{small}}
\psfrag{0.2}{\begin{small}0.2
                \end{small}}
\psfrag{0.4}{\begin{small}0.4
                \end{small}}
\psfrag{0.6}{\begin{small}0.6
                \end{small}}
\psfrag{0.8}{\begin{small}0.8
                \end{small}}
\psfrag{1.0}{\begin{small}1
                \end{small}}
\psfrag{1}{\begin{small}1
                \end{small}}
              \psfrag{2}{\begin{small}2
                \end{small}}
\psfrag{3}{\begin{small}3
                \end{small}}
\psfrag{4}{\begin{small}4
                \end{small}}
\psfrag{5}{\begin{small}5
                \end{small}}
\psfrag{6}{\begin{small}6            \end{small}}
\psfrag{7}{\begin{small}7
                \end{small}}
\psfrag{f}{\begin{small}\!\!\!\!\!\!$f(Q^2,k)$
           \end{small}}
\psfrag{k}{\begin{small}$k$~(GeV)
           \end{small}}
\psfrag{Q20}{\begin{small}  $Q^2=0~\mathrm{GeV}^2$        \end{small}}
\psfrag{Q202}{\begin{small}  $Q^2=0.2~\mathrm{GeV}^2$       \end{small}}
\psfrag{Q205}{\begin{small}  $Q^2=0.5~\mathrm{GeV}^2$        \end{small}}
\includegraphics[width=11cm]{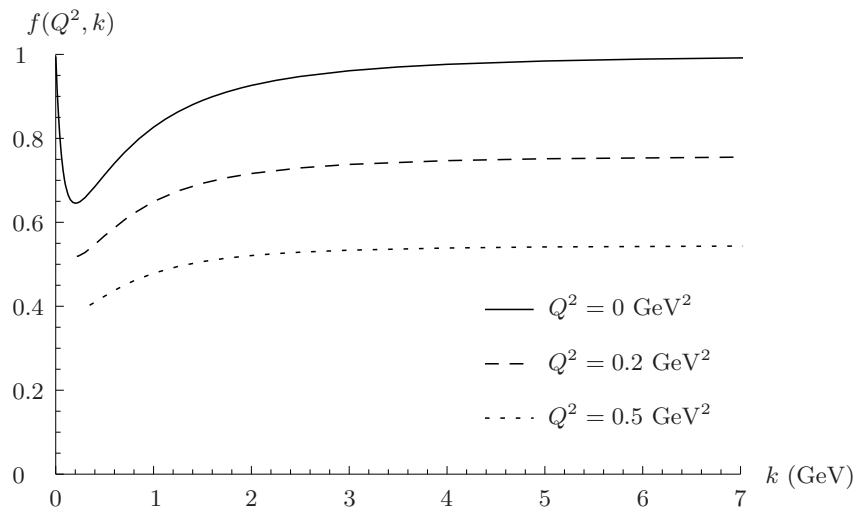}
\caption{Dependence of the physical pion form factor $f$ of Eq.~(\ref{eq:f1}) on
the pion center-of-mass momentum $k$ for different
values of the squared momentum transfer $Q^2$.}\label{fig:ff1}
\end{center}
\end{figure}

\begin{figure}[t!]
\begin{center}

\psfrag{0}{\begin{small}0
                \end{small}}
\psfrag{0.0}{\begin{small}0
                \end{small}}
\psfrag{0.2}{\begin{small}0.2
                \end{small}}
\psfrag{0.4}{\begin{small}0.4
                \end{small}}
\psfrag{0.6}{\begin{small}0.6
                \end{small}}
\psfrag{0.8}{\begin{small}0.8
                \end{small}}
\psfrag{1.0}{\begin{small}1
                \end{small}}
\psfrag{1}{\begin{small}1
                \end{small}}
              \psfrag{2}{\begin{small}2
                \end{small}}
\psfrag{3}{\begin{small}3
                \end{small}}
\psfrag{4}{\begin{small}4
                \end{small}}
\psfrag{5}{\begin{small}5
                \end{small}}
\psfrag{6}{\begin{small}6            \end{small}}
\psfrag{7}{\begin{small}7
                \end{small}}
\psfrag{b}{\begin{small}\!\!\!\!\!\!$\tilde b(Q^2,k)$
           \end{small}}
\psfrag{k}{\begin{small}$k$~(GeV)
           \end{small}}
\psfrag{Q20}{\begin{small}  $Q^2=0~\mathrm{GeV}^2$        \end{small}}
\psfrag{Q202}{\begin{small}  $Q^2=0.2~\mathrm{GeV}^2$       \end{small}}
\psfrag{Q205}{\begin{small}  $Q^2=0.5~\mathrm{GeV}^2$        \end{small}}
\includegraphics[width=11cm]{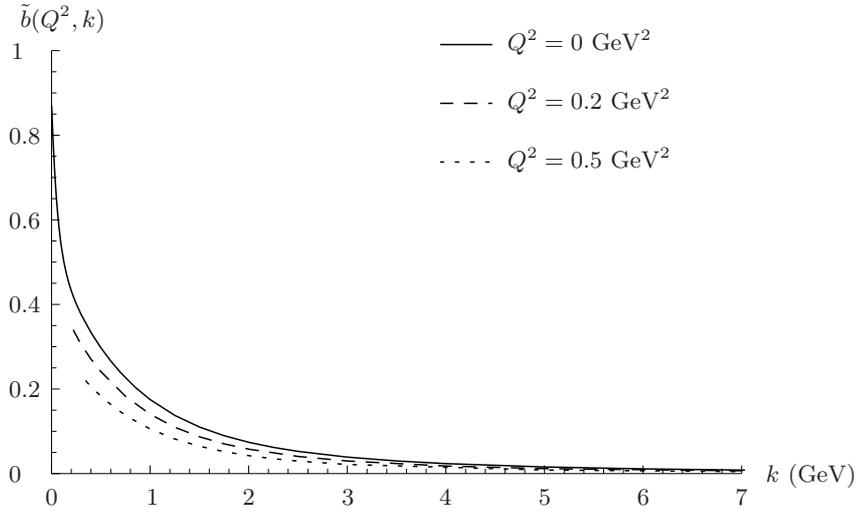}
\caption{Dependence of
the spurious pion form factor $\tilde b$ of Eq.~(\ref{eq:f2}) on
the pion center-of-mass momentum $k$ for different
values of the squared momentum transfer $Q^2$.}\label{fig:ff2}
\end{center}
\end{figure}

Nevertheless, for $k> Q/2$ the 2 form factors $f$ and $\tilde b$ can be separated uniquely by means of Eq.~(\ref{eq:formffPS2}). They are then given by Eqs.~(\ref{eq:f1}) and~(\ref{eq:f2}) and plotted in Figs.~\ref{fig:ff1} and~\ref{fig:ff2} versus $k$ for fixed $Q^2$.
One observes that both, the physical form factor $f$ and the spurious form factor $\tilde b$ depend on $k$. The $k$-dependence of $f$ vanishes rather quickly with increasing $k$. At the same time the spurious form factor $\tilde b$ is seen to
vanish for large $k$. This means by taking the limit $k\rightarrow
\infty$ effects from the violation of cluster separability are removed. In this limit we also ended up with the simple analytic expression for the form factor $F$ given in Eq.~(\ref{eq:pionformfactor}). This expression has been proved in Sec.~\ref{sec:comparestandardff} to be equivalent with the standard front-form expression for the pion form factor of Ref.~\cite{ Chung:1988mu} obtained from a spectator current in the $q^+=0$
frame. 

From Fig.~\ref{fig:ff1} we also read off the correct charge of the pion, i.e. $F(0)=\lim_{k\rightarrow
\infty}f(0,k)=1$ in units of the elementary charge $\mathrm e$. This is only the case for the form factor one gets from the decomposition of the pion current given by Eq.~(\ref{eq:formffPS}), which justifies to call $f$ the physical and $\tilde b$ the unphysical form factors of the pion~\cite{Carbonell:1998rj}. In addition, we note that the fact that we end up with only one form factor is also confirmed by the observation that $J_{\pi}^0(\boldsymbol k_\mathrm C',\boldsymbol k_\mathrm C)\rightarrow J_{\pi}^3(\boldsymbol k_\mathrm C',\boldsymbol k_\mathrm C)$ in the limit $k\rightarrow
\infty$. 

\begin{figure}[h!]
\begin{center}
\psfrag{1.}{\begin{small}1                  \end{small}}
\psfrag{0.3}{\begin{small}0.3              \end{small}}
\psfrag{0.5}{\begin{small}\!0.5              \end{small}}
\psfrag{0.05}{\begin{small}0.05              \end{small}}
\psfrag{0.15}{\,\,\begin{small}0.15              \end{small}}
\psfrag{0.20}{\begin{small}0.2                 \end{small}}
\psfrag{0.50}{\begin{small}0.5              \end{small}}
\psfrag{0}{\begin{small}0             \end{small}}
\psfrag{0}{\begin{small}0             \end{small}}
\psfrag{1.00}{\,\,\,\,\begin{small}1              \end{small}}
\psfrag{5.00}{\,\,\,\,\begin{small}5              \end{small}}
\psfrag{10.00}{\,\,\,\,\begin{small}10               \end{small}}
\psfrag{F2}{\begin{small}$F^2(Q^2)$                     \end{small}          }
\psfrag{Q2}{\begin{small}$Q^2 (\mathrm {GeV}^2)$            \end{small}         }
\psfrag{Sn}{\begin{scriptsize} $\mathcal S \neq 1$                                    \end{scriptsize}}
\psfrag{S}{\begin{scriptsize} $\mathcal S = 1$                                    \end{scriptsize}}
\psfrag{Bebek1978}{\begin{small}\cite{Bebek:1977pe}                                      \end{small}}
\psfrag{Bebek1976}{\begin{small}\cite{Bebek:1974ww}                                      \end{small}}
\psfrag{Bebek1974}{\begin{small} \cite{Bebek:1974iz}                                      \end{small}}
\psfrag{Brown1973}{ \begin{small}\cite{Brown:1973wr}                                      \end{small}}
\psfrag{Amendolia1986}{\begin{scriptsize}\cite{Amendolia:1986wj}                                              \end{scriptsize}}
\psfrag{CC}{\begin{small}present approach\end{small}}
\psfrag{PFSM}{\begin{small}point-form spectator model\end{small}}
\includegraphics[clip=7cm,width=80mm]{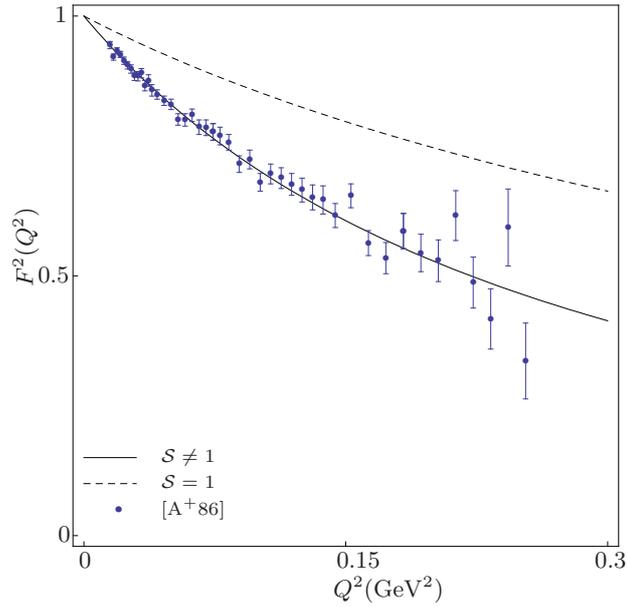}
\caption{\label{fig:pionffspin_low}$Q^2$-dependence of the electromagnetic pion form factor squared $F^2\left(Q^2\right)$ in
the low-$Q^2$ region, as evaluated by means of Eq.~(\ref{eq:pionformfactor}) with parameters $a=350$~MeV$ $ and $m_\mathrm q=210$~MeV with and
without spin-rotation factor $\mathcal S$~\cite{Biernat:2009my}. Data are taken from Ref.~\cite{Amendolia:1986wj}.}
\end{center} 
\end{figure}

\begin{figure}[h!]
\begin{center}
\psfrag{1.6}{\begin{small}\!1.6                 \end{small}}
\psfrag{0.8}{\begin{small}0.8              \end{small}}
\psfrag{5}{\begin{small}5              \end{small}}
\psfrag{10}{\begin{small}10            \end{small}}
\psfrag{0.10}{\,\,\begin{small}0.1              \end{small}}
\psfrag{0.20}{\begin{small}0.2                 \end{small}}
\psfrag{0.50}{\begin{small}0.5              \end{small}}
\psfrag{0}{\begin{small}0             \end{small}}
\psfrag{0.2}{\begin{small}0.2             \end{small}}
\psfrag{1.00}{\,\,\,\,\begin{small}1              \end{small}}
\psfrag{5.00}{\,\,\,\,\begin{small}5              \end{small}}
\psfrag{10.00}{\,\,\,\,\begin{small}10               \end{small}}
\psfrag{Q2F}{\begin{small}$Q^2F(Q^2)(\mathrm {GeV}^2)$                     \end{small}          }
\psfrag{Q2}{\begin{small}$Q^2 (\mathrm {GeV}^2)$            \end{small}         }
\psfrag{Sn}{\begin{scriptsize} $\mathcal S = 1$                                    \end{scriptsize}}
\psfrag{S}{\begin{scriptsize} $\mathcal S \neq 1$                                    \end{scriptsize}}
\psfrag{F}{\begin{small}$F(Q^2)$                     \end{small}          }
\psfrag{Q}{\begin{small}$Q^2 (\mathrm {GeV}^2)$            \end{small}         }
\psfrag{Huber2008}{\begin{scriptsize} \cite{Huber:2008id}                                      \end{scriptsize}}
\psfrag{Bebek1978}{\begin{scriptsize}\cite{Bebek:1977pe}                                      \end{scriptsize}}
\psfrag{Bebek1976}{\begin{scriptsize}\cite{Bebek:1974ww}                                      \end{scriptsize}}
\psfrag{Bebek1974}{\begin{scriptsize} \cite{Bebek:1974iz}                                      \end{scriptsize}}
\psfrag{Brown1973}{\begin{scriptsize}\cite{Brown:1973wr}                                      \end{scriptsize}}
\psfrag{Amendolia1986}{\begin{small}\cite{Amendolia:1986wj}                                              \end{small}}
\psfrag{CC}{\begin{small}present approach\end{small}}
\psfrag{PFSM}{\begin{small}point-form spectator model\end{small}}
\includegraphics[clip=7cm,width=80mm]{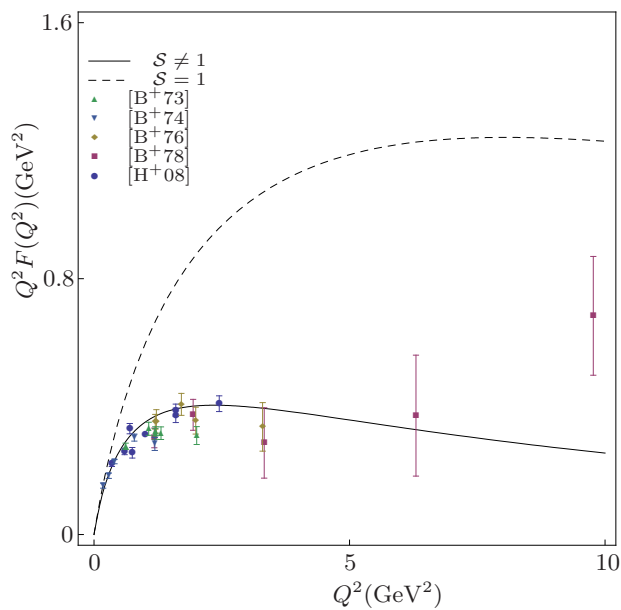}
\caption{\label{fig:pionffspin_high}$Q^2$-dependence of the electromagnetic pion form factor scaled by $Q^2$ as evaluated by means of Eq.~(\ref{eq:pionformfactor}) with the same parameters as in Fig.~\ref{fig:pionffspin_low} with and
without spin-rotation factor $\mathcal S$~\cite{Biernat:2009my}. Data are taken from Refs.~\cite{Brown:1973wr,Bebek:1974ww,Bebek:1974iz,Bebek:1977pe,Huber:2008id}.}
\end{center} 
\end{figure}

In Figs.~\ref{fig:pionffspin_low} and~\ref{fig:pionffspin_high} we show the $Q^2$-dependence of the electromagnetic pion form factor $F\left(Q^2\right)$ as evaluated by means of Eq.~(\ref{eq:pionformfactor}) along with experimental data. Also shown is the role of the spin-rotation
factor $\mathcal S$. The quark spin obviously has a substantial effect on the electromagnetic form factor over nearly the whole
momentum-transfer range and thus cannot be neglected~\cite{Biernat:2009my}.

\subsection{Results and Comparisons}
\begin{figure}[t]
\begin{center}
\psfrag{0.01}{\begin{small}0.01                  \end{small}}
\psfrag{0.02}{\begin{small}0.02              \end{small}}
\psfrag{0.05}{\begin{small}0.05              \end{small}}
\psfrag{0.05}{\begin{small}0.05              \end{small}}
\psfrag{0.10}{\,\,\begin{small}0.1              \end{small}}
\psfrag{0.20}{\begin{small}0.2                 \end{small}}
\psfrag{0.50}{\begin{small}0.5              \end{small}}
\psfrag{0.2}{\begin{small}0.2             \end{small}}
\psfrag{0.2}{\begin{small}0.2             \end{small}}
\psfrag{1.00}{\,\,\,\,\begin{small}1              \end{small}}
\psfrag{5.00}{\,\,\,\,\begin{small}5              \end{small}}
\psfrag{10.00}{\,\,\,\,\begin{small}10               \end{small}}
\psfrag{F}{\begin{small}$F(Q^2)$                     \end{small}          }
\psfrag{Q}{\begin{small}$Q^2 (\mathrm {GeV}^2)$            \end{small}         }
\psfrag{Huber2008}{\!\begin{scriptsize} \cite{Huber:2008id}                                      \end{scriptsize}}
\psfrag{Bebek1978}{\!\begin{scriptsize}\cite{Bebek:1977pe}                                      \end{scriptsize}}
\psfrag{Bebek1976}{\!\begin{scriptsize}\cite{Bebek:1974ww}                                      \end{scriptsize}}
\psfrag{Bebek1974}{\begin{scriptsize} \cite{Bebek:1974iz}                                      \end{scriptsize}}
\psfrag{Brown1973}{ \begin{scriptsize}\cite{Brown:1973wr}                                      \end{scriptsize}}
\psfrag{Amendolia1986}{\begin{scriptsize}\cite{Amendolia:1986wj}                                              \end{scriptsize}}
\psfrag{CC}{\begin{scriptsize}present approach\end{scriptsize}}
\psfrag{PFSM}{\begin{scriptsize}point-form spectator model\end{scriptsize}}
\includegraphics[clip=7cm,width=100mm]{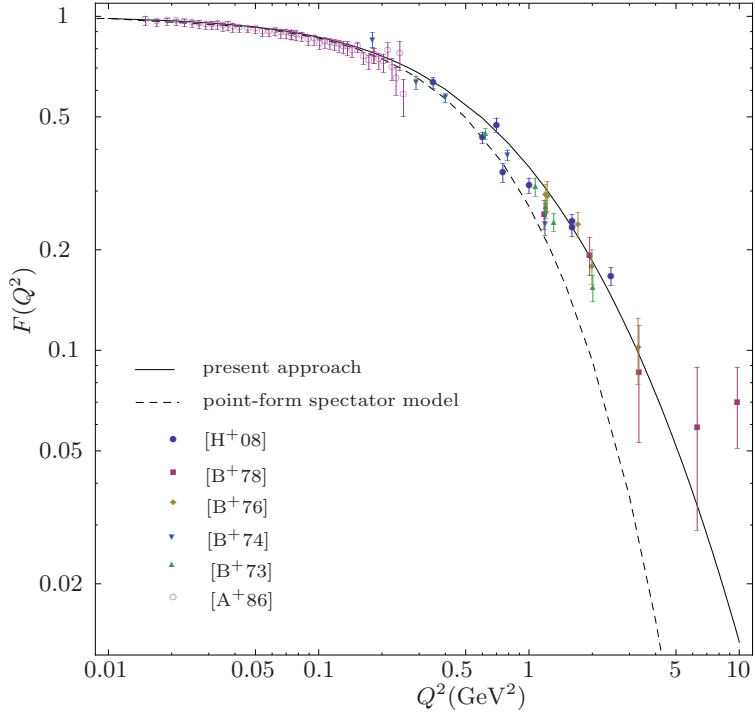}
\caption{\label{fig:pionff}The electromagnetic pion form factor
  $F\left(Q^2\right)$ as evaluated by means of
  Eq.~(\ref{eq:pionformfactor}) with parameters $m_\mathrm q=340\, \mathrm {MeV}$ and $a=312\, \mathrm {MeV}$ in comparison
  with the outcome of the point-form spectator model with the same parameters and $m_{00}=770\, \mathrm {MeV}$~\cite{Biernat:2009my}. Data are taken from Refs.~\cite{Amendolia:1986wj,Brown:1973wr,Bebek:1974ww,Bebek:1974iz,Bebek:1977pe,Huber:2008id}.}
\end{center} 
\end{figure}

We have already mentioned previously that the pion form-factor expression obtained in the point-form spectator model~\cite{Wagenbrunn:2000es,Boffi:2001zb,Melde:2007zz} is 
not only determined by the pion wave function, but exhibits also an explicit dependence on the pion mass. Therefore, the spectrum of the mass operator is directly connected with the electromagnetic structure of its eigenstates. This makes it somewhat delicate to
compare our form-factor results for the simple harmonic-oscillator
confinement potential with corresponding point-form model predictions~\cite{Biernat:2009my}. Whereas the wave function is solely determined by the oscillator parameter $a$ (cf. Eq.~(\ref{eq:u00})), another free constant $c_0$ can be
added to the confinement potential (cf. Eq.~(\ref{eq:spectrumho})) to shift the eigenvalue spectrum. Unlike our results, which do not depend on $c_0$, the point-form spectator model predictions exhibit a strong dependence on $c_0$. The parameters $a=350~\mathrm{MeV}$ and $m_\mathrm q=210~\mathrm{MeV}$ as proposed in Ref.\cite{Chung:1988mu} give a reasonable fit of our pion form factor with the data. Taking these parameters together with $c_0<0$, such that the harmonic-oscillator ground state were to coincide with the physical mass of the pion of $m_\pi=140$~MeV, the fall-off of the point-form spectator model form factor would be unreasonably strong. With $c_0=0$, on the other hand, the
pion ground-state mass would be larger than $1$~GeV and the fall-off
of the form factor would be much too slow~\cite{Biernat:2009my}. We have therefore tried
to take the set of parameters of $a=312$~MeV, $c_0=-1.04$~GeV$^2$ and
$m_\mathrm q=340$~MeV, which have been already discussed in Sec.~\ref{sec:homodelformesons}. These are reasonable values for the purpose of a qualitative comparison between our approach and the point-form spectator model predictions. In Fig.~\ref{fig:pionff} we compare our result with the point-form spectator model prediction and experimental data.
Both results become
comparable at small momentum transfers. Above $Q^2\approx 1$~GeV$^2$, however, significant differences can
be observed. These differences resemble the situation for the
electromagnetic nucleon form factors. In the latter case the
stronger fall-off produced by the point-form spectator model is a welcome feature which
brings the theoretical predictions from constituent quark models
close to experiments~\cite{Melde:2007zz}. For the usual front-form
spectator current in the $q^+=0$ frame agreement with experiment is
 achieved only by introducing electromagnetic form factors
for the constituent quarks~\cite{Simula:2001wx}. It remains to be
seen whether the situation for the electromagnetic pion form factor
could also change in favor of the point-form spectator model if a more sophisticated quark-antiquark
 potential is employed which reproduces the mass of the
pion and its lowest excitations sufficiently accurately~\cite{Biernat:2009my}.

\section{Electromagnetic $\rho$-Meson Form Factors}
\subsection{Discussion and Predictions}
A charged $\rho^\pm$ meson in a constituent quark model is considered to be  a quark-antiquark bound state with total spin $j=1$. As for the pion the confining interaction between the quarks is modeled by the simple harmonic-oscillator potential of Sec.~\ref{sec:homodelformesons}. We have fixed the 3 parameters $a=312$~MeV, $c_0=-1.04$~GeV$^2$ and
$m_\mathrm q=340$~MeV via the vector-meson spectrum, therefore the mass of the ground state $m_{00}=770~\mathrm {MeV}$ coincides already with the correct $\rho$-meson mass $m_\rho$. 

\begin{figure}[t!]
\begin{center}
\psfrag{-}{\begin{small}             \end{small}}
\psfrag{0.2}{\begin{small}-0.2                 \end{small}}
\psfrag{0.4}{\begin{small}-0.4              \end{small}}
\psfrag{0.6}{\begin{small}-0.6               \end{small}}
\psfrag{0.8}{\begin{small}-0.8               \end{small}}
\psfrag{1.0}{\begin{small}\!-1              \end{small}}
\psfrag{2}{\begin{small}2                 \end{small}}
\psfrag{4}{\begin{small}4                  \end{small}}
\psfrag{6}{\begin{small}6                  \end{small}}
\psfrag{2}{\begin{small}2                 \end{small}}
\psfrag{4}{\begin{small}4                  \end{small}}
\psfrag{6}{\begin{small}6                  \end{small}}
\psfrag{8}{\begin{small}8                \end{small}}
\psfrag{FF1}{\begin{small}$F_1(Q^2)$                \end{small}}
\psfrag{Q2}{\begin{small}$Q^2~(\mathrm{GeV}^2)$               \end{small}}
\psfrag{FF}{\begin{small}form factors                 \end{small}}
\psfrag{FF}{\begin{small}form factors                 \end{small}}
\psfrag{FF}{\begin{small}form factors                 \end{small}}
\includegraphics[width=10cm]{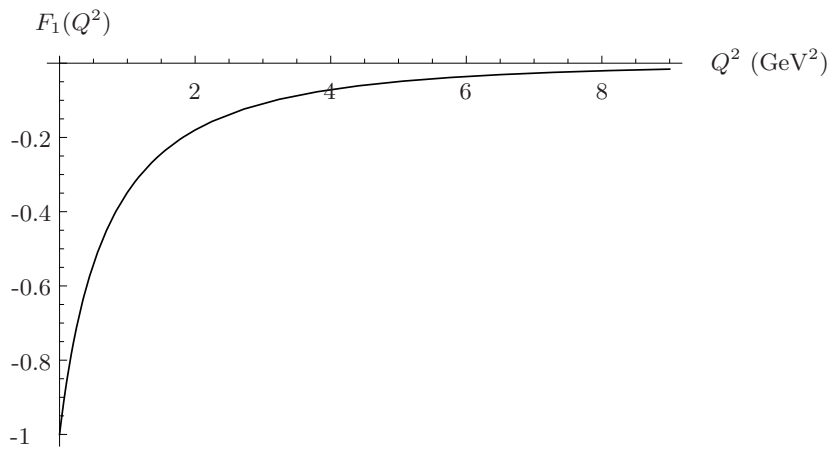}
\caption{The electromagnetic $\rho$-meson form factor
  $F_1\left(Q^2\right)$ evaluated by means of Eq.~(\ref{eq:ff1me}) with $F_1^1(Q^2)+F_1^2(Q^2)=1$ and $F_2^1(Q^2)=F_2^2(Q^2)=0$ and parameters $m_\mathrm q=340\, \mathrm {MeV}$, $a=312\, \mathrm {MeV}$ and $c_0=-1.04\, \mathrm {GeV}^2$ such that $m_{00}=770\, \mathrm {MeV}$.
}\label{fig:rhoFF1}                   \end{center}
\end{figure}

\begin{figure}[t!]
\begin{center}
\psfrag{0.7}{\begin{small}0.7                  \end{small}}
\psfrag{0.6}{\begin{small}0.6                  \end{small}}
\psfrag{0.5}{\begin{small}0.5                \end{small}}
\psfrag{0.4}{\begin{small}0.4                  \end{small}}
\psfrag{0.3}{\begin{small}0.3                 \end{small}}
\psfrag{0.2}{\begin{small}0.2                 \end{small}}
\psfrag{0.1}{\begin{small}0.1                \end{small}}
\psfrag{1.0}{\begin{small}1               \end{small}}
\psfrag{1.5}{\begin{small}1.5                 \end{small}}
\psfrag{2.0}{\begin{small}2                 \end{small}}
\psfrag{2}{\begin{small}2                 \end{small}}
\psfrag{4}{\begin{small}4                  \end{small}}
\psfrag{6}{\begin{small}6                  \end{small}}
\psfrag{2}{\begin{small}2                 \end{small}}
\psfrag{4}{\begin{small}4                  \end{small}}
\psfrag{6}{\begin{small}6                  \end{small}}
\psfrag{8}{\begin{small}8                \end{small}}
\psfrag{FF2}{\begin{small}$F_2(Q^2)$                 \end{small}}
\psfrag{Q2}{\begin{small}$Q^2~(\mathrm{GeV}^2)$               \end{small}}
\psfrag{FF}{\begin{small}form factors                 \end{small}}
\psfrag{FF}{\begin{small}form factors                 \end{small}}
\psfrag{FF}{\begin{small}form factors                 \end{small}}
\includegraphics[width=10cm]{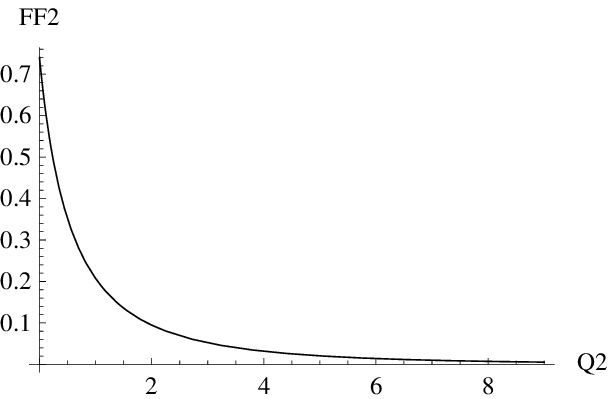}
\caption{The electromagnetic $\rho$-meson form factor
  $F_2\left(Q^2\right)$ evaluated by means of Eq.~(\ref{eq:ff2me}) with $F_1^1(Q^2)+F_1^2(Q^2)=1$ and $F_2^1(Q^2)=F_2^2(Q^2)=0$ and with the same parameters as
in Fig.~\ref{fig:rhoFF1}.
}\label{fig:rhoFF2}                   \end{center}
\end{figure}

\begin{figure}[t!]
\begin{center}
\psfrag{0.5}{\begin{small}0.5                  \end{small}}
\psfrag{1.0}{\begin{small}1               \end{small}}
\psfrag{1.5}{\begin{small}1.5                 \end{small}}
\psfrag{2.0}{\begin{small}2                 \end{small}}
\psfrag{2}{\begin{small}2                 \end{small}}
\psfrag{4}{\begin{small}4                  \end{small}}
\psfrag{6}{\begin{small}6                  \end{small}}
\psfrag{2}{\begin{small}2                 \end{small}}
\psfrag{4}{\begin{small}4                  \end{small}}
\psfrag{6}{\begin{small}6                  \end{small}}
\psfrag{8}{\begin{small}8                \end{small}}
\psfrag{GM}{\begin{small}$G_\mathrm M(Q^2)$                 \end{small}}
\psfrag{Q2}{\begin{small}$Q^2~(\mathrm{GeV}^2)$               \end{small}}
\psfrag{FF}{\begin{small}form factors                 \end{small}}
\psfrag{FF}{\begin{small}form factors                 \end{small}}
\psfrag{FF}{\begin{small}form factors                 \end{small}}
\includegraphics[width=10cm]{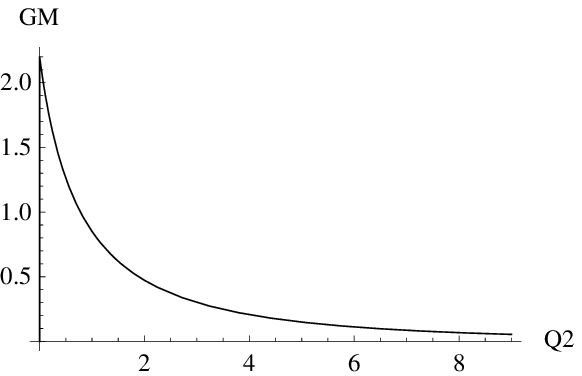}
\caption{The magnetic $\rho$-meson form factor
  $G_\mathrm M\left(Q^2\right)$ evaluated by means of Eq.~(\ref{eq:ffGMme}) with $F_1^1(Q^2)+F_1^2(Q^2)=1$ and $F_2^1(Q^2)=F_2^2(Q^2)=0$ and with the same parameters as
in Fig.~\ref{fig:rhoFF1}.
}\label{fig:rhoGM}                   \end{center}
\end{figure}

\begin{figure}[t!]
\begin{center}
\psfrag{0.2}{\begin{small}0.2                 \end{small}}
\psfrag{0.4}{\begin{small}0.4              \end{small}}
\psfrag{0.6}{\begin{small}0.6               \end{small}}
\psfrag{0.8}{\begin{small}0.8               \end{small}}
\psfrag{1.0}{\begin{small}1              \end{small}}
\psfrag{2}{\begin{small}2                 \end{small}}
\psfrag{4}{\begin{small}4                  \end{small}}
\psfrag{6}{\begin{small}6                  \end{small}}
\psfrag{3}{\begin{small}3                 \end{small}}
\psfrag{5}{\begin{small}5                  \end{small}}
\psfrag{1}{\begin{small}1                  \end{small}}
\psfrag{8}{\begin{small}8                \end{small}}
\psfrag{GC}{\begin{small}$G_\mathrm C(Q^2)$                \end{small}}
\psfrag{Q2}{\begin{small}$Q^2~(\mathrm{GeV}^2)$               \end{small}}
\includegraphics[width=10cm]{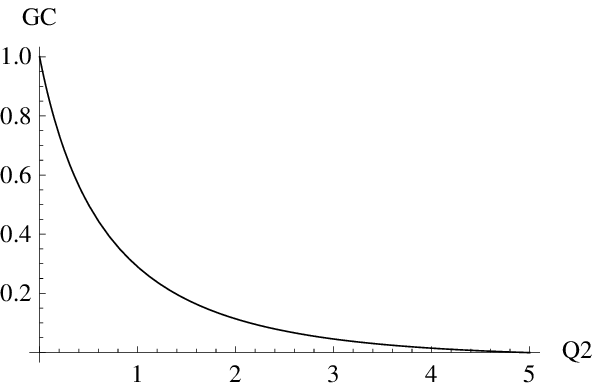}
\caption{The electric charge $\rho$-meson form factor
  $G_\mathrm C\left(Q^2\right)$ calculated from Eq.~(\ref{eq:rhoGC}) with the same parameters as
in Fig.~\ref{fig:rhoFF1}. 
}  \label{fig:rhoGC}                    \end{center}
\end{figure}

\begin{figure}
\begin{center}
\psfrag{-}{\begin{small}\,\,-              \end{small}}
\psfrag{0.2}{\begin{small}0.2                 \end{small}}
\psfrag{0.4}{\begin{small}0.4              \end{small}}
\psfrag{0.1}{\begin{small}0.1               \end{small}}
\psfrag{0.3}{\begin{small}0.3               \end{small}}
\psfrag{1.0}{\begin{small}\!1              \end{small}}
\psfrag{2}{\begin{small}2                 \end{small}}
\psfrag{4}{\begin{small}4                  \end{small}}
\psfrag{6}{\begin{small}6                  \end{small}}
\psfrag{1}{\begin{small}1                 \end{small}}
\psfrag{3}{\begin{small}3                  \end{small}}
\psfrag{5}{\begin{small}5                  \end{small}}
\psfrag{8}{\begin{small}8                \end{small}}
\psfrag{GQ}{\begin{small}$G_\mathrm Q(Q^2)$                \end{small}}
\psfrag{Q2}{\begin{small}$Q^2~(\mathrm{GeV}^2)$               \end{small}}
\includegraphics[width=10cm]{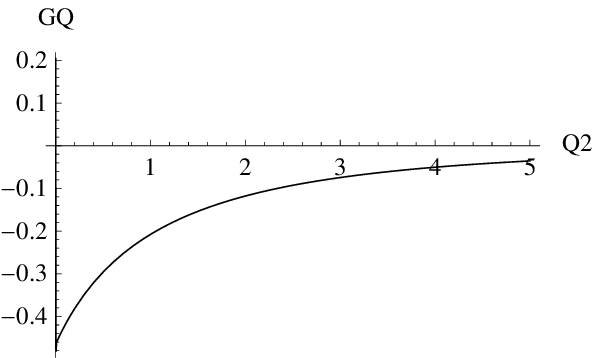}
\caption{The electric quadrupole $\rho$-meson form factor
  $G_\mathrm Q\left(Q^2\right)$ calculated from Eq.~(\ref{eq:rhoGQ}) with the same parameters as
in Fig.~\ref{fig:rhoFF1}. 
}  \label{fig:rhoGQ}                    \end{center}
\end{figure}

The $\rho$-meson electromagnetic form factors are computed by simply plugging the ground-state harmonic-oscillator wave function $u_{00}$ and the ground-state mass $m_{00}$ into the form factor expressions Eqs.~(\ref{eq:ff1me}),~(\ref{eq:ff2me}) 
and~(\ref{eq:ffGMme}) with $F_1^1(Q^2)+F_1^2(Q^2)=1$ and $F_2^1(Q^2)=F_2^2(Q^2)=0$ (since the quarks are considered as point-like).
The results are depicted in Figs.~\ref{fig:rhoFF1}-\ref{fig:rhoGM}.
By means of Eqs.~(\ref{eq:rhoGC}) and~(\ref{eq:rhoGQ}) we obtain from  $F_1\left(Q^2\right)$, $F_2\left(Q^2\right)$ and $G_\mathrm M\left(Q^2\right)$ the electric charge and quadrupole form factors of the $\rho$ meson, $G_\mathrm C\left(Q^2\right)$ and $G_\mathrm Q\left(Q^2\right)$, respectively.
They are plotted in Figs.~\ref{fig:rhoGC} and~\ref{fig:rhoGQ}.
From Fig.~\ref{fig:rhoGC} we read off the correct $\rho^+$-meson charge $G_\mathrm C(0)=1$ in units of the fundamental charge $|\,\mathrm e\,|$. This is ensured only by the decomposition of the $\rho$-meson current given in Eq.~(\ref{eq:covstructurephysdeutcurr}), which justifies this particular way of separating the physical from the unphysical contributions. Our predictions for the magnetic dipole and the electric quadrupole moment given by the $Q^2\rightarrow 0$ limits of $G_\mathrm M\left(Q^2\right)$ and $G_\mathrm Q\left(Q^2\right)$, cf. Eqs.~(\ref{eq:murho}) and~(\ref{eq:qrho}), are $\mu_\rho=2.2$ and $Q_\rho=-0.47$ (in units $|\,\mathrm e\,|/2m_{\rho}$ and $|\,\mathrm e\,|/m_{\rho}^2$), respectively. 

\begin{figure}
\begin{center}
\psfrag{0.0}{\begin{small}0                \end{small}}
\psfrag{0}{\begin{small}0            \end{small}}
\psfrag{0.2}{\begin{small}0.2                 \end{small}}
\psfrag{0.4}{\begin{small}0.4              \end{small}}
\psfrag{0.1}{\begin{small}0.1               \end{small}}
\psfrag{0.3}{\begin{small}0.3               \end{small}}
\psfrag{0.6}{\begin{small}0.6               \end{small}}
\psfrag{0.5}{\begin{small}0.5              \end{small}}
\psfrag{0.8}{\begin{small}0.8               \end{small}}
\psfrag{1.0}{\begin{small}\!1              \end{small}}
\psfrag{2}{\begin{small}2                 \end{small}}
\psfrag{4}{\begin{small}4                  \end{small}}
\psfrag{6}{\begin{small}6                  \end{small}}
\psfrag{1}{\begin{small}1                 \end{small}}
\psfrag{3}{\begin{small}3                  \end{small}}
\psfrag{5}{\begin{small}5                  \end{small}}
\psfrag{8}{\begin{small}8                \end{small}}
\psfrag{A}{\begin{small}$A(Q^2)$                \end{small}}
\psfrag{Q2}{\begin{small}$Q^2~(\mathrm{GeV}^2)$               \end{small}}
\includegraphics[width=10cm]{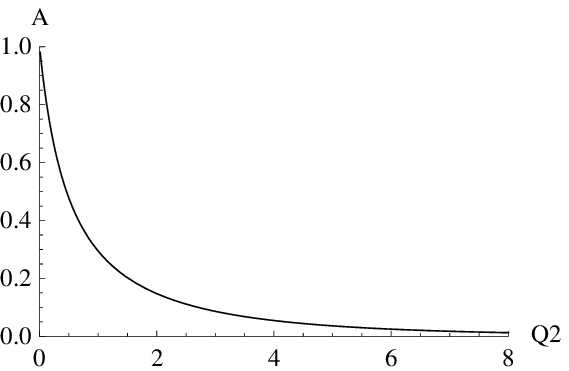}
\caption{The elastic scattering observable $A(Q^2)$ from Eq.~(\ref{eq:Aobservable}) for the $\rho$ meson calculated with the same parameters as
in Fig.~\ref{fig:rhoFF1}.
}  \label{fig:Arho}                    \end{center}
\end{figure}

\begin{figure}
\begin{center}
\psfrag{0.0}{\begin{small}0                \end{small}}
\psfrag{0}{\begin{small}0            \end{small}}
\psfrag{0.2}{\begin{small}0.2                 \end{small}}
\psfrag{0.4}{\begin{small}0.4              \end{small}}
\psfrag{0.1}{\begin{small}0.1               \end{small}}
\psfrag{0.3}{\begin{small}0.3               \end{small}}
\psfrag{0.6}{\begin{small}0.6               \end{small}}
\psfrag{0.5}{\begin{small}0.5              \end{small}}
\psfrag{0.7}{\begin{small}0.7               \end{small}}
\psfrag{1.0}{\begin{small}\!1              \end{small}}
\psfrag{2}{\begin{small}2                 \end{small}}
\psfrag{4}{\begin{small}4                  \end{small}}
\psfrag{6}{\begin{small}6                  \end{small}}
\psfrag{1}{\begin{small}1                 \end{small}}
\psfrag{3}{\begin{small}3                  \end{small}}
\psfrag{5}{\begin{small}5                  \end{small}}
\psfrag{8}{\begin{small}8                \end{small}}
\psfrag{B}{\begin{small}$B(Q^2)$                \end{small}}
\psfrag{Q2}{\begin{small}$Q^2~(\mathrm{GeV}^2)$               \end{small}}
\includegraphics[width=10cm]{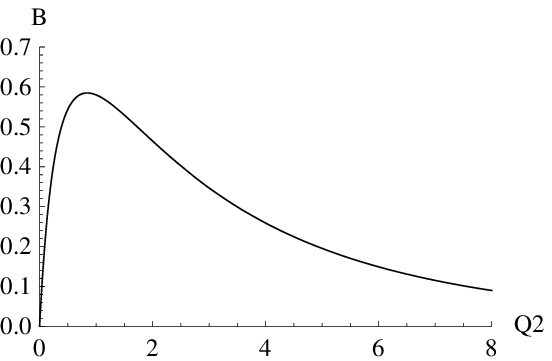}
\caption{The elastic scattering observable $B(Q^2)$ from Eq.~(\ref{eq:Bobservable}) for the $\rho$ meson calculated with the same parameters as
in Fig.~\ref{fig:rhoFF1}. 
}  \label{fig:Brho}                    \end{center}
\end{figure}

\begin{figure}
\begin{center}
\psfrag{-}{\begin{small}              \end{small}}
\psfrag{0.0}{\begin{small}0                \end{small}}
\psfrag{0}{\begin{small}0            \end{small}}
\psfrag{0.2}{\begin{small}-0.2                 \end{small}}
\psfrag{0.4}{\begin{small}-0.4              \end{small}}
\psfrag{0.1}{\begin{small}-0.1               \end{small}}
\psfrag{0.3}{\begin{small}-0.3               \end{small}}
\psfrag{0.6}{\begin{small}-0.6               \end{small}}
\psfrag{0.5}{\begin{small}-0.5              \end{small}}
\psfrag{0.7}{\begin{small}0.7               \end{small}}
\psfrag{1.0}{\begin{small}\!1              \end{small}}
\psfrag{2}{\begin{small}2                 \end{small}}
\psfrag{4}{\begin{small}4                  \end{small}}
\psfrag{6}{\begin{small}6                  \end{small}}
\psfrag{1}{\begin{small}1                 \end{small}}
\psfrag{3}{\begin{small}3                  \end{small}}
\psfrag{5}{\begin{small}5                  \end{small}}
\psfrag{8}{\begin{small}8                \end{small}}
\psfrag{T20}{\begin{small}$T_{20}(Q^2)$                \end{small}}
\psfrag{Q2}{\begin{small}$Q^2~(\mathrm{GeV}^2)$               \end{small}}
\includegraphics[width=10cm]{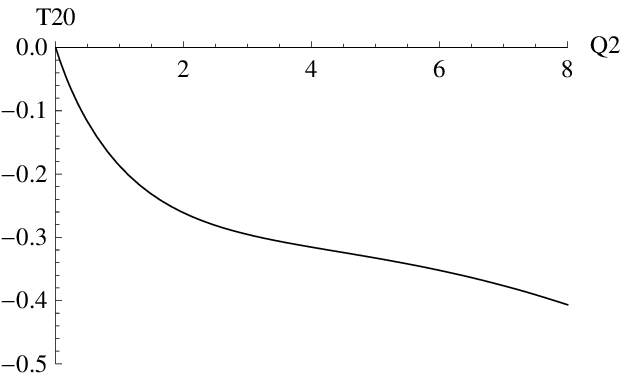}
\caption{The elastic scattering observable $T_{20}(Q^2)$ from Eq.~(\ref{eq:T20}) for the $\rho$ meson calculated with the same parameters as
in Fig.~\ref{fig:rhoFF1}.
}  \label{fig:T20rho}                    \end{center}
\end{figure}

From the form factors $G_\mathrm C(Q^2)$, $G_\mathrm M(Q^2)$ and $G_\mathrm Q(Q^2)$ we obtain the elastic scattering observables $A(Q^2)$, $B(Q^2)$ and $T_{20}(Q^2)$ by means of Eqs.~(\ref{eq:Aobservable}), (\ref{eq:Bobservable}) 
and (\ref{eq:T20}), respectively. The corresponding results are 
depicted in Figs.~\ref{fig:Arho},~\ref{fig:Brho} and~\ref{fig:T20rho}.     

\begin{figure}[h!]
\begin{center}
\psfrag{-}{\begin{small}\;-              \end{small}}
\psfrag{0.0}{\begin{small}0                \end{small}}
\psfrag{0.2}{\begin{small}0.2                 \end{small}}
\psfrag{0.4}{\begin{small}0.4              \end{small}}
\psfrag{0.5}{\begin{small}0.5              \end{small}}
\psfrag{0.1}{\begin{small}0.1               \end{small}}
\psfrag{0.3}{\begin{small}0.3               \end{small}}
\psfrag{1.0}{\begin{small}\!1              \end{small}}
\psfrag{20}{\begin{small}20                \end{small}}
\psfrag{40}{\begin{small}40                  \end{small}}
\psfrag{6}{\begin{small}6                  \end{small}}
\psfrag{10}{\begin{small}10                 \end{small}}
\psfrag{30}{\begin{small}30                  \end{small}}
\psfrag{5}{\begin{small}5                  \end{small}}
\psfrag{8}{\begin{small}8                \end{small}}
\psfrag{J1p0}{\begin{small}$J^1_{10}(Q^2)$                \end{small}}
\psfrag{Q2}{\begin{small}$Q^2~(\mathrm{GeV}^2)$               \end{small}}

\includegraphics[width=10cm]{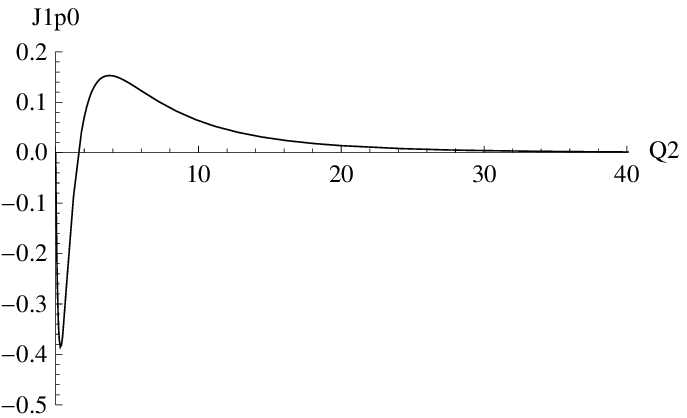}
\caption{\label{fig:currentconsviolation}The violation of conservation of the $\rho$-meson current given by $J^1_{10}(Q^2)$  with our standard kinematics of momentum transfer in the 1-direction calculated calculated with the same parameters as
in Fig.~\ref{fig:rhoFF1} (here depicted in units of the elementary charge $|\,\mathrm e\,|$). 
}                    \end{center}
\end{figure}

\begin{figure}[h!]
\begin{center}
\psfrag{0.05}{\begin{small}0.05                \end{small}}
\psfrag{0.20}{\begin{small}0.2                 \end{small}}
\psfrag{0.4}{\begin{small}0.4              \end{small}}
\psfrag{0.15}{\begin{small}0.15              \end{small}}
\psfrag{0.10}{\begin{small}0.1               \end{small}}
\psfrag{0.3}{\begin{small}0.3               \end{small}}
\psfrag{1.0}{\begin{small}\!1              \end{small}}
\psfrag{2}{\begin{small}2                \end{small}}
\psfrag{4}{\begin{small}4                  \end{small}}
\psfrag{6}{\begin{small}6                  \end{small}}
\psfrag{10}{\begin{small}10                 \end{small}}
\psfrag{30}{\begin{small}30                  \end{small}}
\psfrag{5}{\begin{small}5                  \end{small}}
\psfrag{8}{\begin{small}8                \end{small}}
\psfrag{mB5B7}{\begin{small}\!\!\!\!\!\!\!\!\!\!\!\!\!\!\!\!\!$-(B_5(Q^2)+B_7(Q^2))$                \end{small}}
\psfrag{Q2}{\begin{small}$Q^2~(\mathrm{GeV}^2)$               \end{small}}
\includegraphics[width=10cm]{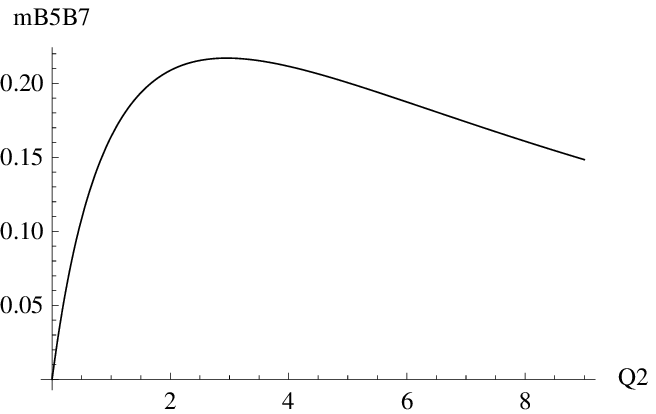}
\caption{The violation of the angular condition for the $\rho$-meson given by Eq.~(\ref{eq:angularconditionviolation}) calculated with the same parameters as
in Fig.~\ref{fig:rhoFF1}. \label{fig:ACrho}  
}                    \end{center}
\end{figure}

\begin{figure}[h!]
\begin{center}
\psfrag{0.08}{\begin{small}0.08                \end{small}}
\psfrag{0.06}{\begin{small}0.06                \end{small}}
\psfrag{0.04}{\begin{small}0.04                \end{small}}
\psfrag{0.02}{\begin{small}0.02                \end{small}}
\psfrag{0.20}{\begin{small}0.2                 \end{small}}
\psfrag{0.4}{\begin{small}0.4              \end{small}}
\psfrag{0.15}{\begin{small}0.15              \end{small}}
\psfrag{0.10}{\begin{small}0.1               \end{small}}
\psfrag{0.3}{\begin{small}0.3               \end{small}}
\psfrag{1.0}{\begin{small}\!1              \end{small}}
\psfrag{2}{\begin{small}2                \end{small}}
\psfrag{4}{\begin{small}4                  \end{small}}
\psfrag{6}{\begin{small}6                  \end{small}}
\psfrag{10}{\begin{small}10                 \end{small}}
\psfrag{30}{\begin{small}30                  \end{small}}
\psfrag{5}{\begin{small}5                  \end{small}}
\psfrag{8}{\begin{small}8                \end{small}}
\psfrag{B6}{\begin{small}$B_6(Q^2)$                \end{small}}
\psfrag{Q2}{\begin{small}$Q^2~(\mathrm{GeV}^2)$               \end{small}}
\includegraphics[width=10cm]{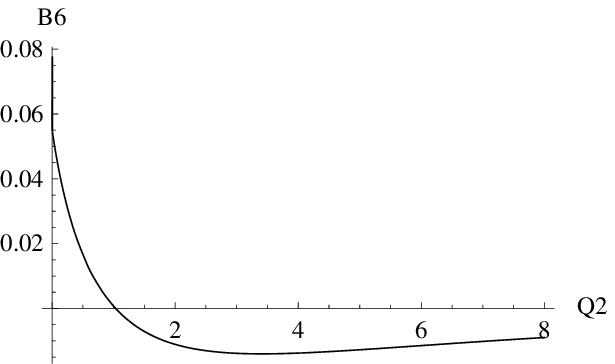}
\caption{The spurious form factor $B_6(Q^2)$ of the $\rho$-meson calculated with the same parameters as
in Fig.~\ref{fig:rhoFF1}. \label{fig:rhoB6}  
}                    \end{center}
\end{figure}

It is also interesting to see how large the spurious contributions to the $\rho$-meson current are. In particular we shall concentrate on the violation of current conservation, the violation of the angular condition and the spurious form factor $B_6(Q^2)$, cf. Sec.~\ref{sec:spuriousconstrib}. The $\rho$-meson current is not conserved due to a non-vanishing current component $J^1_{10}(Q^2)$. For our standard kinematics, Eq.~(\ref{eq:kM}), its magnitude is depicted in Fig.~\ref{fig:currentconsviolation}.
We also give our result for the violation of the angular condition, which is equal to the sum of the spurious form factors $B_5$ and $B_7$ (cf. Eq.~(\ref{eq:angularconditionviolation})).
The result is plotted in Fig.~\ref{fig:ACrho}.
Finally we calculate the spurious form factor 
$B_6(Q^2)$ from the difference of $G_\mathrm M$ and the expression Eq.~(\ref{eq:GMB6}). It is shown in Fig.~\ref{fig:rhoB6}.
As one observes, the spurious contributions can contribute significantly to the current matrix elements and their separation is crucial for the extraction of meaningful physical form factors.
\subsection{Comparisons}
In order to make reasonable comparisons with other approaches to $\rho$-meson form factors we look, in particular, at calculations that have also used the simple harmonic-oscillator potential of Sec.~\ref{sec:homodelformesons}:
the light-front approach of Ref.~\cite{Choi:2004ww}, the covariant extension of the light-front formalism of Ref.~\cite{Jaus:2002sv}, the covariant light-front prescription of Ref.~\cite{Carbonell:1998rj} (the results from this prescription using a Gaussian wave function can be found in Ref.~\cite{Jaus:2002sv}) and the standard light-front prescription of Ref.~\cite{Chung:1988my} (the results from this prescription can be found in Ref.~\cite{Jaus:2002sv}). 
In order to end up with qualitative comparable results we adopt the values for the 2 parameters $m_\mathrm q$ and $a$ of each approach and use them in our prescription. 
The predictions for the magnetic dipole moment $\mu_\rho$ and the electric quadrupole moment $Q_\rho$ are compared in Tab.~\ref{tab:rhomuQ}.
\begin{table*} 
\begin{center}
\begin{tabular}
{|l|c|c|c|c|}\hline
Ref.&$m_\mathrm q$ (MeV)&$a$ (MeV)&$\mu_\rho$&$Q_\rho $ \\ \hline
this work&340&312&2.20&-0.47\\\hline
Choi et al.~\cite{Choi:2004ww} &220&365.9&1.92&-0.43\\ 
this work&220&365.9&2.33&-0.33\\ \hline
Jaus~\cite{Jaus:2002sv}&250&280&1.83&-0.33\\ 
this work&250&280&2.25&-0.33\\ \hline
Carbonell et al.~\cite{Carbonell:1998rj}&250&262&2.23& -0.005\\
this work&250&262&2.231&-0.0058\\\hline 
Chung et al.~\cite{Chung:1988my}&250&316&2.23& -0.19\\
this work &250&316& 2.27344&-0.253915 \\\hline 
\end{tabular}             \end{center}\caption{\label{tab:rhomuQ}
Comparison of the magnetic dipole moment  
$\mu_\rho$ (in units $|\,\mathrm e\,|/2m_{\rho}$) and the electric quadrupole moment $Q_\rho$ (in units $|\,\mathrm e\,|/m_{\rho}^2$) in different approaches with a harmonic oscillator confining potential.
}\end{table*}
It turns out that our results for $\mu_\rho$ and $Q_\rho$ show agreement with the results of Carbonell et al., Ref.~\cite{Carbonell:1998rj}. This is just what we expected due to the similarities and correspondences in both approaches. Our value for $Q_\rho$ also coincides with the approach of Jaus, Ref.~\cite{Jaus:2002sv}, however the values for $\mu_\rho$ differ significantly. Quite good agreement is found with Chung et al., Ref.~\cite{Chung:1988my}. In particular, their value for $\mu_\rho$ is about $0.0434$ units of $|\,\mathrm e\,|/2m_{\rho}$ lower than ours which coincides exactly with the value of $B_6(0)-B_5(0)-B_7(0)$. This difference between our value for the magnetic moment and theirs can be understood as the spurious term $-B_6(Q^2)+(B_5(Q^2)+B_7(Q^2))/(1+\eta))$ which is added to the magnetic form factor when using the standard light-front prescription of Ref.~\cite{Chung:1988my}  (cf. Ref.~\cite{Carbonell:1998rj}).

In Tab.~\ref{tab:rhomuQpred} we also give our predictions for $\mu_\rho$ and $Q_\rho$ in comparison with predictions of other approaches which use different model wave functions.
\begin{table*}[h!]
\begin{center}
\begin{tabular}
{|l|c|c|}\hline
Ref.&$\mu_\rho$&$Q_\rho$\\ \hline
this work&2.20&-0.47\\
Bagdasaryan et al.~\cite{Bagdasaryan:1984kz}&2.30&-0.45\\ 
Samsonov~\cite{Samsonov:2003hs}&2.00$\pm$0.3&-\\ 
Aliev et al.~\cite{Aliev:2004uj}&2.30&-\\ 
Cardarelli et al.~\cite{Cardarelli:1994yq}&2.23&-0.61\\ 
Bhagwat et al.~\cite{Bhagwat:2006pu}&2.01&-0.41\\ 
Hawes et al.~\cite{Hawes:1998bz}&2.69&-0.84\\ 
De Melo et al.~\cite{deMelo:1997hh}&2.14&-0.79\\ \hline
\end{tabular}             \end{center}\caption{\label{tab:rhomuQpred}
Predictions of the magnetic dipole moment 
$\mu_\rho$ (in units $|\,\mathrm e\,|/2m_{\rho}$) and the electric quadrupole moment $Q_\rho$ (in units $|\,\mathrm e\,|/m_{\rho}^2$) in different approaches.
}\end{table*}
Our predicted $\mu_\rho$ and $Q_\rho$ have quite reasonable values which are lying between those of other approaches. In Tab.~\ref{tab:ffsrho} we compare our approach with the work by Choi et al., Ref.~\cite{Choi:2004ww} using the harmonic-oscillator potential with the same parameters as therein. Here it turns out that best agreement is achieved for the magnetic form factor $G_\mathrm M$. 
\begin{table*} 
\begin{center}
\begin{tabular}
{|c|c|c|c|}
\hline
$Q^2(\mathrm{GeV}^2)$&&\cite{Choi:2004ww}&this work \\ \hline
&$G_\mathrm C$&0.38&0.29\\
$Q^2=1$&$G_\mathrm M$&0.93&0.93\\
&$G_\mathrm Q$&-0.23&-0.21\\\hline
&$G_\mathrm C$&0.18&0.12\\
$Q^2=2$&$G_\mathrm M$&0.59&0.58\\
&$G_\mathrm Q$&-0.15&-0.14\\\hline
&$G_\mathrm C$&0.08&0.05\\
$Q^2=3$&$G_\mathrm M$&0.41&0.41\\
&$G_\mathrm Q$&-0.10&-0.10\\\hline
\end{tabular}                                             \end{center}
\caption{\label{tab:ffsrho}
Comparison of the form factors of our approach with the work by Choi et al., Ref.~\cite{Choi:2004ww} for some fixed values of $Q^2$.
In both approaches the same harmonic-oscillator potential has been used with the parameters $m_\mathrm q=220~\mathrm{MeV}$ and $a=365.9~\mathrm{MeV}$.  
}\end{table*}

In Tab.~\ref{tab:ffsrhopred} we compare our prediction for the $\rho$-meson form factors with those of other approaches which use different model wave functions.
\begin{table*} 
\begin{center}
\begin{tabular}
{|c|c|c|c|c|c|c|}
\hline
$Q^2(\mathrm{GeV}^2)$&&this work&\cite{Bhagwat:2006pu}&\cite{Hawes:1998bz}&\cite{Aliev:2004uj}&\cite{Braguta:2004kx}\\ \hline
&$G_\mathrm C$&0.29&0.22&0.17&0.25&0.10\\
$Q^2=1$&$G_\mathrm M$&0.85&0.57&0.85&0.58&0.46\\
&$G_\mathrm Q$&-0.21&-0.11&-0.51&-0.49&-0.16\\\hline
&$G_\mathrm C$&0.11&0.08&0.04&0.13&0.16\\
$Q^2=2$&$G_\mathrm M$&0.47&0.27&0.45&0.28&0.27\\
&$G_\mathrm Q$&-0.12&-0.05&-0.32&-0.24&-0.11\\\hline
&$G_\mathrm C$&0.05&&0.11&0.08&-0.03\\
$Q^2=3$&$G_\mathrm M$&0.30&&0.25&0.17&0.18\\
&$G_\mathrm Q$&-0.07&&-0.23&-0.15&-0.10\\\hline
\end{tabular}                                             \end{center}
\caption{\label{tab:ffsrhopred}
Comparison of the $\rho$-meson form factors in various approaches.}\end{table*}
First we observe that our electric charge form factor $G_\mathrm C$ lies for small values of $Q^2$ above the predictions of other approaches. For higher $Q^2$ 
our values for $G_\mathrm C$ lie somewhere between the values of the other approaches. Second, our magnetic form factor $G_\mathrm M$ lies above the values of other approaches in the whole range of $Q^2$. Third, our prediction for the electric quadrupole form factor $G_\mathrm Q$ lies between the ones of other approaches.

To conclude, in view of the simplicity of our harmonic-oscillator confining potential these are altogether quite reasonable results for our electromagnetic $\rho$-meson form factor.
\section{Electromagnetic Deuteron Form Factors}
\begin{figure}
\begin{center}
\psfrag{0}{\begin{small}0               \end{small}}
\psfrag{0.1}{\begin{small}0.1                \end{small}}
\psfrag{0.01}{\begin{small}0.01                \end{small}}
\psfrag{0.001}{\begin{small}0.001                \end{small}}
\psfrag{0.0001}{\begin{small}  \;$10^{-4}$                \end{small}}
\psfrag{0.00001}{\begin{small}\quad$10^{-5}$                \end{small}}
\psfrag{0.06}{\begin{small}0.06                \end{small}}
\psfrag{0.04}{\begin{small}0.04                \end{small}}
\psfrag{0.02}{\begin{small}0.02                \end{small}}
\psfrag{0.20}{\begin{small}0.2                 \end{small}}
\psfrag{0.4}{\begin{small}0.4              \end{small}}
\psfrag{0.15}{\begin{small}0.15              \end{small}}
\psfrag{0.10}{\begin{small}0.1               \end{small}}
\psfrag{0.3}{\begin{small}0.3               \end{small}}
\psfrag{1}{\begin{small}\!1              \end{small}}
\psfrag{2}{\begin{small}2                \end{small}}
\psfrag{4}{\begin{small}4                  \end{small}}
\psfrag{6}{\begin{small}6                  \end{small}}
\psfrag{10}{\begin{small}10                 \end{small}}
\psfrag{3}{\begin{small}3                  \end{small}}
\psfrag{5}{\begin{small}5                  \end{small}}
\psfrag{7}{\begin{small}7               \end{small}}
\psfrag{MMD}{\begin{small}\cite{Mergell:1995bf}              \end{small}}
\psfrag{GK}{\begin{small}\cite{Gari:1992qw}              \end{small}}
\psfrag{pl}{\begin{small}point-like nucleons             \end{small}}
\psfrag{AbsF1}{\begin{small}$\vert F_1(Q^2)\vert$                \end{small}}
\psfrag{Q2}{\begin{small}$Q^2~(\mathrm{GeV}^2)$               \end{small}}
\includegraphics[width=10cm]{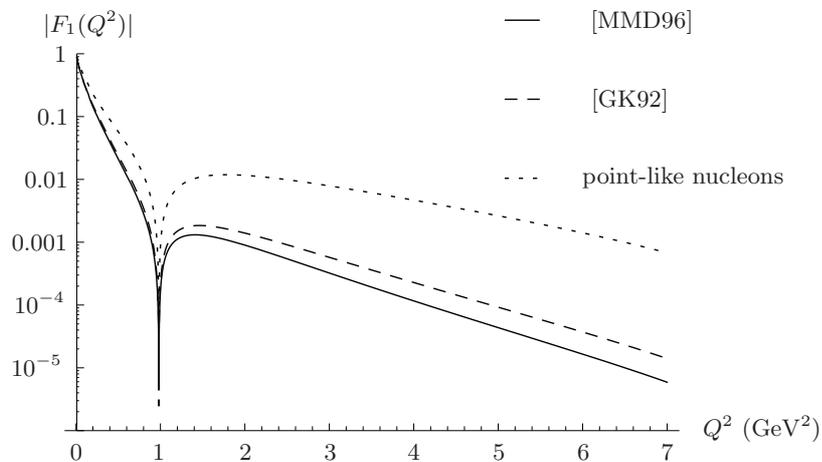}
\caption{The deuteron form factor $F_1$ evaluated within the Walecka-type model together with 2 different parametrizations for the nucleon form factors taken from Refs.~\cite{Gari:1992qw} and~\cite{Mergell:1995bf} and for point-like nucleons. 
 \label{fig:f1deuteron} }                    \end{center}
\end{figure}

\begin{figure}
\begin{center}
\psfrag{0}{\begin{small}0                \end{small}}
\psfrag{0.01}{\begin{small}0.01                \end{small}}
\psfrag{0.0001}{\begin{small}  \;$10^{-4}$                \end{small}}
\psfrag{0.00001}{\begin{small}\quad$10^{-5}$                \end{small}}
\psfrag{0.001}{\begin{small}0.001                \end{small}}
\psfrag{0.0001}{\begin{small}0.0001                \end{small}}
\psfrag{0}{\begin{small}0               \end{small}}
\psfrag{0.1}{\begin{small}0.1                \end{small}}
\psfrag{0.01}{\begin{small}0.01                \end{small}}
\psfrag{0.001}{\begin{small}0.001                \end{small}}
\psfrag{0.0001}{\begin{small}  \;$10^{-4}$                \end{small}}
\psfrag{0.00001}{\begin{small}\quad$10^{-5}$                \end{small}}
\psfrag{0.06}{\begin{small}0.06                \end{small}}
\psfrag{0.04}{\begin{small}0.04                \end{small}}
\psfrag{0.02}{\begin{small}0.02                \end{small}}
\psfrag{0.20}{\begin{small}0.2                 \end{small}}
\psfrag{0.4}{\begin{small}0.4              \end{small}}
\psfrag{0.15}{\begin{small}0.15              \end{small}}
\psfrag{0.10}{\begin{small}0.1               \end{small}}
\psfrag{0.3}{\begin{small}0.3               \end{small}}
\psfrag{1}{\begin{small}\!1              \end{small}}
\psfrag{2}{\begin{small}2                \end{small}}
\psfrag{4}{\begin{small}4                  \end{small}}
\psfrag{6}{\begin{small}6                  \end{small}}
\psfrag{10}{\begin{small}10                 \end{small}}
\psfrag{3}{\begin{small}3                  \end{small}}
\psfrag{5}{\begin{small}5                  \end{small}}
\psfrag{8}{\begin{small}8               \end{small}}
\psfrag{MMD}{\begin{small}\cite{Mergell:1995bf}              \end{small}}
\psfrag{GK}{\begin{small}\cite{Gari:1992qw}              \end{small}}
\psfrag{pl}{\begin{small}point-like nucleons             \end{small}}
\psfrag{AbsF2}{\begin{small}$\vert F_2(Q^2)\vert$                \end{small}}
\psfrag{Q2}{\begin{small}$Q^2~(\mathrm{GeV}^2)$               \end{small}}
\includegraphics[width=10cm]{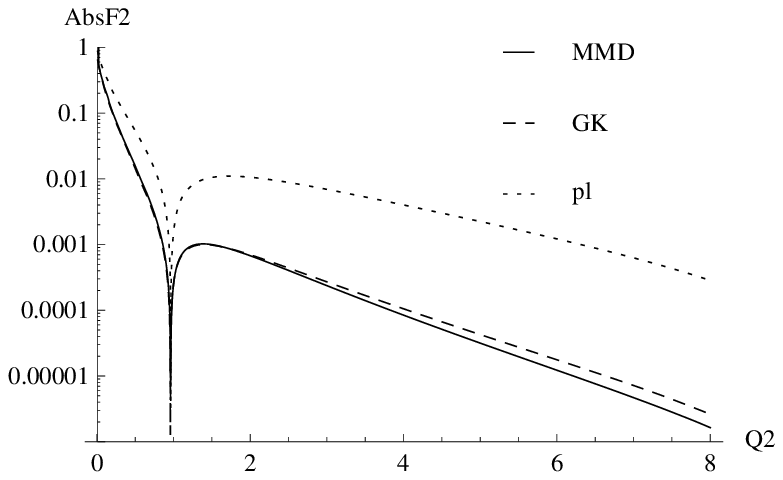}
\caption{The deuteron form factor $F_2$ evaluated within the Walecka-type model together with 2 different parametrizations for the nucleon form factors taken from Refs.~\cite{Gari:1992qw} and~\cite{Mergell:1995bf} and for point-like nucleons.\label{fig:f2deuteron}  
}                     \end{center}
\end{figure}

\begin{figure}
\begin{center}
\psfrag{0}{\begin{small}0                \end{small}}
\psfrag{0.1}{\begin{small}0.1                \end{small}}
\psfrag{0.01}{\begin{small}0.01                \end{small}}
\psfrag{0.0001}{\begin{small}  \;$10^{-4}$                \end{small}}
\psfrag{0.01}{\begin{small}0.01                \end{small}}
\psfrag{0.001}{\begin{small}0.001                \end{small}}
\psfrag{0.06}{\begin{small}0.06                \end{small}}
\psfrag{0.04}{\begin{small}0.04                \end{small}}
\psfrag{0.02}{\begin{small}0.02                \end{small}}
\psfrag{0.20}{\begin{small}0.2                 \end{small}}
\psfrag{0.4}{\begin{small}0.4              \end{small}}
\psfrag{0.15}{\begin{small}0.15              \end{small}}
\psfrag{0.10}{\begin{small}0.1               \end{small}}
\psfrag{0.3}{\begin{small}0.3               \end{small}}
\psfrag{1}{\begin{small}\!1              \end{small}}
\psfrag{2}{\begin{small}2                \end{small}}
\psfrag{4}{\begin{small}4                  \end{small}}
\psfrag{6}{\begin{small}6                  \end{small}}
\psfrag{10}{\begin{small}10                 \end{small}}
\psfrag{3}{\begin{small}3                  \end{small}}
\psfrag{5}{\begin{small}5                  \end{small}}
\psfrag{7}{\begin{small}7              \end{small}}
\psfrag{MMD}{\begin{small}\cite{Mergell:1995bf}              \end{small}}
\psfrag{GK}{\begin{small}\cite{Gari:1992qw}              \end{small}}
\psfrag{pl}{\begin{small}point-like nucleons             \end{small}}
\psfrag{AbsF1}{\begin{small}$\vert G_\mathrm M(Q^2)\vert$                \end{small}}
\psfrag{Q2}{\begin{small}$Q^2~(\mathrm{GeV}^2)$               \end{small}}
\includegraphics[width=10cm]{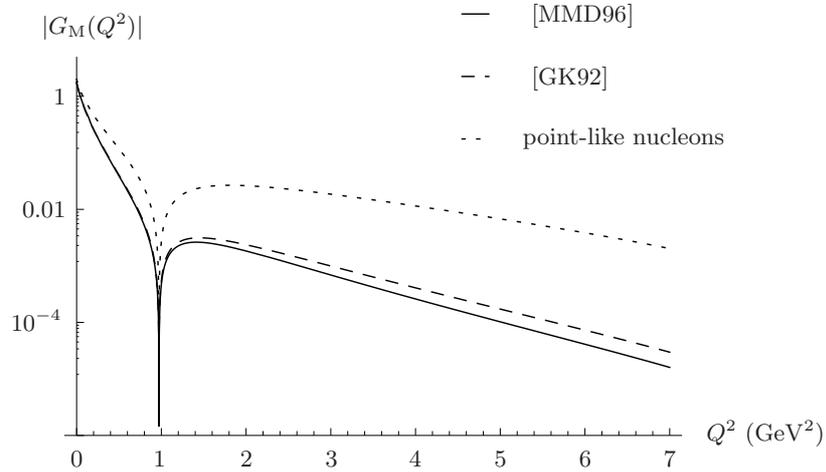}
\caption{The magnetic dipole form factor $G_\mathrm M$ of the deuteron evaluated within the Walecka-type model together with 2 different parametrizations for the nucleon form factors taken from Refs.~\cite{Gari:1992qw} and~\cite{Mergell:1995bf} and for point-like nucleons.
\label{fig:gmdeuteron} }                     \end{center}
\end{figure}

In the present work we model the nucleon-nucleon interaction by the Pauli-Villars regularized Walecka-type potential which we have proposed in Sec.~\ref{sec:waleckatypemodel}. For simplicity we have restricted ourselves to the static limits for the meson-exchange potential. However, we have refined our bound-state problem by replacing the non-relativistic by the relativistic kinetic-energy operator. To get numerical results for the deuteron form factors we plug the solution $u_{00}$ and the mass eigenvalue $m_{00}=m_\mathrm D$ of the Walecka-model mass eigenvalue equation~(\ref{eq:schroedingermomspace}) into the expressions for the spin-1 form factors
Eqs.~(\ref{eq:ff1me}),~(\ref{eq:ff2me}) 
and~(\ref{eq:ffGMme}). In the present work we do not derive expressions for the nucleon form factors $F_1^\mathrm p(Q^2)$, $F_1^\mathrm n(Q^2)$, $F_2^\mathrm p(Q^2)$ and $F_2^\mathrm n(Q^2)$ but we rather use the 2 parametrizations for the nucleon form factors proposed in Refs.~\cite{Gari:1992qw} and~\cite{Mergell:1995bf}. For comparison, we also consider point-like nucleons, i.e. $F_1^\mathrm p(Q^2)=1$ and $F_1^\mathrm n(Q^2)=F_2^\mathrm p(Q^2)=F_2^\mathrm n(Q^2)=0$. The electromagnetic deuteron form factors $F_1\left(Q^2\right)$, $F_2\left(Q^2\right)$ and $G_\mathrm M\left(Q^2\right)$ are then evaluated with these 3 parametrizations. They are depicted in Figs.~\ref{fig:f1deuteron},~\ref{fig:f2deuteron} and~\ref{fig:gmdeuteron}, respectively. 
With the help of Eqs.~(\ref{eq:rhoGC}) and~(\ref{eq:rhoGQ}) we obtain from  $F_1\left(Q^2\right)$, $F_2\left(Q^2\right)$ and $G_\mathrm M\left(Q^2\right)$ the electric charge and quadrupole form factors of the deuteron, $G_\mathrm C(Q^2)$ and $G_\mathrm Q(Q^2)$, respectively. The correct charge $G_\mathrm C(0)=-F_1(0)=1$ can be read off from Fig.~\ref{fig:f1deuteron}, which is, in general, guaranteed only by the particular decomposition of the deuteron current given by Eq.~(\ref{eq:covstructurephysdeutcurr}).
%
Our predictions for the magnetic dipole moment 
$\mu_\mathrm D$ and the electric quadrupole moment $Q_\mathrm D$ defined by the $Q^2\rightarrow 0$ limits of $G_\mathrm M\left(Q^2\right)$ and $G_\mathrm Q\left(Q^2\right)$ are summarized in Tab.~\ref{tab:muQdeutpredicts}.
\begin{table*} 
\begin{center}
\begin{tabular}
{|l|c|c|}\hline
nucleon form factor parametrizations &$\mu_\mathrm D$&$Q_\mathrm D$\\ \hline
Gari et al., Ref.~\cite{Gari:1992qw}&1.76354&-0.013501\\\hline
Mergell et al., Ref.~\cite{Mergell:1995bf}& 1.76585& -0.013519\\\hline
point-like nucleons&2.00472& -0.018137 \\ \hline
\end{tabular}             \end{center}\caption{\label{tab:muQdeutpredicts}
The magnetic dipole moment 
$\mu_\mathrm D$ (in units $|\,\mathrm e\,|/2m_{\mathrm D}$) and the electric quadrupole moment $Q_\mathrm D$ (in units $|\,\mathrm e\,|/m_{\mathrm D}^2$) for the deuteron 
evaluated with the Walecka-type wave function and 2 different parametrizations for the nucleon form factors taken from Refs.~\cite{Gari:1992qw} and~\cite{Mergell:1995bf} and for point-like nucleons.
}\end{table*}

Due to the simplicity of our Walecka-type model we do not expect to get form factor results that are close to experimental data. Therefore we do not compare our results with data, which is anyway not the purpose of this work. We should mention, however, that we find a quite reasonable value for the magnetic dipole moment $\mu_\mathrm D$. The value for point-like nucleons, e.g. comes close to the value $\mu_\mathrm D=2$ for point-like spin-1 particles. Our results for the electric quadrupole moment $Q_\mathrm D$ are quite small as a consequence of the absence of a d-wave contribution. In the absence of a d-wave contribution the non-vanishing of $Q_\mathrm D$ is a pure relativistic effect.    

\begin{figure}
\begin{center}
\psfrag{0}{\begin{small}0              \end{small}}
\psfrag{0.001}{\begin{small}0.001                \end{small}}
\psfrag{0.0001}{\begin{small}$10^-4$                \end{small}}
\psfrag{0.00001}{\begin{small}\quad$10^{-5}$                \end{small}}
\psfrag{E7}{\begin{small}\!\!\!\!\!\!\!$10^{-7}$                \end{small}}
\psfrag{E9}{\begin{small}\!\!\!\!\!\!\!$10^{-9}$                \end{small}}
\psfrag{0.06}{\begin{small}0.06                \end{small}}
\psfrag{0.04}{\begin{small}0.04                \end{small}}
\psfrag{0.02}{\begin{small}0.02                \end{small}}
\psfrag{0.20}{\begin{small}0.2                 \end{small}}
\psfrag{0.4}{\begin{small}0.4              \end{small}}
\psfrag{0.15}{\begin{small}0.15              \end{small}}
\psfrag{0.10}{\begin{small}0.1               \end{small}}
\psfrag{0.3}{\begin{small}0.3               \end{small}}
\psfrag{1}{\begin{small}\!1              \end{small}}
\psfrag{2}{\begin{small}2                \end{small}}
\psfrag{4}{\begin{small}4                  \end{small}}
\psfrag{6}{\begin{small}6                  \end{small}}
\psfrag{10}{\begin{small}10                 \end{small}}
\psfrag{9}{\begin{small}9                 \end{small}}
\psfrag{3}{\begin{small}3                  \end{small}}
\psfrag{5}{\begin{small}5                  \end{small}}
\psfrag{7}{\begin{small}7              \end{small}}
\psfrag{MMD}{\begin{small}\cite{Mergell:1995bf}              \end{small}}
\psfrag{GK}{\begin{small}\cite{Gari:1992qw}              \end{small}}
\psfrag{pointlike}{\begin{small}point-like nucleons             \end{small}}
\psfrag{A}{\begin{small}\!\!\!\!\!\!$\vert A(Q^2)\vert$                \end{small}}
\psfrag{Q2}{\begin{small}$Q^2~(\mathrm{GeV}^2)$               \end{small}}
\includegraphics[width=10cm]{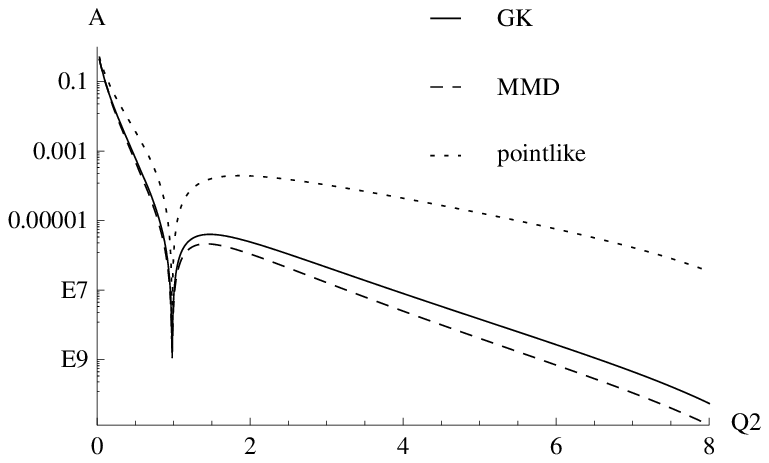}
\caption{\label{fig:Adeuteron}
The elastic scattering observable $A(Q^2)$ of the deuteron evaluated within the Walecka-type model together with
2 different parametrizations for the nucleon form factors taken from Refs.~\cite{Gari:1992qw} and~\cite{Mergell:1995bf} and for point-like nucleons. 
}                     \end{center}
\end{figure}

\begin{figure}
\begin{center}
\psfrag{0}{\begin{small}0               \end{small}}
\psfrag{0.001}{\begin{small}\!\!\!0.001                \end{small}}
\psfrag{0.0001}{\begin{small}$10^-4$                \end{small}}
\psfrag{E5}{\begin{small}\!\!\!\!\!\!\!$10^{-5}$                \end{small}}
\psfrag{E7}{\begin{small}\!\!\!\!\!\!\!$10^{-7}$                \end{small}}
\psfrag{E9}{\begin{small}\!\!\!\!\!\!\!$10^{-9}$                \end{small}}
\psfrag{0.06}{\begin{small}0.06                \end{small}}
\psfrag{0.04}{\begin{small}0.04                \end{small}}
\psfrag{0.02}{\begin{small}0.02                \end{small}}
\psfrag{0.20}{\begin{small}0.2                 \end{small}}
\psfrag{0.4}{\begin{small}0.4              \end{small}}
\psfrag{0.15}{\begin{small}0.15              \end{small}}
\psfrag{0.10}{\begin{small}0.1               \end{small}}
\psfrag{0.3}{\begin{small}0.3               \end{small}}
\psfrag{1}{\begin{small}\!1              \end{small}}
\psfrag{2}{\begin{small}2                \end{small}}
\psfrag{4}{\begin{small}4                  \end{small}}
\psfrag{6}{\begin{small}6                  \end{small}}
\psfrag{10}{\begin{small}10                 \end{small}}
\psfrag{8}{\begin{small}8                 \end{small}}
\psfrag{3}{\begin{small}3                  \end{small}}
\psfrag{5}{\begin{small}5                  \end{small}}
\psfrag{7}{\begin{small}7              \end{small}}
\psfrag{MMD}{\begin{small}\cite{Mergell:1995bf}              \end{small}}
\psfrag{GK}{\begin{small}\cite{Gari:1992qw}              \end{small}}
\psfrag{pointlike}{\begin{small}point-like nucleons             \end{small}}
\psfrag{B}{\begin{small}\!\!\!\!\!\!$\vert B(Q^2)\vert$                \end{small}}
\psfrag{Q2}{\begin{small}$Q^2~(\mathrm{GeV}^2)$               \end{small}}
\includegraphics[width=8.0cm]{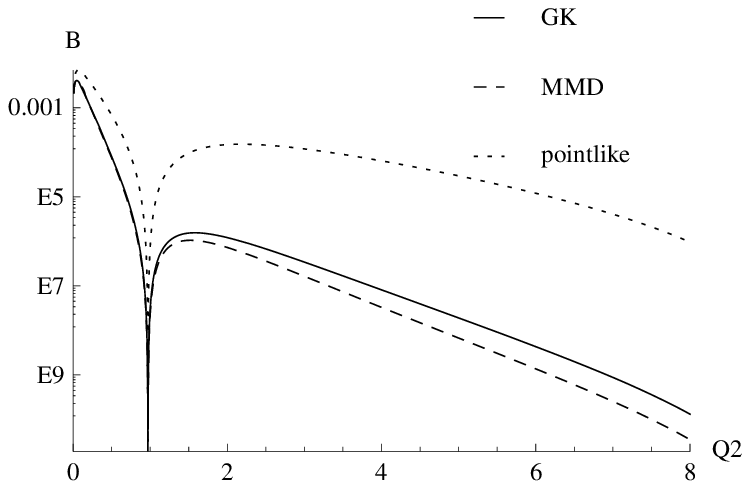}
\caption{\label{fig:Bdeuteron}
The elastic scattering observable $B(Q^2)$ of the deuteron evaluated within the Walecka-type model together with
2 different parametrizations for the nucleon form factors taken from Refs.~\cite{Gari:1992qw} and~\cite{Mergell:1995bf} and for point-like nucleons. 
}                     \end{center}
\end{figure}

\begin{figure}
\begin{center}
\psfrag{0}{\begin{small}0               \end{small}}
\psfrag{0.001}{\begin{small}\!\!\!0.001                \end{small}}
\psfrag{0.0001}{\begin{small}$10^-4$                \end{small}}
\psfrag{E5}{\begin{small}\!\!\!\!\!\!\!$10^{-5}$                \end{small}}
\psfrag{E7}{\begin{small}\!\!\!\!\!\!\!$10^{-7}$                \end{small}}
\psfrag{E9}{\begin{small}\!\!\!\!\!\!\!$10^{-9}$                \end{small}}
\psfrag{0.1}{\begin{small}\!\!\!\!-0.1               \end{small}}
\psfrag{0.2}{\begin{small}\!\!\!\!-0.2                \end{small}}
\psfrag{0.3}{\begin{small}\!\!\!\!-0.3                \end{small}}
\psfrag{0.4}{\begin{small}\!\!\!\!-0.4               \end{small}}
\psfrag{-}{\begin{small}            \end{small}}
\psfrag{0.5}{\begin{small}\!\!\!\!-0.5              \end{small}}
\psfrag{0.0}{\begin{small}0              \end{small}}
\psfrag{0.10}{\begin{small}0.1               \end{small}}
\psfrag{1}{\begin{small}\!1              \end{small}}
\psfrag{2}{\begin{small}2                \end{small}}
\psfrag{4}{\begin{small}4                  \end{small}}
\psfrag{6}{\begin{small}6                  \end{small}}
\psfrag{10}{\begin{small}10                 \end{small}}
\psfrag{8}{\begin{small}8                 \end{small}}
\psfrag{3}{\begin{small}3                  \end{small}}
\psfrag{5}{\begin{small}5                  \end{small}}
\psfrag{7}{\begin{small}7              \end{small}}
\psfrag{MMD}{\begin{small}\cite{Mergell:1995bf}              \end{small}}
\psfrag{GK}{\begin{small}\cite{Gari:1992qw}              \end{small}}
\psfrag{pointlike}{\begin{small}\; point-like nucleons             \end{small}}
\psfrag{T20}{\begin{small}\!\!\!\!\!\!$T_{20} (Q^2)$                \end{small}}
\psfrag{Q2}{\begin{small}$Q^2~(\mathrm{GeV}^2)$               \end{small}}
\includegraphics[width=10cm]{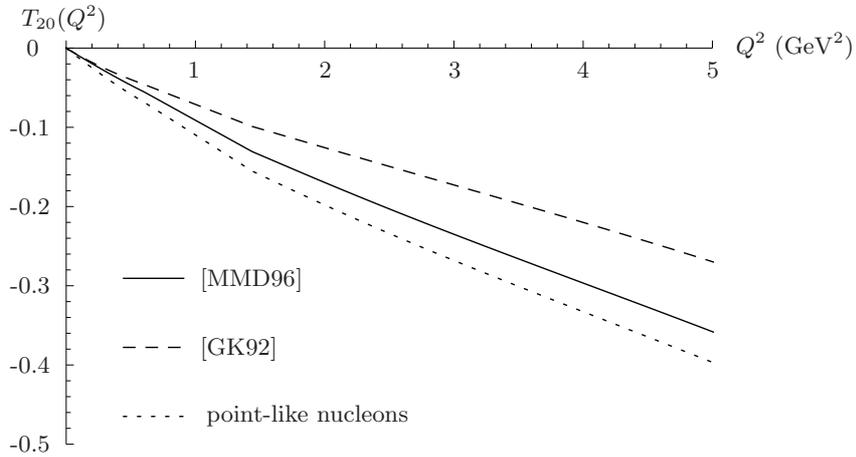}
\caption{\label{fig:T20deuteron}
The elastic scattering observable $T_{20}(Q^2)$ of the deuteron evaluated within the Walecka-type model together with 2 different parametrizations for the nucleon form factors taken from Refs.~\cite{Gari:1992qw} and~\cite{Mergell:1995bf} and for point-like nucleons. 
}                     \end{center}
\end{figure}

From the form factors $G_\mathrm C$, $G_\mathrm M$ and $G_\mathrm Q$ we obtain the elastic scattering observables $A(Q^2)$, $B(Q^2)$ and $T_{20}(Q^2)$ by means of Eqs.~(\ref{eq:Aobservable}), (\ref{eq:Bobservable}) 
and (\ref{eq:T20}), respectively. We compare the results for different parametrizations of the nucleon form factors in Figs.~\ref{fig:Adeuteron},~\ref{fig:Bdeuteron} and~\ref{fig:T20deuteron}.

\begin{figure}
\begin{center}
\psfrag{-}{\begin{small}\;-              \end{small}}
\psfrag{0.0002}{\begin{small}0.0002               \end{small}}
\psfrag{0.0004}{\begin{small}0.0004                \end{small}}
\psfrag{0.0006}{\begin{small}0.0006              \end{small}}
\psfrag{0.5}{\begin{small}0.5              \end{small}}
\psfrag{0.1}{\begin{small}0.1               \end{small}}
\psfrag{0.3}{\begin{small}0.3               \end{small}}
\psfrag{1.0}{\begin{small}\!1              \end{small}}
\psfrag{20}{\begin{small}20                \end{small}}
\psfrag{40}{\begin{small}40                  \end{small}}
\psfrag{6}{\begin{small}6                  \end{small}}
\psfrag{10}{\begin{small}10                 \end{small}}
\psfrag{15}{\begin{small}15                  \end{small}}
\psfrag{5}{\begin{small}5                  \end{small}}
\psfrag{8}{\begin{small}8                \end{small}}
\psfrag{J1p0}{\begin{small}$J^1_{10}(Q^2)$                \end{small}}
\psfrag{Q2}{\begin{small}$Q^2~(\mathrm{GeV}^2)$               \end{small}}
\psfrag{MMD}{\begin{small}\!\!\!\cite{Mergell:1995bf}              \end{small}}
\psfrag{GK}{\begin{small}\cite{Gari:1992qw}              \end{small}}
\psfrag{pointlike}{\begin{small}point-like nucleons             \end{small}}
\includegraphics[width=10cm]{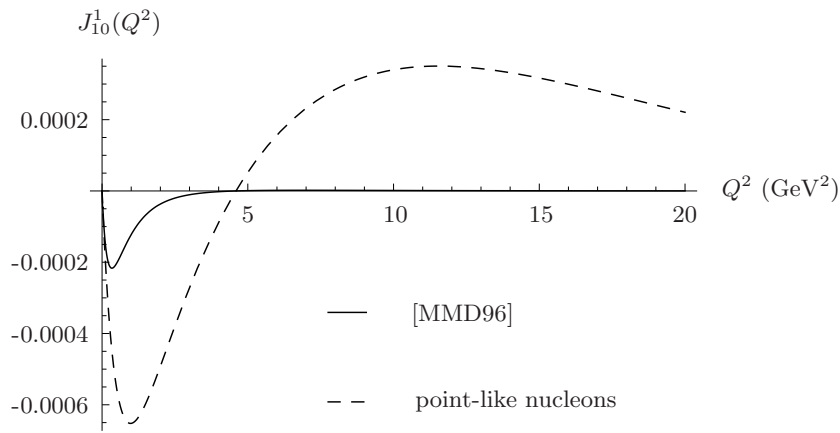}
\caption{\label{fig:currentconsviolationDeut}The violation of current conservation given by $J^1_{10}(Q^2)$ with our standard kinematics of momentum transfer in the 1-direction calculated within the Walecka-type model together with the nucleon form factor parametrization of Ref.~\cite{Mergell:1995bf} and for point-like nucleons (here depicted in units of the elementary charge $\mathrm e$) . 
}                   \end{center}
\end{figure}

\begin{figure}
\begin{center}
\psfrag{-}{\begin{small}\;-              \end{small}}
\psfrag{0.0002}{\begin{small}0.0002               \end{small}}
\psfrag{0.0004}{\begin{small}0.0004                \end{small}}
\psfrag{0.0006}{\begin{small}0.0006              \end{small}}
\psfrag{0.0008}{\begin{small}0.0008             \end{small}}
\psfrag{0.0010}{\begin{small}0.001              \end{small}}
\psfrag{0.05}{\begin{small}0.05                \end{small}}
\psfrag{0.20}{\begin{small}0.2                 \end{small}}
\psfrag{0.4}{\begin{small}0.4              \end{small}}
\psfrag{0.15}{\begin{small}0.15              \end{small}}
\psfrag{0.10}{\begin{small}0.1               \end{small}}
\psfrag{0.3}{\begin{small}0.3               \end{small}}
\psfrag{1.0}{\begin{small}\!1              \end{small}}
\psfrag{2}{\begin{small}2                \end{small}}
\psfrag{5}{\begin{small}5                  \end{small}}
\psfrag{6}{\begin{small}6                  \end{small}}
\psfrag{10}{\begin{small}10                 \end{small}}
\psfrag{15}{\begin{small}15                 \end{small}}
\psfrag{5}{\begin{small}5                  \end{small}}
\psfrag{8}{\begin{small}8                \end{small}}
\psfrag{B5B7}{\begin{small}\!\!\!\!\!\!\!\!\!\!\!\!\!\!\!\!\!$-(B_5(Q^2)+B_7(Q^2))$                \end{small}}
\psfrag{Q2}{\begin{small}$Q^2~(\mathrm{GeV}^2)$               \end{small}}
\psfrag{MMD}{\begin{small}\!\!\!\cite{Mergell:1995bf}              \end{small}}
\psfrag{GK}{\begin{small}\cite{Gari:1992qw}              \end{small}}
\psfrag{pointlike}{\begin{small}point-like nucleons             \end{small}}
\includegraphics[width=10cm]{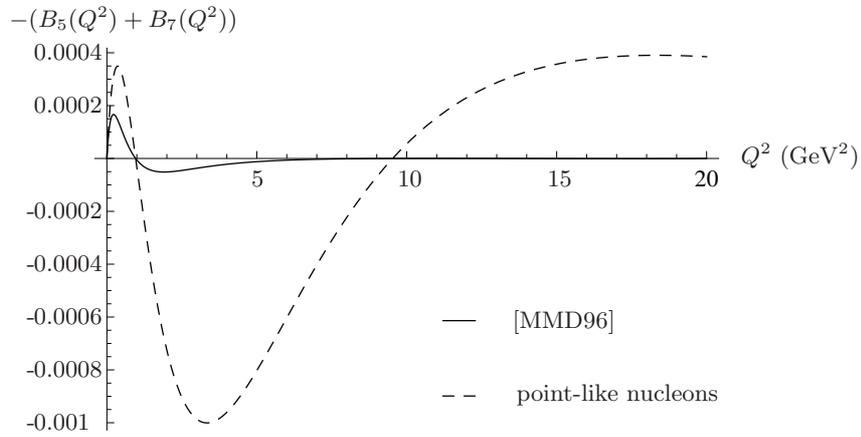}
\caption{The violation of the angular condition for the $\rho$ meson given by the sum of the spurious form factors  $B_5(Q^2)$ and $B_7(Q^2)$. Here it is calculated within the Walecka-type model together with the nucleon form factor parametrization of Ref.~\cite{Mergell:1995bf} and for point-like nucleons. \label{fig:ACdeut}  
}                    \end{center}
\end{figure}

\begin{figure}
\begin{center}
\psfrag{E4}{\begin{small}\!\!\!\!\!\!$10^{-4}$               \end{small}}
\psfrag{E5}{\begin{small}\!\!\!\!\!\!$10^{-5}$             \end{small}}
\psfrag{E6}{\begin{small}\!\!\!\!\!\!$10^{-6}$             \end{small}}
\psfrag{E7}{\begin{small}\!\!\!\!\!\!$10^{-7}$               \end{small}}
\psfrag{E8}{\begin{small}\!\!\!\!\!\!$10^{-8}$             \end{small}}
\psfrag{E9}{\begin{small}\!\!\!\!\!\!$10^{-9}$             \end{small}}
\psfrag{0.001}{\begin{small}0.001               \end{small}}
\psfrag{1}{\begin{small}1                \end{small}}
\psfrag{2}{\begin{small}2             \end{small}}
\psfrag{3}{\begin{small}3              \end{small}}
\psfrag{4}{\begin{small}4             \end{small}}
\psfrag{5}{\begin{small}5               \end{small}}
\psfrag{1.0}{\begin{small}\;\;1              \end{small}}
\psfrag{2.0}{\begin{small}\;\;2                \end{small}}
\psfrag{4}{\begin{small}4                  \end{small}}
\psfrag{6}{\begin{small}6                  \end{small}}
\psfrag{10}{\begin{small}10                 \end{small}}
\psfrag{30}{\begin{small}30                  \end{small}}
\psfrag{5.0}{\begin{small}\;\;5                  \end{small}}
\psfrag{8}{\begin{small}8                \end{small}}
\psfrag{B6}{\begin{small}\!\!\!\!\!\!$|B_6(Q^2)|$                \end{small}}
\psfrag{Q2}{\begin{small}$Q^2~(\mathrm{GeV}^2)$               \end{small}}
\psfrag{MMD}{\begin{small}\!\!\!\cite{Mergell:1995bf}              \end{small}}
\psfrag{GK}{\begin{small}\cite{Gari:1992qw}              \end{small}}
\psfrag{pl}{\begin{small}point-like nucleons             \end{small}}
\includegraphics[width=10cm]{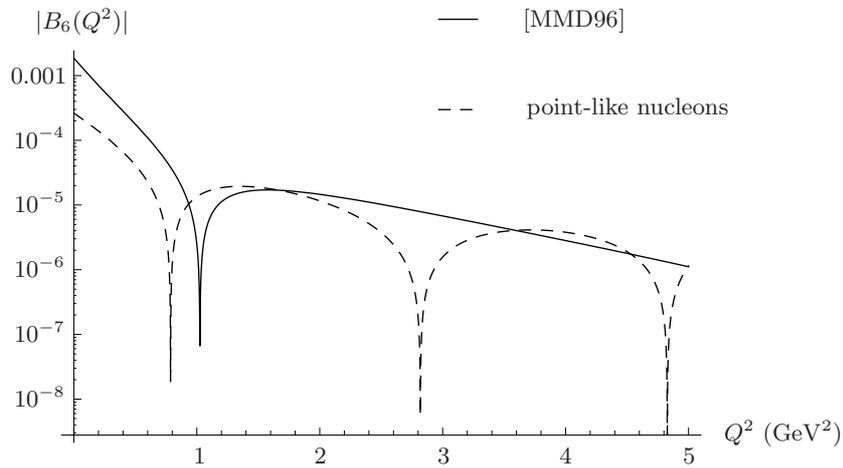}
\caption{The spurious form factor $B_6(Q^2)$ of the deuteron calculated within the Walecka-type model together with the nucleon form factor parametrization of Ref.~\cite{Mergell:1995bf} and for point-like nucleons.  \label{fig:DeutB6}  
}                    \end{center}
\end{figure}

Finally, it is quite interesting to look at the spurious contributions to the deuteron current causing violation of continuity, violation of the angular condition and the spurious form factor $B_6(Q^2)$. To this end we use the parametrization for the nucleon form factors of Ref.~\cite{Mergell:1995bf} and compare it to the point-like nucleon case. The continuity-violating matrix element of the current is $J^1_{10}(Q^2)$ (with our standard kinematics). It is depicted in Fig.~\ref{fig:currentconsviolationDeut}.
In Fig.~\ref{fig:ACdeut} we show the violation of the angular condition which is equal to the sum of the spurious form factors $B_5(Q^2)+B_7(Q^2)$.
The spurious form factor $B_6(Q^2)$, which is contained in the magnetic form factor extracted from the plus component of the current in the standard light-front prescription~\cite{Chung:1988my,Carbonell:1998rj}, is plotted in Fig.~\ref{fig:DeutB6}.
What we observe is that the relative importance of spurious contributions diminishes if one goes from the strongly bound $\mathrm q\bar{\mathrm q}$ system to the weakly bound two-nucleon system. The influence of spurious contributions is further reduced by the nucleon form factors.

We should stress once more that the purpose of the above computation of the deuteron form factors within the simple Walecka-type model is not to find best agreement with experimental data. The calculation is rather part of the benchmark about relativistic effects in few-body physics, where different approaches to deuteron form factors are compared on equal grounds starting with the same fully covariant Walecka-model Lagrangean. Therefore, in order to make concise statements about our deuteron form factor results the findings of the other approaches participating in the benchmark are required.
\chapter{Summary and Outlook}\label{chap:8}
\section{Summary}
Of the various forms of relativistic dynamics given by Dirac, the point form is the least utilized, although it has several advantageous properties. The key benefit of the point form is the property that the Lorentz group is its kinematic group implying that only the Abelian group of space-time translations is affected by interactions. This natural way of separating the kinematic from the dynamic Poincar\'{e} generators allows for a manifest Lorentz covariant formulation of dynamical equations and makes for a simple behavior under Lorentz transformations. 

In the present work we exploit the virtues of point-form relativistic quantum mechanics to analyze the electromagnetic properties of relativistic two-body systems. To this end we treat the elastic electromagnetic scattering of an electron by a bound-state as a two-channel problem for a Bakamjian-Thomas type mass operator. The 2 channels are coupled by appropriately defined electromagnetic vertex operators, which are constructed from quantum field theoretical Lagrangean densities. These vertex interactions describe the emission and absorption of a photon by the electron or the constituents of the bound state and they have the Lorentz structure of a field theoretical vertex. In this way the dynamics of the exchanged photon is taken explicitly into account. These vertices are, however, non-local due to the conservation of the overall four-velocity of the electron-bound-state(-photon) system that is inherent to the point-form Bakamjian-Thomas framework. Four-velocity conservation at the vertex is a necessary approximation in order to be able to formulate electron-bound-state scattering as a simple eigenvalue problem for a Bakamjian-Thomas type
mass operator. 

In this work we consider, for simplicity, only instantaneous interactions between the constituents forming the bound state. Instantaneous interactions can be used within the Bakamjian-Thomas framework without losing Poincar\'{e} invariance. Therefore, our approach preserves Poincar\'{e} invariance by construction. What we violate, however, within a Bakamjian-Thomas construction for more than 2 particles is macroscopic locality (i.e. cluster separability).

By a Feshbach reduction the eigenvalue problem for the coupled-channel mass operator can be brought into the form of a non-linear eigenvalue problem which defines the one-photon-exchange optical potential. As expected, the optical potential is a contraction of the (point-like) electron current with a bound-state current times the covariant
photon propagator. The resulting bound-state current, which is expressed in terms of bound-state wave functions and constituent currents, is Hermitean and transforms covariantly under Lorentz transformations. Its Lorentz structure is, however, not fully determined by the incoming and outgoing bound-state momenta and spins. It turns out that additional, unphysical Lorentz covariant terms are needed to parametrize the current. These spurious contributions depend on the momentum of the electron and are the consequence of the violation of cluster separability in the Bakamjian-Thomas framework. Furthermore, both the physical and the spurious form factors (associated with the spurious contributions) depend not only on Mandelstam $\mathrm t$, i.e. the four-momentum transfer squared, but also exhibit an additional dependence on Mandelstam $\mathrm s$, i.e. the total invariant mass squared of the electron-bound-state system. These are the only independent Lorentz invariants that can be constructed from the incoming and outgoing bound-state and electron momenta.
 
In the present work we consider, in particular, spin-0 bound systems like the pion, and spin-1 bound systems like the $\rho$ meson and the deuteron.
The current we obtain for the case of a spin-0 bound state is conserved. It can be parametrized by 2 form factors, a physical and a spurious form factor. The structure of our current reveals an interesting correspondence between our approach and the covariant light-front approach~\cite{Karmanov:1994ck,Carbonell:1998rj}. Their current also contains a spurious contribution which is, in their case, associated with an arbitrary light-like four-vector $\omega$ that describes the orientation of the light front. The dependence on this light-front orientation is a consequence of retaining explicit Lorentz covariance in the light-front approach. 

We observe that the $\mathrm s$-dependence of our physical form factor vanishes rather quickly with increasing $\mathrm s$, which suggests taking the limit $\mathrm s\rightarrow\infty$ to obtain a sensible physical form factor that depends only on Mandelstam $\mathrm t$. At the same time the spurious form factor is seen to vanish, implying that taking the limit $\mathrm s\rightarrow\infty$ removes all cluster-separability-violating effects. The current in the limit is a one-body current. The limit of letting the bound-state momentum go to infinity can be understood as considering only the subprocess where the photon is absorbed or emitted by the bound state in the infinite-momentum frame of the bound state. Therefore, it is not too surprising that in this limit we have been able to prove that our analytical formula for the form factor coincides with the standard front-form expression that is extracted from the plus component of a spectator current in the $q^+= 0$ frame~\cite{Chung:1988mu}. With regard to the similarities of our approach and the covariant light-front approach this is quite obvious: in the standard-light front dynamics, with the fixed light front vector $\omega=(1,0,0,-1)$, the plus component of the current does not contain any spurious contributions (since $\omega^+=0$).

As in the scalar case, the Lorentz structure of our spin-1 bound-state current (which contains 3 physical and 8 spurious contributions) resembles the corresponding current of the covariant light-front dynamics if the sum of the incoming and outgoing electron momenta is identified with the light-front vector $\omega$. 
However, unlike the spin-0 case, by taking the limit $\mathrm s\rightarrow\infty$, only 4 spurious contributions proportional to $\omega$ are removed. The remaining 4 contributions are responsible for the violation of current conservation and for the violation of the angular condition. This again resembles the situation of covariant light-front dynamics after fixing $\omega$ to the standard value $\omega=(1,0,0,-1)$. The plus component of the spin-1 spectator current (from which the form factors are usually extracted) then still contains 4 spurious contributions leading to the violation of the angular condition. There is, however, an unambiguous prescription proposed by Karmanov and Smirnov~\cite{Karmanov:1994ck} for how to separate the physical from the unphysical contributions, which can be directly applied to our case.

For mesons, we model the confining interaction between the quark and the antiquark by a simple harmonic-oscillator potential. Our numerical results for the $\rho$-meson magnetic dipole moment and electric quadrupole moment are, as expected, in agreement with the results obtained from the covariant light-front prescription (using the same model). The results for the magnetic dipole moment obtained from the standard front-form prescription (using the same model) differ from our results exactly by the value of the spurious contribution that is contained in the magnetic form factor obtained from the standard, non-covariant light-front prescription.

For the deuteron we propose a simple Walecka-type model based on a Lagrangean density for a field theoretical description of nucleons that interact via scalar $\sigma$- and vector $\omega$-exchange. This model is intended to serve as a common starting point for a benchmark calculation on \lq relativistic effects in few-body physics'. The aim is to identify and compare such effects in different relativistic approaches to few-body systems by means of the deuteron structure. We have fixed the free parameters of the model such that the experimental values for the binding energy of the deuteron and the triplet scattering length are reasonably well reproduced in the non-relativistic limit. In order to go beyond the non-relativistic limit we have included Pauli-Villars regulators. The simplest quantity we can derive is the so-called \textit{body form factor}, which is just the Fourier transform of the charge distribution. All calculations within this thesis use the static approximation to this Walecka-type one-boson exchange nucleon-nucleon interaction. As a consequence the deuteron is a pure s-wave. Nevertheless it is possible to study several relativistic effects. The first is the change of the body form factor that occurs if the non-relativistic kinetic energy is replaced by the corresponding relativistic expression. This affects the wave function and in the sequel the body form factor. As a next step we have plugged this \lq\lq relativized'' deuteron wave function into our point-form expression for the deuteron current and the form factors. Here Wigner rotations and further relativistic kinematical factors come into play. Due to the covariance of the deuteron current not only one, but 3 form factors are found. The body form factor then corresponds to the electric monopole form factor $G_\mathrm C$. We observe that cluster-separability violating effects are less important in the weakly bound deuteron than in the strongly bound $\rho$ meson.  

To summarize, we have presented a relativistic formalism which makes it possible to derive the electromagnetic currents and form factors of bound few-body systems consistent with the binding forces. One advantage of this approach is that the bound-state current is uniquely determined by the interaction dynamics which is responsible for the binding. Furthermore, the current can directly be extracted from the one-photon-exchange optical potential. However, in using a Bakamjian-Thomas approach, problems involving cluster separability arise which manifest themselves by unphysical contributions to the current. Nevertheless, we have found an unambiguous procedure for how to separate them from the physical contributions to get meaningful results for the form factors.
\section{Outlook}
Our coupled-channel formalism for calculating form factors is quite general and can be applied immediately to other few-body systems. In Refs.~\cite{Rocha:2010uu,Rocha:2010wm} for example, the present formalism has been applied to calculate  electromagnetic and weak form factors of heavy-light systems. Therein, a simple analytical expression for the Isgur-Wise function~\cite{Isgur:1989vq,Isgur:1989ed} has been found in the heavy-quark limit.

The above formalism can also be generalized in different directions. A next step to full relativity would be to account for dynamical particle-exchange interactions by adding additional channels. In particular, for the Walecka-type model one could make the $\sigma$- and $\omega$-exchange dynamical. This would add meson-exchange contributions to the spectator current that results from instantaneous binding forces. Corresponding extensions of the point-form approach are under investigation. In general, the construction of such many-body currents is a highly nontrivial task~\cite{Gross:1987bu}. It remains to be seen how large the effects of such additional exchange-current contributions on the form factors are. Furthermore, due to the generality of our approach, applications to bound systems with half-integer spins such as baryons or other nuclei seem to be straightforward.

Altogether, the approach of this thesis offers the possibility to be further refined and generalized and to be applied to various problems of relativistic few-body physics.

\begin{appendix}

\chapter{Minkowski Space}\label{app:A}
\section{Conventions and Basic Relations}
\label{app:Minkowskispace}
We define a contravariant four-vector by 
\begin{equation}
 x^\mu:=(x^0,x^1,x^2,x^3)=(x^0,\boldsymbol x)\,.
\end{equation}
The corresponding covariant vector is defined by 
\begin{equation}
 x_\mu:=(x_0,x_1,x_2,x_3)=(x^0,-\boldsymbol x)=\mathrm g_{\mu\nu}x^\nu
\end{equation}
with the metric tensor
\begin{equation}
 \mathrm {diag}\,(\mathrm g_{\mu\nu})=(1,-1,-1,-1)\,.
\end{equation}
Throughout the present thesis we use Einstein's sum convention meaning that a sum over repeated indices is always assumed. Greek indices like $\mu,\nu,\tau,\ldots$ on four-vectors $\{x^\mu\}\equiv x$ vary from 0 to 3 and Roman indices like $i,j,k,\ldots$ on three-vectors $\{x^i\}\equiv\boldsymbol x$ vary from 1 to 3.

The invariant scalar product between 2 four-vectors $x^\mu$ and $y^\mu$ is defined by
 \begin{equation}
 x\cdot y:=x^\mu \mathrm g_{\mu\nu} y^\nu=x^0y^0-\boldsymbol x\cdot\boldsymbol y\,.
\end{equation}
This scalar product is invariant under the Lorentz transformation
 \begin{equation}
 x^\mu\stackrel{\varLambda}{\longrightarrow}x'^\mu=\varLambda^\mu_{\,\,\nu}x^\nu
\end{equation}
if the matrices $\varLambda^\mu_{\,\,\nu}$ satisfy a generalized orthogonality relation
\begin{equation}\label{eq:lambdaortho}
 \varLambda^\mu_{\,\,\nu} \varLambda^\sigma_{\,\,\tau} \mathrm g^{\nu\tau}= \mathrm g^{\mu\sigma}.
\end{equation}
This relation can be rewritten as
\begin{equation}\label{eq:lambdaortho2}
 \varLambda^\mu_{\,\,\nu} \varLambda_\sigma^{\,\,\nu}= \mathrm g^{\mu}_{\sigma}\equiv\delta^{\mu}_{\sigma}
\end{equation}
which implies that 
\begin{equation}
\varLambda_\sigma^{\,\,\nu} =(\varLambda^{-1})^\nu_{\,\,\sigma}\,.
\end{equation}

\section[Matrix Representation of the Poincar\'e Group]{Matrix Representation of the\\ Poincar\'e Group}
\label{app:MinkowskispaceMatrixrepPmu}
In order to find a matrix representation for the inhomogeneous part of the Poincar\'e group, we go over to homogeneous coordinates defined by the five-component vectors~\cite{Scheck:2001}
\begin{eqnarray}
 \xi^\eta:=\left(\xi^4x^0,\xi^4x^1,\xi^4x^2,\xi^4x^3,\xi^4\right)\,.
\end{eqnarray}
Then a Poincar\'e group element is given by the $(5\times5)$-matrix
\begin{eqnarray}
\varPi\left(\varLambda,a\right):=
\left(%
\begin{array}{cc}
  \varLambda^\mu_{\,\,\nu}&a^\mu\\
  0^{\mathrm T\nu}&1
\end{array}
\right)
\end{eqnarray}
and a Poincar\'e transformation can be written in the homogeneous form
\begin{eqnarray}
\xi'^\eta= \varPi^\eta_{\,\,\theta}\xi^\theta \quad \text{with}\quad \eta,\theta=0,\ldots,4\,.
\end{eqnarray}
Consequently, a Lorentz transformation and a space-time translation are given by 
\begin{eqnarray}
\varPi\left(\varLambda,0\right):=
\left(%
\begin{array}{cc}
  \varLambda^\mu_{\,\nu}&0^\mu\\
  0^{\mathrm T\nu}&1\\
\end{array}
\right)\quad\text{and}\quad T\left(a\right):=\varPi\left(1_{4\times 4},a\right)\,,
\end{eqnarray}
respectively.
Writing $T\left(a\right)$ in exponential form as $
T\left(a\right)=\exp\left(-\mathrm i\,a_\mu P^\mu\right)
$
the generators $P^\mu$ for infinitesimal space-time translations, being $(5\times5)$-matrices, are defined by
 \begin{eqnarray}
P^\mu:=\left.\mathrm i\,\mathrm g^{\mu\nu}\frac{\partial T\left(a\right)}{\partial a^\nu}\right\vert_{a^\nu=0 }\,.
\end{eqnarray}

\chapter[Spinors and Polarization Vectors]{Spinors and \\Polarization Vectors}\label{app:B}
\section{Spin-1/2 Dirac Particles}\label{app:Diracrepresentation}
\subsection{Dirac Representation} \label{app:standardrepresentation}
We use the standard (Dirac) representation of the Dirac matrices given by
\begin{eqnarray}&&\label{eq:standardrepr}
\gamma^0=\left(%
 \begin{array}{cc}
  1_2&0 \\
   0 &-1_2
 \end{array}
 \right)\,,\quad\boldsymbol{\gamma}=\left(%
 \begin{array}{cc}
  0&\boldsymbol{\sigma} \\
   -\boldsymbol{\sigma} &0
 \end{array}
 \right)=\gamma^0 \boldsymbol{\alpha}\,
\end{eqnarray}
and 
\begin{eqnarray}
\gamma^5=\left(%
 \begin{array}{cc}
  0&1_2 \\
   1_2 &0
 \end{array}
 \right)\,.
\end{eqnarray}
Here
\begin{eqnarray}
\sigma_1=-\sigma^1=\left(%
 \begin{array}{cc}
  0&1 \\
   1 &0
 \end{array}
 \right)\,,\quad \sigma_2=-\sigma^2=\left(%
 \begin{array}{cc}
  0&-\mathrm{i} \\
    \mathrm{i}&0
 \end{array}
 \right)\, \nonumber
\end{eqnarray}
and
\begin{eqnarray}
\sigma_3=-\sigma^3=\left(%
 \begin{array}{cc}
  1&0 \\
   0 &-1
 \end{array}
 \right)
\end{eqnarray}
are the usual Pauli matrices. A useful relation for the Pauli matrices is 
\begin{eqnarray}
\label{eq:paulimatricesproduct}
\left(\boldsymbol{\sigma}\cdot\boldsymbol{a}\right)
\left(\boldsymbol{\sigma}\cdot\boldsymbol{b}
\right)=1_2\,\boldsymbol{a}\cdot\boldsymbol{b}+\mathrm i\,\boldsymbol{\sigma}\cdot\left(\boldsymbol{a}\times
\boldsymbol{b}\right)\,.
\end{eqnarray}
Using this standard representation a canonical boost in the $4\times4$ matrix representation of the SL$(2,\mathbb C$), Eq.~(\ref{eq:boost4dimsl2c}), has the form 
\begin{eqnarray}&&
S[\underline B_{\mathrm c}(\boldsymbol v)]=
\sqrt{\frac{p^0+m}{2m}}
\left(%
 \begin{array}{cc}
  1_2&\frac{\boldsymbol{\sigma}\cdot\boldsymbol{p}}{p^0+m} \\
   \frac{\boldsymbol{\sigma}\cdot\boldsymbol{p}}{p^0+m} &1_2
 \end{array}
 \right)\quad\text{with}\quad v:=\frac{p}{m}\,.%
\end{eqnarray}
The Dirac spinors for an (anti)particle at rest are given by
\begin{eqnarray}
u_{\sigma}(\boldsymbol 0):=\sqrt{2m}\left(%
 \begin{array}{c}
   \varsigma_{\sigma} \\
   0
 \end{array}
 \right)\quad\text{and}\quad
v_{\sigma}(\boldsymbol 0):=-\sqrt{2m}\left(%
 \begin{array}{c}
   0 \\
  \varepsilon \varsigma_{\sigma}
 \end{array}
 \right)\,,\quad \sigma=\pm\frac12\,,\nonumber\\
\end{eqnarray}
where $\varsigma_{\sigma}$ is the two-component Pauli spinor
\begin{eqnarray}\label{eq:metricPaulispinors}
 \varsigma_{\sigma}=\left(%
 \begin{array}{c}
   \frac12+\sigma \\
   \frac12-\sigma
 \end{array}
 \right)%
 \quad \text{and} \quad
\varepsilon=\mathrm i\,\sigma_2\,.
  \end{eqnarray}
Here $\sigma$ denotes the canonical spin projection on the 3-axis.
The Pauli spinors are normalized 
according to $\varsigma_{\sigma'}^\dag\varsigma_{\sigma}=\delta_{\sigma'\sigma}$.
The spinors for an (anti)particle with momentum $p=mv$ are given by 
 \begin{eqnarray}\label{eq:Diracspinors1}
&&u_{\sigma}(\boldsymbol p)_{\alpha}=S_{\alpha\beta}\left[\underline B_{\mathrm c}\left(\boldsymbol v\right)\right] u_{\sigma}(\boldsymbol{0})_{\beta}=\sqrt{p^0+m}\left(%
 \begin{array}{c}
   \varsigma_{\sigma} \\
   \frac{\boldsymbol{\sigma}\cdot\boldsymbol{p}}{p^0+m}\varsigma_{\sigma}
 \end{array}
 \right)_\beta\,,\\\label{eq:Diracspinors2}
&&v_{\sigma}(\boldsymbol p)_{\alpha}=S_{\alpha\beta}\left[\underline B_{\mathrm c}\left(\boldsymbol v\right)\right] v_{\sigma}(\boldsymbol{0})_{\beta}=-\sqrt{p^0+m}\left(%
 \begin{array}{c}
   \frac{\boldsymbol{\sigma}\cdot\boldsymbol{p}}{p^0+m}\varepsilon\varsigma_{\sigma} \\
   \varepsilon\varsigma_{\sigma}
 \end{array}
 \right)_\beta\,,\nonumber\\
\end{eqnarray}
with $\alpha,\beta =1,\ldots,4$. We use the Greek letters $\alpha$ and $\beta$ to denote Dirac indices.
For equal particle and antiparticle masses the spinors $u_{\sigma}\left(\boldsymbol p\right)$ and $v_{\sigma}\left(\boldsymbol p\right)$  are related by~\cite{Bjorken:1964,Nachtmann:1986}
\begin{eqnarray}&&
v_{\sigma}\left(\boldsymbol p\right)=C\bar{u}_{\sigma}^{\mathrm{T}}\left(\boldsymbol  p\right)\quad\text{and}\quad u_{\sigma}\left(\boldsymbol p\right)=C \bar{v}_{\sigma}^{\mathrm{T}}\left(\boldsymbol p\right)\,,
\end{eqnarray} where
$C:=i\gamma^2\gamma^0=-C^{-1}=-C^{\dag}=-C^{\mathrm{T}}$ is the charge conjugation operator. With $C\gamma^{\mu\,\mathrm{T}}C^{-1}=-\gamma^{\mu}$ it follows immediately that
\begin{eqnarray}\label{eq:ugammauevgammav}
\bar{v}_{\sigma}\left(\boldsymbol p\right)\gamma^{\mu} v_{\sigma'}\left(\boldsymbol p'\right)=\left[\bar{u}_{\sigma'}\left(\boldsymbol p' \right)\gamma^{\mu} u_{\sigma}\left(\boldsymbol p\right)\right]^{\mathrm{T}}=\bar{u}_{\sigma'}\left(\boldsymbol p'\right)\gamma^{\mu}u_{\sigma}\left(\boldsymbol p \right)
\end{eqnarray}
since $\bar{u}_{\sigma'}\left(\boldsymbol p'\right)\gamma^{\mu}u_{\sigma}\left(\boldsymbol p \right)$ is a scalar quantity.
\subsection{Covariance Properties}
\label{app:covarianceproperties}
Under a Lorentz transformation $\varLambda$ the spinor $u_{\sigma}(\boldsymbol p)$ and its adjoint $\bar u_{\sigma}(\boldsymbol p)$ transform according to~\cite{Keister:1991sb}
\begin{eqnarray}\label{eq:spinortransformationprop}&&
u_{\sigma}(\boldsymbol p)\stackrel{\varLambda}{\longrightarrow}u_{\bar{\sigma}}( \boldsymbol \varLambda p)=D_{\sigma\bar \sigma}^{\frac12}\left(\underline{R}_\mathrm{W_{\!c}}^{-1}(v,\varLambda)\right)S(\underline \varLambda)u_{\sigma}(\boldsymbol p)\,,\nonumber\\&&
\bar u_{\sigma}(\boldsymbol p)\stackrel{\varLambda}{\longrightarrow}\bar u_{\bar{\sigma}}( \boldsymbol \varLambda p)=\bar u_{\sigma}(\boldsymbol p) S^{-1}(\underline \varLambda)D_{\sigma\bar\sigma}^{\frac12\ast}\left(\underline{R}_\mathrm{W_{\!c}}^{-1}(v,\varLambda)\right)\,,\nonumber\\
\end{eqnarray}
with $\sigma,\bar \sigma=\pm\frac12$. Here we have used
\begin{eqnarray}\label{eq:Sgamm0}
 S^\dag(\underline \varLambda)\gamma^0=\gamma^0S^{-1}(\underline \varLambda)\,.
\end{eqnarray}
The aim of this section is to find the transformation properties of the current of an extended massive spin-1/2 particle such as a nucleon. Such a current is given by
\begin{eqnarray}\label{eq:nuclcurrent2}\lefteqn{
 J^{\mu}(\boldsymbol p^\prime,\sigma^\prime;\boldsymbol p,\sigma)}\nonumber\\&&=
|\, \mathrm e\,|\,
\bar{u}_{\sigma'}(\boldsymbol p^\prime)\left[F_1((p-p')^2) \gamma^\mu+F_2((p-p')^2)\frac{\mathrm i (p'-p)_\nu \sigma^{\mu\nu}}{2m}\right]
u_{\sigma}(\boldsymbol p)\nonumber\\
\end{eqnarray} 
where $\sigma^{\mu\nu}=\frac{\mathrm i}{2} (\gamma^\mu\gamma^\nu-\gamma^\nu\gamma^\mu)$ and $\mathrm e$ is the elementary charge. In our notation the anomalous magnetic moment $\kappa$ of the spin-1/2 constituent (in units of the Bohr magneton) is absorbed into $F_2$ such that $F_2(0)=\kappa$. For point-like particles, such as quarks (electrons), the current simplifies by setting the Dirac form factor $F_1((p-p')^2)=Q_{\mathrm q (\mathrm e)}$ and the Pauli form factor $F_2((p-p')^2)=0$, where $Q_{\mathrm q (\mathrm e)}$ is the electric charge of the corresponding quark (electron).
For the first term in Eq.~(\ref{eq:nuclcurrent2}) we find, using the transformation properties of the spinors, Eq.~(\ref{eq:spinortransformationprop}), together with 
\begin{eqnarray}
\label{eq:covgammamatr}
 S^{-1}(\underline \varLambda)\gamma^{\mu}S(\underline \varLambda)=\varLambda_{\;\nu}^{\mu}\gamma^{\nu}\,,
\end{eqnarray}
 the following transformation properties under a Lorentz transformation:
\begin{eqnarray}\lefteqn{
\bar u_{\sigma'}(\boldsymbol p')\gamma^{\mu}u_{\sigma}(\boldsymbol p)\stackrel{\varLambda}{\longrightarrow}
\bar u_{\bar \sigma'}(\boldsymbol \varLambda p')\gamma^{\mu}u_{\bar \sigma}(\boldsymbol \varLambda p)}\nonumber\\&=&\varLambda^{\mu}_{\;\nu}\bar u_{\sigma'}(\boldsymbol p')\gamma^{\nu}u_{\sigma}(\boldsymbol p)D_{\sigma'\bar\sigma'}^{\frac12\ast}\left(\underline R_\mathrm{W_{\!c}}^{-1}(v',\varLambda)\right)
D_{\sigma\bar\sigma}^{\frac12}\left(\underline R_\mathrm{W_{\!c}}^{-1}(v,\varLambda)\right)\,.
\end{eqnarray}
For the $\sigma^{\mu\nu}$-part we use the transformation properties of the Dirac matrices $S^{-1}(\underline \varLambda)\gamma^\mu\gamma^\nu S(\underline \varLambda)=
\varLambda^\mu_{\;\rho}\gamma^\rho \varLambda^\nu_{\;\sigma} \gamma^\sigma$ to find
\begin{eqnarray}\lefteqn{
\bar u_{\sigma'}(\boldsymbol p')\sigma^{\mu\nu}u_{\sigma}(\boldsymbol p)\stackrel{\varLambda}{\longrightarrow}
 \bar u_{\bar\sigma'}(\boldsymbol \varLambda p')\sigma^{\mu\nu} u_{\bar \sigma}(\boldsymbol \varLambda p)}\nonumber\\&&= \varLambda^\mu_{\;\rho}\varLambda^\nu_{\;\tau}\bar u_{\sigma'}( \boldsymbol p')\sigma^{\rho\tau} u_{\sigma}(\boldsymbol p)D_{\sigma'\bar\sigma'}^{\frac12\ast}\left(\underline R_\mathrm{W_{\!c}}^{-1}(v',\varLambda)\right)
D_{\sigma\bar\sigma}^{\frac12}\left(\underline R_\mathrm{W_{\!c}}^{-1}(v,\varLambda)\right)\,.\nonumber\\
\end{eqnarray}
The whole term proportional to $\sigma^{\mu\nu}$ transforms in the same way as the $\gamma^\mu$-part:
\begin{eqnarray}\lefteqn{
 \bar u_{\bar\sigma'}(\boldsymbol \varLambda p')\frac{\mathrm i(\varLambda p'-\varLambda p)_\nu \sigma^{\mu\nu}} {2m}u_{\bar\sigma}(\boldsymbol \varLambda p)}\nonumber\\&=&
\varLambda^\mu_{\,\,\rho}\varLambda^\nu_{\,\,\tau}\bar u_{\sigma'}( \boldsymbol p'))\frac{\mathrm i (\varLambda^{-1})^\lambda_{\,\,\nu} ( p'- p)_\lambda  \sigma^{\rho\tau}}{2m} u_{\sigma}( \boldsymbol p)
\nonumber\\&&\times D_{\sigma'\bar\sigma'}^{\frac12\ast}\left(\underline R_\mathrm{W_{\!c}}^{-1}(v',\varLambda)\right)
D_{\sigma\bar\sigma}^{\frac12}\left(\underline R_\mathrm{W_{\!c}}^{-1}(v,\varLambda)\right)
\nonumber\\&=&
 \varLambda^\mu_{\,\,\rho}\bar u_{\sigma'}(\boldsymbol p'))\frac{\mathrm i ( p'- p)_\lambda  \sigma^{\rho\lambda}}{2m} u_{\sigma}(\boldsymbol p)\nonumber\\&&\times 
D_{\sigma'\bar\sigma'}^{\frac12\ast}\left(\underline R_\mathrm{W_{\!c}}^{-1}(v',\varLambda)\right)
D_{\sigma\bar\sigma}^{\frac12}\left(\underline R_\mathrm{W_{\!c}}^{-1}(v,\varLambda)\right)\,,
\end{eqnarray}
where we have used (by rewriting Eq.~(\ref{eq:lambdaortho}))
\begin{eqnarray}&&\label{eq:inverseLambda}
 \varLambda^\mu_{\,\,\nu} \varLambda_\sigma^{\,\,\nu} = \mathrm g^{\mu}_{\sigma}=\delta^{\mu}_{\sigma}\quad \Rightarrow\quad \varLambda_\sigma^{\,\,\nu}=(\varLambda^{-1})_{\,\,\sigma}^{\nu}\,.
\end{eqnarray}
With these findings we are finally able to write down the transformation properties of a spin-1/2 current under Lorentz transformations:
\begin{eqnarray}\label{eq:transformpropconstcurrent}
 J^{\mu}(\boldsymbol p^\prime,\sigma^\prime;\boldsymbol p,\sigma)&\stackrel{\varLambda}{\longrightarrow}&
J^{\mu}(\boldsymbol \varLambda p^\prime,\bar \sigma^\prime;\boldsymbol \varLambda p,\bar \sigma)\nonumber\\&=&
\varLambda^\mu_{\,\,\nu}J^{\nu}(\boldsymbol p^\prime,\sigma^\prime;\boldsymbol p, \sigma)D_{\sigma'\bar\sigma'}^{\frac12\ast}\left(\underline R_\mathrm{W_{\!c}}^{-1}(v',\varLambda)\right)
D_{\sigma\bar\sigma}^{\frac12}\left(\underline R_\mathrm{W_{\!c}}^{-1}(v,\varLambda)\right)\,.\nonumber\\
\end{eqnarray}
\subsection{Hermiticity}
With the help of the relations
\begin{eqnarray}
\gamma^0\gamma^{\mu\dag}\gamma^0=\gamma^\mu\quad\text{and}\quad \gamma^0\sigma^{\mu\nu\dag}\gamma^0 =\sigma^{\mu\nu}
\end{eqnarray}
Hermiticity of the spin-1/2 current~(\ref{eq:nuclcurrent2}) is easily proved.
To this end we look at the Hermitean conjugate of the current matrix:
\begin{eqnarray}\label{eq:nuclcurrenthermiticity}
\lefteqn{\left[J^{\mu}(\boldsymbol p^\prime,\sigma^\prime;\boldsymbol p,\sigma)\right]^\dag}\nonumber\\&=&\left[J^{\mu\ast}(\boldsymbol p^\prime,\sigma^\prime;\boldsymbol p,\sigma)\right]^\mathrm T\nonumber\\&=&|\,\mathrm e\,|\,
u_{\sigma}^\dag(\boldsymbol p)\left[F_1((p-p')^2) \gamma^{\mu\dag}-F_2((p-p')^2)\frac{\mathrm i (p'-p)_\nu \sigma^{\mu\nu\dag}}{2m}\right]
\gamma^0u_{\sigma'}(\boldsymbol p^\prime)\nonumber\\&=&
J^{\mu}(\boldsymbol p,\sigma;\boldsymbol p^\prime,\sigma^\prime)\nonumber\\&=&J^{\mu\ast}(\boldsymbol p^\prime,\sigma^\prime;\boldsymbol p,\sigma)\,,
\end{eqnarray} 
where we have used in the last step $J^{\mu\ast}(\boldsymbol p^\prime,\sigma^\prime;\boldsymbol p,\sigma) =\left[J^{\mu\ast}(\boldsymbol p^\prime,\sigma^\prime;\boldsymbol p,\sigma)\right]^\mathrm T$ since it is a scalar quantity.
\section{Massless Spin-1 Particles}\label{app:spin-1polarizationvectors}
 Appropriately orthonormalized photon polarization vectors $\epsilon_{\lambda}^\mu(\boldsymbol p)$, which are most convenient for our purpose are given by the components of the helicity boost matrix 
$\epsilon_{\lambda}^\mu(\boldsymbol p):=B_\mathrm h(\boldsymbol p)^\mu_{\;\lambda}$~\cite{Klink:2000pq}. They satisfy the usual completeness relation
\begin{equation}
 \label{eq:polcomp}
\sum_{\lambda=0}^3
\epsilon^{\mu}_{\lambda}(\boldsymbol {p})(-\mathrm g^{\lambda\lambda})\, \epsilon^{\ast \nu}_{\lambda}(\boldsymbol{p})
= - \mathrm g^{\mu \nu}\, .
\end{equation}
These photon polarization vectors transform under a Lorentz transformation according to
\begin{eqnarray}\label{eq:phpolarvectortransfprop}
&&\epsilon_\lambda^\mu(\boldsymbol p)\stackrel{\varLambda}{\longrightarrow}\epsilon^\mu_{\bar{\lambda}}(\boldsymbol \varLambda p)=
 R_{\mathrm{W_h}}^{-1}(p,\varLambda)_{\lambda\bar\lambda}\varLambda^\mu_{\;\nu}\epsilon_\lambda^\nu(\boldsymbol p)\,.
\end{eqnarray}

\section{Massive Spin-1 Particles}
\subsection{Polarization Vectors}
\label{app:massivespin1polarizvecs}
The polarization vectors $\epsilon^{\nu}_{\sigma}(\boldsymbol0)$  of a spin-1 particle at rest with $\sigma=\pm 1,0$ are usually defined by~\cite{Aitchison:2003,Scadron}
\begin{eqnarray}
\epsilon^{\nu}_{1}(\boldsymbol 0)&:=&-\frac{1}{\sqrt{2}}(0,1,\mathrm i,0)^{\nu}\,,\\
\epsilon^{\nu}_{-1}(\boldsymbol 0)&:=&\frac{1}{\sqrt{2}}(0,1,-\mathrm i,0)^{\nu}\,,\\
\epsilon^{\nu}_{0}(\boldsymbol 0)&:=&(0,0,0,1)^{\nu}\,.
                                           \end{eqnarray}
The boosted polarization vectors are obtained from the rest-frame vectors by a canonical boost with velocity $\boldsymbol v:=
\boldsymbol p/m$:
 \begin{eqnarray}
\epsilon^{\mu}_{\sigma}(\boldsymbol p)=B_\mathrm c(\boldsymbol v)^\mu_{\,\,\nu}\epsilon^{\nu}_{\sigma}(\boldsymbol 0)\,.\end{eqnarray}
The transversality property then reads
\begin{eqnarray} \label{eq:transversalityprop}      
\epsilon_{\sigma}(\boldsymbol p)\cdot  p=0\,.
      \end{eqnarray} 
The transformation properties of the polarization vectors under Lorentz transformations are given by
\begin{eqnarray}
\epsilon^\mu_{\sigma}(\boldsymbol p)\stackrel{\varLambda}{\longrightarrow}\epsilon^\mu_{\bar{\sigma}}(\bar{\boldsymbol  p})=D_{\sigma\bar\sigma}^{1}\left(\underline R_\mathrm{W_{\!c}}^{-1}(v,\varLambda)\right)\varLambda^\mu_{\,\,\nu}\epsilon^\nu_{\sigma}(\boldsymbol p)\,,
\end{eqnarray}
with $\sigma,\bar \sigma=\pm 1,0$.

Polarization vectors of spin-1 particles which are described by velocity states $\vert V;\ldots;\boldsymbol k_j,\mu_j;\ldots\rangle$ transform under a Lorentz transformation with a Wigner rotation according to
\begin{eqnarray}
\epsilon^\mu_{\mu_j}(\boldsymbol k_j)\stackrel{\varLambda}{\longrightarrow}\epsilon^\mu_{\bar{\mu}_j}(\boldsymbol R_\mathrm{W_{\!c}}(V,\varLambda) k_j)=D_{\mu_j\bar\mu_j}^{1}\left(\underline R_\mathrm{W_{\!c}}^{-1}(V,\varLambda)\right)R_\mathrm{W_{\!c}}(V,\varLambda)^\mu_{\,\,\nu}\epsilon^\nu_{\mu_j}(\boldsymbol k_j)\,.\nonumber\\
\end{eqnarray}
The relation between $\epsilon^\mu_{\sigma_j}(\boldsymbol p_j=\boldsymbol B_\mathrm c(\boldsymbol V)k_j)$ and $\epsilon^\mu_{\mu_j}(\boldsymbol k_j)$ is then given by
\begin{eqnarray}&&\label{eq:polphysunphy}
 \epsilon^\mu_{\mu_j}(\boldsymbol k_j)D_{\mu_j\sigma_j}^{1}\left(\underline R_\mathrm{W_{\!c}}^{-1}(w_j,B_\mathrm c(\boldsymbol V))\right)= B_\mathrm c^{-1}(\boldsymbol V)^\mu_{\,\,\nu}\epsilon^\nu_{\sigma_j}(\boldsymbol p_j)
\end{eqnarray}
with $w_j:=k_j/m_j$.
\subsection{Clebsch-Gordan Coefficients}
The Clebsch-Gordan coefficient $C_{\frac12\tilde\mu_1\frac12\tilde\mu_2}^{1\mu_j}$ of Eq.~(\ref{eq:TrafopropCG}) can be written in terms of the polarization vectors as~\cite{Buck:1979ff}
\begin{eqnarray}\label{eq:CG}
   C_{\frac12\tilde\mu_1\frac12\tilde\mu_2}^{1\mu_j}=\left(\sigma_\mu\epsilon_{\mu_j}^\mu(\boldsymbol 0)\frac{\mathrm i \,\sigma_2}{\sqrt 2}\right)_{\tilde\mu_1\tilde\mu_2}
  \end{eqnarray}
with 
\begin{eqnarray}
  \sigma_\mu\epsilon^{\mu}_{\mu_j}(\boldsymbol 0)=\boldsymbol \sigma\cdot \boldsymbol\epsilon_{\mu_j}(\boldsymbol 0)\,.
  \end{eqnarray}
Explicitly, they are given in matrix form by
\begin{eqnarray}
 C_{\frac12\tilde\mu_1\frac12\tilde\mu_2}^{11}&=&\delta_{\frac12\tilde\mu_1}\delta_{\frac12\tilde\mu_2}=
\left(
\begin{array}{ll}
1 & 0\\
0 & 0
\end{array}
\right)_{\tilde\mu_1\tilde\mu_2}\,,\\
C_{\frac12\tilde\mu_1\frac12\tilde\mu_2}^{1-1}&=&\delta_{-\frac12\tilde\mu_1}\delta_{-\frac12\tilde\mu_2}=
\left(
\begin{array}{ll}
0 & 0\\
0 & 1
\end{array}
\right)_{\tilde\mu_1\tilde\mu_2}\,,\\
C_{\frac12\tilde\mu_1\frac12\tilde\mu_2}^{10}&=&\frac{1}{\sqrt{2}}\delta_{\tilde\mu_1-\tilde\mu_2}=
\frac{1}{\sqrt{2}}\left(
\begin{array}{ll}
0 & 1\\
1 & 0
\end{array}
\right)_{\tilde\mu_1\tilde\mu_2}\,.
  \end{eqnarray}
Since the Clebsch-Gordan coefficients are real we can write 
 \begin{eqnarray}\label{eq:CGprimedast}
  C_{\frac12\tilde\mu_1'\frac12\tilde\mu'_2}^{1\mu_j'}=C_{\frac12\tilde\mu_1'\frac12\tilde\mu'_2}^{1\mu_j'\ast} =-
\left(\sigma^\ast_\mu\epsilon^{\ast\mu}_{\mu_j'}(\boldsymbol 0)\frac{\mathrm i\,\sigma_2^\ast}{\sqrt 2}\right)_{\tilde\mu_1'\tilde\mu_2'}\,.
  \end{eqnarray}
Moreover, due to the symmetry of the Clebsch-Gordan coefficients under $\tilde\mu_1'\leftrightarrow\tilde\mu'_2$ exchange and the hermiticity of the Pauli matrices we obtain
\begin{eqnarray}\label{eq:CGprimeddag}
  C_{\frac12\tilde\mu_1'\frac12\tilde\mu'_2}^{1\mu_j'}&=&C_{\frac12\tilde\mu_1'\frac12\tilde\mu'_2}^{1\mu_j'\dag} \nonumber\\&=&
-\epsilon^{\ast\mu}_{\mu_j'}(\boldsymbol 0)\left(\frac{\mathrm i\,\sigma_2^\dag}{\sqrt 2}\sigma^\dag_\mu\right)_{\tilde\mu_2'\tilde\mu_1'}
\nonumber\\&=&
-\epsilon^{\ast\mu}_{\mu_j'}(\boldsymbol 0)\left(\frac{\mathrm i\,\sigma_2}{\sqrt 2}\sigma_\mu\right)_{\tilde\mu_2'\tilde\mu_1'}\,.\nonumber\\
\end{eqnarray}
Since the contraction $\sigma_\mu\epsilon_{\mu_j}^\mu(\boldsymbol 0)$ is Lorentz invariant we can write
\begin{eqnarray}\label{eq:edotsigma}
 \sigma_\mu\epsilon_{\mu_j}^\mu(\boldsymbol 0)&=&
\epsilon^{\mu}_{\mu_j}(\boldsymbol k_{\mathrm C})
\mathrm g_{\mu\nu} B_\mathrm c(\boldsymbol w_{\mathrm C})^\nu_{\,\,\rho}\sigma^\rho\nonumber\\&
=&\epsilon^{\mu}_{\mu_j}(\boldsymbol k_{\mathrm C})
\mathrm g_{\mu\nu} B_\mathrm c(\boldsymbol w_{\mathrm C})^\nu_{\,\,\rho} \mathrm g^{\rho\lambda} \sigma_\lambda\nonumber\\&
=&\epsilon^{\nu}_{\mu_j}(\boldsymbol k_{\mathrm C})
 B_\mathrm c(\boldsymbol w_{\mathrm C})_\nu^{\,\,\lambda} \sigma_\lambda\nonumber\\&
=&\epsilon^{\nu}_{\mu_j}(\boldsymbol k_{\mathrm C})B_\mathrm c(-\boldsymbol w_{\mathrm C})_{\,\,\nu}^{\lambda} \sigma_\lambda\,,\quad w_{\mathrm C}:=\frac{k_{\mathrm C}}{m_{\mathrm C}}\,,
  \end{eqnarray}
were we have used Eq.~(\ref{eq:inverseLambda}).

\chapter[Useful Relations in the SL$\left(2,\mathbb C\right)$ and SU$\left(2\right)$]{Useful Relations in the SL$\left(2,\mathbb C\right)$ and  SU$\left(2\right)$}\label{app:C}
\section{Canonical Boosts in the SL$\left(2,\mathbb C\right)$} \label{app:boostSL2C}
In Eq.~(\ref{eq:boostSL2C}) a Hermitean SL$\left(2,\mathbb C\right)$-matrix representation of canonical boosts is introduced by
 \begin{eqnarray}&&
\underline{B}_{\mathrm c}\left(\boldsymbol v\right) :=
\sqrt{\frac{v^0+1}{2}}\sigma^0+\frac{ \boldsymbol\sigma\cdot \boldsymbol v}{\sqrt{2\left(v^0+1 \right) }}\,.
\end{eqnarray}
One can prove by direct calculation that 
\begin{eqnarray}
\underline{B}_{\mathrm c}\left( \boldsymbol v\right)\underline{B}_{\mathrm c}\left( \boldsymbol v\right)= \sigma^0 v^0+\sigma_i v^i =\sigma^0 v^0+\boldsymbol \sigma\cdot\boldsymbol v=\sigma_\mu v^\mu\,.\end{eqnarray}
This can be equivalently written as \begin{eqnarray}
\underline{B}_{\mathrm c}\left( \boldsymbol v\right)\underline{B}_{\mathrm c}\left( \boldsymbol v\right)=\underline{B}_{\mathrm c}\left( \boldsymbol v\right)\sigma_\mu (1,\boldsymbol 0)^\mu \underline{B}_{\mathrm c}\left( \boldsymbol v\right)=\sigma_\mu B_{\mathrm c}(\boldsymbol v)^\mu_{\,\,\nu}(1,\boldsymbol 0)^\nu\,. 
\end{eqnarray}
Generalizing to arbitrary vectors $w$ gives
\begin{eqnarray}
\underline{B}_{\mathrm c}\left( \boldsymbol v\right)\sigma_\mu w^\mu \underline{B}_{\mathrm c}\left( \boldsymbol v\right)=\sigma_\mu B_{\mathrm c}(\boldsymbol v)^\mu_{\,\,\nu} w^\nu\,. 
\end{eqnarray}
\section{Useful Relations of Wigner $D$-Functions}\label{app:WignerDf}
The Wigner $D$-functions defined by Eq.~(\ref{eq:wignerDfuntion})
 satisfy the following useful relations:
\begin{eqnarray}\label{eq:RWprop1}
D^{j\ast}_{\sigma\sigma'}(\underline R )
&=& D^{j\dag}_{\sigma'\sigma}(\underline R )=D^{j}_{\sigma'\sigma}(\underline R^{-1} )\,;\\
\sum_{\sigma''}D^{j}_{\sigma\sigma''}(\underline R )D^{j}_{\sigma''\sigma'}(\underline R' )
&=&D^{j}_{\sigma\sigma'}(\underline R \,\underline R' )\,;\\
D^{j}_{\sigma\sigma'}(1_2 )&=&\delta_{\sigma\sigma'}\,;\label{eq:D-functdelta}\\
D^{j}_{\sigma\sigma'}(\underline R)&=&(-1)^{\sigma'-\sigma}D^{j\ast}_{-\sigma-\sigma'}(\underline R)\,.\label{eq:RWprop4}
\end{eqnarray}
For the spin-1/2 case we need the relation 
\begin{eqnarray}\label{eq:sigam2Dsigma2}
\sigma_2\,
D^{\frac12\ast}(\underline R)\,\sigma_2&=&\left(
\begin{array}{cc}
D^{\frac12\ast}_{-\frac12-\frac12}(\underline R) & -D^{\frac12\ast}_{-\frac12\frac12}(\underline R) \\
-D^{\frac12\ast}_{\frac12-\frac12}(\underline R) & D^{\frac12\ast}_{\frac12\frac12}(\underline R)
                                                                                                      \end{array}                                                                                                      
\right)\nonumber\\&=&\left(
\begin{array}{cc}
D^{\frac12}_{\frac12\frac12}(\underline R) & D^{\frac12}_{\frac12-\frac12}(\underline R) \\
D^{\frac12}_{-\frac12\frac12}(\underline R) & D^{\frac12}_{-\frac12-\frac12}(\underline R)
                                                                                                      \end{array}                                                                                                      
\right)\nonumber\\&=&D^\frac12(\underline R)\,,
\end{eqnarray}
where we have used Eq.~(\ref{eq:RWprop4}).
\section[Transformation Properties of the Clebsch-Gordan Coefficients]{Transformation Properties of the\\Clebsch-Gordan Coefficients}
\label{app:TrafopropCG}
In order to find the transformation properties of the Clebsch Gordan coefficients under Lorentz transformations (for $l=0$ bound states) we look at the transformed wave function of Eq.~(\ref{eq:transfwavefunction}):
 \begin{eqnarray}\label{eq:transfomredwf}
\lefteqn{\varPsi _{nj\bar \mu_j \bar \mu_1'\bar \mu_2'}(\bar{\tilde{\boldsymbol k}}')}\nonumber\\&=&
\varPsi _{nj\mu_j \mu_1'\mu_2'}(\tilde{\boldsymbol k}')  D_{\bar\mu_1'\mu_1'}^{j_1}\left[\underline R_\mathrm{W_{\!c}}(V,\varLambda)\right] D_{\bar\mu_2'\mu_2'}^{j_2}\left[\underline R_\mathrm{W_{\!c}}(V,\varLambda)\right]
 D_{\bar\mu_j\mu_j}^{j\ast}\left[\underline R_\mathrm{W_{\!c}}(V,\varLambda)\right]
\nonumber\\&=&
C^{j\mu_j}_{j_1\tilde \mu_1j_2\tilde \mu_2} u_{n0}(\tilde{k}') D_{\bar\mu_j\mu_j}^{j\ast}\left[\underline R_\mathrm{W_{\!c}}(V,\varLambda)\right]
\nonumber\\&&\times
 D^{j_1}_{\mu_1'\tilde \mu_1}\left[\underline R_\mathrm{W_{\!c}}\left(\tilde w_1',B_{\mathrm c}(\boldsymbol w_{12}')\right)\right]
D^{j_2}_{\mu_2'\tilde \mu_2}\left[\underline R_\mathrm{W_{\!c}}\left(\tilde w_2',B_{\mathrm c}(\boldsymbol w_{12}')\right)\right]
\nonumber\\&&\times D_{\bar\mu_1'\mu_1'}^{j_1}\left[\underline R_\mathrm{W_{\!c}}(V,\varLambda)\right] D_{\bar\mu_2'\mu_2'}^{j_2}\left[\underline R_\mathrm{W_{\!c}}(V,\varLambda)\right]\,.
 \end{eqnarray}
Here we have inserted Eq.~(\ref{eq:2bodywf}) for the wave function $\varPsi _{nj\mu_j \mu_1'\mu_2'}(\tilde{\boldsymbol k}')$.
On the other hand, from Eq.~(\ref{eq:2bodywf}), we have 
\begin{eqnarray}\label{eq:transfomredwf2}
\lefteqn{\varPsi _{nj\bar \mu_j \bar \mu_1'\bar \mu_2'}(\bar{\tilde{\boldsymbol k}}')}\nonumber\\&=&
C^{j\bar \mu_j}_{j_1\bar {\tilde \mu}_1j_2\bar{\tilde \mu}_2}u_{n0}(\tilde{k}')
\nonumber\\&&\times
 D^{j_1}_{\bar \mu_1'\bar{\tilde \mu}_1}\left[\underline R_\mathrm{W_{\!c}}\left(R_\mathrm{W_{\!c}}(V,\varLambda)\tilde w_1', B_{\mathrm c}(\boldsymbol R_\mathrm{W_{\!c}}(V,\varLambda)w_{12}')\right)\right]
\nonumber\\&&\times D^{j_2}_{\bar \mu_2'\bar{\tilde \mu}_2}\left[\underline R_\mathrm{W_{\!c}}\left(R_\mathrm{W_{\!c}}(V,\varLambda)\tilde w_2', B_{\mathrm c}(\boldsymbol R_\mathrm{W_{\!c}}(V,\varLambda) w_{12}')\right)\right]\,.
 \end{eqnarray}
Here, we can further simplify the argument of the Wigner $D$-functions containing the transformed spins (using the short-hand notation $R_\mathrm{W_{\!c}}:=R_\mathrm{W_{\!c}}(V,\varLambda)$): 
\begin{eqnarray}
\label{eq:3WignerDfucntions}
\lefteqn{
 \underline R_\mathrm{W_{\!c}}\left(R_\mathrm{W_{\!c}}\tilde w_1',B_{\mathrm c}(\boldsymbol R_\mathrm{W_{\!c}}w_{12}')\right)}\nonumber\\&=&
\underline B^{-1}_\mathrm c( \boldsymbol B_{\mathrm c}( R_\mathrm{W_{\!c}}w_{12}') R_\mathrm{W_{\!c}}\tilde w_1')\,\underline B_{\mathrm c}(\boldsymbol R_\mathrm{W_{\!c}}w_{12}') \,
\underline B_\mathrm c(\boldsymbol R_\mathrm{W_{\!c}}\tilde w_1')\nonumber\\&=&
\underline R_\mathrm{W_{\!c}}(\tilde w_1',\underbrace{B_{\mathrm c}(\boldsymbol R_\mathrm{W_{\!c}} w_{12}') R_\mathrm{W_{\!c}}}_{=R_\mathrm{W_{\!c}} B_{\mathrm c}(\boldsymbol w_{12}') } )\,\underline B^{-1}_\mathrm c (\tilde {\boldsymbol w}_1') \,\underline R_\mathrm{W_{\!c}} ^{-1} 
\nonumber\\&&\times \underline B_{\mathrm c}^{-1}(\boldsymbol R_\mathrm{W_{\!c}} w_{12}')\,  
\underline B_{\mathrm c}(\boldsymbol R_\mathrm{W_{\!c}}w_{12}')\, 
\underline B_\mathrm c(\boldsymbol R_\mathrm{W_{\!c}}\tilde w_1')\nonumber\\&=&
\underline R_\mathrm{W_{\!c}}(\tilde w_1',R_\mathrm{W_{\!c}} B_{\mathrm c}(\boldsymbol w_{12}') ) \,\underline B^{-1}_\mathrm c (\tilde {\boldsymbol w}_1')\, \underline R_\mathrm{W_{\!c}} ^{-1} \,
\underline B_\mathrm c(\boldsymbol R_\mathrm{W_{\!c}}\tilde w_1')\nonumber\\&=&
 \underline B_\mathrm c^{-1}(\boldsymbol R_\mathrm{W_{\!c}} B_{\mathrm c}(\boldsymbol w_{12}')\tilde w_1')\,\underline R_\mathrm{W_{\!c}}\, \underline B_{\mathrm c}(\boldsymbol w_{12}')\, \underline B_\mathrm c(\tilde {\boldsymbol w}_1')\,
\underline B^{-1}_\mathrm c (\tilde {\boldsymbol w}_1') \,\underline R_\mathrm{W_{\!c}} ^{-1} \nonumber\\&&\times 
\underline R_\mathrm{W_{\!c}}\,\underline B_\mathrm c( \tilde {\boldsymbol w}_1')\,\underline R_\mathrm{W_{\!c}}^{-1}
\nonumber\\&=&
 \underline R_\mathrm{W_{\!c}}\,\underline B_\mathrm c^{-1}(\boldsymbol B_{\mathrm c}(\boldsymbol w_{12}')\tilde w_1')\,\underline R_\mathrm{W_{\!c}}^{-1} \,\underline R_\mathrm{W_{\!c}}\, \underline B_{\mathrm c}(\boldsymbol w_{12}')\, \underline B_\mathrm c(\tilde {\boldsymbol w}_1')\,
\underline R_\mathrm{W_{\!c}}^{-1}
\nonumber\\&=&
\underline R_\mathrm{W_{\!c}}\,
\underline B_\mathrm c^{-1}(\boldsymbol B_{\mathrm c}(\boldsymbol w_{12}')\tilde w_1')\, \underline B_{\mathrm c}(\boldsymbol w_{12}')\, \underline B_\mathrm c(\tilde {\boldsymbol w}_1')\,
\underline R_\mathrm{W_{\!c}}^{-1}
\nonumber\\&=&\underline R_\mathrm{W_{\!c}}\,
\underline R_\mathrm{W_{\!c}}(\tilde w_1',B_{\mathrm c}(\boldsymbol w_{12}'))\,
\underline R_\mathrm{W_{\!c}}^{-1}\,.
\end{eqnarray}
Here we have made use of the property of canonical Wigner rotations, Eq.~(\ref{eq:propWc}).
A similar calculation applies to constituent 2.
Consequently, we can split the Wigner $D$-functions in Eq.~(\ref{eq:transfomredwf2}) into 3 $D$-functions:
\begin{eqnarray}\lefteqn{
 D^{j_1}_{\bar \mu_1'\bar{\tilde \mu}_1'}\left[\underline R_\mathrm{W_{\!c}}\left(R_\mathrm{W_{\!c}}(V,\varLambda)\tilde w_1',B_{\mathrm c}(\boldsymbol R_\mathrm{W_{\!c}}(V,\varLambda)w_{12}')\right)\right]}\nonumber\\&&=
D^{j_1}_{\bar \mu_1'\mu_1'}[\underline R_\mathrm{W_{\!c}}(V,\varLambda)]
D^{j_1}_{\mu_1'\tilde\mu_1}[
\underline R_\mathrm{W_{\!c}}(\tilde w_1',B_{\mathrm c}(\boldsymbol w_{12}'))] D^{j_1}_{\tilde\mu_1\bar{\tilde\mu}_1}[\underline R_\mathrm{W_{\!c}}^{-1}(V,\varLambda)]\nonumber\\
 \end{eqnarray} 
and similarly for constituent 2.
Comparing Eqs.~(\ref{eq:transfomredwf}) and~(\ref{eq:transfomredwf2}) gives the desired transformation properties of the
Clebsch-Gordan coefficients under Lorentz transformations:
\begin{eqnarray}\lefteqn{C^{j\mu_j}_{j_1\tilde \mu_1j_2\tilde \mu_2}\stackrel{\varLambda}{\rightarrow}
 C^{j\bar \mu_j}_{j_1\bar {\tilde \mu}_1j_2\bar{\tilde \mu}_2}}\nonumber\\&&=C^{j\mu_j}_{j_1\tilde \mu_1j_2\tilde \mu_2}D_{\bar\mu_j\mu_j}^{j\ast}\left[\underline R_\mathrm{W_{\!c}}(V,\varLambda)\right]
D^{j_1}_{\bar{\tilde\mu}_1\tilde\mu_1}[\underline R_\mathrm{W_{\!c}}(V,\varLambda)]D^{j_2}_{\bar{\tilde\mu}_2\tilde\mu_2}[\underline R_\mathrm{W_{\!c}}(V,\varLambda)]\,.
\nonumber\\
\end{eqnarray}

\chapter{Electromagnetic Vertices}\label{app:D}
In this chapter we derive the phenomenological currents from matrix elements of the interaction Lagrangean density and investigate their covariance properties. We also take a closer look at the covariance and continuity properties of the microscopic bound-state current.

\section{Vertex Interaction}
\label{app:vertexint}
In this section we present the explicit calculation of  Eq.~(\ref{eq:vertexinteractionlagrangedens}). In the present work we consider two kinds of bound systems: charged mesons ($\pi^+$ and $\rho^+$) consisting of a quark and an antiquark of different flavors ($\mathrm u$ and $\bar{\mathrm d}$) and a charged nucleus (deuteron) consisting of 2 nucleons (proton and neutron). It should be noted that the creation and annihilation operators associated with quark and anti-quark of different flavors anti-commute. In this section we will derive the constituent currents for the point-like case of quarks and then generalize the result to non-pointlike constituents like nucleons. 

We start with the left-hand side of Eq.~(\ref{eq:vertexinteractionlagrangedens}) and insert for the interaction Lagrangean density Eq.~(\ref{eq:lagrangeandens}) at $x=0$, where constituent 1 is a quark and constituent 2 is an antiquark:
\begin{eqnarray}
 \lefteqn{-\langle \boldsymbol k_\mathrm e',\mu_\mathrm e';\boldsymbol k_1',\mu_1';\boldsymbol k_2',\mu_2';\boldsymbol k'_\gamma,\mu'_\gamma\vert \hat{\mathcal L}_{\mathrm {int}} (0)\vert\boldsymbol k_\mathrm e,\mu_\mathrm e;\boldsymbol k_1,\mu_1;\boldsymbol k_2,\mu_2\rangle}
\nonumber\\&=& \,|\,\mathrm e\,|\,\langle \boldsymbol k_\mathrm e',\mu_\mathrm e';\boldsymbol k_1',\mu_1';\boldsymbol k_2',\mu_2';\boldsymbol k'_\gamma,\mu'_\gamma\vert\left[ 
Q_{\mathrm e}:\hat{\bar \psi}_{\mathrm e}(0)\gamma^\mu\hat \psi_{\mathrm e} (0) \hat A_{\mu}(0):\right.\nonumber\\&&\left.\;\;+Q_1:\hat{\bar \psi}_{1}(0)\gamma^\mu\hat \psi_{1} (0) \hat A_{\mu}(0):+Q_{2}:\hat{\bar \psi}_{2}(0)\gamma^\mu\hat \psi_{2} (0) \hat A_{\mu}(0):\right]\nonumber\\&&\times\vert\boldsymbol k_\mathrm e,\mu_\mathrm e;\boldsymbol k_1,\mu_1;\boldsymbol k_2,\mu_2\rangle\,.\nonumber\\
\end{eqnarray}
First we concentrate on the second term of this expression:
\begin{eqnarray}\label{eq:vertexterm2}
 \lefteqn{\langle \boldsymbol k_\mathrm e',\mu_\mathrm e';\boldsymbol k_1',\mu_1';\boldsymbol k_2',\mu_2';\boldsymbol k'_\gamma,\mu'_\gamma\vert 
:\hat{\bar \psi}_{1}(0)\gamma^\mu\hat \psi_{1} (0) \hat A_{\mu}(0):\vert\boldsymbol k_\mathrm e,\mu_\mathrm e;\boldsymbol k_1,\mu_1;\boldsymbol k_2,\mu_2\rangle}\nonumber\\&
=&
\frac{1}{(2\pi)^3}\int\frac{\mathrm d^3p_1}{2p_1^0}\sum_{\sigma_1=\pm\frac12}\frac{1}{(2\pi)^3}\int\frac{\mathrm d^3p'_1}{2p_1'^0}\sum_{\sigma_1'=\pm\frac12}
\frac{1}{(2\pi)^3}\int\frac{\mathrm d^3p_\gamma}{2\vert\boldsymbol p_\gamma\vert}\sum_{\lambda_\gamma=0}^3(-\mathrm g^{\lambda_\gamma\lambda_\gamma})
\nonumber\\&&
\times\bar u_{\sigma_1}(\boldsymbol p_1)\gamma_{\nu} u_{\sigma'_1}(\boldsymbol p'_1) B_\mathrm h(\boldsymbol p)^\nu_{\;\lambda_{\gamma}} 
\langle0\vert c_{\mu_{\mathrm e}'}(\boldsymbol k_{\mathrm e}') d_{\mu_2'}(\boldsymbol k_2')c_{\mu_1'}(\boldsymbol k_1')a_{\mu'_\gamma}(\boldsymbol k'_\gamma)\nonumber\\&&\;\;\times
:\hat c^\dag_{\sigma_1}(\boldsymbol p_1)\hat c_{\sigma'_1}(\boldsymbol p'_1) \hat a^\dag_{\lambda_{\gamma}}(\boldsymbol p_\gamma):  d_{\mu_2}^\dag(\boldsymbol k_2)c_{\mu_{\mathrm e}}^\dag(\boldsymbol k_{\mathrm e})
c_{\mu_1}^\dag(\boldsymbol k_1) \vert 0\rangle
 \nonumber\\&
=&
\frac{1}{(2\pi)^3}\int\frac{\mathrm d^3p_1}{2p_1^0}\sum_{\sigma_1=\pm\frac12}\frac{1}{(2\pi)^3}\int\frac{\mathrm d^3p'_1}{2p_1'^0}\sum_{\sigma_1'=\pm\frac12}
\frac{1}{(2\pi)^3}\int\frac{\mathrm d^3p_\gamma}{2\vert\boldsymbol p_\gamma\vert}\sum_{\lambda_\gamma=0}^3(-\mathrm g^{\lambda_\gamma\lambda_\gamma})
\nonumber\\&&
\times\bar u_{\sigma_1}(\boldsymbol p_1)\gamma_{\nu} u_{\sigma'_1}(\boldsymbol p'_1) B_\mathrm h(\boldsymbol p)^\nu_{\;\lambda_{\gamma}} 
\nonumber\\&&\times\langle0\vert c_{\mu_1'}(\boldsymbol k_1')\hat c^\dag_{\sigma_1}(\boldsymbol p_1)\vert 0\rangle\langle0\vert 
\hat c_{\sigma'_1}(\boldsymbol p'_1) c_{\mu_1}^\dag(\boldsymbol k_1) \vert 0\rangle\langle0\vert  d_{\mu_2'}(\boldsymbol k_2')d_{\mu_2}^\dag(\boldsymbol k_2)
\vert 0\rangle\nonumber\\&&
\times\langle0\vert c_{\mu_{\mathrm e}'}(\boldsymbol k_{\mathrm e}') c_{\mu_{\mathrm e}}^\dag(\boldsymbol k_{\mathrm e})\vert 0\rangle\langle0\vert 
a_{\mu'_\gamma}(\boldsymbol k'_\gamma)\hat a^\dag_{\lambda_{\gamma}}(\boldsymbol p_\gamma) \vert 0\rangle   \nonumber\\&
= &
\frac{1}{(2\pi)^3}\int\frac{\mathrm d^3p_1}{2p_1^0}\sum_{\sigma_1=\pm\frac12}\frac{1}{(2\pi)^3}\int\frac{\mathrm d^3p'_1}{2p_1'^0}\sum_{\sigma_1'=\pm\frac12}
\frac{1}{(2\pi)^3}\int\frac{\mathrm d^3p_\gamma}{2\vert\boldsymbol p_\gamma\vert}\sum_{\lambda_\gamma=0}^3(-\mathrm g^{\lambda_\gamma\lambda_\gamma})
\nonumber\\&&
\times\bar u_{\sigma_1}(\boldsymbol p_1)\gamma_{\nu} u_{\sigma'_1}(\boldsymbol p'_1) B_\mathrm h(\boldsymbol p)^\nu_{\;\lambda_{\gamma}} 
(2\pi)^3 \delta_{\sigma_1\mu_1'}2 p_1^0\delta^3(\boldsymbol p_1-\boldsymbol k_1')\nonumber\\&&\times(2\pi)^3 \delta_{\sigma_1'\mu_1} 2 p_1'^0\delta^3(\boldsymbol p_1'-\boldsymbol k_1)
(2\pi)^3 \delta_{\mu_{\mathrm e}\mu_{\mathrm e}'}2 k_{\mathrm e}^0 \delta^3(\boldsymbol{k}_{\mathrm e}^\prime - \boldsymbol{k}_{\mathrm e})
\nonumber\\&&\times  (2\pi)^3\delta_{\mu_2\mu_2'} 2  k_2^0 \delta^3(\boldsymbol{k}_2^\prime - \boldsymbol{k}_2) 
 (2\pi)^3(-\mathrm g_{\mu'_\gamma \lambda_{\gamma}})2  k_\gamma^0 \delta^3(\boldsymbol{k}_\gamma^\prime - \boldsymbol{p}_\gamma)
\nonumber\\&
= &
\bar u_{\mu'_1}(\boldsymbol k_1')\gamma_{\nu} u_{\mu_1}(\boldsymbol k_1) B_\mathrm h(\boldsymbol k_\gamma')^\nu_{\;\mu'_\gamma } \nonumber\\&&
\times
(2\pi)^3 \delta_{\mu_{\mathrm e}\mu_{\mathrm e}'}2 k_{\mathrm e}^0 \delta^3(\boldsymbol{k}_{\mathrm e}^\prime - \boldsymbol{k}_{\mathrm e})
 (2\pi)^3\delta_{\mu_2\mu_2'} 2  k_2^0 \delta^3(\boldsymbol{k}_2^\prime - \boldsymbol{k}_2).\nonumber\\
\end{eqnarray}
Here we have used Wick's theorem (see, e.g., Ref.~\cite{Aitchison:2003}), the anti-commutation and commutation relations, Eqs.~(\ref{eq:anticommrel}) and~(\ref{eq:commrel}), and that the annihilation operators annihilate the vacuum,
Eqs.~(\ref{eq:vac}) and~(\ref{eq:vac1}).
The calculation of the first term is equivalent to the one for the second term:
\begin{eqnarray}
 && \langle \boldsymbol k_\mathrm e',\mu_\mathrm e';\boldsymbol k_1',\mu_1';\boldsymbol k_2',\mu_2';\boldsymbol k'_\gamma,\mu'_\gamma\vert 
:\hat{\bar \psi}_{\mathrm e}(0)\gamma^\mu\hat \psi_{\mathrm e} (0) \hat A_{\mu}(0):\vert\boldsymbol k_\mathrm e,\mu_\mathrm e;\boldsymbol k_1,\mu_1;\boldsymbol k_2,\mu_2\rangle
\nonumber\\&&\;\;
= 
\bar u_{\mu'_\mathrm e}(\boldsymbol k_\mathrm e')\gamma_{\nu} u_{\mu_\mathrm e}(\boldsymbol k_\mathrm e) B_\mathrm h(\boldsymbol k_\gamma')^\nu_{\;\mu'_\gamma } \nonumber\\&&\;\;\;\;\times
(2\pi)^3 \delta_{\mu_{1}\mu_{1}'}2 k_{1}^0 \delta^3(\boldsymbol{k}_1^\prime - \boldsymbol{k}_1)
 (2\pi)^3\delta_{\mu_2\mu_2'} 2  k_2^0 \delta^3(\boldsymbol{k}_2^\prime - \boldsymbol{k}_2)\,.
\end{eqnarray}
For the third term we obtain a similar result, except for a relative minus sign which emerges from the normal ordering of the anti-particle operators,
$:\hat d_{\sigma_1}(\boldsymbol p_1)\hat d^\dag_{\sigma'_1}(\boldsymbol p'_1) :=-\hat d^\dag_{\sigma'_1}(\boldsymbol p'_1)\hat d_{\sigma_1}(\boldsymbol p_1)$, i.e.
\begin{eqnarray}&&
 \langle \boldsymbol k_\mathrm e',\mu_\mathrm e';\boldsymbol k_1',\mu_1';\boldsymbol k_2',\mu_2';\boldsymbol k'_\gamma,\mu'_\gamma\vert 
:\hat{\bar \psi}_{2}(0)\gamma^\mu\hat \psi_{2} (0) \hat A_{\mu}(0):\vert\boldsymbol k_\mathrm e,\mu_\mathrm e;\boldsymbol k_1,\mu_1;\boldsymbol k_2,\mu_2\rangle\nonumber\\&&\;\;=
-\bar v_{\mu_2}(\boldsymbol k_1)\gamma_{\nu} v_{\mu_2'}(\boldsymbol k_2') B_\mathrm h(\boldsymbol k_\gamma')^\nu_{\;\mu'_\gamma } \nonumber\\&&\;\;\;\;\times
(2\pi)^3 \delta_{\mu_{\mathrm e}\mu_{\mathrm e}'}2 k_{\mathrm e}^0 \delta^3(\boldsymbol{k}_{\mathrm e}^\prime - \boldsymbol{k}_{\mathrm e})
 (2\pi)^3\delta_{\mu_1\mu_1'} 2  k_2^0 \delta^3(\boldsymbol{k}_1^\prime - \boldsymbol{k}_1)\,.
\end{eqnarray} 
The matrix elements of the Lagrangean density derived above for point-like spin-$\frac12$ particles can be immediately generalized to current matrix elements for extended objects, such as nucleons, mesons or the deuteron. This is done by simply replacing $|\,\mathrm e\,|\,Q_i\bar u_{\mu'_i}(\boldsymbol k_i')\gamma^{\mu} u_{\mu_i}(\boldsymbol k_i)$ by the appropriate expression for the current matrix elements of an extended particle, $J_{i}^\mu (\boldsymbol k_i',\mu_i';\boldsymbol k_i,\mu_i)$ and $J_{\mathrm C}^\mu (\boldsymbol k_\mathrm C',\mu_j';\boldsymbol k_\mathrm C,\mu_j;K_\mathrm e)$.
\section{Covariance Properties}\label{app:transfpropcurrents}
\subsection{Phenomenological Currents}\label{app:covphenomenocurrents}
In this section we prove the correct transformation properties of the currents $J_{i}^\mu( \boldsymbol p_i^\prime, \sigma_i^\prime; \boldsymbol p_i,\sigma_i)$ defined by Eq.~(\ref{eq:physicalphenocurrent}). Note that the currents $J_{i}^\mu( \boldsymbol p_i^\prime, \sigma_i^\prime; \boldsymbol p_i,\sigma_i)$ are defined via the center-of-momentum currents $J_{i}^\mu( \boldsymbol k_i^\prime, \mu_i^\prime; \boldsymbol k_i,\mu_i)$ whose transformation properties are given by Eq.~(\ref{eq:transformpropcomcurrents}). We start our analysis by looking at the Lorentz transformed current:
\begin{eqnarray}\lefteqn{
J_{i}^\mu( \boldsymbol p_i^\prime, \sigma_i^\prime; \boldsymbol p_i,\sigma_i)\stackrel{\varLambda}{\longrightarrow}J_{i}^\mu( \boldsymbol\varLambda \boldsymbol p_i^\prime, \bar \sigma_i^\prime; \boldsymbol \varLambda p_i,\bar\sigma_i)}\nonumber\\&=&
B_\mathrm c(\boldsymbol \varLambda V)^\mu_{\,\,\nu}J_{i}^{\nu}( \boldsymbol R_\mathrm{W_{\!c}}(V,\varLambda) k_i',\bar \mu_i';\boldsymbol R_\mathrm{W_{\!c}}(V,\varLambda) k_i,\bar{\mu}_i)
\nonumber\\&&\times D_{\bar{\mu}_i'\bar\sigma_i'}^{j_i\ast}\left[\underline R_\mathrm{W_{\!c}}^{-1}(R_\mathrm{W_{\!c}}(V,\varLambda)w_i',B_\mathrm c(\boldsymbol \varLambda V))\right]
D_{\bar{\mu}_i\bar\sigma_i}^{j_i}\left[\underline R_\mathrm{W_{\!c}}^{-1}(R_\mathrm{W_{\!c}}(V,\varLambda)w_i,B_\mathrm c(\boldsymbol \varLambda V))\right]
\nonumber\\&=&
\underbrace{B_\mathrm c(\boldsymbol \varLambda V)^\mu_{\,\,\nu}R_\mathrm{W_{\!c}}(V,\varLambda)^\nu_{\,\,\rho}}_{=(\varLambda B_\mathrm c(\boldsymbol V))^\mu_{\,\,\rho}}
J_{i}^{\rho}( \boldsymbol k_i',\mu_i';\boldsymbol  k_i,\mu_i)\nonumber\\&&\times
D^{j_i\ast}_{\mu_i'\bar \mu_i '}\left[\underline R_\mathrm{W_{\!c}}^{-1}(V,\varLambda)\right]  
D^{j_i}_{\mu_i\bar \mu_i}\left[\underline R_\mathrm{W_{\!c}}^{-1}(V,\varLambda)\right]\nonumber\\&&\times
 D_{\bar{\mu}_i'\bar\sigma_i'}^{j_i\ast}\left[\underline R_\mathrm{W_{\!c}}^{-1}(R_\mathrm{W_{\!c}}(V,\varLambda)w_i',B_\mathrm c(\boldsymbol \varLambda V))\right]
\nonumber\\&&\times
D_{\bar{\mu}_i\bar\sigma_i}^{j_i}\left[\underline R_\mathrm{W_{\!c}}^{-1}(R_\mathrm{W_{\!c}}(V,\varLambda)w_i,B_\mathrm c(\boldsymbol \varLambda V))\right]
\nonumber\\&=&
\varLambda^\mu_{\,\,\nu}
J_{i}^{\nu}( \boldsymbol p_i',\sigma_i';\boldsymbol  p_i,\sigma_i)
D^{j_i\ast}_{\sigma_i'\mu_i ' }\left[\underline R_\mathrm{W_{\!c}}(w_i',B_\mathrm c(\boldsymbol V))\right]
D^{j_i}_{ \sigma_i \mu_i}\left[\underline R_\mathrm{W_{\!c}}(w_i,B_\mathrm c(\boldsymbol V))\right]\nonumber\\&&\times
D^{j_i\ast}_{\mu_i'\bar\sigma_i'}\left[\underline R_\mathrm{W_{\!c}}^{-1}(V,\varLambda)\underline R_\mathrm{W_{\!c}}^{-1}(R_\mathrm{W_{\!c}}(V,\varLambda)w_i',B_\mathrm c(\boldsymbol \varLambda V))\right]
\nonumber\\&&\times D^{j_i}_{\mu_i\bar\sigma_i}\left[\underline R_\mathrm{W_{\!c}}^{-1}(V,\varLambda)\underline R_\mathrm{W_{\!c}}^{-1}(R_\mathrm{W_{\!c}}(V,\varLambda)w_i,B_\mathrm c(\boldsymbol \varLambda V))
\right]
\nonumber\\&=&
\varLambda^\mu_{\,\,\nu}
J_{i}^{\nu}( \boldsymbol p_i',\sigma_i';\boldsymbol  p_i,\sigma_i)
\nonumber\\&&\times
D^{j_i\ast}_{\sigma_i'\bar\sigma_i'}\left[\underline R_\mathrm{W_{\!c}}(w_i',B_\mathrm c(\boldsymbol V))
\underline R_\mathrm{W_{\!c}}^{-1}(V,\varLambda)\underline R_\mathrm{W_{\!c}}^{-1}(R_\mathrm{W_{\!c}}(V,\varLambda)w_i',B_\mathrm c(\boldsymbol \varLambda V))
\right]\nonumber\\&&\times
D^{j_i}_{ \sigma_i \bar\sigma_i}\left[\underline R_\mathrm{W_{\!c}}(w_i,B_\mathrm c(\boldsymbol V))\underline R_\mathrm{W_{\!c}}^{-1}(V,\varLambda)\underline R_\mathrm{W_{\!c}}^{-1}(R_\mathrm{W_{\!c}}(V,\varLambda)w_i,B_\mathrm c(\boldsymbol \varLambda V))
\right]
\nonumber\\&=&
\varLambda^\mu_{\,\,\nu}
J_{i}^{\nu}( \boldsymbol p_i',\sigma_i';\boldsymbol  p_i,\sigma_i)
D^{j_i\ast}_{\sigma_i'\bar\sigma_i'}\left[\underline R_\mathrm{W_{\!c}}^{-1}(v_i',\varLambda)
\right]
D^{j_i}_{ \sigma_i \bar\sigma_i}\left[\underline R_\mathrm{W_{\!c}}^{-1}(v_i,\varLambda)
\right]\,.
\end{eqnarray}
Here we have used in the last step
\begin{eqnarray}\lefteqn{
 \underline R_\mathrm{W_{\!c}}(w_i,B_\mathrm c(\boldsymbol V))\underline R_\mathrm{W_{\!c}}^{-1}(V,\varLambda)\underline R_\mathrm{W_{\!c}}^{-1}(R_\mathrm{W_{\!c}}(V,\varLambda)w_i,B_\mathrm c(\boldsymbol \varLambda V))}\nonumber\\&=&
\underline B_\mathrm c^{-1}(\boldsymbol v_i) \underline B_\mathrm c(\boldsymbol V) \underbrace{\underline B_\mathrm c(\boldsymbol w_i) \underline B_\mathrm c^{-1}(\boldsymbol V)\underline \varLambda^{-1} \underline B_\mathrm c(\boldsymbol \varLambda V)  \underline B_\mathrm c^{-1}(\boldsymbol R_\mathrm{W_{\!c}}(V,\varLambda)w_i)}_
{=\underline R_\mathrm{W_{\!c}}^{-1}\left[w_i,R_\mathrm{W_{\!c}}(V,\varLambda)\right]=\underline R_\mathrm{W_{\!c}}^{-1}(V,\varLambda),\,\text{cf.~Eq.~(\ref{eq:propWc})}}\nonumber\\&&\times
\underline B_\mathrm c^{-1}(\boldsymbol \varLambda V) \underline B_\mathrm c( \boldsymbol B_\mathrm c^{-1}(\boldsymbol \varLambda V) R_\mathrm{W_{\!c}}(V,\varLambda)w_i)\nonumber\\&=&
\underline B_\mathrm c^{-1}(\boldsymbol v_i) \underline B_\mathrm c(\boldsymbol V) \underline R_\mathrm{W_{\!c}}^{-1}(V,\varLambda)
\underline B_\mathrm c^{-1}(\boldsymbol \varLambda V) \underline B_\mathrm c( \boldsymbol B_\mathrm c^{-1}(\boldsymbol \varLambda V) R_\mathrm{W_{\!c}}(V,\varLambda)w_i)\nonumber\\&=&
\underline B_\mathrm c^{-1}(\boldsymbol v_i) \underline \varLambda^{-1} \underline B_\mathrm c( \boldsymbol \varLambda v_i)\,
\end{eqnarray}
 and a similar relation for the primed momenta.
We see, by comparison with Eq.~(\ref{eq:transformpropconstcurrent}), that $J_{i}^\mu( \boldsymbol p_i^\prime, \sigma_i^\prime; \boldsymbol p_i,\sigma_i)$ transforms under Lorentz transformations like a contravariant four-vector.
\subsection{Microscopic Bound-State Current}\label{app:covmicrocurrents}
In this section we prove the transformation properties of the bound-state current, Eq.~(\ref{eq:trafopropcombscurrent}), under Lorentz transformations for the case of equal constituent masses and for a pure s-wave ($l=l'=0$). 
First we note that, under a Lorentz transformation $\varLambda$, the center-of-mass momenta of the electron-bound-state system, $k_\mathrm C^{(\prime)}$, $k_1^{(\prime)}$ and $k_2^{(\prime)}$, undergo a Wigner rotation $R_\mathrm{W_{\!c}}(V,\varLambda)$. Consequently, their zero components
 are not affected by a Lorentz transformation, i.e. $\bar k_\mathrm C^{(\prime)0}= k_\mathrm C^{(\prime)0}$, $\bar k_1^{(\prime)0}=k_1^{(\prime)0}$ and $\bar k_2^{(\prime)0}=k_2^{(\prime)0}$. Note that under such a Lorentz transformation also the cluster center-of-mass momenta $\tilde k_1$ and $\tilde k_1'$ undergo the same Wigner rotation, cf. Eq.~(\ref{eq:ktildebar1}).
Therefore, the three-dimensional integration measure as well as quantities involving the magnitude of the cluster center-of-mass three-momenta remain invariant under Lorentz transformations, i.e. $\mathrm{d}^3\bar{\tilde{k}}'_1=\mathrm{d}^3\tilde{k}'_1$,  $\bar{\tilde{\boldsymbol k}}_1^{(\prime)2}=\tilde{\boldsymbol k}_1^{(\prime)2}$ and $\bar{m}_{12}^{(\prime)}=m_{12}^{(\prime)}$.
With these findings at hand we look at the Lorentz transformed current of Eq.~(\ref{eq:bscurrentequalmasses}):
\begin{eqnarray} 
\lefteqn{J^\mu_{\mathrm C}(\boldsymbol R_\mathrm{W_{\!c}}(V,\varLambda) k_{\mathrm C}',\bar \mu_j';\boldsymbol R_\mathrm{W_{\!c}}(V,\varLambda)k_{\mathrm C},\bar \mu_j;R_\mathrm{W_{\!c}}(V,\varLambda)K_\mathrm e )}\nonumber\\&=&
\frac{\sqrt{ k^0_{\mathrm C} k'^0_{\mathrm C}}}{4\pi}\int \frac{\mathrm d^3 \tilde k_{1}'}{ k_{1}^0}
\sqrt{\frac{m_{12}}{m_{12}'}}
\sqrt{\frac{ k^0_{12}}{k'^0_{12}}}
u^\ast_{n0}(\tilde{k}')
 u_{n0}(\tilde{k})
\nonumber\\&&\times
D^{\frac12}_{\bar {\tilde \mu}_1'\bar \mu_1'}\left[\underline R_\mathrm{W_{\!c}}^{-1}\left(R_\mathrm{W_{\!c}} (V,\varLambda)\tilde w_1', B_{\mathrm c}(\boldsymbol R_\mathrm{W_{\!c}} (V,\varLambda)w_{12}')\right)\right]
C^{j\bar \mu_j'\ast}_{\frac12\bar {\tilde \mu}_1'\frac12\bar{\tilde \mu}_2'} \nonumber\\&&\times D^{\frac12}_{\bar{\tilde \mu}_2'\bar{\tilde \mu}_2}\left[\underline R_\mathrm{W_{\!c}}\left(R_\mathrm{W_{\!c}} (V,\varLambda)\tilde w_2, B^{-1}_{\mathrm c}(\boldsymbol R_\mathrm{W_{\!c}} (V,\varLambda)w_{12}') B_{\mathrm c}(\boldsymbol R_\mathrm{W_{\!c}} (V,\varLambda)w_{12})\right)\right] 
\nonumber\\&&\times C^{j\bar \mu_j}_{\frac12\bar{\tilde \mu}_1\frac12\bar{\tilde \mu}_2}D^{\frac12}_{\bar \mu_1\bar{\tilde \mu}_1}\left[\underline R_\mathrm{W_{\!c}}\left(R_\mathrm{W_{\!c}} (V,\varLambda)\tilde w_1,B_{\mathrm c}(\boldsymbol R_\mathrm{W_{\!c}} (V,\varLambda)w_{12})\right)\right]
\nonumber\\&&\times|\,\mathrm e\,|\,\bar u_{\bar \mu_1'}(\boldsymbol R_\mathrm{W_{\!c}} (V,\varLambda)k_1')(\varGamma_1+\varGamma_2)^\mu u_{\bar \mu_1}(\boldsymbol R_\mathrm{W_{\!c}} (V,\varLambda)k_1)
 \,.
\end{eqnarray}
Next we look at the product of the two Clebsch-Gordan coefficients between the three Wigner $D$-functions. Using the transformation properties of the Clebsch-Gordans derived previously, cf. Eq.~(\ref{eq:TrafopropCG}), together with Eq.~(\ref{eq:3WignerDfucntions}), which allows to split the transformed $D$-function into the product of three $D$-functions, we obtain the following equation:
\begin{eqnarray}&&\label{eq:DCDCD}
\lefteqn{D^{\frac12}_{\bar {\tilde \mu}_1'\bar \mu_1'}\left[\underline R_\mathrm{W_{\!c}}^{-1}\left(R_\mathrm{W_{\!c}}\tilde w_1',B_{\mathrm c}(\boldsymbol R_\mathrm{W_{\!c}}w_{12}')\right)\right]
C^{j\bar \mu_j'\ast}_{\frac12\bar {\tilde \mu}_1'\frac12\bar{\tilde \mu}_2'} }\nonumber\\&&\times D^{\frac12}_{\bar{\tilde \mu}_2'\bar{\tilde \mu}_2}\left[\underline R_\mathrm{W_{\!c}}\left(R_\mathrm{W_{\!c}}\tilde w_2,B^{-1}_{\mathrm c}(\boldsymbol R_\mathrm{W_{\!c}}w_{12}')B_{\mathrm c}(\boldsymbol R_\mathrm{W_{\!c}}w_{12})\right)\right] 
\nonumber\\&&\times C^{j\bar \mu_j}_{\frac12\bar{\tilde \mu}_1\frac12\bar{\tilde \mu}_2}D^{\frac12}_{\bar \mu_1\bar{\tilde \mu}_1}\left[\underline R_\mathrm{W_{\!c}}\left(R_\mathrm{W_{\!c}}\tilde w_1,B_{\mathrm c}(\boldsymbol R_\mathrm{W_{\!c}}w_{12})\right)\right]
\nonumber\\&&\;\;\;\;=
 D^{\frac12}_{\tilde \mu_1'\mu_1'}\left[\underline R_\mathrm{W_{\!c}}^{-1}\left(\tilde w_1',B_{\mathrm c}(\boldsymbol w_{12}')\right)\right]
C^{j\mu_j'\ast}_{\frac12 {\tilde \mu}_1'\frac12\tilde \mu_2'} \nonumber\\&&\;\;\;\;\;\;\;\;\times D^{\frac12}_{\tilde \mu_2'\tilde \mu_2}\left[\underline R_\mathrm{W_{\!c}}\left(\tilde w_2, B^{-1}_{\mathrm c}(\boldsymbol w_{12}')B_{\mathrm c}(\boldsymbol w_{12})\right)\right] 
\nonumber\\&&\;\;\;\;\;\;\;\;\times C^{j \mu_j}_{\frac12\tilde \mu_1\frac12\tilde \mu_2}D^{\frac12}_{ \mu_1\tilde \mu_1}\left[\underline R_\mathrm{W_{\!c}}\left(\tilde w_1,B_{\mathrm c}(\boldsymbol w_{12})\right)\right]
\nonumber\\&&\;\;\;\;\;\;\;\;\times D_{\bar \mu_1'\mu_1'}^{\frac12\ast}(\underline R_\mathrm{W_{\!c}})D_{\bar \mu_1\mu_1}^{\frac12}(\underline R_\mathrm{W_{\!c}})
D_{\bar\mu_j\mu_j}^{j\ast}(\underline R_\mathrm{W_{\!c}}) D_{\bar\mu_j'\mu_j'}^{j}(\underline R_\mathrm{W_{\!c}})\,.
\end{eqnarray}
In this calculation the 4 Wigner $D$-functions $D_{\tilde \mu_1\bar{\tilde\mu}_1}^{\frac12}$, $D_{\tilde \mu_2\bar{\tilde\mu}_2}^{\frac12}$,
$D_{\tilde \mu_1'\bar{\tilde\mu}_1'}^{\frac12\ast}$ and $D_{\tilde \mu_2'\bar{\tilde\mu}_2'}^{\frac12\ast}$ have been canceled with the corresponding inverse ones from the Clebsch-Gordan coefficients.
From the transformation properties of the constituent currents, Eq.~(\ref{eq:transformpropcomcurrents}), we get 2 more Wigner $D$-functions which cancel $D_{\bar \mu_1\mu_1}^{\frac12}$ and $D_{\bar \mu_1'\mu_1'}^{\frac12\ast}$ in Eq.~(\ref{eq:DCDCD}). With these findings we finally end up with the transformation properties of the microscopic bound-state current given by Eq.~(\ref{eq:trafopropcombscurrent}).  
\section{Current Conservation}
\subsection{Pseudoscalar-Bound-State Current}\label{app:currentconservation}
In this appendix we prove the conservation of the electromagnetic current, Eq.~(\ref{eq:pscurrent2}), for pseudoscalar bound states.
For simplicity we restrict our considerations to point-like constituents, like quarks, for which the vertex reads $(\varGamma_1+\varGamma_2)^\mu=(Q_1+Q_2)\gamma^\mu$.
We start with the right-hand side of Eq.~(\ref{eq:currentconservps}): 
\begin{eqnarray}\label{eq:currentconserv2}
\lefteqn{(k_\mathrm C'-k_\mathrm C)_{\mu}J^\mu_{\mathrm {PS}}(\boldsymbol k_{\mathrm C}',\boldsymbol k_{\mathrm C},K_\mathrm e)}\nonumber\\&=&
-(\boldsymbol k_\mathrm C'-\boldsymbol k_\mathrm C)\cdot \boldsymbol J_{\mathrm {PS}}(\boldsymbol k_{\mathrm C}',\boldsymbol k_{\mathrm C},K_\mathrm e)\nonumber\\&=&
-|\,\mathrm e\,|\,(Q_1+Q_2)\frac{1}{4\pi}\frac{\sqrt{ k^0_{\mathrm C} k'^0_{\mathrm C}}}{2}\int \frac{\mathrm d^3 \tilde k_{1}'}{ k_{1}^0}
\sqrt{\frac{m_{12}}{m_{12}'}}
\sqrt{\frac{ k^0_{12}}{k'^0_{12}}}
u^\ast_{n0}(\tilde{k}')
 u_{n0}(\tilde{k})
\nonumber\\&&\times
\sum_{\mu_1'\mu_1}D^{\frac12}_{\mu_1\mu_1'}\left[\underline B_{\mathrm c}\left(\boldsymbol w_1\right)
   \sigma_\tau 
\left(
\begin{array}{c}
w_{12}^{0}\\
-\boldsymbol w_{12}
\end{array}
\right)^\tau
\underline w_{2}\,
  \sigma_\nu
\left(
\begin{array}{c}
w_{12}'^{0}\\
-\boldsymbol w_{12}'
\end{array}
\right)^\nu\underline B_{\mathrm c}\left(\boldsymbol w'_1\right)\right]
\nonumber\\&&\times\bar u_{\mu_1'}(\boldsymbol k_1')(\boldsymbol k_1'-\boldsymbol k_1)\cdot \boldsymbol \gamma u_{\mu_1}(\boldsymbol k_1)
 \,,
\end{eqnarray}
since $k_\mathrm C'^0=k_\mathrm C^0$ and $\boldsymbol q=\boldsymbol k_\mathrm C'-\boldsymbol k_\mathrm C=\boldsymbol k_1'-\boldsymbol k_1=\boldsymbol q_1$.
In the following calculation we will suppress the argument of the Wigner $D$-function for better readability. Let us first concentrate on the contraction of the spatial part of the constituent current with $\boldsymbol q_1$. Using the standard representation of Dirac spinors, Eq.~(\ref{eq:Diracspinors1}),
we obtain after a short calculation
\begin{eqnarray}\label{eq:cuurentconsercalc}\lefteqn{
 \sum_{\mu_1'\mu_1}\bar u_{\mu_1'}(\boldsymbol k_1')(\boldsymbol k_1'-\boldsymbol k_1)\cdot \boldsymbol \gamma\, u_{\mu_1}(\boldsymbol k_1) D^{\frac12}_{\mu_1\mu_1'}}\nonumber\\&=&\!\!\!\!
\sqrt{k_1^0+m}\sqrt{k_1'^0+m}\sum_{\mu_1'\mu_1}\left[\left( \frac{\boldsymbol k_1'^2}{k_1'^0+m}+\frac{\boldsymbol k_1\cdot\boldsymbol k_1'}{k_1^0+m }-\frac{\boldsymbol k_1^2}{k_1^0+m }-
\frac{\boldsymbol k_1\cdot\boldsymbol k_1'}{k_1'^0+m }\right) \delta_{\mu_1\mu_1'}\right.\nonumber\\&&\left.+\mathrm{i}\,\boldsymbol k_1'\times
\boldsymbol k_1\cdot \varsigma^{\mathrm T}_{\mu_1'}\boldsymbol{\sigma}\,\varsigma_{\mu_1}\left( \frac{1}{k_1^0+m }-\frac{1}{k_1'^0+m }\right)  \right]
D^{\frac12}_{\mu_1\mu_1'}\,,
\end{eqnarray}
where we have used Eq.~(\ref{eq:paulimatricesproduct}).
On the right-hand side of Eq.~(\ref{eq:cuurentconsercalc}) the terms between the first round brackets are multiplied by $\delta_{\mu_1\mu_1'}$ and the Wigner $D$-function. Summation over the spins gives for this product
\begin{eqnarray}\label{eq:reDfunct}
 \sum_{\mu_1'\mu_1} \delta_{\mu_1\mu_1'}D^{\frac12}_{\mu_1\mu_1'}=\sum_{\mu_1} D^{\frac12}_{\mu_1\mu_1}=2\,\Re ( D_{\frac12\frac12}^{\frac12})\,,
\end{eqnarray}
where we have used Eq.~(\ref{eq:RWprop4}).
Let us now concentrate on the first two terms between the first round brackets, $\frac{\boldsymbol k_1'^2}{k_1'^0+m}+\frac{\boldsymbol k_1\cdot\boldsymbol k_1'}{k_1^0+m }$.
If we replace all primed by unprimed momenta and vice versa we obtain simply $\frac{\boldsymbol k_1^2}{k_1^0+m}+\frac{\boldsymbol k_1'\cdot\boldsymbol k_1}{k_1'^0+m }$. Further, we observe that this expression has just the opposite sign of the second two terms between the first round brackets. Consequently, if we can show that the remaining expressions under the integral in Eq.~(\ref{eq:currentconserv2}) are symmetric under the interchange of primed and unprimed momenta, these 4 terms between the first round brackets cancel upon integration.
To this aim we transform the integration measure in Eq.~(\ref{eq:currentconserv2}) according to Eq.~(\ref{eq:integrationmeausremanip}).
Then we obtain for the kinematical factors under the integral of Eq.~(\ref{eq:currentconserv2}) the following equality:
\begin{eqnarray}\label{eq:currentconserv22}
\frac{\mathrm d^3 \tilde k_{1}'}{ k_{1}^0}
\sqrt{\frac{m_{12}}{m_{12}'}}
\sqrt{\frac{ k^0_{12}}{k'^0_{12}}}=
\frac {\mathrm{d}^3 \tilde{k}_1}{k_1'^0}
\sqrt{\frac{m_{12}'}{m_{12}}}
\sqrt{\frac{ k'^0_{12}}{k^0_{12}}}\,.
\end{eqnarray}
It is evident that this expression is symmetric under the interchange of primed and unprimed momenta. The Wigner $D$-function in Eq.~(\ref{eq:pscurrent2}) transforms under the interchange of primed and unprimed momenta and spins by complex conjugation:
  \begin{eqnarray}\lefteqn{
   D^{\frac12}_{\mu_1\mu_1'}\left[\underline R_\mathrm{W_{\!c}}\left(\tilde w_1,B_{\mathrm c}(\boldsymbol w_{12})\right)
\underline R_\mathrm{W_{\!c}}^{-1}\left(\tilde w_2,B^{-1}_{\mathrm c}(\boldsymbol w_{12}')B_{\mathrm c}(\boldsymbol w_{12})\right)
\underline R_\mathrm{W_{\!c}}^{-1}\left(\tilde w_1',B_{\mathrm c}(\boldsymbol w_{12}')\right)\right]}\nonumber\\&
\stackrel{ \{k_i,\mu_1\}\leftrightarrow \{k_i',\mu_1'\}}{\longrightarrow}&
D^{\frac12}_{\mu_1'\mu_1}\left[\underline R_\mathrm{W_{\!c}}\left(\tilde w'_1,B_{\mathrm c}(\boldsymbol w_{12}')\right)
\underline R_\mathrm{W_{\!c}}^{-1}\left(\tilde w_2, B^{-1}_{\mathrm c}(\boldsymbol w_{12})B_{\mathrm c}(\boldsymbol w_{12}')\right)
\right.\nonumber\\&&\left.\;\;\;\;\times\underline R_\mathrm{W_{\!c}}^{-1}\left(\tilde w_1,B_{\mathrm c}(\boldsymbol w_{12})\right)\right]\nonumber\\&=&
D^{\frac12\ast }_{\mu_1\mu_1'}\left[\underline R_\mathrm{W_{\!c}}\left(\tilde w_1,B_{\mathrm c}(\boldsymbol w_{12})\right)
\underline R_\mathrm{W_{\!c}}^{-1}\left(\tilde w_2,B^{-1}_{\mathrm c}(\boldsymbol w_{12}')B_{\mathrm c}(\boldsymbol w_{12})\right)
\right.\nonumber\\&&\left.\;\;\;\;\times\underline R_\mathrm{W_{\!c}}^{-1}\left(\tilde w_1',B_{\mathrm c}(\boldsymbol w_{12}')\right)\right]\,.\nonumber\\
  \end{eqnarray}
Further we have
\begin{eqnarray}
 \sum_{\mu_1'\mu_1} \delta_{\mu_1\mu_1'}D^{\frac12\ast}_{\mu_1\mu_1'}=2\,\Re ( D_{\frac12\frac12}^{\frac12})\,,
\end{eqnarray}
where we have used Eq.~(\ref{eq:RWprop4}). Thus, by comparison with Eq.~(\ref{eq:reDfunct}), we see that this expression is also symmetric under the interchange of primed and unprimed momenta. 
For real radial wave functions $u_{n0}(\tilde{k})$ the product $u^\ast_{n0}(\tilde{k}')
 u_{n0}(\tilde{k})$ is also symmetric under interchange of primed and unprimed momenta.  
Putting the pieces together we have 
\begin{eqnarray}\lefteqn{\sqrt{ k^0_{\mathrm C} k'^0_{\mathrm C}}\int \frac{\mathrm d^3 \tilde k_{1}'}{ k_{1}^0}
\sqrt{\frac{m_{12}}{m_{12}'}}
\sqrt{\frac{ k^0_{12}}{k'^0_{12}}}
u^\ast_{n0}(\tilde{k}')\,
 u_{n0}(\tilde{k})}
\nonumber\\&&\times
\sum_{\mu_1'\mu_1}D^{\frac12}_{\mu_1\mu_1'}\delta_{\mu_1\mu_1'}
\sqrt{k_1^0+m}\sqrt{k_1'^0+m}\left( \frac{\boldsymbol k_1'^2}{k_1'^0+m}+\frac{\boldsymbol k_1\cdot\boldsymbol k_1'}{k_1^0+m }\right)
\nonumber\\&&\;\;=
\sqrt{ k^0_{\mathrm C} k'^0_{\mathrm C}}\int \frac{\mathrm d^3 \tilde k_{1}}{ k_{1}'^0}
\sqrt{\frac{m_{12}'}{m_{12}}}
\sqrt{\frac{ k'^0_{12}}{k^0_{12}}}
u^\ast_{n0}(\tilde{k})\,
 u_{n0}(\tilde{k}')
\nonumber\\&&\;\;\;\;\times
\sum_{\mu_1'\mu_1}D^{\frac12}_{\mu_1'\mu_1}\delta_{\mu_1\mu_1'}
\sqrt{k_1^0+m}\sqrt{k_1'^0+m}\left( \frac{\boldsymbol k_1'^2}{k_1'^0+m}+\frac{\boldsymbol k_1\cdot\boldsymbol k_1'}{k_1^0+m }\right)
\nonumber\\
&&\;\;=\sqrt{ k^0_{\mathrm C} k'^0_{\mathrm C}}\int \frac{\mathrm d^3 \tilde k_{1}'}{ k_{1}^0}
\sqrt{\frac{m_{12}}{m_{12}'}}
\sqrt{\frac{ k^0_{12}}{k'^0_{12}}}
u^\ast_{n0}(\tilde{k}')\,
 u_{n0}(\tilde{k})
\nonumber\\&&\;\;\;\;\times
\sum_{\mu_1'\mu_1}D^{\frac12}_{\mu_1\mu_1'}\delta_{\mu_1\mu_1'}
\sqrt{k_1^0+m}\sqrt{k_1'^0+m}\left(\frac{\boldsymbol k_1^2}{k_1^0+m}+\frac{\boldsymbol k_1'\cdot\boldsymbol k_1}{k_1'^0+m }\right)\,.\nonumber\\
\end{eqnarray} 
Here we have in the last step just renamed the integration variable. The right-hand side of this equation is just the negative of the second 2 terms between the first round brackets in Eq.~(\ref{eq:cuurentconsercalc}). Therefore, the first 2 terms cancel the second two terms between the first round brackets due to the relative minus sign.

Next we consider the first term between the second round brackets in Eq.~(\ref{eq:cuurentconsercalc}), $\frac{1}{k_1^0+m }$, which has just the opposite sign of the second term if all primed and unprimed momenta are interchanged.
Therefore, in an similar manner as above, we have to show that all other factors are symmetric under the interchange of primed and unprimed momenta in order that these two terms cancel.
To this aim we start with the terms proportional to $ \varsigma^{\mathrm T}_{\mu_1'}\sigma_3\,\varsigma_{\mu_1}$ in Eq.~(\ref{eq:cuurentconsercalc}).
The relevant factors to be considered are
\begin{eqnarray}\label{eq:relfactors1}
  (\boldsymbol k_1'\times\boldsymbol k_1)^3\sum_{\mu_1'\mu_1}\varsigma^{\mathrm T}_{\mu_1}\sigma_3\,\varsigma_{\mu_1'}D_{\mu_1'\mu_1}^{\frac12}\,.
 \end{eqnarray}
Performing the sums over the spins gives
 \begin{eqnarray}
  \sum_{\mu_1'\mu_1}\varsigma^{\mathrm T}_{\mu_1'}\sigma_3\,\varsigma_{\mu_1}D_{\mu_1\mu_1'}^{\frac12}=2\,\mathrm i \,\Im( D_{\frac12\frac12}^{\frac12})\,.
 \end{eqnarray}
Under interchange of all primed and unprimed momenta this expression picks up a minus sign:
\begin{eqnarray}
  \sum_{\mu_1'\mu_1}\varsigma^{\mathrm T}_{\mu_1}\sigma_3\,\varsigma_{\mu_1'}D_{\mu_1'\mu_1}^{\frac12}=2\,\mathrm i \,\Im( D_{\frac12\frac12}^{\frac12\ast})=-2\,\mathrm i\,\Im( D_{\frac12\frac12}^{\frac12})
\,.
 \end{eqnarray}
Since this expression is multiplied by $(\boldsymbol k_1'\times\boldsymbol k_1)^3$, the product, 
Eq.~(\ref{eq:relfactors1}), is again symmetric under primed-unprimed momenta interchange. Therefore, the two terms proportional to $ \varsigma^{\mathrm T}_{\mu_1'}\sigma_3\,\varsigma_{\mu_1}$ cancel.

The proof for the $\varsigma^{\mathrm T}_{\mu_1'}\sigma_2\,\varsigma_{\mu_1}$-terms is similar. Carrying out the spin sums we get
\begin{eqnarray}
\sum_{\mu_1'\mu_1}\varsigma^{\mathrm T}_{\mu_1'}\sigma_2\,\varsigma_{\mu_1}D_{\mu_1\mu_1'}^{\frac12}=-2\,\mathrm i\,\Re( D_{-\frac12\frac12}^{\frac12})\,.
\end{eqnarray}
If we interchange again all primed by unprimed variables we find also a relative minus sign:
\begin{eqnarray}
\sum_{\mu_1'\mu_1}\varsigma^{\mathrm T}_{\mu_1}\sigma_2\,\varsigma_{\mu_1'}D_{\mu_1'\mu_1}^{\frac12}=2\,\mathrm i\,\Re( D_{-\frac12\frac12}^{\frac12})\,.
\end{eqnarray}
Therefore, 
\begin{eqnarray}\label{eq:relfactors2}
  (\boldsymbol k_1'\times\boldsymbol k_1)^2\sum_{\mu_1'\mu_1}\varsigma^{\mathrm T}_{\mu_1}\sigma_2\,\varsigma_{\mu_1'}D_{\mu_1'\mu_1}^{\frac12}
 \end{eqnarray}
is symmetric under primed-unprimed momenta interchange and the two terms proportional to $ \varsigma^{\mathrm T}_{\mu_1'}\sigma_2\,\varsigma_{\mu_1}$ cancel.

Finally we are left with the $\varsigma^{\mathrm T}_{\mu_1'}\sigma_1\varsigma_{\mu_1}$-terms:
\begin{eqnarray}&&
\sum_{\mu_1'\mu_1}\varsigma^{\mathrm T}_{\mu_1'}\sigma_1\varsigma_{\mu_1}D_{\mu_1\mu_1'}^{\frac12}=2\,\mathrm i\,\Im( D_{-\frac12\frac12}^{\frac12})\,.
\end{eqnarray}
Interchanging all primed by unprimed variables we find also a relative minus sign:
\begin{eqnarray}&&
\sum_{\mu_1'\mu_1}\varsigma^{\mathrm T}_{\mu_1}\sigma_1\varsigma_{\mu_1'}D_{\mu_1'\mu_1}^{\frac12}=-2\,\mathrm i\,\Im( D_{-\frac12\frac12}^{\frac12})\,.
\end{eqnarray}
Therefore, also the product
\begin{eqnarray}\label{eq:relfactors3}
  (\boldsymbol k_1'\times\boldsymbol k_1)^1\sum_{\mu_1'\mu_1}\varsigma^{\mathrm T}_{\mu_1}\sigma_1\varsigma_{\mu_1'}D_{\mu_1'\mu_1}^{\frac12}
 \end{eqnarray}
is invariant under primed-unprimed interchange and the two terms proportional to $ \varsigma^{\mathrm T}_{\mu_1'}\sigma_1\varsigma_{\mu_1}$ cancel.

Since all contributions cancel each other upon integration over $\tilde{\boldsymbol k}_1$ the conservation of our pseudoscalar-bound-state current is proven. 

\subsection{Vector-Bound-State Current}\label{app:currentnonconservation}
For the vector-bound-state current, Eq.~(\ref{eq:bscurrentVectorphys}), current conservation does, unlike the pseudoscalar case, in general not hold. This statement is  expressed by Eq.~(\ref{eq:currentnonconserv}) and will be justified numerically later on. As we will see in this section, an analytical proof of current conservation along the lines of the pseudoscalar case does not work, since the product of the 3 Wigner $D$-functions together with the 2 Clebsch-Gordan coefficients in Eq.~(\ref{eq:physicalmicrocurrent}) \textit{cannot} be written as one single Wigner $D$-function. Therefore, it is not possible to use symmetry arguments for the spin-dependent factors under the interchange of primed and unprimed momenta and spins. To see this it suffices to consider the constituents as point-like particles. For better readability we use the following abbreviation for the product of the 3 Wigner $D$-functions and the 2 Clebsch-Gordan coefficients:
 \begin{eqnarray}
\mathcal S^{\mu_j'\mu_j}_{\mu_1\mu_1'}&:=&D^{\frac12}_{\tilde \mu_1'\mu_1'}\left[\underline R_\mathrm{W_{\!c}}^{-1}\left(\tilde w_1',B_{\mathrm c}(\boldsymbol w_{12}')\right)\right]
C^{1'\mu_j'\ast}_{\frac12\tilde \mu_1'\frac12\tilde \mu_2'} \nonumber\\&&\times D^{\frac12}_{\tilde \mu_2'\tilde \mu_2}
\left[\underline R_\mathrm{W_{\!c}}\left(\tilde w_2, B^{-1}_{\mathrm c}(\boldsymbol w_{12}')B_{\mathrm c}(\boldsymbol w_{12})\right)\right] 
\nonumber\\&&\times C^{1\mu_j}_{\frac12\tilde \mu_1\frac12\tilde \mu_2}D^{\frac12}_{\mu_1\tilde \mu_1}\left[\underline R_\mathrm{W_{\!c}}\left(\tilde w_1,B_{\mathrm c}(\boldsymbol w_{12})\right)\right]\,.
 \end{eqnarray}
On the right-hand side of Eq.~(\ref{eq:currentnonconserv}) we have the contraction of the spatial part of the (point-like) constituent current with $\boldsymbol q_1$ (cf. Eq.~(\ref{eq:bscurrent2}))
\begin{eqnarray}\label{eq:cuurentnonconsercalc2}
 \lefteqn{\sum_{\mu_1'\mu_1}\bar u_{\mu_1'}(\boldsymbol k_1')(\boldsymbol k_1'-\boldsymbol k_1)\cdot \boldsymbol \gamma \,u_{\mu_1}(\boldsymbol k_1) \,\mathcal S^{\mu_j'\mu_j}_{\mu_1\mu_1'}}\nonumber\\&=&\!\!\!\!\!
\sqrt{k_1^0+m}\sqrt{k_1'^0+m}\sum_{\mu_1'\mu_1}\left[\left( \frac{\boldsymbol k_1'^2}{k_1'^0+m}+\frac{\boldsymbol k_1\cdot\boldsymbol k_1'}{k_1^0+m }-\frac{\boldsymbol k_1^2}{k_1^0+m }-
\frac{\boldsymbol k_1\cdot\boldsymbol k_1'}{k_1'^0+m }\right) \delta_{\mu_1\mu_1'}\right.\nonumber\\&&\left.+\mathrm{i}\,\boldsymbol k_1'\times
\boldsymbol k_1\cdot \varsigma^{\mathrm T}_{\mu_1'}\boldsymbol{\sigma}\,\varsigma_{\mu_1}\left( \frac{1}{k_1^0+m }-\frac{1}{k_1'^0+m }\right)  \right]
\mathcal S^{\mu_j'\mu_j}_{\mu_1\mu_1'}\,.
\end{eqnarray}
For the first term proportional to $\delta_{\mu_1\mu_1'}$ we can perform the spin sum
\begin{eqnarray}
   \sum_{\mu_1'\mu_1}\delta_{\mu_1\mu_1'} \mathcal S^{\mu_j'\mu_j}_{\mu_1\mu_1'}= \sum_{\mu_1}\mathcal S^{\mu_j'\mu_j}_{\mu_1\mu_1}\,.
\end{eqnarray}
The important observation is that this expression is, in general, complex and not just a real number like in the pseudoscalar case. On the other hand, by interchanging all primed and unprimed momenta and spins in $\mathcal S^{\mu_j'\mu_j}_{\mu_1\mu_1'}$ that are integrated and summed over, respectively, we find
\begin{eqnarray}
\mathcal S^{\mu_j'\mu_j}_{\mu_1'\mu_1}&=&D^{\frac12}_{\tilde \mu_1\mu_1}\left[\underline R_\mathrm{W_{\!c}}^{-1}\left(\tilde w_1,B_{\mathrm c}(\boldsymbol w_{12})\right)\right]
C^{1\mu_j'\ast}_{\frac12\tilde \mu_1\frac12\tilde \mu_2} \nonumber\\&&\times D^{\frac12}_{\tilde \mu_2\tilde \mu_2'}
\left[\underline R_\mathrm{W_{\!c}}\left(\tilde w_2',B^{-1}_{\mathrm c}(\boldsymbol w_{12})B_{\mathrm c}(\boldsymbol w_{12}')\right)\right] 
\nonumber\\&&\times C^{1\mu_j}_{\frac12\tilde \mu_1'\frac12\tilde \mu_2'}D^{\frac12}_{\mu_1'\tilde \mu_1'}
\left[\underline R_\mathrm{W_{\!c}}\left(\tilde w_1',B_{\mathrm c}(\boldsymbol w_{12}')\right)\right]\nonumber\\&=&
\mathcal S^{\mu_j'\mu_j\ast}_{\mu_1\mu_1'}\,.
 \end{eqnarray}
Carrying out the spin sum of this expression gives
\begin{eqnarray}
   \sum_{\mu_1'\mu_1}\delta_{\mu_1\mu_1'} \mathcal S^{\mu_j'\mu_j\ast}_{\mu_1\mu_1'}= \sum_{\mu_1}\mathcal S^{\mu_j'\mu_j\ast}_{\mu_1\mu_1}\neq \sum_{\mu_1}\mathcal S^{\mu_j'\mu_j}_{\mu_1\mu_1}\,,
\end{eqnarray}
since this is, in general, a complex number. Consequently, cancellation of the 4 terms between the first round brackets of Eq.~(\ref{eq:cuurentnonconsercalc2}) cannot be shown by simply renaming primed and unprimed momenta and spins that are integrated and summed over.

With a similar argumentation we proceed for the other terms between the second round brackets of Eq.~(\ref{eq:cuurentnonconsercalc2}). For the term proportional to $\varsigma^{\mathrm T}_{\mu_1'}\sigma_3\,\varsigma_{\mu_1}$ we can perform the spin sums:
\begin{eqnarray}
   \sum_{\mu_1'\mu_1}\varsigma^{\mathrm T}_{\mu_1'}\sigma_3\,\varsigma_{\mu_1}\mathcal S^{\mu_j'\mu_j}_{\mu_1\mu_1'}=\mathcal S^{\mu_j'\mu_j}_{\frac12\frac12}-\mathcal S^{\mu_j'\mu_j}_{-\frac12-\frac12}\, .
\end{eqnarray}
On the other hand, interchanging all primed and unprimed variables that are summed and integrated over we get
\begin{eqnarray}
   \sum_{\mu_1'\mu_1}\varsigma^{\mathrm T}_{\mu_1}\sigma_3\,\varsigma_{\mu_1'}\mathcal S^{\mu_j'\mu_j\ast}_{\mu_1\mu_1'}=(\mathcal S^{\mu_j'\mu_j}_{\frac12\frac12}-\mathcal S^{\mu_j'\mu_j}_{-\frac12-\frac12})^\ast\,. 
\end{eqnarray}
Since $\mathcal S^{\mu_j'\mu_j}_{\frac12\frac12}-\mathcal S^{\mu_j'\mu_j}_{-\frac12-\frac12}$ is a complex number, we have in general
 \begin{eqnarray}
  \mathcal S^{\mu_j'\mu_j}_{\frac12\frac12}-\mathcal S^{\mu_j'\mu_j}_{-\frac12-\frac12}\neq -(\mathcal S^{\mu_j'\mu_j}_{\frac12\frac12}-\mathcal S^{\mu_j'\mu_j}_{-\frac12-\frac12})^\ast\, 
\end{eqnarray}
and therefore the terms between the round brackets do not cancel. 

Similarly, we proceed for the  $\varsigma^{\mathrm T}_{\mu_1'}\sigma_2\,\varsigma_{\mu_1}$-term:
\begin{eqnarray}
   \sum_{\mu_1'\mu_1}\varsigma^{\mathrm T}_{\mu_1'}\sigma_2\,\varsigma_{\mu_1}\mathcal S^{\mu_j'\mu_j}_{\mu_1\mu_1'}=\mathrm i\,(\mathcal S^{\mu_j'\mu_j}_{\frac12-\frac12}-\mathcal S^{\mu_j'\mu_j}_{-\frac12\frac12})\,. 
\end{eqnarray}
Interchanging all primed and unprimed variables that are summed and integrated over gives
\begin{eqnarray}
   \sum_{\mu_1'\mu_1}\varsigma^{T}_{\mu_1}\sigma_2\,\varsigma_{\mu_1'}\mathcal S^{\mu_j'\mu_j\ast}_{\mu_1\mu_1'}=\mathrm i\,(\mathcal S^{\mu_j'\mu_j\ast}_{-\frac12\frac12}-\mathcal S^{\mu_j'\mu_j\ast}_{\frac12-\frac12})\,. 
\end{eqnarray}
Since, in general 
\begin{eqnarray}
 (\mathcal S^{\mu_j'\mu_j}_{-\frac12\frac12}-\mathcal S^{\mu_j'\mu_j}_{\frac12-\frac12})^\ast\neq-(\mathcal S^{\mu_j'\mu_j}_{\frac12-\frac12}-\mathcal S^{\mu_j'\mu_j}_{-\frac12\frac12})
\end{eqnarray}
the corresponding terms do not cancel.

Finally, for the $\varsigma^{\mathrm T}_{\mu_1'}\sigma_1\,\varsigma_{\mu_1}$-term we have
\begin{eqnarray}
   \sum_{\mu_1'\mu_1}\varsigma^{\mathrm T}_{\mu_1'}\sigma_1\,\varsigma_{\mu_1}\mathcal S^{\mu_j'\mu_j}_{\mu_1\mu_1'}=\mathcal S^{\mu_j'\mu_j}_{\frac12-\frac12}+\mathcal S^{\mu_j'\mu_j}_{-\frac12\frac12}\,. 
\end{eqnarray}
Interchanging all primed and unprimed variables and summing over gives
\begin{eqnarray}
   \sum_{\mu_1'\mu_1}\varsigma^{\mathrm T}_{\mu_1}\sigma_1\,\varsigma_{\mu_1'}\mathcal S^{\mu_j'\mu_j\ast}_{\mu_1\mu_1'}=\mathcal S^{\mu_j'\mu_j\ast}_{-\frac12\frac12}+\mathcal S^{\mu_j'\mu_j\ast}_{\frac12-\frac12}\,. 
\end{eqnarray}
Since, in general 
\begin{eqnarray}
 (\mathcal S^{\mu_j'\mu_j}_{-\frac12\frac12}+\mathcal S^{\mu_j'\mu_j}_{\frac12-\frac12})^\ast\neq-(\mathcal S^{\mu_j'\mu_j}_{\frac12+\frac12}-\mathcal S^{\mu_j'\mu_j}_{-\frac12\frac12})
\end{eqnarray}
the corresponding terms do also not cancel.

Since, in general, the integrals will not vanish by symmetry arguments we do not expect the vector-bound-state current,  Eq.~(\ref{eq:bscurrentVectorphys}), to be conserved.
\section{Covariant Structure} \label{app:covstructure}
The explicit analysis of the covariant structure of the vector bound-state current $J^\mu_{\mathrm {V}}(\boldsymbol p_{\mathrm C}',\sigma_j';\boldsymbol p_{\mathrm C},\sigma_j;P_{\mathrm e})$ is carried out as follows: by multiplying the tensor $\epsilon^{\mu\ast}_{\sigma_j'}(\boldsymbol p_\mathrm C')\epsilon^{\nu}_{\sigma_j}(\boldsymbol p_\mathrm C)$ with the available covariants $P_\mathrm C^\mu$, $d^\mu$ and/or $P_\mathrm e^\mu$ together with $\mathrm g^{\mu\nu}$ we build all independent Hermitean covariants.
From the transversality property, Eq.~(\ref{eq:transversalityprop}), follow the relations \begin{eqnarray}
       \epsilon^\ast_{\sigma'_j}(\boldsymbol p'_\mathrm C)\cdot P_\mathrm C=-\epsilon^\ast_{\sigma'_j}(\boldsymbol p'_\mathrm C)\cdot d\quad\text{and} \quad
\epsilon_{\sigma_j}(\boldsymbol p_\mathrm C)\cdot P_\mathrm C=\epsilon_{\sigma_j}(\boldsymbol p_\mathrm C)\cdot d\,.
      \end{eqnarray}
Then, we can immediately construct the following two independent Hermitean covariants:
\begin{eqnarray}\label{eq:cov1}
  &&\epsilon^{\mu\ast}_{\sigma'_j}(\boldsymbol p'_\mathrm C)[\epsilon_{\sigma_j}(\boldsymbol p_\mathrm C)\cdot d]-
\epsilon_{\sigma_j}^\mu (\boldsymbol p_\mathrm C)[\epsilon^\ast_{\sigma'_j}(\boldsymbol p'_\mathrm C)\cdot d]\,,
\\ \label{eq:cov2}
&&\epsilon^{\mu\ast}_{\sigma'_j}(\boldsymbol p'_\mathrm C)[\epsilon_{\sigma_j}(\boldsymbol p_\mathrm C)\cdot P_\mathrm e]+
\epsilon_{\sigma_j}^\mu (\boldsymbol p_\mathrm C)[\epsilon^\ast_{\sigma'_j}(\boldsymbol p'_\mathrm C)\cdot P_\mathrm e]\,.
\end{eqnarray}
In order to set up the other covariants we construct all independent scalars where the two polarizations vectors appear either symmetric or antisymmetric under mutual exchange:
\begin{eqnarray}&&
\epsilon^\ast_{\sigma_j'}(\boldsymbol p'_\mathrm C)\cdot\epsilon_{\sigma_j}(\boldsymbol p_\mathrm C)\,,\quad [\epsilon^\ast_{\sigma'_j}(\boldsymbol p'_\mathrm C)\cdot d]
[\epsilon_{\sigma_j}(\boldsymbol p_\mathrm C)\cdot d]
\,,\quad [\epsilon^\ast_{\sigma'_j}(\boldsymbol p'_\mathrm C)\cdot P_\mathrm e][\epsilon_{\sigma_j}(\boldsymbol p_\mathrm C)\cdot P_\mathrm e]\,,\nonumber \\&&
[\epsilon^\ast_{\sigma_j'}(\boldsymbol p'_\mathrm C)\cdot d][\epsilon_{\sigma_j}(\boldsymbol p_\mathrm C)\cdot P_\mathrm e]\pm 
[\epsilon^\ast_{\sigma_j'}(\boldsymbol p'_\mathrm C)\cdot P_\mathrm e][\epsilon_{\sigma_j}(\boldsymbol p_\mathrm C)\cdot d] \,.\nonumber
\end{eqnarray}
These scalars are then multiplied with the available covariants $P_\mathrm C^\mu$, $d^\mu$ and $P_\mathrm e^\mu$. Now we exclude all 
anti-Hermitean combinations such that only the following Hermitean covariants remain:
\begin{eqnarray}&&
[\epsilon^\ast_{\sigma_j'}(\boldsymbol p'_\mathrm C)\cdot\epsilon_{\sigma_j}(\boldsymbol p_\mathrm C)]P_\mathrm C^\mu\,,\qquad
[\epsilon^\ast_{\sigma_j'}(\boldsymbol p'_\mathrm C)\cdot\epsilon_{\sigma_j}(\boldsymbol p_\mathrm C)]P_\mathrm e^\mu\,,\nonumber \\&&
 [\epsilon^\ast_{\sigma'_j}(\boldsymbol p'_\mathrm C)\cdot d]
[\epsilon_{\sigma_j}(\boldsymbol p_\mathrm C)\cdot d]P_\mathrm C^\mu\,,\qquad  [\epsilon^\ast_{\sigma'_j}(\boldsymbol p'_\mathrm C)\cdot d]
[\epsilon_{\sigma_j}(\boldsymbol p_\mathrm C)\cdot d]P_\mathrm e^\mu\,,\nonumber \\&&
 [\epsilon^\ast_{\sigma'_j}(\boldsymbol p'_\mathrm C)\cdot P_\mathrm e][\epsilon_{\sigma_j}(\boldsymbol p_\mathrm C)\cdot P_\mathrm e]P_\mathrm C^\mu\,,\qquad
 [\epsilon^\ast_{\sigma'_j}(\boldsymbol p'_\mathrm C)\cdot P_\mathrm e][\epsilon_{\sigma_j}(\boldsymbol p_\mathrm C)\cdot P_\mathrm e]P_\mathrm e^\mu\,,\nonumber \\
&& \{[\epsilon^\ast_{\sigma_j'}(\boldsymbol p'_\mathrm C)\cdot d][\epsilon_{\sigma_j}(\boldsymbol p_\mathrm C)\cdot P_\mathrm e]- 
[\epsilon^\ast_{\sigma_j'}(\boldsymbol p'_\mathrm C)\cdot P_\mathrm e][\epsilon_{\sigma_j}(\boldsymbol p_\mathrm C)\cdot d] \} P_\mathrm C^\mu\,,\nonumber \\
&& \{[\epsilon^\ast_{\sigma_j'}(\boldsymbol p'_\mathrm C)\cdot d][\epsilon_{\sigma_j}(\boldsymbol p_\mathrm C)\cdot P_\mathrm e]- 
[\epsilon^\ast_{\sigma_j'}(\boldsymbol p'_\mathrm C)\cdot P_\mathrm e][\epsilon_{\sigma_j}(\boldsymbol p_\mathrm C)\cdot d] \} P_\mathrm e^\mu\,,\nonumber \\
&& \{[\epsilon^\ast_{\sigma_j'}(\boldsymbol p'_\mathrm C)\cdot d][\epsilon_{\sigma_j}(\boldsymbol p_\mathrm C)\cdot P_\mathrm e]+ 
[\epsilon^\ast_{\sigma_j'}(\boldsymbol p'_\mathrm C)\cdot P_\mathrm e][\epsilon_{\sigma_j}(\boldsymbol p_\mathrm C)\cdot d] \} d^\mu\,
.\label{eq:cov3}
\end{eqnarray}
These 9 covariants together with~(\ref{eq:cov1}) and (\ref{eq:cov2}) are then 11 Hermitean covariants. Consequently, we can parametrize the current
$J^\mu_{\mathrm {V}}(\boldsymbol p_{\mathrm C}',\sigma_j';\boldsymbol p_{\mathrm C},\sigma_j;P_\mathrm e)$ in terms of 11 form factors, the 3 physical form factors $f_1$, $f_2$ and $g_\mathrm M$ and 8 spurious form factors $b_1,\ldots, b_8$. 
\chapter{Matrix Elements of the Optical Potential}
\label{app:1phoptpotmatrixelemnts}
\section{Constituent-Level Calculation}
\begin{figure}
\begin{center}
\psfrag{e}{$\mathrm e$                }
\psfrag{C}{$\mathrm C$          }
\psfrag{c1}{$\mathrm c_1$         }
\psfrag{c2}{$\mathrm c_2$          }
\psfrag{gamma}{\;$\gamma$           }
\includegraphics[clip=7cm,width=50mm]{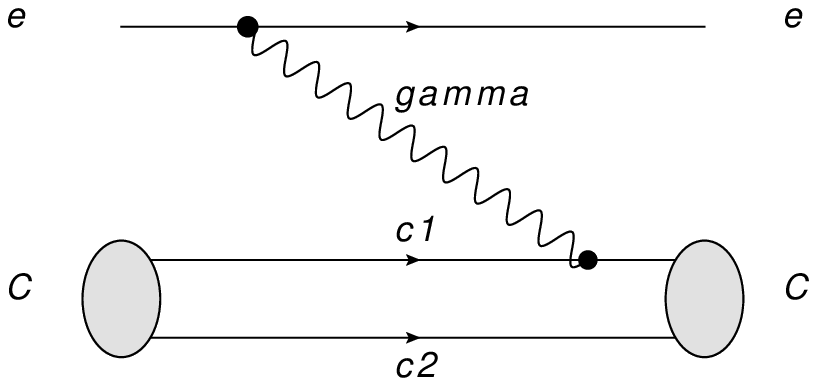}
\qquad\quad
\includegraphics[clip=7cm,width=50mm]{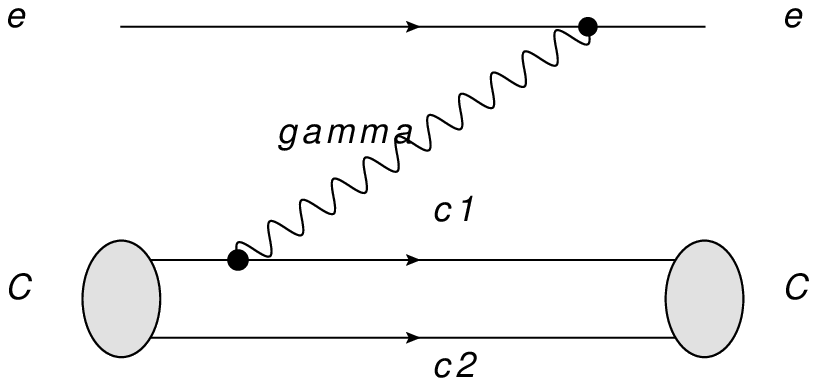}\vspace{6ex}
\includegraphics[clip=7cm,width=50mm]{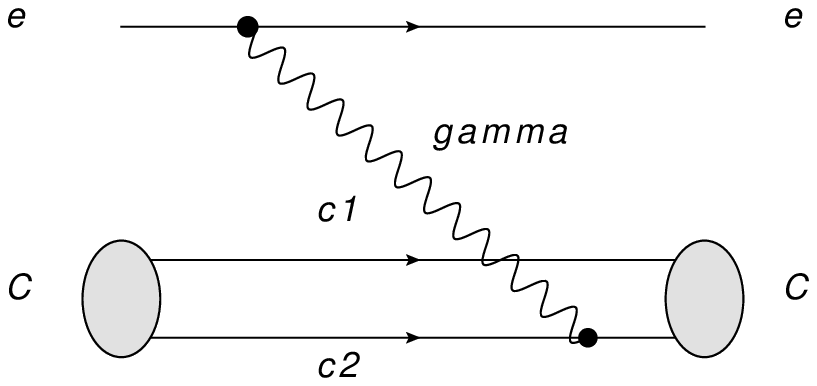}
\qquad\quad
\includegraphics[clip=7cm,width=50mm]{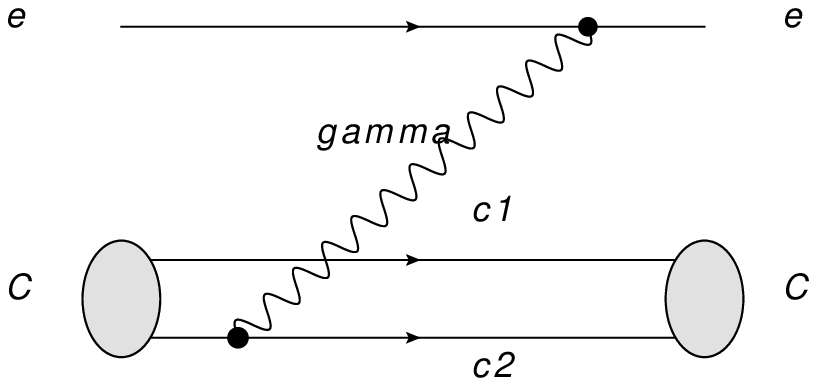}
\caption{\label{fig:graphec1}
The graphical representation of the 4 possibilities $\varGamma_{\mathrm e\rightarrow1}$, $\varGamma_{1\rightarrow\mathrm e}$, $\varGamma_{\mathrm e\rightarrow2}$ and $\varGamma_{2\rightarrow \mathrm e}$ to exchange the photon between the electron and one of the constituents.
}
\end{center} 
\end{figure}

In this appendix we present the rather lengthy calculation of the on-shell velocity-state matrix elements of the one-photon-exchange optical potential for elastic electron-bound-state scattering. A similar and quite detailed calculation for the case of spinless constituents can be found in Ref.~\cite{Fuchsberger:2007}. In the same manner we start with Eq.~(\ref{eq:1gammaamplit}) and insert completeness relations for free and
clustered velocity states, Eqs.~(\ref{eq:vcompl}) and~(\ref{eq:completenesseigenstates4part}), at the appropriate places:
\begin{eqnarray}\lefteqn{\langle  \underline{V}^\prime; \underline{\boldsymbol k}_{\mathrm e}^\prime,
\underline{\mu}_{\mathrm e}^\prime; \underline{\boldsymbol k}_{12}^\prime, n,j,\mu_j'\vert \hat{K} \left(\hat{M}_{\mathrm {e C} \gamma}
-M\right)^{-1} \hat{K}^\dag }\nonumber \\&&\;\;\times\vert\underline{V};
\underline{\boldsymbol k}_{\mathrm e}, \underline{\mu}_{\mathrm e};\underline{\boldsymbol k}_{12}, n,
j,\mu_j \rangle\nonumber\\&&=\langle  \underline{V}^\prime; \underline{\boldsymbol k}_{\mathrm e}^\prime,
\underline{\mu}_{\mathrm e}^\prime; \underline{\boldsymbol k}_{12}^\prime, n,j,\mu_j'\vert\hat 1_{12\mathrm e}' \,\hat{K}\,\hat 1_{12\mathrm e\gamma}''' \left(\hat{M}_{\mathrm {e C} \gamma}
-M\right)^{-1} \hat 1_{\mathrm {Ce}\gamma}'' \nonumber \\&&\;\;\times \hat 1_{12\mathrm e\gamma}''\,\hat{K}^\dag\,\hat 1_{12\mathrm e} \vert\, \underline{V};
\underline{\boldsymbol k}_{\mathrm e}, \underline{\mu}_{\mathrm e};\underline{\boldsymbol k}_{12}, n,
j,\mu_j \rangle\,\nonumber
 \end{eqnarray}
\\
\begin{eqnarray}\label{eq:optpottotal}
\lefteqn{=\frac{1}{(2\pi)^{51}}
\int \frac{\mathrm d^3 V'}{V'^0}\frac{\mathrm d^3 k_{\mathrm e}'}{2 k_{\mathrm e}'^0}\frac{\mathrm d^3 k_{1}'}{2 k_{1}'^0}\frac{(k_{1}'^0+k_{2}'^0+k_{\mathrm e}'^0)^3}{2k_{2}'^0}\sum_{\mu_{\mathrm e}'\mu_1'\mu_2'}}
 \nonumber\\&&\times
\int \frac{\mathrm d^3 V'''}{V'''^0}\frac{\mathrm d^3 k_{\mathrm e}'''}{2 k_{\mathrm e}'''^0}\frac{\mathrm d^3 k_{1}'''}{2 k_{1}'''^0}\frac{\mathrm d^3 k_{2}'''}{2 k_{2}'''^0}\frac{(k_{1}'''^0+k_{2}'''^0+k_{\mathrm e}'''^0+k'''^0_\gamma)^3}{2k_{\gamma}'''^0}\sum_{\mu_{\mathrm e}'''\mu_1'''\mu_2'''\mu_\gamma'''}(-\mathrm g^{\mu_\gamma'''\mu_\gamma'''})
 \nonumber\\&&\times 
\int \frac{\mathrm d^3 \underline{V}''}{\underline{V}''^0}\frac{\mathrm d^3 \underline{k}''_{12}}{2 k_{\mathrm C}''^0}\frac{\mathrm d^3 \underline{k}_{\gamma}''}{2 \underline{k}_{\gamma}''^0}\frac{(k_{\mathrm C}''^0+\underline{k}_{\mathrm e}''^0+\underline{k}''^0_\gamma)^3}{2\underline{k}_{\mathrm e}''^0}\sum_{n''j''\mu_j''\underline{\mu}_{\mathrm e}''\underline{\mu}_\gamma''}
(-\mathrm g^{\underline{\mu}_\gamma''\underline{\mu}_\gamma''})
\nonumber\\&&\times
\int \frac{\mathrm d^3 V''}{V''^0}\frac{\mathrm d^3 k_{\mathrm e}''}{2 k_{\mathrm e}''^0}\frac{\mathrm d^3 k_{1}''}{2 k_{1}''^0}\frac{\mathrm d^3 k_{2}''}{2 k_{2}''^0}
\frac{(k_{1}''^0+k_{2}''^0+k_{\mathrm e}''^0+k''^0_\gamma)^3}{2k_{\gamma}''^0}\sum_{\mu_{\mathrm e}''\mu_1''\mu_2''\mu_\gamma''}(-\mathrm g^{\mu_\gamma''\mu_\gamma''})
 \nonumber\\&&\times
\int \frac{\mathrm d^3 V}{V^0}\frac{\mathrm d^3 k_{\mathrm e}}{2 k_{\mathrm e}^0}\frac{\mathrm d^3 k_{1}}{2 k_{1}^0}
\frac{(k_{1}^0+k_{2}^0+k_{\mathrm e}^0)^3}{2k_2^0}\sum_{\mu_{\mathrm e}\mu_1\mu_2}
 \nonumber\\&&\times 
\langle\underline{V}^\prime; \underline{\boldsymbol k}_{\mathrm e}^\prime,
\underline{\mu}_{\mathrm e}^\prime; \underline{\boldsymbol k}_{12}^\prime, n,j,\mu_j' \vert V';\boldsymbol k_{\mathrm e}',\mu_{\mathrm e}';\boldsymbol k_{1}',\mu_{1}';
\boldsymbol k_{2}',\mu_{2}'\rangle \nonumber\\&&\times 
\langle V';\boldsymbol k_{\mathrm e}',\mu_{\mathrm e}';\boldsymbol k_{1}',\mu_{1}';
\boldsymbol k_{2}',\mu_{2}'\vert \hat{K}\vert V''';\boldsymbol k_{\mathrm e}''',\mu_{\mathrm e}''';\boldsymbol k_{1}''',\mu_{1}''';
\boldsymbol k_{2}''',\mu_{2}''';\boldsymbol k_{\gamma}''',\mu_{\gamma}'''\rangle
\nonumber\\&&\times
 \langle V''';\boldsymbol k_{\mathrm e}''',\mu_{\mathrm e}''';\boldsymbol k_{1}''',\mu_{1}''';
\boldsymbol k_{2}''',\mu_{2}''';\boldsymbol k_{\gamma}''',\mu_{\gamma}'''\vert\nonumber\\&&\;\;\;\;\times\vert\underline{V}''; \underline{\boldsymbol k}_{\mathrm e}'',
\underline{\mu}_{\mathrm e}''; \underline{\boldsymbol k}_{12}'', n'',j'',\mu_j'';\underline{\boldsymbol k}_\gamma'',\underline{\mu}_\gamma''\rangle
\nonumber\\&&\times  \left( k_{\mathrm e}''^0+k_{\mathrm C}''^0+ k_{\gamma}''^0
-M\right)^{-1}\nonumber\\&&\times
\langle \underline{V}''; \underline{\boldsymbol k}_{\mathrm e}'',
\underline{\mu}_{\mathrm e}''; \underline{\boldsymbol k}_{12}'', n'',j'',\mu_j'';\underline{\boldsymbol k}_\gamma'',\underline{\mu}_\gamma''\vert 
V'';\boldsymbol k_{\mathrm e}'',\mu_{\mathrm e}'';\boldsymbol k_{1}'',\mu_{1}'';
\boldsymbol k_{2}'',\mu_{2}'';\boldsymbol k_{\gamma}'',\mu_{\gamma}''\rangle
\nonumber\\&&\times
\langle V'';\boldsymbol k_{\mathrm e}'',\mu_{\mathrm e}'';\boldsymbol k_{1}'',\mu_{1}'';
\boldsymbol k_{2}'',\mu_{2}'';\boldsymbol k_{\gamma}'',\mu_{\gamma}''\vert \hat K^\dag \vert V;\boldsymbol k_{\mathrm e},\mu_{\mathrm e};\boldsymbol k_{1},\mu_{1};
\boldsymbol k_{2},\mu_{2}\rangle
\nonumber\\&&\times
\langle V;\boldsymbol k_{\mathrm e},\mu_{\mathrm e};\boldsymbol k_{1},\mu_{1};
\boldsymbol k_{2},\mu_{2}\vert\, \underline{V};
\underline{\boldsymbol k}_{\mathrm e}, \underline{\mu}_{\mathrm e};\underline{\boldsymbol k}_{12}, n,
j,\mu_j\rangle\,.
\end{eqnarray}
The matrix elements we need to know are typically scalar products between free and clustered velocity states as well as matrix elements of the vertex operators between free velocity states, which have been already derived previously. The former define the bound-state wave functions on the three- and four-body Hilbert spaces, cf. Eqs.~(\ref{eq:norm3partcl}) and (\ref{eq:norm4partcl}). The latter consist of 3 terms and are defined by Eq.~(\ref{eq:vertexoper1}). Inserting the latter gives altogether 9 terms which are depicted in Figs.~\ref{fig:graphec1} and \ref{fig:grapheee}. There the blobs that connect the $\mathrm c_1$ and $\mathrm c_2$ lines stand for integrals over wave functions of the incoming and outgoing bound state.

\begin{figure}
\begin{center}
\psfrag{e}{$\mathrm e$                }
\psfrag{C}{$\mathrm C$          }
\psfrag{c1}{$\mathrm c_1$         }
\psfrag{c2}{$\mathrm c_2$          }
\psfrag{gamma}{\;$\gamma$           }
\end{center} 
\end{figure}
\begin{figure}
\begin{center}
\psfrag{e}{$\mathrm e$                }
\psfrag{C}{$\mathrm C$          }
\psfrag{c1}{$\mathrm c_1$         }
\psfrag{c2}{$\mathrm c_2$          }
\psfrag{gamma}{\quad $\gamma$           }
\includegraphics[clip=7cm,width=50mm]{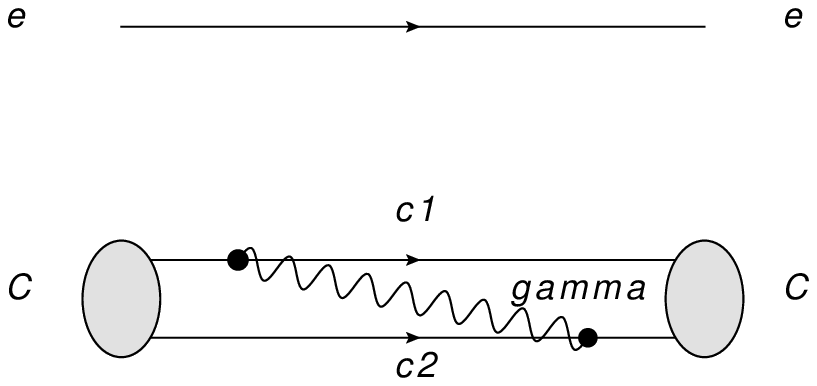}
\qquad\quad
\includegraphics[clip=7cm,width=50mm]{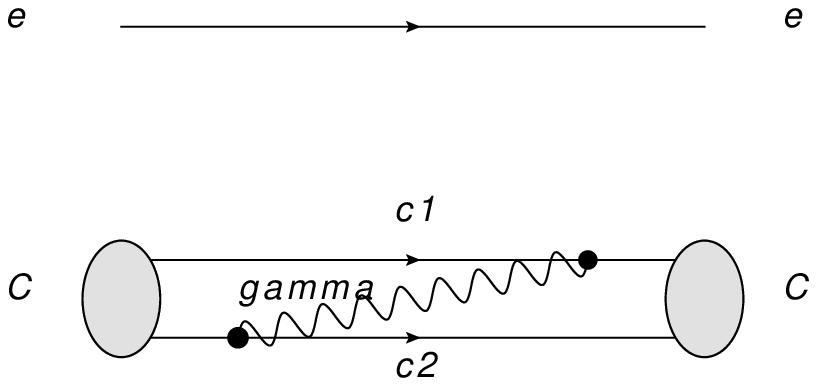}\vspace{6ex}\qquad\quad
\includegraphics[clip=7cm,width=50mm]{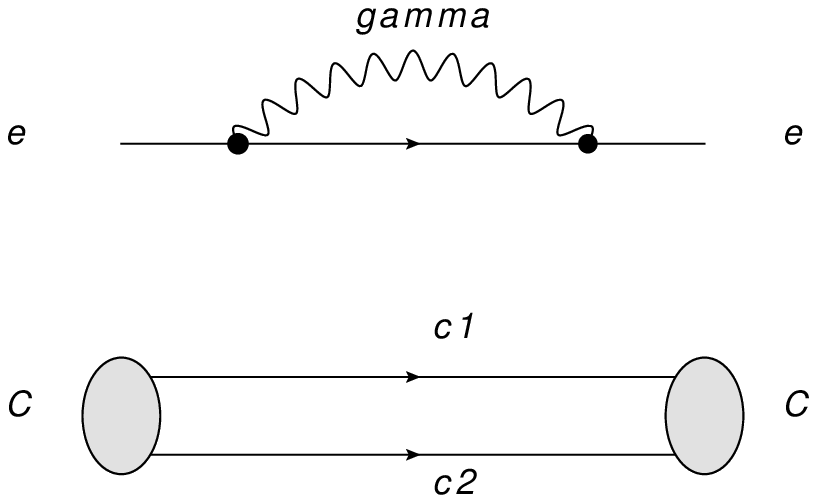}\qquad\quad
\includegraphics[clip=7cm,width=50mm]{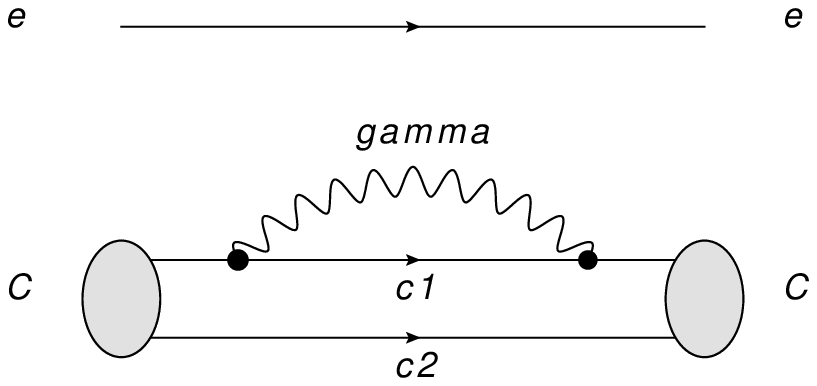}\vspace{6ex}
\includegraphics[clip=7cm,width=50mm]{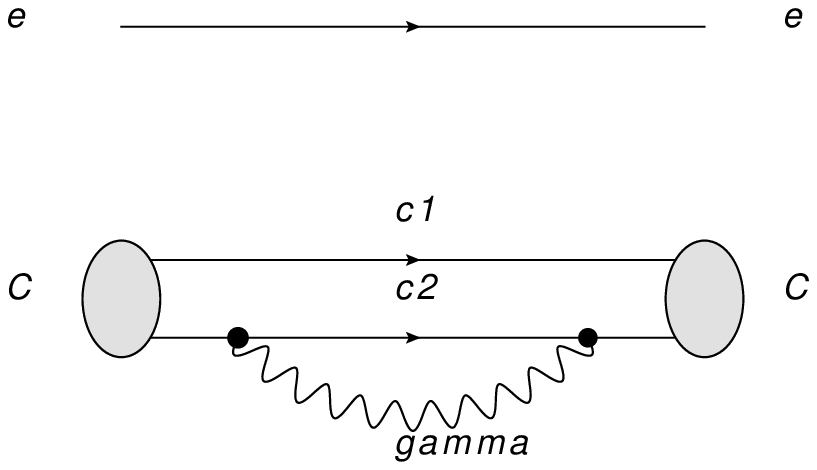}
\caption{\label{fig:grapheee}The graphical representation of the 5 self-energy contributions $\varGamma_{1\rightarrow2}$, $\varGamma_{2\rightarrow1}$, $\varGamma_{\mathrm e\rightarrow\mathrm e}$, $\varGamma_{1\rightarrow 1}$ and $\varGamma_{2\rightarrow 2}$ to the optical potential, where the photon is emitted and absorbed by the same particle.
}
\end{center} 
\end{figure}
We are interested only in the 4 contributions of Fig.~\ref{fig:graphec1}, since the remaining 5 of Fig.~\ref{fig:grapheee} are self-energy contributions in which the photon is emitted and absorbed by the same particle. The latter contribute to the renormalized masses of the electron and the bound state and will therefore be neglected. 
The first contribution of Fig.~\ref{fig:graphec1}, $\varGamma_{e\rightarrow1}$, where the photon is emitted by the electron and absorbed by constituent 1 is given by the rather lengthy expression
\begin{eqnarray} \label{eq:optpot2}
\varGamma_{e\rightarrow1}&=&(2\pi)^3
\int \frac{\mathrm d^3 V'}{V'^0}\frac{\mathrm d^3 k_{\mathrm e}'}{2 k_{\mathrm e}'^0}\frac{\mathrm d^3 k_{1}'}{2 k_{1}'^0}\frac{(k_{1}'^0+k_{2}'^0+k_{\mathrm e}'^0)^3}{2k_{2}'^0}\sum_{\mu_{\mathrm e}'\mu_1'\mu_2'}\sum_{\mu_{\mathrm e}'''\mu_1'''\mu_2'''\mu_\gamma'''}(-\mathrm g^{\mu_\gamma'''\mu_\gamma'''})
 \nonumber\\&&\times
\int \frac{\mathrm d^3 V'''}{V'''^0}\frac{\mathrm d^3 k_{\mathrm e}'''}{2 k_{\mathrm e}'''^0}\frac{\mathrm d^3 k_{1}'''}{2 k_{1}'''^0}\frac{\mathrm d^3 k_{2}'''}{2 k_{2}'''^0}\frac{(k_{1}'''^0+k_{2}'''^0+k_{\mathrm e}'''^0+k'''^0_\gamma)^3}{2k_{\gamma}'''^0}
 \nonumber\\&&\times 
\int \frac{\mathrm d^3 \underline{V}''}{\underline{V}''^0}\frac{\mathrm d^3 \underline{k}''_{\mathrm e}}{2 \underline{k}_{\mathrm e}''^0}\frac{\mathrm d^3 \underline{k}_{\gamma}''}{2 \underline{k}_{\gamma}''^0}\frac{(k_{\mathrm C}''^0+\underline{k}_{\mathrm e}''^0+\underline{k}''^0_\gamma)^3}{2k_{\mathrm C}''^0}\sum_{n''j''\mu_j''\underline{\mu}_{\mathrm e}''\underline{\mu}_\gamma''}
(-\mathrm g^{\underline{\mu}_\gamma''\underline{\mu}_\gamma''})
\nonumber\\&&\times
\int \frac{\mathrm d^3 V''}{V''^0}\frac{\mathrm d^3 k_{\mathrm e}''}{2 k_{\mathrm e}''^0}\frac{\mathrm d^3 k_{1}''}{2 k_{1}''^0}\frac{\mathrm d^3 k_{2}''}{2 k_{2}''^0}
\frac{(k_{1}''^0+k_{2}''^0+k_{\mathrm e}''^0+k''^0_\gamma)^3}{2k_{\gamma}''^0}\sum_{\mu_{\mathrm e}''\mu_1''\mu_2''\mu_\gamma''}(-\mathrm g^{\mu_\gamma''\mu_\gamma''})
 \nonumber\\&&\times
\int \frac{\mathrm d^3 V}{V^0}\frac{\mathrm d^3 k_{\mathrm e}}{2 k_{\mathrm e}^0}\frac{\mathrm d^3 k_{1}}{2 k_{1}^0}
\frac{(k_{1}^0+k_{2}^0+k_{\mathrm e}^0)^3}{2k_2^0}\sum_{\mu_{\mathrm e}\mu_1\mu_2}
 \nonumber\\&&\times
 \sqrt{\frac{2\tilde k'^0_12\tilde k'^0_2}{2(\tilde k'^0_1+\tilde k'^0_2)}}\sqrt{\frac{2 k'^0_{\mathrm C}2\underline{k}'^0_{\mathrm e}}{(k'^0_{\mathrm C}+\underline k'^0_{\mathrm e})^3}}
\sqrt{\frac{2 k'^0_{12}2k'^0_{\mathrm e}}{(k'^0_{12}+k'^0_{\mathrm e})^3}} V'^0\nonumber\\&&\times\delta ^3(\boldsymbol V'-\underline{\boldsymbol V}') 
\delta_{\underline{\mu}_\mathrm e'\mu_\mathrm e'} \delta^3\left(\underline{\boldsymbol k}_{\mathrm e}'-\boldsymbol k_{\mathrm e}'\right) 
\varPsi^\ast _{nj\mu_j'\mu_1'\mu_2'}(\tilde{\boldsymbol k}')
\nonumber\\&&\times
\frac{(-1)}{\sqrt{M_{\mathrm e 12}'^3 M_{\mathrm e 12\gamma}'''^3}}V'^0\delta^3(\boldsymbol V'-\boldsymbol V''') J_{1\mu}(\boldsymbol{k}_1',\mu_1';\boldsymbol{k}_1''',\mu_1''')
\nonumber\\&&\times
\epsilon_{\mu_{\gamma}'''}^\mu(\boldsymbol{k}_{\gamma}''')\, \delta_{\mu_{\mathrm e}'\mu_{\mathrm e}'''} 2 k_{\mathrm e}'^0
 \delta^3(\boldsymbol{k}_e' - \boldsymbol{k}_e''')  
 \delta_{\mu_2'\mu_2'''} 2  k_2'^0 \delta^3(\boldsymbol{k}_2' - \boldsymbol{k}_2''')\nonumber\\&&\times
 \sqrt{\frac{2\tilde k'''^0_12\tilde k'''^0_2}{2(\tilde k'''^0_1+\tilde k'''^0_2)}}\sqrt{\frac{2 k''^0_{\mathrm C}2\underline{k}''^0_{\mathrm e}2\underline{k}''^0_{\gamma}}{(k''^0_{\mathrm C}+\underline{k}''^0_{\mathrm e}+\underline{k}''^0_{\gamma})^3}}
\sqrt{\frac{2 k'''^0_{12}2k'''^0_{\mathrm e}2k'''^0_{\gamma}}{(k'''^0_{12}+k'''^0_{\mathrm e}+k'''^0_{\gamma})^3}}V''^0\nonumber\\&&\times\delta ^3(\underline{\boldsymbol V}''-\boldsymbol V''') 
\delta_{\underline{\mu}_\mathrm e''\mu_\mathrm e'''} \delta^3\left(\underline{\boldsymbol k}_{\mathrm e}''-\boldsymbol k_{\mathrm e}'''\right) 
(-\mathrm g_{\underline{\mu}_\gamma''\mu_\gamma'''}) \delta^3\left(\underline{\boldsymbol k}_{\gamma}''-\boldsymbol k_{\gamma}'''\right)\nonumber\\&&\times\varPsi _{n''j''\mu_j''\mu_1'''\mu_2'''}(\tilde{\boldsymbol k}''')
 \left( k_{\mathrm e}''^0+k_{\mathrm C}''^0+ k_{\gamma}''^0
-M\right)^{-1}\nonumber\\&&\times
\sqrt{\frac{2\tilde k''^0_12\tilde k''^0_2}{2(\tilde k''^0_1+\tilde k''^0_2)}}\sqrt{\frac{2 k''^0_{\mathrm C}2\underline{k}''^0_{\mathrm e}2\underline{k}''^0_{\gamma}}{(k''^0_{\mathrm C}+\underline{k}''^0_{\mathrm e}+\underline{k}''^0_{\gamma})^3}}
\sqrt{\frac{2 k''^0_{12}2k''^0_{\mathrm e}2k''^0_{\gamma}}{(k''^0_{12}+k''^0_{\mathrm e}+k''^0_{\gamma})^3}}V''^0\nonumber\\&&\times\delta ^3(\underline{\boldsymbol V}''-\boldsymbol V'') 
\delta_{\underline{\mu}_\mathrm e''\mu_\mathrm e''} \delta^3\left(\underline{\boldsymbol k}_{\mathrm e}''-\boldsymbol k_{\mathrm e}''\right) 
(-\mathrm g_{\underline{\mu}_\gamma''\mu_\gamma''}) \delta^3\left(\underline{\boldsymbol k}_{\gamma}''-\boldsymbol k_{\gamma}''\right)
 \nonumber\\&&\times\varPsi^\ast_{n''j''\mu_j''\mu_1''\mu_2''}(\tilde{\boldsymbol k}'')
\frac{(-1)}{\sqrt{M_{\mathrm e 12}^3 M_{\mathrm e 12\gamma}''^3}}V^0\delta^3(\boldsymbol V-\boldsymbol V'') 
\nonumber\\&&\times
|\,\mathrm e\,|\,Q_{\mathrm e}\,
\bar{u}_{\mu_{\mathrm e}''}(\boldsymbol{k}_{\mathrm e}^\prime)\gamma_\nu
u_{\mu_{\mathrm e}}(\boldsymbol{k}_e)\,
\epsilon_{\mu_{\gamma}''}^\nu(\boldsymbol{k}_{\gamma}'')\,  \delta_{\mu_1\mu_1''}2 k_1^0 \delta^3(\boldsymbol{k}_1'' - \boldsymbol{k}_1)   
\nonumber\\&&\times
 \delta_{\mu_2\mu_2''}2 k_2^0 \delta^3(\boldsymbol{k}_2'' - \boldsymbol{k}_2) \sqrt{\frac{2\tilde k^0_12\tilde k^0_2}{2(\tilde k^0_1+\tilde k^0_2)}}\sqrt{\frac{2 k^0_{\mathrm C}2\underline{k}^0_{\mathrm e}}{(k^0_{\mathrm C}+\underline k^0_{\mathrm e})^3}}
\sqrt{\frac{2 k^0_{12}2k^0_{\mathrm e}}{(k^0_{12}+k^0_{\mathrm e})^3}} V^0\nonumber\\&&\times\delta ^3(\boldsymbol V-\underline{\boldsymbol V}) 
\delta_{\underline{\mu}_\mathrm e\mu_\mathrm e} \delta^3\left(\underline{\boldsymbol k}_{\mathrm e}-\boldsymbol k_{\mathrm e}\right) 
\varPsi _{nj\mu_j\mu_1\mu_2}(\tilde{\boldsymbol k})\,.
\end{eqnarray}
The integrations over  $\boldsymbol V,\, \boldsymbol V',\, \boldsymbol V'',\, \boldsymbol V''',\, \underline {\boldsymbol V}'',\, \boldsymbol k_{\mathrm e},\,\boldsymbol k_{\mathrm e}',\,\boldsymbol k_{\mathrm e}'',\,\boldsymbol k_{\mathrm e}''',\,
\underline{\boldsymbol k}_{\mathrm e}'',\,\underline{\boldsymbol k}_{\gamma}'',\,\boldsymbol k_{\gamma}''',$ $\boldsymbol k_1'',\, \boldsymbol k_2'',\, \boldsymbol k_{2}'''$  together with the sums over $\mu_{\mathrm e}, \, \mu_{\mathrm e}',\, \mu_{\mathrm e}'',\, \mu_{\mathrm e}''',\,
\underline{ \mu}_{\mathrm e}'',\,\mu_\gamma''',\,\underline \mu_\gamma'',\,\mu_\gamma'',$ $\mu_{1}'',\,  \, \mu_{2}'', \, \mu_{2}''' $ can be carried out by means of the corresponding
Dirac delta functions and Kronecker deltas, respectively. Then the contribution simplifies to
\begin{eqnarray} 
\varGamma_{e\rightarrow1}&=&(2\pi)^3\underline V^0\delta ^3(\underline {\boldsymbol V}-\underline {\boldsymbol V}')
\nonumber\\&&\times\int \frac{\mathrm d^3 k_{1}'}{2 k_{1}'^0}\frac{1}{2k_{2}'^0}\sum_{\mu_1'\mu_2'}
 \frac{1}{2 k_{1}'''^0}\sum_{\mu_1'''}
 \frac{1}{2\vert\boldsymbol{k}_{\mathrm e}-\boldsymbol{k}_{\mathrm e}'\vert }\sum_{n''j''\mu_j''}
\int\frac{\mathrm d^3 k_{1}}{2 k_{1}^02k_{2}^0}\sum_{\mu_1\mu_2}
 \nonumber\\&&\times
 \sqrt{\frac{2\tilde k'^0_12\tilde k'^0_2}{2(\tilde k'^0_1+\tilde k'^0_2)}}\sqrt{\frac{2 k'^0_{12}2 k'^0_{\mathrm C}}{(k'^0_{\mathrm C}+\underline k'^0_{\mathrm e})^3}}
\varPsi^\ast _{nj\mu_j'\mu_1'\mu_2'}(\tilde{\boldsymbol k}')
\nonumber\\&&\times
 J_{1\nu}(\boldsymbol{k}_1',\mu_1';\boldsymbol{k}_1''',\mu_1''')
 \sqrt{\frac{2\tilde k'''^0_12\tilde k'''^0_2}{2(\tilde k'''^0_1+\tilde k'''^0_2)}}
\sqrt{2 k'''^0_{12}}\nonumber\\&&\times
\varPsi _{n''j''\mu_j''\mu_1'''\mu_2'}(\tilde{\boldsymbol k}''')
 \left( \underline{k}_{\mathrm e}'^0+k_{\mathrm C}''^0+ \vert\boldsymbol{k}_{\mathrm e}-\boldsymbol{k}_{\mathrm e}'\vert
-M\right)^{-1}\nonumber\\&&\times
2 k^0_{12}\varPsi^\ast_{n''j''\mu_j''\mu_1\mu_2}(\tilde{\boldsymbol k})
 (-\mathrm g^{\mu\nu})\,
|\,\mathrm e\,|\,Q_{\mathrm e}\,
\bar{u}_{\underline{\mu}_{\mathrm e}'}(\boldsymbol{k}_{\mathrm e}^\prime)\gamma_\mu
u_{\underline{\mu}_{\mathrm e}}(\boldsymbol{k}_e)\,
\nonumber\\&&\times
 \frac{2\tilde k^0_12\tilde k^0_2}{2(\tilde k^0_1+\tilde k^0_2)}\sqrt{\frac{2 k^0_{\mathrm C}}{(k^0_{\mathrm C}+\underline k^0_{\mathrm e})^3}}
\varPsi _{nj\mu_j\mu_1\mu_2}(\tilde{\boldsymbol k})\,.
\end{eqnarray}
Here we have used the completeness relation for the photon polarization vectors Eq.~(\ref{eq:polcomp}). Furthermore, the delta function $\delta^3(\boldsymbol{k}_2'' - \boldsymbol{k}_2)$ of Eq.~(\ref{eq:optpot2}) has been rewritten according to $\delta^3(\boldsymbol{k}_2'' - \boldsymbol{k}_2)=\delta^3(\boldsymbol{k}_\gamma'' - (\boldsymbol{k}_{\mathrm e}-\boldsymbol{k}_{\mathrm e}'))$
by using $\underline {\boldsymbol k}_{\mathrm e}'+\underline {\boldsymbol k}_{\gamma}'+\underline {\boldsymbol k}_{1}=-\underline {\boldsymbol k}_{2}''$ and
$\underline {\boldsymbol k}_{\mathrm e}+\underline {\boldsymbol k}_{1}=-\underline {\boldsymbol k}_{2}$. For better readability we have kept $\boldsymbol k_1'''=\boldsymbol k_1'-\underline{\boldsymbol k}_\mathrm e+\underline{\boldsymbol k}_\mathrm e'$ and $\boldsymbol k_2'''=\boldsymbol k_2'$. In Ref.~\cite{Fuchsberger:2007} the Jacobian for the variable transformation between the integration measures $\mathrm d^3 k_1$ and $\mathrm d^3\tilde  k_1$ has been derived:
 \begin{eqnarray} \label{eq:momchangekktilde}
  \mathrm d^3 k_1=\mathrm d^3\tilde  k_1 \frac{2 k_1^02k_2^0}{2 \tilde k_1^02\tilde k_2^0}\frac{2( \tilde k_1^0+\tilde k_2^0)}{2( k_1^0+k_2^0)}\,.
 \end{eqnarray}
Subsequently, the integration over $\mathrm d^3\tilde  k_1$ can be carried out with the help of the normalization condition
of the wave function, Eq.~(\ref{eq:normeigenfucntions1}). For the special case of a pure central potential as used, e.g., in Ref.~\cite{Krassnigg:2003gh}, the normalization condition can be proved with the help of the appropriate orthogonality relations of the Wigner $D$-functions, the Clebsch-Gordan coefficients and the spherical harmonics, Eqs.~(\ref{eq:D-functdelta}),~(\ref{eq:cgortho}) and~(\ref{eq:sphharmotho}):  
\begin{eqnarray}\lefteqn{
 \int \mathrm d^3\tilde  k_1 \sum_{\mu_1\mu_2} \varPsi^\ast_{n''j''\mu_j''(l''s'')\mu_1\mu_2}(\tilde{\boldsymbol k})\varPsi _{nj\mu_j(ls)\mu_1\mu_2}(\tilde{\boldsymbol k})}
 \nonumber\\&=&
\int \mathrm d^3\tilde  k_1 
\sum_{\mu_l''\mu_s''\tilde \mu_1''\tilde \mu_2''\mu_l\mu_s\tilde \mu_1\tilde \mu_2}
Y^\ast_{l''\mu_l''}(\hat{\tilde{\boldsymbol{ k}}})
C^{s''\mu_s''}_{j_1\tilde\mu_1''j_2\tilde \mu_2''}C^{j''\mu_j''}_{l''\mu_l''s''\mu_s''} u^\ast_{n''l''}(\tilde{k})\nonumber\\&&\times
Y_{l\mu_l}(\hat{\tilde{\boldsymbol{ k}}})
C^{s\mu_s}_{j_1\tilde\mu_1j_2\tilde \mu_2}C^{j\mu_j}_{l\mu_ls\mu_s} u_{nl}(\tilde{k})
\nonumber\\&&\times
 \underbrace{\sum_{\mu_1} D^{j_1}_{\tilde \mu_1''\mu_1}\left[\underline R_\mathrm{W_{\!c}}^{-1}\left(\tilde w_1,B_{\mathrm c}(\boldsymbol w_{12})\right)\right]
D^{j_1}_{\mu_1\tilde \mu_1}\left[\underline R_\mathrm{W_{\!c}}\left(\tilde w_1,B_{\mathrm c}(\boldsymbol w_{12})\right)\right]}_{=\delta_{\tilde \mu_1\tilde \mu_1''},\text{ Eq.~(\ref{eq:D-functdelta})}}\nonumber\\&&\times
\underbrace{\sum_{\mu_2}
D^{j_2}_{\tilde \mu_2''\mu_2}\left[\underline R_\mathrm{W_{\!c}}^{-1}\left(\tilde w_2,B_{\mathrm c}(\boldsymbol w_{12})\right)\right]
D^{j_2}_{\mu_2\tilde \mu_2}\left[\underline R_\mathrm{W_{\!c}}\left(\tilde w_2,B_{\mathrm c}(\boldsymbol w_{12})\right)\right]}_{=\delta_{\tilde \mu_2\tilde \mu_2''},\text{ Eq.~(\ref{eq:D-functdelta})}}\nonumber
\end{eqnarray}
\begin{eqnarray}&
=&\int \mathrm d\tilde  k\, \tilde  k^2 
\sum_{\mu_l''\mu_s''\mu_l\mu_s}
\underbrace{\sum_{\tilde \mu_1\tilde \mu_2}C^{s''\mu_s''}_{j_1\tilde\mu_1j_2\tilde \mu_2}C^{s\mu_s}_{j_1\tilde\mu_1j_2\tilde \mu_2}}_{=\delta_{ss''}\delta_{\mu_s\mu_s''}\,, \text{ Eq.~(\ref{eq:cgortho})}}C^{j''\mu_j''}_{l''\mu_l''s''\mu_s''} u^\ast_{n''l''}(\tilde{k})\nonumber\\&&\times
\underbrace{\int \mathrm d\Omega(\hat{\tilde{\boldsymbol{ k}}})Y^\ast_{l''\mu_l''}(\hat{\tilde{\boldsymbol{ k}}})Y_{l\mu_l}(\hat{\tilde{\boldsymbol{ k}}})}_{=\delta_{ll''}\delta_{\mu_l\mu_l''}\,, \text{ Eq.~(\ref{eq:sphharmotho})}}
C^{j\mu_j}_{l\mu_ls\mu_s} u_{nl}(\tilde{k})
\nonumber\\&=&
\sum_{\mu_l\mu_s}
C^{j''\mu_j''}_{l\mu_ls\mu_s} C^{j\mu_j}_{l\mu_ls\mu_s}\int \mathrm d\tilde  k\,\tilde  k^2 u^\ast_{n''l}(\tilde{k})
 u_{nl}(\tilde{k})
\nonumber\\&=&\delta_{jj''}\delta_{\mu_j\mu_j''} \delta_{nn''}\,.
\end{eqnarray}
Here we have assumed that the radial wave functions are normalized to unity:
\begin{eqnarray}
\int \mathrm d\tilde  k\,\tilde  k^2 u^\ast_{n''l}(\tilde{k})
 u_{nl}(\tilde{k})=\delta_{nn''}\,.
\end{eqnarray}
Finally we denote all incoming momenta and spins by unprimed symbols and all outgoing momenta and spins by primed symbols. Then the result reads
\begin{eqnarray}
\varGamma_{e\rightarrow1}&=&(2\pi)^3\underline V^0\delta ^3(\underline {\boldsymbol V}-\underline {\boldsymbol V}')\frac{1}{2\vert\boldsymbol{k}_{\mathrm e}-\boldsymbol{k}_{\mathrm e}'\vert }
\sqrt{\frac{2 k'^0_{\mathrm C}}{(k'^0_{\mathrm C}+\underline k'^0_{\mathrm e})^3}}\sqrt{\frac{2 k^0_{\mathrm C}}{(k^0_{\mathrm C}+\underline k^0_{\mathrm e})^3}}
 \nonumber\\&&\times\left( \underline{k}_{\mathrm e}'^0+k_{\mathrm C}^0+ \vert\boldsymbol{k}_{\mathrm e}-\boldsymbol{k}_{\mathrm e}'\vert
-M\right)^{-1}\nonumber\\&&\times
\int \frac{\mathrm d^3 k_{1}'}{2 k_{1}'^0}\frac{1}{2k_{2}'^0}\frac{1}{2 k_{1}^0}\nonumber\\&&\times
\sqrt{2 k'^0_{12}}\sqrt{2 k^0_{12}}
\sqrt{\frac{2\tilde k'^0_12\tilde k'^0_2}{2(\tilde k'^0_1+\tilde k'^0_2)}}\sqrt{\frac{2\tilde k^0_12\tilde k^0_2}{2(\tilde k^0_1+\tilde k^0_2)}}
\nonumber\\&&\times\sum_{\mu_1\mu_1'\mu_2'}\varPsi^\ast _{nj\mu_j'\mu_1'\mu_2'}(\tilde{\boldsymbol k}')\varPsi _{nj\mu_j\mu_1\mu_2'}(\tilde{\boldsymbol k})
\nonumber\\&&\times
 J_{1\nu}(\boldsymbol{k}_1',\mu_1';\boldsymbol{k}_1,\mu_1) 
(-\mathrm g^{\mu\nu})
\,|\,\mathrm e\,|\,Q_{\mathrm e}\,
\bar{u}_{\underline{\mu}_{\mathrm e}'}(\boldsymbol{k}_{\mathrm e}^\prime)\gamma_\mu
u_{\underline{\mu}_{\mathrm e}}(\boldsymbol{k}_e)\,
.
\end{eqnarray}
In this expression the momenta satisfy the relations $\boldsymbol k_1=\boldsymbol k_1'-(\boldsymbol k_{\mathrm e}-\boldsymbol k_{\mathrm e}')$ since 
$\boldsymbol k_2'=\boldsymbol k_2\Rightarrow-\boldsymbol k_1'-\underline{\boldsymbol k}_{\mathrm e}'=-\boldsymbol k_1-\underline{\boldsymbol k}_{\mathrm e}'-\underline{\boldsymbol k}_{\gamma}''$.
The other three contributions from Fig.~\ref{fig:graphec1} are obtained in an analogous manner and will not be calculated here explicitly. They read
\begin{eqnarray}
\varGamma_{e\rightarrow2}&=&(2\pi)^3\underline V^0\delta ^3(\underline {\boldsymbol V}-\underline {\boldsymbol V}')\frac{1}{2\vert\boldsymbol{k}_{\mathrm e}-\boldsymbol{k}_{\mathrm e}'\vert }
\sqrt{\frac{2 k'^0_{\mathrm C}}{(k'^0_{\mathrm C}+\underline k'^0_{\mathrm e})^3}}\sqrt{\frac{2 k^0_{\mathrm C}}{(k^0_{\mathrm C}+\underline k^0_{\mathrm e})^3}}
 \nonumber\\&&\times\left( \underline{k}_{\mathrm e}'^0+k_{\mathrm C}^0+ \vert\boldsymbol{k}_{\mathrm e}-\boldsymbol{k}_{\mathrm e}'\vert
-M\right)^{-1}\nonumber\\&&\times
\int \frac{\mathrm d^3 k_{2}'}{2 k_{1}'^0}\frac{1}{2k_{2}'^0}\frac{1}{2 k_{2}^0}\nonumber\\&&\times
\sqrt{2 k'^0_{12}}\sqrt{2 k^0_{12}}
\sqrt{\frac{2\tilde k'^0_12\tilde k'^0_2}{2(\tilde k'^0_1+\tilde k'^0_2)}}\sqrt{\frac{2\tilde k^0_12\tilde k^0_2}{2(\tilde k^0_1+\tilde k^0_2)}}
\nonumber\\&&\times\sum_{\mu_2\mu_1'\mu_2'}\varPsi^\ast _{nj\mu_j'\mu_1'\mu_2'}(\tilde{\boldsymbol k}')\varPsi _{nj\mu_j\mu_1'\mu_2}(\tilde{\boldsymbol k})
\nonumber\\&&\times
 J_{2\nu}(\boldsymbol{k}_2',\mu_2';\boldsymbol{k}_2,\mu_2) 
(-\mathrm g^{\mu\nu})
\,|\,\mathrm e\,|\,Q_{\mathrm e}\,
\bar{u}_{\underline{\mu}_{\mathrm e}'}(\boldsymbol{k}_{\mathrm e}^\prime)\gamma_\mu
u_{\underline{\mu}_{\mathrm e}}(\boldsymbol{k}_e)\,
\end{eqnarray}
where $\boldsymbol k_2=\boldsymbol k_2'-(\boldsymbol k_{\mathrm e}-\boldsymbol k_{\mathrm e}')$ and $ \boldsymbol k_1'=\boldsymbol k_1$,
\begin{eqnarray} 
\varGamma_{1\rightarrow\mathrm e}&=&(2\pi)^3\underline V^0\delta ^3(\underline {\boldsymbol V}-\underline {\boldsymbol V}')\frac{1}{2\vert\boldsymbol{k}_{\mathrm e}'-\boldsymbol{k}_{\mathrm e}\vert }
\sqrt{\frac{2 k'^0_{\mathrm C}}{(k'^0_{\mathrm C}+\underline k'^0_{\mathrm e})^3}}\sqrt{\frac{2 k^0_{\mathrm C}}{(k^0_{\mathrm C}+\underline k^0_{\mathrm e})^3}}
 \nonumber\\&&\times\left( \underline{k}_{\mathrm e}^0+k_{\mathrm C}'^0+ \vert\boldsymbol{k}_{\mathrm e}'-\boldsymbol{k}_{\mathrm e}\vert
-M\right)^{-1}\nonumber\\&&\times
\int \frac{\mathrm d^3 k_{1}}{2 k_{1}^0}\frac{1}{2k_{2}^0}\frac{1}{2 k_{1}'^0}\nonumber\\&&\times
\sqrt{2 k'^0_{12}}\sqrt{2 k^0_{12}}
\sqrt{\frac{2\tilde k'^0_12\tilde k'^0_2}{2(\tilde k'^0_1+\tilde k'^0_2)}}\sqrt{\frac{2\tilde k^0_12\tilde k^0_2}{2(\tilde k^0_1+\tilde k^0_2)}}
\nonumber\\&&\times\sum_{\mu_1'\mu_1\mu_2}\varPsi^\ast _{nj\mu_j'\mu_1'\mu_2}(\tilde{\boldsymbol k}')\varPsi _{nj\mu_j\mu_1\mu_2}(\tilde{\boldsymbol k})
\nonumber\\&&\times
 J_{1\nu}(\boldsymbol{k}_1',\mu_1';\boldsymbol{k}_1,\mu_1) 
(-\mathrm g^{\mu\nu})
\,|\,\mathrm e\,|\,Q_{\mathrm e}\,
\bar{u}_{\underline{\mu}_{\mathrm e}'}(\boldsymbol{k}_{\mathrm e}^\prime)\gamma_\mu
u_{\underline{\mu}_{\mathrm e}}(\boldsymbol{k}_e)\,
\end{eqnarray}
where $\boldsymbol k_1'=\boldsymbol k_1-(\boldsymbol k_{\mathrm e}'-\boldsymbol k_{\mathrm e})$ and $ \boldsymbol k_2'=\boldsymbol k_2$ and finally 
\begin{eqnarray}
\varGamma_{2\rightarrow\mathrm e}&=&(2\pi)^3\underline V^0\delta ^3(\underline {\boldsymbol V}-\underline {\boldsymbol V}')\frac{1}{2\vert\boldsymbol{k}_{\mathrm e}'-\boldsymbol{k}_{\mathrm e}\vert }
\sqrt{\frac{2 k'^0_{\mathrm C}}{(k'^0_{\mathrm C}+\underline k'^0_{\mathrm e})^3}}\sqrt{\frac{2 k^0_{\mathrm C}}{(k^0_{\mathrm C}+\underline k^0_{\mathrm e})^3}}
 \nonumber\\&&\times\left( \underline{k}_{\mathrm e}^0+k_{\mathrm C}'^0+ \vert\boldsymbol{k}_{\mathrm e}'-\boldsymbol{k}_{\mathrm e}\vert
-M\right)^{-1}\nonumber\\&&\times
\int \frac{\mathrm d^3 k_{2}}{2 k_{1}^0}\frac{1}{2k_{2}^0}\frac{1}{2 k_{2}'^0}\nonumber\\&&\times
\sqrt{2 k'^0_{12}}\sqrt{2 k^0_{12}}
\sqrt{\frac{2\tilde k'^0_12\tilde k'^0_2}{2(\tilde k'^0_1+\tilde k'^0_2)}}\sqrt{\frac{2\tilde k^0_12\tilde k^0_2}{2(\tilde k^0_1+\tilde k^0_2)}}
\nonumber\\&&\times\sum_{\mu_2'\mu_1\mu_2}\varPsi^\ast _{nj\mu_j'\mu_1\mu_2'}(\tilde{\boldsymbol k}')\varPsi _{nj\mu_j\mu_1\mu_2}(\tilde{\boldsymbol k})
\nonumber\\&&\times
 J_{2\nu}(\boldsymbol{k}_2',\mu_2';\boldsymbol{k}_2,\mu_2) 
(-\mathrm g^{\mu\nu})
\,|\,\mathrm e\,|\,Q_{\mathrm e}\,
\bar{u}_{\underline{\mu}_{\mathrm e}'}(\boldsymbol{k}_{\mathrm e}^\prime)\gamma_\mu
u_{\underline{\mu}_{\mathrm e}}(\boldsymbol{k}_e)\,
\end{eqnarray}
where $\boldsymbol k_2'=\boldsymbol k_2-(\boldsymbol k_{\mathrm e}'-\boldsymbol k_{\mathrm e})$ and $ \boldsymbol k_1'=\boldsymbol k_1$.\\
\section{Combining the Time Orderings}
The next step towards a simplification of the sum of the four contributions we have obtained above is to combine the 2 possible time orderings for each exchange process to one covariant contribution. 
To this end we use 
for $\varGamma_{1\rightarrow\mathrm e}$ the spectator condition $k_2^0=k_2'^0$ and the equality of integration measures $\mathrm d^3 k_1=\mathrm d^3 k_1'$ since $\boldsymbol k_{\mathrm e}'-\boldsymbol k_{\mathrm e}$ is a constant. Moreover, we can simply rename the spin indices that are summed over according to the spectator condition $\mu_2=\mu_2'$ and by the replacement  $\mu_1\leftrightarrow\mu_1'$.
These rewritings make $\varGamma_{1\rightarrow\mathrm e}$ easier comparable with $\varGamma_{\mathrm e\rightarrow1}$. An analogous manipulation applies to $\varGamma_{2\rightarrow\mathrm e}$. Consequently, after neglecting the self-energy contributions, the optical potential, Eq.~(\ref{eq:optpottotal}), becomes
\begin{eqnarray} \label{eq:1gammaamplitres1}
\lefteqn{
 \varGamma_{\mathrm e\rightarrow1}+\varGamma_{1\rightarrow\mathrm e}+\varGamma_{\mathrm e\rightarrow2}+\varGamma_{2\rightarrow \mathrm e}}\nonumber\\&=&
(2\pi)^3\underline V^0\delta ^3(\underline {\boldsymbol V}-\underline {\boldsymbol V}')\frac{1}{2\vert\boldsymbol{k}_{\mathrm e}-\boldsymbol{k}_{\mathrm e}'\vert }
\sqrt{\frac{2 k'^0_{\mathrm C}}{(k'^0_{\mathrm C}+\underline k'^0_{\mathrm e})^3}}\sqrt{\frac{2 k^0_{\mathrm C}}{(k^0_{\mathrm C}+\underline k^0_{\mathrm e})^3}}
 \nonumber\\&&\times\left[\left( \underline{k}_{\mathrm e}'^0+k_{\mathrm C}^0+ \vert\boldsymbol{k}_{\mathrm e}-\boldsymbol{k}_{\mathrm e}'\vert
-M\right)^{-1}+ \left( \underline{k}_{\mathrm e}^0+k_{\mathrm C}'^0+ \vert\boldsymbol{k}'_{\mathrm e}-\boldsymbol{k}_{\mathrm e}\vert
-M\right)^{-1}\right]\nonumber\\&&\times
\sum_{\mu_1'\mu_2'}\left[\int \frac{\mathrm d^3 k_{1}'}{2 k_{1}'^0}\frac{1}{2k_{2}'^0}\frac{1}{2 k_{1}^0}\sqrt{\frac{2\tilde k'^0_12\tilde k'^0_2}{2(\tilde k'^0_1+\tilde k'^0_2)}}\sqrt{\frac{2\tilde k^0_12\tilde k^0_2}{2(\tilde k^0_1+\tilde k^0_2)}}
\sqrt{2 k'^0_{12}}\sqrt{2 k^0_{12}}\right.\nonumber\\&&\;\;\;\;\times\left.
\sum_{\mu_1}\varPsi^\ast _{nj\mu_j'\mu_1'\mu_2'}(\tilde{\boldsymbol k}')\varPsi _{nj\mu_j\mu_1\mu_2'}(\tilde{\boldsymbol k})
 J_{1\nu}(\boldsymbol{k}_1',\mu_1';\boldsymbol{k}_1,\mu_1)\right.\nonumber\\&&\;\;+\left.
\int \frac{\mathrm d^3 k_{2}'}{2 k_{2}'^0}\frac{1}{2k_{1}'^0}\frac{1}{2 k_{2}^0}\sqrt{\frac{2\tilde k'^0_12\tilde k'^0_2}{2(\tilde k'^0_1+\tilde k'^0_2)}}\sqrt{\frac{2\tilde k^0_12\tilde k^0_2}{2(\tilde k^0_1+\tilde k^0_2)}}
\sqrt{2 k'^0_{12}}\sqrt{2 k^0_{12}}\right.\nonumber\\&&\;\;\;\;\times\left.
\sum_{\mu_2}\varPsi^\ast _{nj\mu_j'\mu_1'\mu_2'}(\tilde{\boldsymbol k}')\varPsi _{nj\mu_j\mu_1'\mu_2}(\tilde{\boldsymbol k})
 J_{2\nu}(\boldsymbol{k}_2',\mu_2';\boldsymbol{k}_2,\mu_2) 
\right]\nonumber\\&&\times
(-\mathrm g^{\mu\nu})
\,|\,\mathrm e\,|\,Q_{\mathrm e}\,
\bar{u}_{\underline{\mu}_{\mathrm e}'}(\boldsymbol{k}_{\mathrm e}^\prime)\gamma_\mu
u_{\underline{\mu}_{\mathrm e}}(\boldsymbol{k}_e)\,.
\end{eqnarray}
The two terms in the first square brackets correspond to the two possible time orderings, i.e. emission of the photon with subsequent absorption by one of the constituents and vice versa. Since we consider on-shell matrix elements, cf. Eq.~(\ref{eq:onshell}), both time orderings of the photon exchange can be combined to give the covariant photon propagator:
\begin{eqnarray}\label{eq:covphotonprop}&&
\left( \underline{k}_{\mathrm e}'^0+k_{\mathrm C}^0+ \vert\boldsymbol{k}_{\mathrm e}-\boldsymbol{k}_{\mathrm e}'\vert
-M\right)^{-1}+ \left( \underline{k}_{\mathrm e}^0+k_{\mathrm C}'^0+ \vert\boldsymbol{k}'_{\mathrm e}-\boldsymbol{k}_{\mathrm e}\vert
-M\right)^{-1}\nonumber\\&&\;\;\;\;=
\frac{2 \vert\boldsymbol{k}'_{\mathrm e}-\boldsymbol{k}_{\mathrm e}\vert}
{(k_{\mathrm e}^0-k_{\mathrm e}'^0)(k_{\mathrm C}^0-k_{\mathrm C}'^0)+(\boldsymbol{k}_{\mathrm e}-\boldsymbol{k}_{\mathrm e}')^2 }\nonumber\\&&\;\;\;\;=
\frac{2 \vert\boldsymbol{k}'_{\mathrm e}-\boldsymbol{k}_{\mathrm e}\vert}
{Q^2}\,,
\end{eqnarray}
with the four-momentum transfer to the bound-state (or equivalently of the electron) given by
\begin{eqnarray}
 q:=\left(
\begin{array}{c}
k_{\mathrm C}'^0-k_{\mathrm C}^0\\
\boldsymbol k_{12}'-\boldsymbol k_{12}
    \end{array}\right)\equiv \left(
\begin{array}{c}
k_{\mathrm e}^0-k_{\mathrm e}'^0\\
\boldsymbol k_{\mathrm e}-\boldsymbol k_{\mathrm e}'
    \end{array}\right)\quad \text{and}\quad Q^2:=-q^\mu q_\mu\,.
    \end{eqnarray}
Note that the three-momentum of the photon is $\boldsymbol{k}_\gamma^{\prime\prime}=\pm
(\boldsymbol{k}_\mathrm e^\prime - \boldsymbol{k}_\mathrm e)=\pm\boldsymbol q$. Here the sign depends on the time ordering whereas $k_\gamma^{0\prime\prime}=
\vert\,(\boldsymbol{k}_\mathrm e^\prime - \boldsymbol{k}_\mathrm e) \vert=Q\neq q^0$ is independent of the
time ordering. This modification further simplifies Eq.~(\ref{eq:1gammaamplitres1}), which then finally takes the form given by Eq.~(\ref{eq:1gammaamplitres}).
\chapter[Extraction of the Electromagnetic Form Factors]{Extraction of the Electromagnetic\\ Form Factors}\label{app:F}
 \section{Current Matrix Elements}\label{app:ifinitemomeframe}
Similarly as in the pseudoscalar case we extract the electromagnetic form factors of a vector bound state, $f_1$, $f_2$ and $g_\mathrm M$, from Eq.~(\ref{eq:currentphneospin1}) in the limit $k \rightarrow
\infty$, where the form factors become independent of $k$. First we shall
analyze the behavior of the current~(\ref{eq:currentphneospin1}) in this limit. For better readability we use the following abbreviations for the covariants multiplying the form factors. The covariants associated with the physical form factors $f_1$, $f_2$ and $g_\mathrm M$ are defined by
\begin{eqnarray}
  L^\mu_1(\mu_j',\mu_j)&:=&\epsilon^\ast_{\mu_j'}(\boldsymbol k'_\mathrm C)\cdot\epsilon_{\mu_j}(\boldsymbol k_\mathrm C)K_\mathrm C^\mu\,,\\
L^\mu_2(\mu_j',\mu_j)&:=&\frac{[\epsilon^\ast_{\mu'_j}(\boldsymbol k'_\mathrm C)\cdot q][\epsilon_{\mu_j}(\boldsymbol k_\mathrm C)\cdot q]}{2m_{\mathrm C}^2 }K_\mathrm C^\mu\,,\\
L^\mu_\mathrm M(\mu_j',\mu_j)&:=&\left\lbrace\epsilon^{\mu\ast}_{\mu'_j}(\boldsymbol k'_\mathrm C)[\epsilon_{\mu_j}(\boldsymbol k_\mathrm C)\cdot q]-
 \epsilon_{\mu_j}^\mu (\boldsymbol k_\mathrm C)[\epsilon^\ast_{\mu'_j}(\boldsymbol k'_\mathrm C)\cdot q]\right\rbrace\,,
\end{eqnarray}
respectively.
The $K_\mathrm e$-dependent spurious covariants which are multiplied with the spurious form factors $b_1,\ldots,b_8$ are denoted by
 \begin{eqnarray}
C^\mu_1(\mu_j',\mu_j)&:=& \epsilon^\ast_{\mu_j'}(\boldsymbol k'_\mathrm C)\cdot\epsilon_{\mu_j}(\boldsymbol k_\mathrm C)\frac{ m_{\mathrm C}^2K_\mathrm e^\mu}{K_\mathrm e\cdot K_\mathrm C} 
\,,\\
C^\mu_2(\mu_j',\mu_j)&:=&[\epsilon^\ast_{\mu'_j}(\boldsymbol k'_\mathrm C)\cdot q]
[\epsilon_{\mu_j}(\boldsymbol k_\mathrm C)\cdot q]\frac{K_\mathrm e^\mu}{K_\mathrm e\cdot K_\mathrm C}\,,\\ 
C^\mu_3(\mu_j',\mu_j)&:=&4 \left\lbrace[\epsilon^\ast_{\mu'_j}(\boldsymbol k'_\mathrm C)\cdot K_\mathrm e][\epsilon_{\mu_j}(\boldsymbol k_\mathrm C)\cdot K_\mathrm e]\right\rbrace
\frac{ m_{\mathrm C}^4K_\mathrm e^\mu}{(K_\mathrm e\cdot K_\mathrm C)^3} \,,\\
C^\mu_4(\mu_j',\mu_j)&:=&\left\lbrace[\epsilon^\ast_{\mu_j'}(\boldsymbol k'_\mathrm C)\cdot q][\epsilon_{\mu_j}(\boldsymbol k_\mathrm C)\cdot K_\mathrm e]- 
[\epsilon^\ast_{\mu_j'}(\boldsymbol k'_\mathrm C)\cdot K_\mathrm e][\epsilon_{\mu_j}(\boldsymbol k_\mathrm C)\cdot q] \right\rbrace\frac{ m_{\mathrm C}^2K_\mathrm e^\mu}{(K_\mathrm e\cdot K_\mathrm C)^2} 
 \nonumber\,,\\\\
C^\mu_5(\mu_j',\mu_j)&:=&\frac{[\epsilon^\ast_{\mu'_j}(\boldsymbol k'_\mathrm C)\cdot K_\mathrm e][\epsilon_{\mu_j}(\boldsymbol k_\mathrm C)\cdot K_\mathrm e]}{(K_\mathrm e\cdot K_\mathrm C)^2}4m_{\mathrm C}^2K_\mathrm C^\mu
\,,\\C^\mu_6(\mu_j',\mu_j)&:=&\frac{[\epsilon^\ast_{\mu_j'}(\boldsymbol k'_\mathrm C)\cdot q][\epsilon_{\mu_j}(\boldsymbol k_\mathrm C)\cdot K_\mathrm e]- 
[\epsilon^\ast_{\mu_j'}(\boldsymbol k'_\mathrm C)\cdot K_\mathrm e][\epsilon_{\mu_j}(\boldsymbol k_\mathrm C)\cdot q] }{K_\mathrm e\cdot K_\mathrm C} K_\mathrm C^\mu\,,\\
C^\mu_7(\mu_j',\mu_j)&:=& 2 m_{\mathrm C}^2\frac{\epsilon^{\mu\ast}_{\mu'_j}(\boldsymbol k'_\mathrm C)[\epsilon_{\mu_j}(\boldsymbol k_\mathrm C)\cdot K_\mathrm e]+
\epsilon_{\mu_j}^\mu (\boldsymbol k_\mathrm C)[\epsilon^\ast_{\mu'_j}(\boldsymbol k'_\mathrm C)\cdot K_\mathrm e]}{K_\mathrm e\cdot K_\mathrm C} \,,\\
C^\mu_8(\mu_j',\mu_j)&:=&\frac{[\epsilon^\ast_{\mu_j'}(\boldsymbol k'_\mathrm C)\cdot q][\epsilon_{\mu_j}(\boldsymbol k_\mathrm C)\cdot K_\mathrm e]+ 
[\epsilon^\ast_{\mu_j'}(\boldsymbol k'_\mathrm C)\cdot K_\mathrm e][\epsilon_{\mu_j}(\boldsymbol k_\mathrm C)\cdot q] }{K_\mathrm e\cdot K_\mathrm C}q^\mu\,.
\end{eqnarray}
Using our standard kinematics introduced in Sec.~\ref{sec:kinematics} the non-vanishing covariants in the limit $k \rightarrow
\infty$ are given by
\begin{eqnarray}
 L^\mu_2(1,-1)&\stackrel{k \rightarrow
\infty}{\longrightarrow}&-\frac{Q^2}{4m_{\mathrm C}^2}2k(1,0,0,1)^\mu\,,\label{eq:L21-1}\\
L^\mu_\mathrm M(1,-1)&\stackrel{k \rightarrow
\infty}{\longrightarrow}&-\frac{Q^2}{2m_{\mathrm C}}(1,0,0,1)^\mu\,,\label{eq:LM1-1}
\\
L^\mu_1(1,0)&\stackrel{k \rightarrow
\infty}{\longrightarrow}&\left(-\frac{Q (4 (k - m_{\mathrm C}) m_{\mathrm C} + Q^2)}{2 \sqrt{2} m_{\mathrm C}^2}, 0, 0, -\frac{Q (4 (k - m_{\mathrm C}) m_{\mathrm C} + Q^2)}{2 \sqrt{2} m_{\mathrm C}^2}\right)^\mu\,,\nonumber\\\\
L^\mu_2(1,0)&\stackrel{k \rightarrow
\infty}{\longrightarrow}&\left(\frac{4 (k - m_{\mathrm C}) m_\mathrm C Q^3 + Q^5}{8 \sqrt2 m_{\mathrm C}^4}, 0, 0, \frac{4 (k - m_{\mathrm C}) m_{\mathrm C} Q^3 + Q^5}{8 \sqrt2 m_{\mathrm C}^4}\right)^\mu\,,\nonumber\\
\\
L^\mu_\mathrm M(1,0)&\stackrel{k \rightarrow
\infty}{\longrightarrow}&
\left(-\frac{Q (2 k\, m_{\mathrm C}+ Q^2)}{2 \sqrt2 m_{\mathrm C}^2}, 0,\frac{\mathrm i\, Q^2}{2 \sqrt2 m_{\mathrm C}},  -\frac{Q (2 k\, m_{\mathrm C}+ Q^2)}{2 \sqrt2 m_{\mathrm C}^2}\right)^\mu\,,\\C^\mu_5(1,0)&\stackrel{k \rightarrow
\infty}{\longrightarrow}&-\frac{Q}{\sqrt2}\left(1, 0, 0, 1\right)^\mu\,,\\
C^\mu_6(1,0)&\stackrel{k \rightarrow
\infty}{\longrightarrow}&\frac{Q (2 k\, m_{\mathrm C} + Q^2)}{2 \sqrt{2} m_{\mathrm C}^2}\left(1, 0, 0, 1\right)^\mu\,,\\
\label{eq:c7mu10}C^\mu_7(1,0)&\stackrel{k \rightarrow
\infty}{\longrightarrow}&\frac{1}{\sqrt{2}}(- Q,- m_{\mathrm C},\mathrm i\,m_{\mathrm C},- Q)^\mu\,,\\
\label{eq:c8mu10}C^\mu_8(1,0)&\stackrel{k \rightarrow
\infty}{\longrightarrow}&(0,\frac{Q^2}{2\sqrt{2}m_{\mathrm C}},0,0)^\mu\,,
\end{eqnarray}

\begin{eqnarray} 
L^\mu_1(1,1)&\stackrel{k \rightarrow
\infty}{\longrightarrow}&-(2 k + \frac{Q^2}{m_{\mathrm C}})\left(1, 0, 0, 1\right)^\mu\,,\label{eq:L111} \\
 L^\mu_2(1,1)&\stackrel{k \rightarrow
\infty}{\longrightarrow}&\frac{2 k\, m_{\mathrm C} Q^2 + Q^4}{4 m_{\mathrm C}^3}\left(1, 0, 0, 1\right)^\mu\,,\label{eq:L211} \\
L^\mu_\mathrm M(1,1)&\stackrel{k \rightarrow
\infty}{\longrightarrow}&
\left(-\frac{Q^2}{2 m_{\mathrm C}}, 0, \mathrm i\, Q,-\frac{Q^2}{2 m_{\mathrm C}}\right)^\mu \,,\label{eq:LM11} \\
C^\mu_6(1,1)&\stackrel{k \rightarrow
\infty}{\longrightarrow}&\frac{Q^2}{2 m_{\mathrm C}}\left(1, 0, 0, 1\right)^\mu \,,\label{eq:C611}
\\
L^\mu_1(0,0)&\stackrel{k \rightarrow
\infty}{\longrightarrow}&\frac{-2 k\, m_{\mathrm C}^2 + (k - 2 m_{\mathrm C}) Q^2}{m_{\mathrm C}^2}\left(1, 0, 0, 1\right)^\mu\,,\\
L^\mu_2(0,0)&\stackrel{k \rightarrow
\infty}{\longrightarrow}&-\frac{(k - 2 m_{\mathrm C}) Q^4}{4 m_{\mathrm C}^4}\left(1, 0, 0, 1\right)^\mu\,,\\
L^\mu_M(0,0)&\stackrel{k \rightarrow
\infty}{\longrightarrow}&
-\frac{(-k + m_{\mathrm C}) Q^2}{m_{\mathrm C}^2}
(1,0,0,1)^\nu\,,\\
C^\mu_5(0,0)&\stackrel{k \rightarrow
\infty}{\longrightarrow}&k\left(2, 0, 0, 2\right)^\mu\,,\\
C^\mu_6(0,0)&\stackrel{k \rightarrow
\infty}{\longrightarrow}&\frac{(-k + m_{\mathrm C}) Q^2}{2m_{\mathrm C}^2}\left(2, 0, 0, 2\right)^\mu\,,\\
C^\mu_7(0,0)&\stackrel{k \rightarrow
\infty}{\longrightarrow}&k(2,0,0,2)^\mu\,.\label{eq:C700}
\end{eqnarray}
It becomes evident that the current is not conserved, not even in the limit $k \rightarrow
\infty$, due to the non-vanishing contributions of $C^1_7(1,0)$ and
$C^1_8(1,0)$ (note that for the given kinematics the four-momentum transfer is $q^\mu=(0,Q,0,0)^\mu$). In addition, we observe, using our kinematics, that the zeroth and third components of the current become identical in this limit, i.e.
\begin{eqnarray} 
 J^0_{\mu_j'\mu_j}:=\lim_{k \rightarrow
\infty}J^0_{\mathrm V}(\boldsymbol k_\mathrm C',\mu_j';\boldsymbol k_\mathrm C,\mu_j;K_\mathrm e)=\lim_{k \rightarrow
\infty}J^3_{\mathrm V}(\boldsymbol k_\mathrm C',\mu_j';\boldsymbol k_\mathrm C,\mu_j;K_\mathrm e)=J^3_{\mu_j'\mu_j}\,.\nonumber\\
\end{eqnarray}
 This reduces the number of independent matrix elements given by (\ref{eq:indepmatrixelements}) from 11 to 7.
Consequently, we need 7 form factors, the 3 physical $f_1, f_2, g_\mathrm M$ and the 4 spurious form factors $b_5,\ldots,b_8$, to parametrize the current in the limit $k \rightarrow
\infty$. From Eqs.~(\ref{eq:L21-1})-(\ref{eq:C700}) we find the following leading-order contributions to the current matrix elements in the limit $k \rightarrow
\infty$ (using the short-hand notation $\lim_{k \rightarrow
\infty}J^\mu_{\mathrm V}(\boldsymbol k_\mathrm C',\mu_j';\boldsymbol k_\mathrm C,\mu_j;K_\mathrm e)\equiv J^\mu_{\mu_j'\mu_j}$ and $\lim_{k \rightarrow
\infty}f_i(Q^2,k)\equiv F_i(Q^2)$ ):
 \begin{eqnarray}
\frac{1}{|\,\mathrm e\,|}\,J^0_{00}&=&F_1 L_1^0(0,0)+F_2 L_2^0(0,0)+G_\mathrm M L_\mathrm M^0(0,0)\nonumber\\&&+B_6 C_6^0(0,0)+B_5 C_5^0(0,0)+B_7 C_7^0(0,0)+\mathcal O(k^0)\,, \label{eq:J000}\\
\label{eq:J011}
\frac{1}{|\,\mathrm e\,|}\,J^0_{11} &=& F_1 L_1^0(1,1)+F_2 L_2^0(1,1)+\mathcal O(k^0)\,, \\
\frac{1}{|\,\mathrm e\,|}\,J^0_{10} &=&F_1 L_1^0(1,0)+F_2 L_2^0(1,0)+G_\mathrm M L_\mathrm M^0(1,0)+
B_6 C_6^0(1,0)+\mathcal O(k^0)\,,\nonumber\\\label{eq:J010}\\
\label{eq:J0m11}
\frac{1}{|\,\mathrm e\,|}\,J^0_{1-1}&=&F_2 L_2(1,-1)+\mathcal O(k^0)\,, \\
\frac{1}{|\,\mathrm e\,|}\,J^2_{11}&=& G_\mathrm M L_\mathrm M^2(1,1)+\mathcal O(k^{-1})\,, \\
\frac{1}{|\,\mathrm e\,|}\,J^1_{10}&=& B_7 C_7^1(1,0)+B_8 C_8^1(1,0)+\mathcal O(k^{-1})\,,\\
\frac{1}{|\,\mathrm e\,|}\,J^2_{10}&=& B_7 C_7^2(1,0)+\mathcal O(k^{-1})\,.
 \end{eqnarray}
We observe that the current components $J^0_{11}$, $J^0_{1-1}$ and $J^2_{11}$ do not contain any dominant spurious contributions and are therefore \lq\lq good'' components for the extraction of the physical form factors. With the help of the limiting expressions for the kinematical factors and constituent currents, cf. Eqs.~(\ref{eq:kinf}) and~(\ref{eq:nuclcurrlim}), respectively, we find for the limit of the zeroth (and third component) of the current (divided by $2k |\,\mathrm e\,|$) the following overlap integral:
\begin{eqnarray}
\lefteqn{\frac{1}{|\,\mathrm e\,|}\,\lim_{k\rightarrow \infty}\frac{1}{2k}J^0_{\mathrm V}(\mu_j',\mu_j)}\nonumber \\
&=&\frac{1}{4\pi}\int\mathrm{d}^3\tilde{k}'_1\sqrt{\frac{m_{12}}{m'_{12}}}u_{n0}^\ast\left(\tilde{k}'_1 \right)u_{n0}\left(\tilde{k}_1 \right)
\nonumber\\&&\times\left\lbrace\left[F_1^1(Q^2)+F_1^2(Q^2)\right]\mathcal S^{\mu_j'\mu_j}_{1}+\sqrt{\tau}\left[F_2^1(Q^2)+F_2^2(Q^2)\right]\mathcal S^{\mu_j'\mu_j}_{2} \right\rbrace\,.\nonumber\\
\end{eqnarray}
Here $\mathcal S^{\mu_j'\mu_j}_{1}$ and $\mathcal S^{\mu_j'\mu_j}_{2}$ are the spin factors from the Wigner $D$-functions and Clebsch-Gordan coefficients as given by Eqs.~(\ref{eq:S1}) and~(\ref{eq:S2}). 
Then using Eqs.~(\ref{eq:L21-1}),~(\ref{eq:L111}) and~(\ref{eq:L211}) the physical form factors $F_1$ and $F_2$ can be extracted from $J^0_{\mathrm V}(1,1)$ and $J^0_{\mathrm V}(1,-1)$ according to
\begin{eqnarray}
  F_1(Q^2)&:=&\lim_{k\rightarrow\infty}f_1(Q^2,k)=-\frac{1}{|\,\mathrm e\,|}\,\lim_{k\rightarrow\infty} \frac{1}{2k}\left[J^0_{\mathrm V}(1,1)+J^0_{\mathrm V}(1,-1)\right]\,,\nonumber \\
\\
F_2(Q^2)&:=&\lim_{k\rightarrow\infty}f_2(Q^2,k)=-\frac{1}{|\,\mathrm e\,|\,\eta}\,\lim_{k\rightarrow\infty} \frac{1}{2k}J^0_{\mathrm V}(1,-1)\,,
\end{eqnarray}
with $\eta=Q^2/(4 m_{\mathrm C}^2)$.
For the magnetic form factor $G_\mathrm M$ we need the current matrix element $J_{11}^{2}$.
From the expansion of the nucleon current around $k=\infty$, cf. Eq.~(\ref{eq:constcurrentexpansion}), we see that the second component of the nucleon current is suppressed by $1/k$. By pulling the factor $1/k$ 
out of the integral it is seen to cancel with the factor $\sqrt{k_\mathrm C^0 k_\mathrm C'^0}$.
Thus we find the following overlap integral:
\begin{eqnarray}
\lefteqn{ \frac{1}{|\,\mathrm e\,|}\,J_{\mu_j'\mu_j}^{2}(Q^2)}\nonumber\\&=&\frac{1}{4\pi}\int\mathrm{d}^3\tilde{k}'_1\sqrt{\frac{m_{12}}{ m'_{12}}}u_{n0}^\ast\left(\tilde{k}'_1 \right)u_{n0}\left(\tilde{k}_1 \right)
\frac{m_{12}'}{ ( m_{12}' + 2  \tilde k_1'^3)}
\nonumber\\&&\times\,|\,\mathrm e\,|\,\left\lbrace\left[F_1^1(Q^2) +F_1^2(Q^2)\right]\left(
\tilde k_1'^2 \mathcal S^{\mu_j'\mu_j}_{1}+  \frac{\mathrm i \,Q}{2} 
\mathcal S^{\mu_j'\mu_j}_{3}\right)\right.\nonumber\\&&+
\left.
\sqrt{\tau}\left[F_2^1(Q^2)
  +F_2^2(Q^2)\right]\left(\mathrm i\, m\,  \mathcal S^{\mu_j'\mu_j}_{3}+\tilde k_1'^2\mathcal S^{\mu_j'\mu_j}_{2}\right)
\right\rbrace\,,
\end{eqnarray}
with the spin factor $\mathcal S^{\mu_j'\mu_j}_{3}$ given by Eq.~(\ref{eq:S3}). Then the magnetic form factor is extracted using Eq.~(\ref{eq:LM11}) as
\begin{eqnarray}
 G_\mathrm M(Q^2):=\lim_{k\rightarrow\infty}g_\mathrm M(Q^2,k)=-\frac{\mathrm i}{|\,\mathrm e\,|\,Q}\,J^2_{11}(Q^2)\,.
\end{eqnarray}
\section{Projection Tensors}\label{app:projecttensors}
To see explicitly that the expressions (\ref{eq:ff1projection})-(\ref{eq:ffGMprojection}) are identical with the integrals of Eqs.~(\ref{eq:ff1me})-(\ref{eq:ffGMme}) we have to investigate the particular expressions that appear under the integral  after contracting with the projection tensors. Using our standard kinematics we obtain the following expressions of interest in the limit $k\rightarrow\infty$:
\begin{eqnarray} 
\frac{\sqrt{k_{\mathrm C}^0k_{\mathrm C}'^0}}{k_1^0}&\stackrel{k \rightarrow
\infty}{\longrightarrow}& \frac{2 m_{12}'}{m_{12}' + 2 \tilde k_1'^3}\,,\\
B_\mathrm c(-\boldsymbol w_{\mathrm C})q &\stackrel{k \rightarrow
\infty}{\longrightarrow}& \left(\frac{Q^2}{2 m_{\mathrm C}}, Q, 0, -\frac{Q^2}{2 m_{\mathrm C}}\right)\,,\\
B_\mathrm c(-\boldsymbol w_{\mathrm C}')q &\stackrel{k \rightarrow
\infty}{\longrightarrow}&\left(-\frac{Q^2}{2 m_{\mathrm C}}, Q, 0, \frac{Q^2}{2 m_{\mathrm C}}\right)\,,\\
B_\mathrm c(-\boldsymbol w_{\mathrm C})K_\mathrm C&\stackrel{k \rightarrow
\infty}{\longrightarrow}&\left(2 m_{\mathrm C} + \frac{Q^2}{2 m_{\mathrm C}}, Q, 0, -\frac{Q^2}{2 m_{\mathrm C}}\right)\,,\\
B_\mathrm c(-\boldsymbol w_{\mathrm C}')K_\mathrm C &\stackrel{k \rightarrow
\infty}{\longrightarrow}&\left(2 m_{\mathrm C} + \frac{Q^2}{2 m_{\mathrm C}},- Q, 0, -\frac{Q^2}{2 m_{\mathrm C}}\right)\,,\\
2\frac{B_\mathrm c(-\boldsymbol w_{\mathrm C})K_\mathrm e }{( K_\mathrm e\cdot K_\mathrm C)}&\stackrel{k \rightarrow
\infty}{\longrightarrow}&\frac{1}{m_{\mathrm C}}\left(1,0,0,-1\right)\,,\\
2\frac{B_\mathrm c(-\boldsymbol w_{\mathrm C}')K_\mathrm e }{( K_\mathrm e\cdot K_\mathrm C)}&\stackrel{k \rightarrow
\infty}{\longrightarrow}& \frac{1}{m_{\mathrm C}}\left(1,0,0,-1\right)\,,
\end{eqnarray}
\begin{eqnarray} \lefteqn{
B_\mathrm c(-\boldsymbol w_{\mathrm C}')^{\lambda}_{\,\,\sigma} \mathrm g^{\sigma\tau}B_\mathrm c(-\boldsymbol w_{\mathrm C})^{\nu}_{\,\,\tau}}\nonumber\\&\stackrel{k \rightarrow
\infty}{\longrightarrow}&
\left(
\begin{array}{cccc}
1 + Q^2/(2 m_{\mathrm C}^2) & Q/m_{\mathrm C} & 0 & -Q^2/(2 m_{\mathrm C}^2)\\
-Q/m_{\mathrm C} & -1 & 0 & Q/m_{\mathrm C}\\
0 & 0 & -1 & 0\\
-Q^2/(2 m_{\mathrm C}^2) & -Q/m_{\mathrm C} & 0 & -1 + Q^2/(2 m_{\mathrm C}^2)
\end{array}\right)^{\lambda\nu}\,.\nonumber\\
\end{eqnarray}
Furthermore, the expressions involving the constituent currents 
\begin{eqnarray}
 B_\mathrm c(-\boldsymbol w_{\mathrm C}^{(\prime)})^\mu_{\,\,\tau}J_i^{\tau}\left(\boldsymbol{k}_i',\mu_i';\boldsymbol{k}_i,\mu_i \right)\,,\quad q_\tau J_i^{\tau}\left(\boldsymbol{k}_i',\mu_i';\boldsymbol{k}_i,\mu_i \right)\,,\nonumber\\ K_{\mathrm e\tau}J_i^{\tau}\left(\boldsymbol{k}_i',\mu_i';\boldsymbol{k}_i,\mu_i \right)/( K_\mathrm e\cdot K_\mathrm C)\quad \text{and}\quad K_{\mathrm C\tau}J_i^{\tau}\left(\boldsymbol{k}_i',\mu_i';\boldsymbol{k}_i,\mu_i \right)
\end{eqnarray}
are seen to be finite in the limit
$k\rightarrow\infty$. They are rather lengthy and will not be given here explicitly.
Also, the spin matrix 
 \begin{eqnarray}&&
  D^{\frac12}_{\mu_1\tilde \mu_1}\left[\underline R_\mathrm{W_{\!c}}\left(\tilde w_1,B_{\mathrm c}(\boldsymbol w_{12})\right)\right]
(\sigma_\nu)_{\tilde \mu_1\tilde \mu_2}
  D^{\frac12}_{\tilde{\mu}_2\tilde{\mu}'_2}\left[\underline R_\mathrm{W_{\!c}}\left(\tilde{w}_{2}',B_{\mathrm c}^{-1}(\boldsymbol w_{12})B_{\mathrm c} (\boldsymbol w'_{12})\right)\right]
\nonumber\\&&\;\;\;\;\times (\sigma_\lambda)_{\tilde \mu_2'\tilde \mu_1'}  D^{\frac12}_{\tilde \mu_1'\mu_1'}\left[\underline R_\mathrm{W_{\!c}}^{-1}\left(\tilde w_1',B_{\mathrm c}(\boldsymbol w_{12}')\right)\right]
 \end{eqnarray}
is a lengthy but also finite in the limit $k\rightarrow\infty$.
Therefore, the form factors defined in Eqs.~(\ref{eq:ff1projection})-(\ref{eq:ffGMprojection}) are finite in the limit $k\rightarrow\infty$. Putting everything together it can be shown that  Eqs.~(\ref{eq:ff1projection})-(\ref{eq:ffGMprojection}) are actually equivalent with the integrals given in Eqs.~(\ref{eq:ff1me})-(\ref{eq:ffGMme}).\footnote{Due to the complexity of the problem we have carried out the proof using the symbolic {\sc Mathematica}$^{\begin{scriptsize}\textcopyright                                                                                                                                                                                                                                                                                                                                              \end{scriptsize}                                                                                                                                                                                                                                                                                                                                                                                                                                                                                                                                                                                                                                                                                                                                                                                                                                         }$ programming language.}

\chapter{Meson-Exchange Potentials in the Static Limit}
\label{App.A}
Here we use appropriately normalized Dirac spinors given by 
\begin{equation}
\tilde u_{\sigma}(\boldsymbol p)=\frac{1}{\sqrt{2m}}\,u_{\sigma}(\boldsymbol p)\,.
\end{equation}
In the static approximation and taking the limit $m\to\infty$
the Dirac spinor takes the form
\begin{eqnarray}
\tilde u_{\sigma}(\boldsymbol p) \rightarrow u^{\text {nr}}_{\sigma}= \left(
 \begin{array}{c}
\varsigma_{\sigma} \\ 0 \end{array}\right)\,,
\end{eqnarray}
with
\begin{equation}
 \bar{u}^{\text {nr}}_{\sigma'}\, u^{\text {nr}}_{\sigma}=
 \varsigma_{\sigma'}^\dag\,\varsigma_{\sigma}=\delta_{\sigma'\sigma}\,.
\end{equation}
Then is easy to see that the second-order amplitude for the scalar meson
exchange $m_\mathrm{int}^{\mathrm {N}\sigma}$ in Eq.~(\ref{eq.03}) becomes just the static
potential in Eq.~(\ref{eq.04}):
\begin{eqnarray}
 g^2_{\sigma} \,\bar{u}^{\text {nr}}_{\tilde \mu_1'} u^{\text {nr}}_{\tilde \mu_1}\,
 \frac{1}{-(\tilde{\boldsymbol k}'_1 - \tilde{\boldsymbol k}_1)^2 - m^2_\sigma}\,
 \bar{u}^{\text {nr}}_{\tilde \mu_2'} u^{\text {nr}}_{\tilde \mu_2}=-g^2_{\sigma}
 \frac{\delta_{\tilde \mu_1'\tilde \mu_1}\delta_{\tilde \mu_2'\tilde \mu_2}}
 {\tilde{\boldsymbol q}_1^2 + m^2_\sigma}\,,
\end{eqnarray}
where $\tilde{\boldsymbol q}_1=\tilde{\boldsymbol k} '_1 - \tilde{\boldsymbol k} _1 $ is the 
three-momentum transfer.
For the vector meson exchange contribution $m_\mathrm{int}^{\mathrm {N}\omega}$ in 
Eq.~(\ref{eq.03}) we first concentrate on the contributions of the second 
term of the $\omega$-meson propagator which reads
\begin{equation}
 -g^2_{\omega}\bar{\tilde u}_{\tilde \mu_1'}(\tilde{\boldsymbol k}_1') \gamma_\mu \tilde u_{\tilde \mu_1}(\tilde{\boldsymbol k}_1) 
 \frac{ (\tilde k'^\mu_1 - \tilde k^\mu_1)(\tilde k'^\nu_2 - \tilde k^\nu_2)/m_{\omega}^2}
{(\tilde k_1' - \tilde k_1)^2 - m^2_\omega} 
 \bar{\tilde u}_{\tilde \mu_2'}(\tilde{\boldsymbol k}_2') \gamma_\nu \tilde u_{\tilde \mu_2}(\tilde{\boldsymbol k}_2).
\end{equation}
Making use of the Dirac equations for the spinors, Eqs.~(\ref{eq:Diracequation}) and~(\ref{eq:adjDiracequation}), we have
\begin{equation}
 \bar{\tilde u}_{\tilde \mu_1'}(\tilde{\boldsymbol k}_1') \gamma_\mu (\tilde k_1'^\mu - \tilde k^\mu_1) \tilde u_{\tilde \mu_1}(\tilde{\boldsymbol k}_1) 
=
 \bar{\tilde u}_{\tilde \mu_1'}(\tilde{\boldsymbol k}_1') (m_{\text{N}}-m_{\text{N}}) \tilde u_{\tilde \mu_1}(\tilde{\boldsymbol k}_1)=0\,.
\end{equation} 
Thus, we find that the second term of $m_\mathrm{int}^{\mathrm {N}\omega}$ vanishes for on-shell nucleons.
For the contributions of the first term of the $\omega$-meson propagator
\begin{equation}
-g^2_{\omega}\bar{\tilde u}_{\tilde \mu_1'}(\tilde{\boldsymbol k}_1') \gamma_\mu \tilde u_{\tilde \mu_1}(\tilde{\boldsymbol k}_1) 
 \frac{\mathrm g^{\mu\nu}}{(\tilde k_1' - \tilde k_1)^2 - m^2_\omega}
 \bar{\tilde u}_{\tilde \mu_2'}(\tilde{\boldsymbol k}_2') \gamma_\nu \tilde u_{\tilde \mu_2}(\tilde{\boldsymbol k}_2)
\end{equation}
we use the standard representation of the $\gamma$-matrices, cf. Eq.~(\ref{eq:standardrepr}).
%
%
In the static limit the spatial components of the nucleon current, 
being proportional to the three-momentum divided by the mass, vanish:
\begin{equation}
 \bar{\tilde u}_{\tilde \mu_1'}(\tilde{\boldsymbol k}_1') \gamma^i\tilde u_{\tilde \mu_1}(\tilde{\boldsymbol k}_1) \approx \bar{u}_{\tilde \mu_1'}^{\text{nr}}\gamma^i u_{\tilde \mu_1}^{\text{nr}} =0, \quad\forall\, i=1,\ldots,3\,.
\end{equation}
Thus the three-vector part of the contraction of the currents vanishes:
\begin{equation}-g^2_{\omega}\,\bar{u}_{\tilde \mu_1'}^{\text{nr}}\gamma^i u_{\tilde \mu_1}^{\text{nr}}\,
 \frac{1}{\tilde{\boldsymbol q}_1^2 + m^2_\omega}\,
 \bar{u}_{\tilde \mu_2'}^{\text{nr}}\gamma^i u_{\tilde \mu_2}^{\text{nr}}=0\,.
\end{equation} 
We are left only with the temporal components 
$\bar{\tilde u}_{\tilde \mu_1'}(\tilde{\boldsymbol k}_1') \gamma^0\tilde u_{\tilde \mu_1}(\tilde{\boldsymbol k}_1)$ which
become unity in the non-relativistic limit:
\begin{equation}
 \bar{\tilde u}_{\tilde \mu_1'}(\tilde{\boldsymbol k}_1') \gamma^0\tilde u_{\tilde \mu_1}(\tilde{\boldsymbol k}_1) \approx
\bar{u}_{\tilde \mu_1'}^{\text{nr}}\gamma^0 u_{\tilde \mu_1}^{\text{nr}}=
 \delta_{\tilde \mu_1'\tilde \mu_1}.
\end{equation}
Therefore, the second-order contribution in the static limit is
\begin{equation}
 -g^2_{\omega} \bar{u}_{\tilde \mu_1'}^{\text{nr}}\gamma^0 u_{\tilde \mu_1}^{\text{nr}}
 \frac{1}{-\boldsymbol q^2 - m^2_\omega}
 \bar{u}_{\tilde \mu_2'}^{\text{nr}}\gamma^0 u_{\tilde \mu_2}^{\text{nr}}=
 g^2_{\omega} \frac{\delta_{\tilde \mu_1'\tilde \mu_1}\delta_{\tilde \mu_2'\tilde \mu_2}}
{\boldsymbol q^2 + m^2_\omega},
\end{equation}
which is just the static potential given in Eq.~(\ref{eq.05}).
\chapter{Notation}\label{app:0}

\begin{description}
 \item[$\vert0\rangle$] Vacuum state
\item[$ 1_{n}$] $n$-dimensional unit matrix
\item[$\hat 1_{\{n\}}$] Unit operator on the $n$-particle Hilbert space
\item[$a$] Harmonic oscillator length
\item[$\hat{a}^{(\dag)}_\lambda(\boldsymbol p)$] Photon annihilation (creation) operator
\item[$\boldsymbol \alpha$] Dirac matrices
\item[$\alpha_{(\mathrm{rel})}$] Deuteron bound-state pole (for relativistic kinetic energy)
\item[$\hat A^\mu(x)$] Local photon field operator
\item[$a_\mathrm t$] Deuteron scattering length
                     \item[$B_\mathrm c(\boldsymbol v)$] Canonical (rotationless) boost
\item[$B_\mathrm g(\boldsymbol v)$] General Lorentz boost
\item[$B_\mathrm h(\boldsymbol v)$] Helicity boost
\item[$B_i(Q^2)$] Spin-1 spurious form factors for $k\rightarrow\infty$
\item[$b_i(Q^2,k)$] Spin-1 spurious form factors
\item[$B(Q^2)$] Spin-1 elastic scattering observable
\item[$b(Q^2,k)$] Spin-0 spurious form factor
\item[$\bar b(Q^2,k)$] Spin-0 form factor associated with $P_\perp^\mu$ 
\item[$\tilde b(Q^2,k)$] Alternatively normalized spin-0 spurious form factor 
\item[$\mathrm c$] Speed of light
\item[$C$] Charge conjugation operator
\item[$c_0$] Spectral shift constant
\item[$C^{\mu}_i(\mu_j',\mu_j)$] Lorentz structure associated with the spurious form factor $b_i$ 
\item[$C^{j\mu_j}_{j_1\mu_1j_2\mu_2}$] Clebsch-Gordan coefficient
\item[$\hat{c}^{(\dag)}_\sigma(\boldsymbol p)$] Fermion annihilation (creation) operator
\item[$\delta_{ij}$] Kronecker symbol
\item[$\delta^n(x)$] $n$-dimensional Dirac $\delta$-distribution
\item[$d^\mu$] Four-momentum transfer
\item[$D^{j}_{\sigma'\sigma'}(\underline R)$] Wigner $D$-function
\item[$\hat{d}^{(\dag)}_\sigma(\boldsymbol p)$] Anti-fermion annihilation (creation) operator
\item[$\mathrm e$] Elementary charge
\item[$E_\mathrm B$] Deuteron binding energy
\item[$E_\mathrm B^{\mathrm {rel}}$] Relativistic deuteron binding energy
\item[$\varepsilon$] Pauli metric
\item[$\epsilon^\mu_\lambda(\boldsymbol p)$] Spin-1 polarization vectors
\item[$\epsilon^{\mu\nu\sigma\tau}$] Levi-Civita symbol
\item[$F_\mathrm B(\tilde q_1)$] Body form factor
\item[$F_i^j(Q^2)$] Spin-1/2 form factors of constituent $j$
\item[$F_{i\mu}^{\sigma\tau}$] Projection tensor for $F_i(Q^2)$
\item[$F_i(Q^2)$] Spin-1 form factors for $k\rightarrow\infty$
\item[$f_i(Q^2,k)$] Spin-1 form factors
\item[$f_\omega$] Tensor coupling constant of the $\omega$ meson
\item[$F(Q^2)$] Spin-0 form factor for $k\rightarrow\infty$
\item[$f(Q^2,k)$] Spin-0 form factor
\item[$\bar f(Q^2,k)$] Spin-0 form factor of orthogonal decomposition
\item[$f_\mathrm v(\vert M'_{12\mathrm e\gamma}-M_{12\mathrm e}\vert)$] Vertex factor
\item[$\hat G$] Infinitesimal generators of $\hat U (\underline\varLambda,\underline a)$
\item[$\hat G_{12\cdots n}$] Infinitesimal generators of $\hat U_{12\cdots n} (\underline\varLambda,\underline a)$
\item[$\varGamma_{i\rightarrow j}$] Time-ordered contribution to $\hat V_{\mathrm{opt}}$ 
\item[$\gamma^\mu$] Dirac matrices
\item[$\varGamma^\mu_i$] Electromagnetic vertex of constituent $i$
\item[$G_\mathrm C(Q^2)$] Electric monopole (charge) form factor
\item[$G_{\mathrm M\mu}^{\sigma\tau}$] Projection tensor for $G_\mathrm M(Q^2)$
\item[$\mathrm g^{\mu\nu}$] Metric tensor
\item[$G_\mathrm M(Q^2)$] Magnetic dipole form factor for $k\rightarrow\infty$
\item[$g_\mathrm M(Q^2,k)$] Magnetic dipole form factor
\item[$G_\mathrm Q(Q^2)$] Electric quadrupole form factor
\item[$g_{\sigma/\omega}$] Coupling constant of the $\sigma/\omega$ meson
\item[$\hbar$] Planck's constant
\item[$I_\mathrm {PS}^\mu(\boldsymbol k_\mathrm C',\boldsymbol k_\mathrm C)$] Physical part of $J_\mathrm {PS}^\mu$
\item[$I_{\mathrm {PS}}^\mu(\boldsymbol V)$] Physical spin-0 current 
\item[$I_{\sigma_j'\sigma_j}^\mu$] Physical spin-1 current for $\boldsymbol V=0$ 
\item[$I_\mathrm V^\mu(\boldsymbol k_\mathrm C',\mu_j';\boldsymbol k_\mathrm C,\mu_j)$] Physical part of $J_\mathrm V^\mu$ 
\item[$I_{\mathrm V\sigma_j'\sigma_j}^\mu(\boldsymbol V)$] Physical spin-1 current 
\item[$j$] Quantum number of total angular momentum of the cluster
\item[$j_0(x)$] First spherical Bessel function
\item[$\hat{\boldsymbol j}_\mathrm c$] Canonical spin operator of the cluster
\item[$\hat{\boldsymbol J}_\mathrm c$] Canonical spin operator
\item[$J_\mathrm C^\mu(\boldsymbol p_\mathrm C',\sigma_j';\boldsymbol p_\mathrm C,\sigma_j;P_\mathrm e)$] Bound-state current
\item[$\hat{\boldsymbol J}_\mathrm g$] General spin operator
\item[$\hat{\boldsymbol J}_\mathrm h$] Helicity spin operator
\item[$J^i$] Generators for rotations
\item[$\hat{j}_i$] Canonical spin operator of constituent $i$
\item[$J_i^\mu(\boldsymbol p_i',\sigma_i';\boldsymbol p_i,\sigma_i)$] Spin-1/2 current of constituent $i$
\item[$J^\mu_{\mu_j'\mu_j}$] Spin-1 bound-state current for $k\rightarrow\infty$
\item[$J_\mathrm {PS}^\mu(\boldsymbol p_\mathrm C',\boldsymbol p_\mathrm C,P_\mathrm e)$] Spin-0 bound-state current
\item[$J^\mu_\mathrm V(\mu_j',\mu_j)$] Short-hand notation for spin-1 bound-state current
\item[$J_\mathrm V^\mu(\boldsymbol p_\mathrm C',\sigma_j';\boldsymbol p_\mathrm C,\sigma_j;P_\mathrm e)$] Spin-1 bound-state current
\item[$J_\mathrm V(\boldsymbol p_\mathrm C',\boldsymbol p_\mathrm C,P_\mathrm e)^\mu_{\kappa\omega}$] Current tensor 
\item[$k$] Magnitude of $\boldsymbol k_\mathrm C$
\item[$ \boldsymbol k_\perp$] Intrinsic transverse momentum
\item[$ k_{12}^\mu$] Four-momentum of the free cluster in the overall center-of-momentum system
\item[$\kappa$] Anomalous magnetic moment
\item[$\hat K^{(\mathrm C)}$] Vertex operator (on the bound-state level)
\item[$K^\mu_\mathrm{C/e}$] Sum of incoming and outgoing bound-state (electron) momenta in the overall center-of-momentum system
\item[$K^i$] Boost generators
\item[$ k_i^\mu$] Four-momentum of particle $i$ in the overall center-of-momentum system
 \item[$\tilde k_i^\mu$] Four-momentum of constituent $i$ in the cluster center-of-momentum system
\item[$l$] Quantum number of orbital angular momentum of the cluster
\item[$\hat{\boldsymbol l}$] Orbital angular momentum operator of the cluster
\item[$\varLambda$] Lorentz transformation
\item[$(\varLambda,a)$] Poincar\'{e} transformation
\item[$\lambda_i$] Compton wavelength corresponding to mass $m_i$
\item[$\varLambda_i$] Mass of Pauli-Villars particle $i$
\item[$\hat{\mathcal L}_{\mathrm {int}}^{(\mathrm C)}(x)$] Interaction Lagrangean density (on bound-state level)
\item[$L^{\mu}_i(\mu_j',\mu_j)$] Lorentz structure associated with the form factor $f_i$ 
\item[$L^{\mu}_\mathrm M(\mu_j',\mu_j)$] Lorentz structure associated with the form factor $g_\mathrm M$
\item[$L^{m}_n(x)$] Generalized Laguerre polynomials 
\item[$\mathcal M$] Melosh rotation factor
\item[$\hat M$] Total invariant mass operator
\item[$\hat m_{12}$] Invariant mass operator of the free cluster
\item[$\hat m_\mathrm C$] Total mass operator of the cluster
\item[$m_{(i)}$] Invariant mass (of particle $i$)
\item[$\hat m_i^{\mu\nu}$] Lorentz generators of $\hat U_i (\underline \varLambda,\underline a)$
\item[$\hat M_{\mathrm {int}}$] Interaction mass operator of the overall system
\item[$\hat m_{\mathrm {int}}$] Two-body interaction mass operator 
\item[$\hat m^j_{\mathrm {int}}$] Reduced two-body interaction mass operator 
\item[$M_{J}$] 3-projection of $\hat{\boldsymbol J_\mathrm c}$
\item[$\hat M^{\mu\nu}$] Lorentz generators
\item[$m_n$] Eigenvalue of $\hat m_\mathrm C$ associated with the radial quantum number $n$
\item[$\hat M_{n}$] Mass operator of the free $n$-particle system
\item[$m_{nl}$] Mass eigenvalue of the three-dimensional isotropic harmonic oscillator
\item[$\hat M^{\mu\nu}_n$] Lorentz generators of the free $n$-particle system
\item[$m_{\mathrm{red}}$] Reduced mass
\item[$\mu_i$] Spin projection of particle $i$ in the overall center-of-momentum system
\item[$\tilde \mu_i$] Spin projection of constituent $i$ in the cluster center-of-momentum system
\item[$\mu_j$] 3-projection of $\hat{\boldsymbol j}_\mathrm c$ in the overall center-of-momentum system
\item[$\mu_l$] 3-projection of $\hat{\boldsymbol l}$ 
\item[$\mu_{\rho/\mathrm D}$] Magnetic dipole moment of the $\rho$-meson/deuteron
\item[$\mu_s$] 3-projection of $\hat{\boldsymbol s}$
\item[$n$] Radial quantum number  
\item[$\omega^\mu$] Light-front four-vector
\item[$\hat \omega^\mu(x)$] $\omega$ meson field operator 
\item[$\hat \omega^\mu_{\mathrm{PV}}(x)$] $\omega$ Pauli-Villars field operator 
\item[$\vert\boldsymbol p_1,\sigma_1;\boldsymbol p_2,\sigma_2;\ldots;\boldsymbol p_n,\sigma_n\rangle$] $n$-particle tensor-product state
\item[$P^\mu_\perp$] Four-vector orthogonal to $P^\mu_\mathrm C$
\item[$P^\mu_\mathrm{C/e}$] Sum of incoming and outgoing bound-state/electron momenta 
\item[$\hat P^\mu_{\mathrm{int}}$] Interaction four-momentum operator of the overall system
\item[$\vert\{\boldsymbol p_i,\sigma_i\}\rangle$] Short-hand notation for $n$-particle tensor-product state 
\item[$\varPi(\varLambda,a)$] ($5\times5$)-matrix representation of the Poincar\'{e} group
\item[$p^\mu_i$] Four-momentum operator of particle $i$
\item[$\hat P^\mu$] Total four-momentum operator
\item[$\hat P^\mu_n$] Four-momentum operator of the free $n$-particle system
\item[$\vert\varPsi\rangle$] State vector
\item[$\tilde \varPsi_{00}(\boldsymbol r)$] Deuteron wave function in coordinate space
\item[$\varPsi_{nj\mu_j\mu_1\mu_2}(\tilde{\boldsymbol k})$] 2-particle wave function
\item[$\hat \psi(x)$] Local spin-1/2 fermion field operator
\item[$q^\mu_{(\mathrm N)}$] Four-momentum transfer (on the nucleon) in the overall center-of-momentum system
\item[$Q^2_{(\mathrm N)}$] Four-momentum transfer squared (on the nucleon)
\item[$Q_{\mathrm q/\mathrm e}$] Quark/electron charge
\item[$Q_{\rho/\mathrm D}$] Electric quadrupole moment of the $\rho$-meson/deuteron
\item[$\boldsymbol r$] Position three-vector 
\item[$\boldsymbol \rho$] Rapidity 
\item[$R_\mathrm{W_{\!c}}$] Wigner rotation associated with canonical boosts  
\item[$R_\mathrm{W_{\!g}}$] Wigner rotation associated with general boosts  
\item[$R_\mathrm{W_{\!h}}$] Wigner rotation associated with helicity boosts  
\item[$\mathrm s$] Mandelstam variable
\item[$s$] Quantum number of total spin of the cluster
\item[$\mathcal S$] Spin rotation factor for spin-0 bound states
\item[$\hat{\boldsymbol s}$] Total spin operator of the cluster
\item[$\boldsymbol \sigma$] Vector of Pauli matrices
\item[$\sigma_0$] ($2\times 2$)-unit matrix
\item[$\sigma_i$] Spin projection of particle $i$ 
\item[$\sigma_j$] 3-projection of $\hat{\boldsymbol j}_\mathrm c$
\item[$\hat \sigma_{\mathrm{PV}}(x)$] $\sigma$ Pauli-Villars field operator 
\item[$\varsigma_\sigma$] Pauli spinor
\item[$\hat \sigma(x)$] $\sigma$ meson field operator 
\item[$\mathcal S^{\mu_j'\mu_j}_i$] Spin rotation factors for spin-1 bound states
\item[$S(\underline \varLambda)$] Four-spinor representation of the SL$(2,\mathbb C)$
\item[$\mathcal S^{\mu_j'\mu_j}_{\mu_1\mu_1'}$] Short-hand notation for 3 Wigner $D$-functions and 2 spin-1 Clebsch-Gordan coefficients
\item[$\mathrm t$] Mandelstam variable
\item[$T_{20}(Q^2)$] Tensor polarization
\item[$T(a)$] Space-time translation
\item[$\tau$] Proper time
\item[$\boldsymbol \theta$] Rotation parameter
\item[$\theta(x)$] Heavyside step function
\item[$\hat T_i(\underline a)$] Space-time translation operator for particle $i$
\item[$\hat {\mathcal T}_{\mathrm{int}}^{\mu\nu}(x)$] Interaction energy-momentum tensor operator
\item[$u_0(\tilde k)$] Radial S-wave function of the deuteron in momentum space
\item[$\tilde u_0(r)$] Radial S-wave function of the deuteron in configuration space 
\item[$\hat U_{12\cdots n} (\underline \varLambda,\underline a)$
] 
Unitary representation of the ISL$(2,\mathbb C)$ acting on $n$-particle tensor-product states
\item[$\hat U_i (\underline \varLambda,\underline a)$] Unitary irreducible representation of the ISL$(2,\mathbb C)$
\item[$\hat U (\underline \varLambda,\underline a)$] Unitary representation of the ISL$(2,\mathbb C)$
\item[$u_{n0}(\tilde k)$] Radial S-wave function 
\item[$u_{nl}(\tilde k)$] Radial wave function of harmonic oscillator 
\item[$u_{nls}^j(\tilde k)$] Radial wave function  
\item[$u_{\sigma}(\boldsymbol p)$] Dirac spinor for spin-1/2 fermions
\item[$\tilde u_{\sigma}(\boldsymbol p)$] Dirac spinor for spin-1/2 fermions normalized to unity 
\item[$ u_{\sigma}^{\mathrm{nr}}$] Dirac spinor for spin-1/2 fermions in the static approximation 
\item[$ V^\mu$] Four-velocity of the free overall system 
\item[$ v_{12}^\mu$] Four-velocity of the free cluster
\item[$ v_i^\mu$] Four-velocity of particle $i$ 
\item[$\vert V;\boldsymbol k_1,\mu_1;\boldsymbol k_2,\mu_2;\ldots;\boldsymbol k_n,\mu_n\rangle$] $n$-particle velocity state
\item[$\vert V;\{\boldsymbol k_i,\mu_i\}\rangle$] Short-hand notation for $n$-particle velocity state 
\item[$ \hat V^{(\mathrm C)}_{\mathrm{opt}}(M)$] One-photon-exchange optical potential (on the bound-state level)
\item[$V(r)$] Meson-exchange potential in configuration space 
\item[$v_{\sigma}(\boldsymbol p)$] Dirac spinor for spin-1/2 anti-fermions
\item[$w_0(r)$] Reduced S-wave function of the deuteron 
\item[$ w_{12}^\mu$] Four-velocity of the free cluster in the overall center-of-momentum system
\item[$ w_i^\mu$] Four-velocity of particle $i$ in the overall center-of-momentum system
 \item[$\tilde w_i^\mu$] Four-velocity of constituent $i$ in the cluster center-of-momentum system
\item[$\hat x_\mathrm c$] Newton-Wigner position operator of the cluster
\item[$\hat X_\mathrm c$] Newton-Wigner position operator
\item[$x^\mu$] Minkowski-space position four-vector
\item[$Y_{l\mu_l}(\hat {\tilde {\boldsymbol k}})$] Spherical harmonic functions
\item[$z$] Longitudinal momentum fraction
                     \end{description}

%
 \end{appendix}

\clearpage \pagebreak \clearpage
\chapter*{Acknowledgements}\pagestyle{plain}
First of all, I would like to thank my supervisor Wolfgang
Schweiger. He guided and supported me throughout my studies in many different ways. His thorough and careful approach to physics formed my ideas and it was always a great pleasure to work with him. I am also very grateful for the opportunity of participating at the doctoral program \lq Hadrons in Vacuum, Nuclei and Stars'. I want to thank William H. Klink for his constructive suggestions and his helpful ideas, in particular, on the cluster problem. I would like to express my gratitude to Bernard L. G. Bakker for his supportive supervision of our joint project on the Walecka model and for being a referee of this thesis. At this point I would also like to thank the whole particle physics group at the VU University Amsterdam for their warm hospitality. I owe thanks to my mentor Willibald Plessas for his support in many ways. In particular, I would like to mention the numerous animated discussions with him which encouraged my critical thinking. I also want to thank the whole Graz group and, in particular, Thomas Melde for many discussions. I would like to thank Fritz Coester and Wayne Polyzou for the interesting and elucidating discussions and I am much obliged to the University of Iowa for their warm hospitality.

Furthermore, I will always be grateful for the love and support I
have received from my parents and from my family. I would also
like to mention my friends and colleagues. Thank you for always being there for me! 

Finally, I thank Justine for everything.

\paragraph*{Financial Support}

This thesis was financially supported by the Austrian Science Fund FWF under Grant No. DK W1203-N16 (Doctoral Program \lq Hadrons in Vacuum, Nuclei and Stars'), as well as by the 
Province of Styria, Austria with a Ph.D. grant. The research stay at the University of Iowa in Iowa City, USA was also financially supported by a \lq\lq
Mobilit\"atszuschuss der Karl-Franzens Universit\"at Graz'' and by the University of Iowa.    
\begin{flushright}
 E. P. B., March 2011
\end{flushright}

\bibliographystyle{alpha}

\bibliography{Diss} 

\end{document}